\documentclass[structabstract]{aa}  

\usepackage[pdftex]{graphicx}
\usepackage{txfonts}
\usepackage{t1enc}
\usepackage{epsfig}
\usepackage{rotating}
\usepackage{verbatim}
\usepackage{subfig}
\usepackage{longtable}
\usepackage{supertabular}
\usepackage{lscape}
\begin{document}
\newcommand{\Td}    {T_\mathrm{d}}
\newcommand{\Tex}   {T_\mathrm{ex}}
\newcommand{\Trot}  {T_\mathrm{rot}}
\newcommand{\mum}   {$\mu$m}
\newcommand{\kms}   {km~s$^{-1}$}
\newcommand{\cmg}   {cm$^{2}$~g$^{-1}$}
\newcommand{\cmt}   {cm$^{-3}$}
\newcommand{\jpb}   {$\rm Jy~beam^{-1}$}    
\newcommand{\lo}    {$L_{\sun}$}
\newcommand{\mo}    {$M_{\sun}$}
\newcommand{\nh}    {NH$_3$}
\newcommand{\nth}   {N$_2$H$^+$}
\newcommand{\chtoh} {CH$_3$OH}
\newcommand{\water} {H$_2$O}
\newcommand{\et}    {et al.}
\newcommand{\eg}    {e.\,g.,}
\newcommand{\ie}    {i.\,e.,}
\newcommand{\hi}    {\ion{H}{i}}
\newcommand{\hii}   {\ion{H}{ii}}
\newcommand{\uchii} {UC~\ion{H}{ii}}
\newcommand{\hchii} {HC~\ion{H}{ii}}
\newcommand{\raun}  {$^\mathrm{h~m~s}$}
\newcommand{\deun}  {$\mathrm{\degr~\arcmin~\arcsec}$}
\newcommand{\taba}  {\tablefootmark{a}}
\newcommand{\tabb}  {\tablefootmark{b}}
\newcommand{\tabc}  {\tablefootmark{c}}
\newcommand{\tabd}  {\tablefootmark{d}}
\newcommand{\tabe}  {\tablefootmark{e}}
\newcommand{\tabf}  {\tablefootmark{f}}
\newcommand{\tabg}  {\tablefootmark{g}}
\newcommand{\tabh}  {\tablefootmark{h}}
\newcommand{\tabi}  {\tablefootmark{i}}
\newcommand{\tabj}  {\tablefootmark{j}}
\newcommand{\tabk}  {\tablefootmark{k}}
\newcommand{\tabz}  {\phantom{\tablefootmark{z}}}
\newcommand{\supa}  {$^\mathrm{a}$}
\newcommand{\supb}  {$^\mathrm{b}$}
\newcommand{\supc}  {$^\mathrm{c}$}
\newcommand{\supd}  {$^\mathrm{d}$}
\newcommand{\supe}  {$^\mathrm{e}$}
\newcommand{\supf}  {$^\mathrm{f}$}
\newcommand{\supg}  {$^\mathrm{g}$}
\newcommand{\suph}  {$^\mathrm{h}$}
\newcommand{\phn}   {\phantom{0}}
\newcommand{\phnn}  {\phantom{0}\phantom{0}}
\newcommand{\phnnn} {\phantom{0}\phantom{0}\phantom{0}}
\newcommand{\phnnnn}{\phantom{0}\phantom{0}\phantom{0}\phantom{0}}
\newcommand{\phe}   {\phantom{$^\mathrm{c}$}}
\newcommand{\phb}   {\phantom{$>$}}
\newcommand{\phl}   {$<$}
\newcommand{\phm}   {$>$}
\newcommand{\phnm}  {\phantom{0}\phantom{$.$}}
\newcommand{\phbn}  {\phantom{$>$}\phantom{0}}
\newcommand{\phbnn} {\phantom{$>00$}}
\newcommand{\phnb}  {\phantom{0}\phantom{$>$}}
\newcommand{\phmm}  {\phantom{\pm0.0}}
\newcommand{\phmn}  {\phantom{0\pm0.00}}
\def\HII{H{\sc ii}}
\def\Nly{\mbox{$N_{\rm Ly}$}}
\def\Lbol{\mbox{$L_{\rm bol}$}}
\def\ne{\mbox{$n_{\rm e}$}}
\def\Ro{R_{\rm o}}
\def\Rd{R_{\rm d}}
\def\Rh{R_{\rm h}}
\def\zd{z_{\rm d}}
\def\NHII{\mbox{$N_{\rm HII}$}}
\def\d{{\rm d}}
\def\e{{\rm e}}
\def\Log{{\rm Log}}
\def\fvol{\frac{\d N}{\d V}}
\def\flum{\frac{\d N}{\d\Log L}}
\def\fne{\frac{\d N}{\d\Log n_{\rm e}}}
\def\fD{\mbox{$\frac{\d N}{\d\Log D_{\rm HII}}$}}
%
%
   \title{Different evolutionary stages in massive star formation}

   \subtitle{Centimeter continuum and H$_2$O maser emission with ATCA}

   \author{\'A.\ S\'anchez-Monge\inst{1} 
           \and M.~T.\ Beltr\'an\inst{1}
           \and R.\ Cesaroni\inst{1}
           \and F.\ Fontani\inst{1}
           \and J.\ Brand\inst{2}
           \and S.\ Molinari\inst{3}
           \and L.\ Testi\inst{4,1} 
           \and M.\ Burton\inst{5}
          }

   \offprints{\email{asanchez@arcetri.astro.it}}

   \institute{Osservatorio Astrofisico di Arcetri, INAF, Largo Enrico Fermi 5, I-50125, Firenze, Italy
              \and Istituto di Radioastronomia \& Italian ALMA Regional Centre, via P.\ Gobetti 101, I-40129, Bologna, Italy
              \and Istituto di Fisica dello Spazio Interplanetario, INAF, Area di Recerca di Tor Vergata, Via Fosso Cavaliere 100, I-00133 Roma, Italy
              \and ESO, Karl Schwarzschild str.\ 2, 85748 Garching bei Munchen, Germany
              \and School of Physics, University of New South Wales, NSW 2052, Australia
             }

   \date{Received; accepted }

  \abstract
{}
{We present ATCA observations of the H$_2$O maser line and radio continuum at 18.0~GHz and 22.8~GHz, toward a sample of 192 massive star forming regions containing several clumps already imaged at 1.2~mm. The main aim of this study is to investigate the water maser and centimeter continuum emission (likely tracing thermal free-free emission) in sources at different evolutionary stages, using the evolutionary classifications proposed by Palla \et\ (1991) and Molinari \et\ (2008).}
{We used the recently comissioned CABB backend at ATCA obtaining images with $\sim$20~\arcsec\ resolution in the 1.3~cm continuum and H$_2$O maser emission, in all targets. For the evolutionary analysis of the sources we used the millimeter continuum emission from Beltr\'an \et\ (2006) and the infrared emission from the MSX Point Source Catalogue.}
{We detect centimeter continuum emission in 88\% of the observed fields with a typical rms noise level of 0.45~m\jpb. Most of the fields show a single radio continuum source, while in 20\% of them we identify multiple components. A total of 214 centimeter continuum sources have been identified, likely tracing optically thin \hii\ regions, with physical parameters typical of both extended and compact \hii\ regions. Water maser emission was detected in 41\% of the regions, resulting in a total of 85 distinct components. The low angular ($\sim$20~\arcsec) and spectral ($\sim$14~\kms) resolutions do not allow a proper analysis of the water maser emission, but suffice to investigate its association with the continuum sources. We have also studied the detection rate of \hii\ regions in the two types of IRAS sources defined by Palla \et\ (1991) on the basis of the IRAS colours: High and Low. No significant differences are found, with large detection rates (>90\%) for both High and Low sources.}
{We classify the millimeter and infrared sources in our fields in three evolutionary stages following the scheme presented by Molinari \et\ (2008): (type~1) millimeter-only sources, (type~2) millimeter plus infrared sources, (type~3) infrared-only sources. We find that \hii\ regions are mainly associated with type~2 and 3 objects, confirming that these are more evolved than type~1 sources. The \hii\ regions associated with type~3 sources are slightly less dense and larger in size than those associated with type~2 sources, as expected if the \hii\ region expands as it evolves, and type~3 objects are older than type~2 ones. Regarding the maser emission, it is mostly found associated with type~1 and 2 sources, with a higher detection rate toward type~2, consistent with the results of Breen \et\ (2010). Finally, our results on \hii\ region and H$_2$O maser association with different evolutionary types confirm the evolutionary classification proposed by Molinari \et\ (2008).}


   \maketitle
%

\begin{figure}[t]
\begin{center}
\begin{tabular}[b]{c}
 \epsfig{file=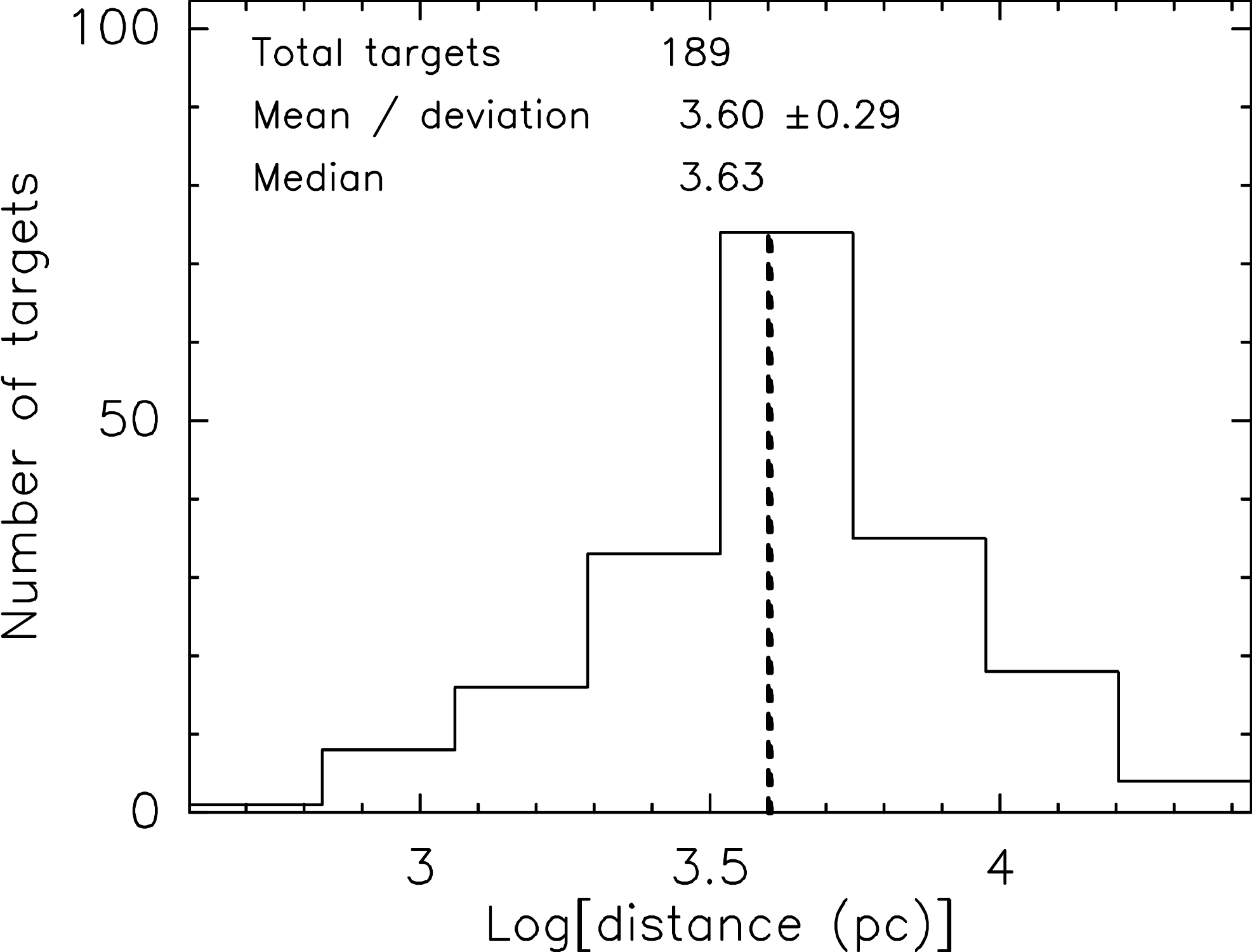, scale=0.32, angle=0} \\
 \noalign{\bigskip}
 \epsfig{file=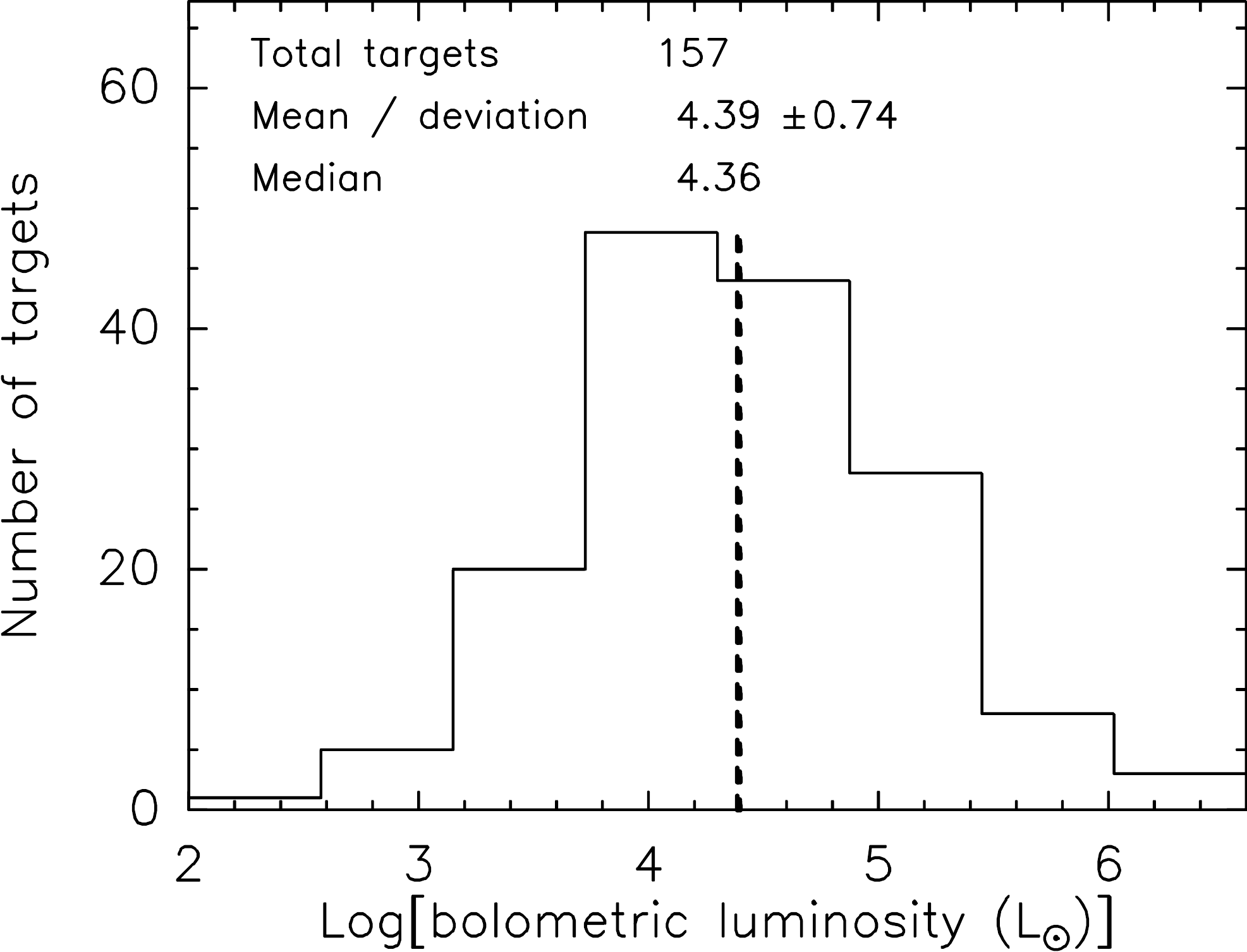, scale=0.32, angle=0} \\
 \noalign{\bigskip}
 \epsfig{file=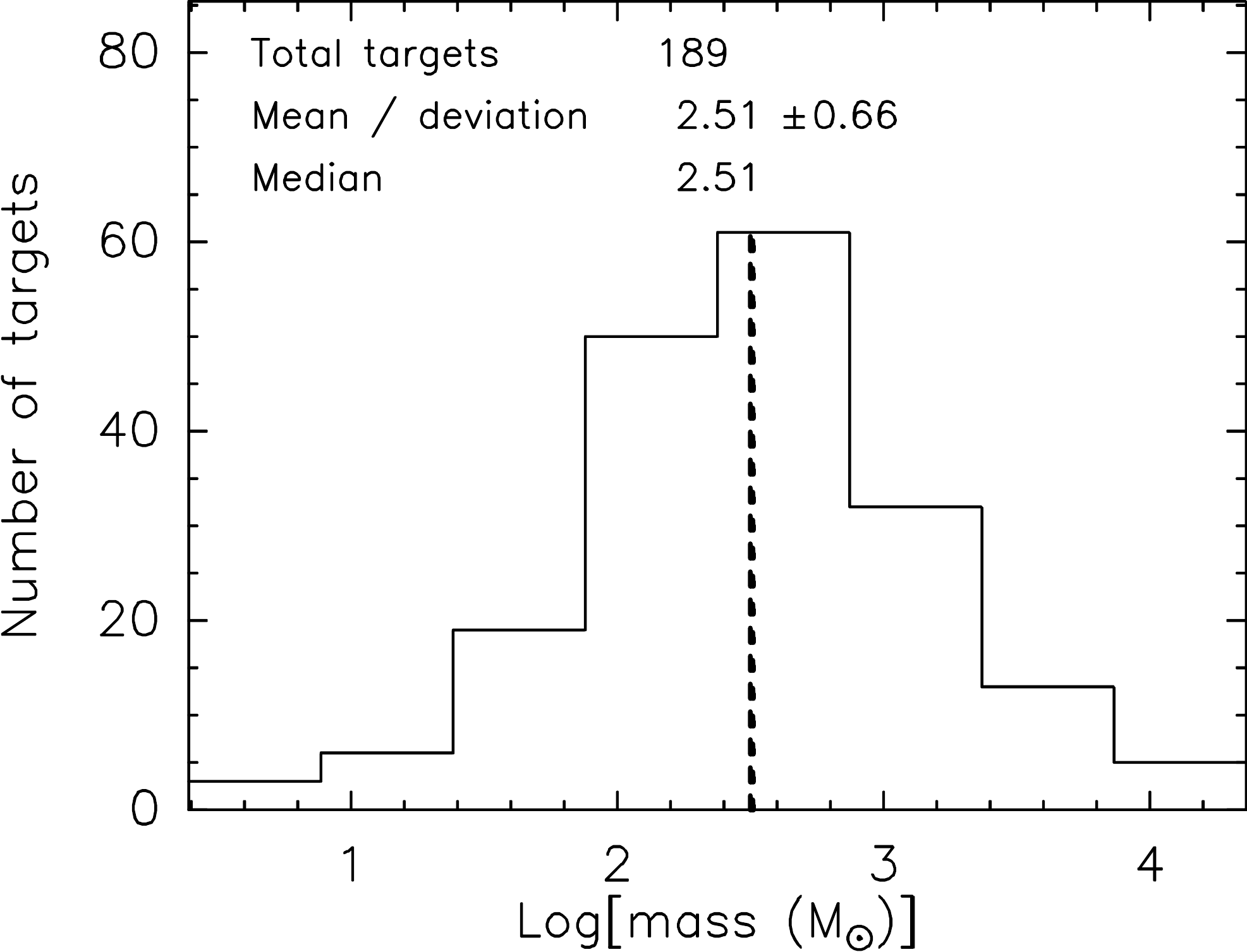, scale=0.32, angle=0} \\
\end{tabular}
\caption{{\bf Top}: Distribution of kinematic distances for the observed fields (see Table~\ref{t:observations} for references). {\bf Middle:} Distribution of bolometric luminosities for the fields centered on an IRAS source. {\bf Bottom:} Distribution of the sum of the masses of the millimeter sources within the observed fields (from Beltr\'an \et\ 2006). At the top of each panel we indicate the total number of sources used in the histogram, the mean, the standard deviation, and the median. The vertical dashed lines indicate the median values.}
\label{f:sampleHisto}
\end{center}
\end{figure}

\section{Introduction\label{s:intro}}

The problem of massive star formation (O \& B stars with masses $\gtrsim$8~\mo) still represents a challenge from both a theoretical and observational point of view. Such stars reach the zero-age main sequence (ZAMS) still undergoing heavy accretion, and their powerful radiation pressure should halt the infalling material, thus inhibiting growth of the stellar mass beyond $\sim$8~\mo\ (\eg\ Palla \& Stahler 1993). Recently, various studies have proposed a solution to this problem based on non-spherical accretion and high accretion rates (\eg\ McKee \& Tan 2003; Bonnell \et\ 2004; Krumholz \et\ 2009; Kuiper \et\ 2010). Together with the theoretical efforts, the comprehension of the massive star formation process requires good observational knowledge of the star-forming environment and of the evolutionary steps through which OB star formation occurs.


In 1991, we started a thorough investigation of a sample of luminous IRAS sources, selected on the basis of their far-infrared colours and with the additional constraint of $\delta$>$-30$\degr\ (Palla \et\ 1991, 1993). The hypothesis that this sample could contain high-mass young stellar objects (YSOs) in different evolutionary phases has been supported by a large number of observations that we have performed both at low- and high-angular resolution (Molinari \et\ 1996, 1998a, 1998b, 2000, 2002; Brand \et\ 2001; Fontani \et\ 2004a, 2004b, 2006; Zhang \et\ 2001, 2005). We started  a similar study in the southern hemisphere, which represents an excellent testbed for these studies, thanks to the large number of (massive) star forming regions that are observable. For this reason, we used the SEST telescope to search for CS-line and 1.2~mm continuum emission in a sample of sources with $\delta$<$-30$\degr\ (Fontani \et\ 2005; Beltr\'an \et\ 2006). Along the same line, other authors (\eg\ Bronfman \et\ 1989; Sridharan \et\ 2002; Beuther \et\ 2002; F\'aundez \et\ 2004; Hill \et\ 2005) have also studied large samples of massive young stellar objects in the millimeter continuum and line emission.

To study a possible evolutionary sequence of the objects, Molinari \et\ (2008) performed a detailed analysis of a small sub-set of our samples from millimeter down to mid-infrared wavelengths. According to these authors, the sources can be divided into three types, in order of increasing age: MM (and MM-S; hereafter \emph{Type~1}) with dominant millimeter emission and possibly close to an infrared MSX source, but not coincident with it; IR-P (hereafter \emph{Type~2}) with both millimeter and infrared emission; and IR-S1/S2 (hereafter \emph{Type~3}) with only infrared emission but located near a millimeter source. Using a simple model, Molinari \et\ (2008) could relate the infrared and millimeter emission to the physical state of the core and explain the distribution of these source types in a mass-luminosity plot in terms of evolution. As illustrated in Fig.~9 of Molinari \et\ (2008), \emph{Type~1} would correspond to high-mass protostars embedded in dusty clumps, possibly surrounded by infrared emission that could originate from more evolved, nearby lower-mass stars (see Faustini \et\ 2009), \emph{Type~2} would be deeply embedded ZAMS OB stars, still undergoing accretion, and \emph{Type~3} would be ZAMS OB stars surrounded only by remnants of their parental clumps.

In this paper, we aim to extend the analysis performed by Molinari \et\ (2008) to a larger sample of sources, and add the information of the centimeter continuum emission (likely tracing \hii\ regions: see \eg\ Kurtz \et\ 1994, S\'anchez-Monge \et\ 2008) and H$_2$O masers (being signposts of the presence of outflows/shocks associated with protostars: \eg\ Felli \et\ 2007; Moscadelli \et\ 2011) to set constraints on the evolutionary scheme. In Sections~\ref{s:sample} and \ref{s:obs} we describe the sample and observations. In Section~\ref{s:results} we present the centimeter continuum and maser line results of the ATCA observations. An analysis of the centimeter continuum sources is presented in Section~\ref{s:analysis}, and the evolutionary scenario is discussed in Section~\ref{s:discussion}. We end in Section~\ref{s:conclusions} with the main conclusions of this work.

\section{Sample\label{s:sample}}

The first step to extend the search for massive YSOs towards the southern hemisphere was to select a sample of possible candidates from the IRAS Point Source Catalogue following the Palla \et\ (1991, 1993) criteria: $F_\mathrm{60~{\mu}m}$$\ge$100~Jy and colours satisfying the criteria established by Richards \et\ (1987) for compact molecular cores.  The sources were divided into two subsamples based on the IRAS colors: sources with [25--12]>0.57 and [60--12]>1.3 were named \emph{High} (with colors characteristic of sources associated with ultracompact \hii\ regions; Wood \& Churchwell 1989), the others were named \emph{Low}. A sample of 235 IRAS sources, containing 142 \emph{Low} and 93 \emph{High} sources, was observed by Beltr\'an \et\ (2006) at 1.2~mm with the SEST, resulting in a total of 667 millimeter clumps detected. In this paper we present observations carried out with the Australia Telescope Compact Array (ATCA; see Sect.~\ref{s:obs}) toward a large sub-sample of these millimeter sources. In total, we observed 192 distinct fields (defined by the full width at half maximum primary beam of ATCA of $\sim$2$\farcm$5; see Sect.~\ref{s:obs}), containing 315 millimeter sources. Out of the total 192 fields, 160 were centered on an IRAS source (81~\emph{Low} and 79~\emph{High} IRAS sources), while the remaining 32 fields were centered on millimeter clumps (detected with the SEST) located $>$150\arcsec\ from an IRAS source.

In Fig.~\ref{f:sampleHisto} we show the distributions of kinematic distances, bolometric luminosities and masses of the sources observed with ATCA. Note that we only have distance determinations for 189 of the 192 observed fields, most of them obtained from CS-line observations (Fontani \et\ 2005; Beltr\'an \et\ 2006; see Table~\ref{t:observations}). For 96 of these fields there is an ambiguity between near and far distances (see Beltr\'an \et\ 2006). Searching the literature we have been able to solve the ambiguity for 36~fields (using \hi\ line observations; see references in Table~\ref{t:observations}), for the remaining fields we have considered the near distance for further analysis\footnote{Most of the distance estimates are obtained from spectral line observations pointing toward an IRAS source. We have assumed that the millimeter sources detected in the SEST fields surrounding an IRAS source (although not coincident) are located at the same kinematic distance (following the analysis carried out by Beltr\'an \et\ 2006).}. The bolometric luminosities have been derived only for the 157 fields centered on an IRAS source with distance determination (see middle panel of Fig.~\ref{f:sampleHisto}). Finally, we have considered the sum of the masses of the millimeter sources within our observed fields, and constructed the distribution of masses. The median values of 4.3~kpc, $2\times10^4$~\lo, and 300~\mo, and the distributions are similar to those obtained for all the sources in Beltr\'an \et\ (2006), confirming that our source selection is representative of the whole sample.

\begin{figure}[t!]
\begin{center}
\begin{tabular}[b]{c}
 \epsfig{file=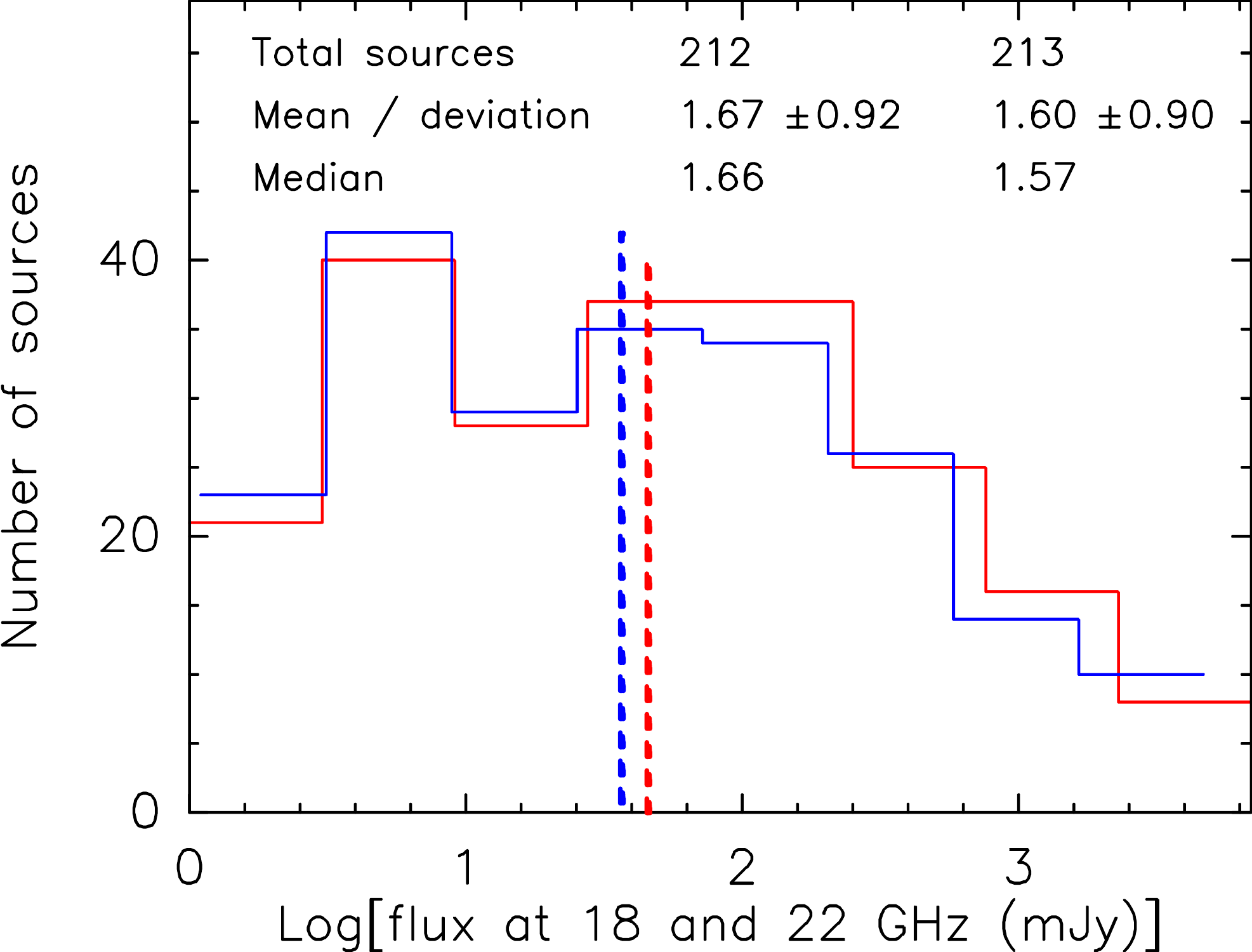, scale=0.32, angle=0} \\
 \noalign{\bigskip}
 \epsfig{file=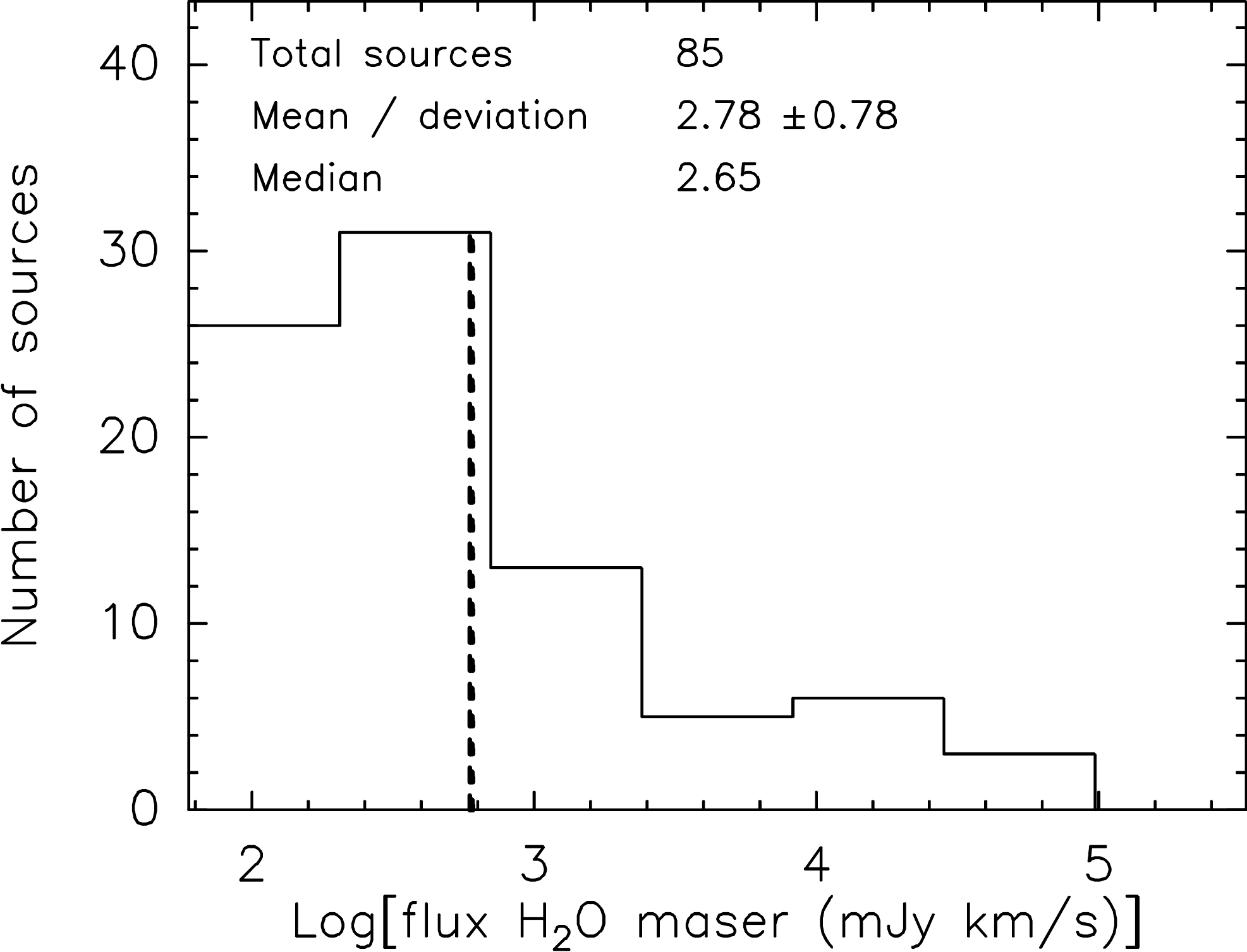, scale=0.32, angle=0} \\
\end{tabular}
\caption{{\bf Top}: Distributions of the flux density of the centimeter continuum sources detected at 18.0~GHz (blue, 212 sources) and 22.8~GHz (red, 213 sources). Note that the number of sources detected at both wavelengths is different, since one source is detected only at 18.0~GHz and two sources are detected only at 22.8~GHz (see Table~\ref{t:results}). {\bf Bottom}: Distribution of the integrated flux of the H$_2$O maser emission. The numbers at the top of each panel correspond to the total number of sources, the mean, the standard deviation, and the median values.}
\label{f:cmmaserHisto}
\end{center}
\end{figure}

\section{Observations and data reduction\label{s:obs}}

\addtocounter{table}{1}  

The 192 selected fields were observed with the Australia Telescope Compact Array (ATCA) interferometer at Narrabri (Australia) on the 25th and 26th of July 2009, and 9th and 10th of October 2009 (project C1804). In Table~\ref{t:observations}, we list the observed fields. Column~1 gives the field name, Col.~2 the source type of the IRAS source (\emph{High} or \emph{Low}), Col.~3 the kinematic distance adopted, Cols.~4 to 7 give the coordinates of the phase center\footnote{Note that in Beltr\'an \et\ (2006) the coordinates of the phase centers reported in their Table~1 for two sources were slightly wrong. For correct coordinates for the SEST observations were $\alpha$=16:12:42.7, $\delta$=$-$51:45:11 for 16085$-$5138; and $\alpha$=17:25:42.0, $\delta$=$-$36:21:57 for 17221$-$3619.} (both in equatorial and galactic systems), Cols.~8 to 13 give the synthesized beams, position angles, and rms noise levels at 18.0~GHz and 22.8~GHz, and Col.~14 indicates the day of observation of the field. Observations were conducted using the recently commissioned Compact Array Broadband Backend (CABB; Wilson \et\ 2011) providing a total bandwidth of 4~GHz divided in two sidebands: 2~GHz in the lower sideband (LSB) centered at a frequency of 18.0~GHz, and 2~GHz in the upper sideband (USB) centered at a frequency of 22.8~GHz, covering the H$_2$O\,($6_{15}$--$5_{23}$) maser transition at 22.235~GHz (although with low spectral resolution: channel width of 1~MHz, corresponding to $\sim$13.5~\kms). The array was in the H75 configuration for the four observing runs, resulting in a synthesized beam of $\sim$25\arcsec at 18.0~GHz and $\sim$20\arcsec\ at 22.8~GHz.

Observations were conducted in snapshot mode. The typical strategy was to observe each target field for 10~minutes. This total on-source time was broken up into a series of 3-minute snapshots in order to get a better coverage of the \emph{uv}-plane. A gain calibrator was observed every 15-minutes to correct for the phase variations. A total of 14 gain calibrators\footnote{The gain calibrators were 0826$-$373, 1004$-$50, 1045$-$62, 1059$-$63, 1129$-$58, J1326$-$5256, 1352$-$63, 1511$-$55, 1613$-$586, 1646$-$5, 1710$-$269, 1714$-$336, 1830$-$210 and 1923$+$210.} were used for the 192 target observed fields. The absolute flux scale was calibrated daily using PKS~B1934$-$638 as a primary calibrator for all the 4 observing runs. The flux of PKS~B1934$-$638 was assumed to be 1.03~Jy at 20~GHz. Daily bandpass calibration was achieved by observing 0537$-$441 (with a flux of 8.35~Jy).

The data reduction was done using the MIRIAD software package (Sault \et\ 1995). In all cases, data from the antenna located at 6~km from the main array (antenna CA06) were not included in the data reduction, to avoid an increase in the rms noise of the final images. Images of each target field were produced using the CLEAN algorithm with natural weighting. Fields with water maser emission at 22~GHz were self-calibrated. The self-calibration process consisted in identifying the channel with the strongest maser component, self-calibrate it (using the task SELFCAL), and apply the gain solutions to the continuum at 22.8~GHz and 18.0~GHz if both continuum maps improved with the new gain solutions. The mean rms noise is 0.45~m\jpb\ at 18.0~GHz and 0.40~m\jpb\ at 22.8~GHz  (see Table~\ref{t:observations} for the noise in individual fields).

\begin{table}
\caption{\label{t:summarycontinuum}Summary of centimeter continuum detections}
\centering
\begin{tabular}{l | r c | r c}
\hline\hline

&\multicolumn{4}{c}{192 observed ATCA fields}
\\
\hline
&\multicolumn{2}{c}{centered on IRAS\supa}
&\multicolumn{2}{c}{centered on mm\supa}
\\
\hline
no cm source			&8		&(--)	&15		&(--)	\\
single cm source		&118		&(118)	&13		&(13)	\\
multiple cm sources	&34		&(74)	&4		&(9)		\\
all					&160		&(192)	&32		&(22)	\\
\hline
\end{tabular}
\begin{list}{}{}
\item[\supa] The first number corresponds to the number of fields, while the second number (in parentheses) corresponds to the number of centimeter sources detected in the corresponding fields.
\end{list}
\end{table}

\section{Results\label{s:results}}

\subsection{Centimeter continuum emission\label{s:continuumres}}

In Fig.~\ref{f:ATCAmaps}, we present a panel for each field showing the centimeter continuum emission at 18.0~GHz (color scale) and 22.8~GHz (contours). In Fig.~\ref{f:globalATCAmaps}, we present three panels (in a row) for each field showing all possible comparisons between maps of the radio continuum emission, the 1.2~mm continuum from SEST (Beltr\'an \et\ 2006) and the 21.8~$\mu$m emission from the MSX (Midcourse Space Experiment; Price \et\ 1999). We detect centimeter continuum emission above the 5$\sigma$ level in 169 out of 192 fields, corresponding to a detection rate of 88\%. In most of the maps, the centimeter continuum emission comes from a single object, and only a few of them show extended complex emission with multiple peaks. In order to identify the centimeter sources and derive their properties, we use the following criterion: if the contour at half-maximum of the intensity peak (over 5$\sigma$) is closed, we consider it as an independent source. With this method we identified a total of 214 centimeter continuum sources. There are multiple sources in 38 fields, and a single detection (with an angular resolution of $\sim$20\arcsec) in 131 fields. Therefore, in 12\% of the 192 observed fields no centimeter continuum emission is detected, in 68\% we find a single component, and in 20\% of the targets there are multiple sources. In Table~\ref{t:summarycontinuum} we list a summary of the detections in the observed fields, distinguishing between fields centered or not on IRAS sources. We can see that most of the fields with no continuum detection (65\%) correspond to fields centered on a millimeter source located far from the IRAS source. 

\begin{figure*}[t!]
\begin{center}
\begin{tabular}[b]{c c c c c c c}
 \epsfig{file=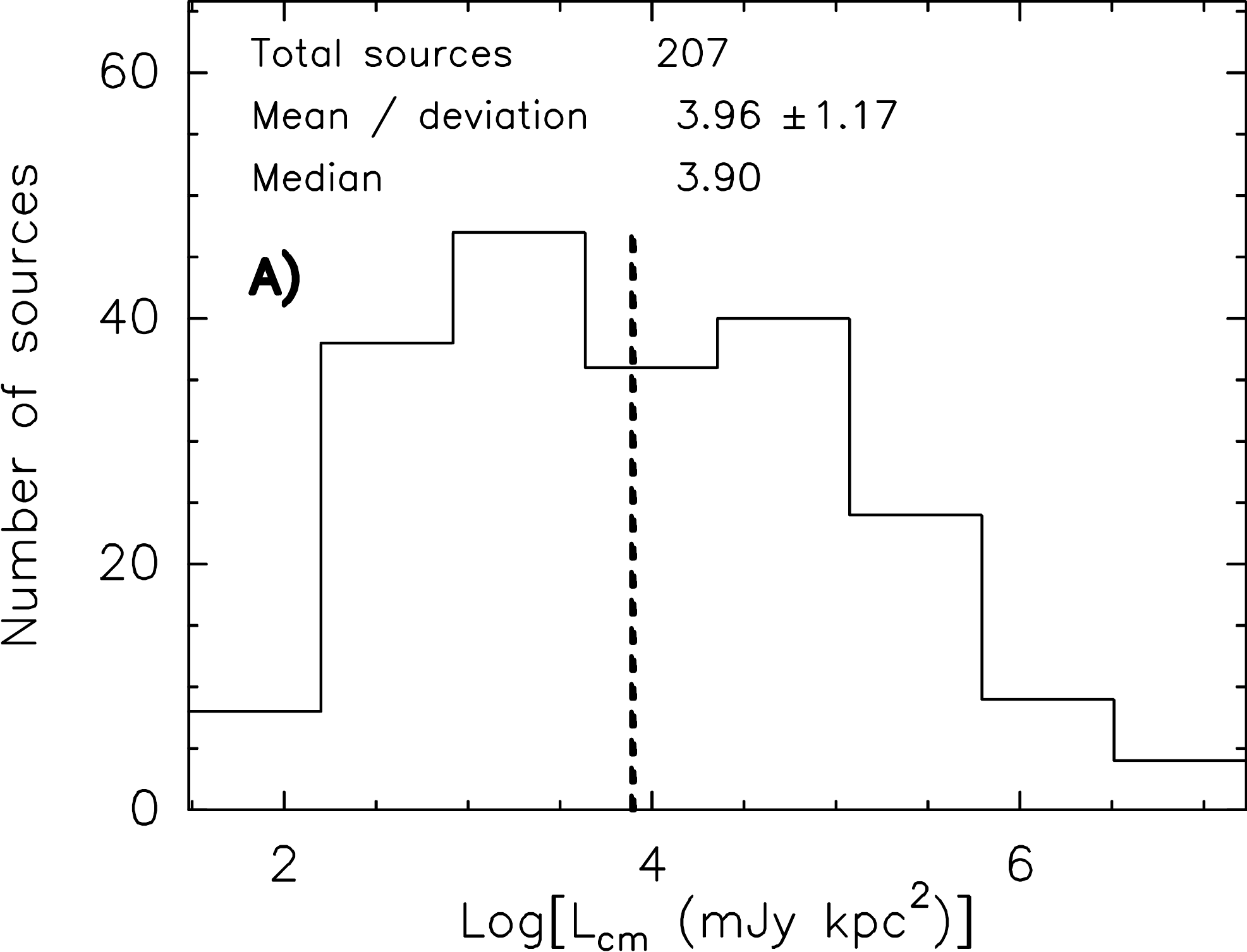, scale=0.27, angle=0} &&
 \epsfig{file=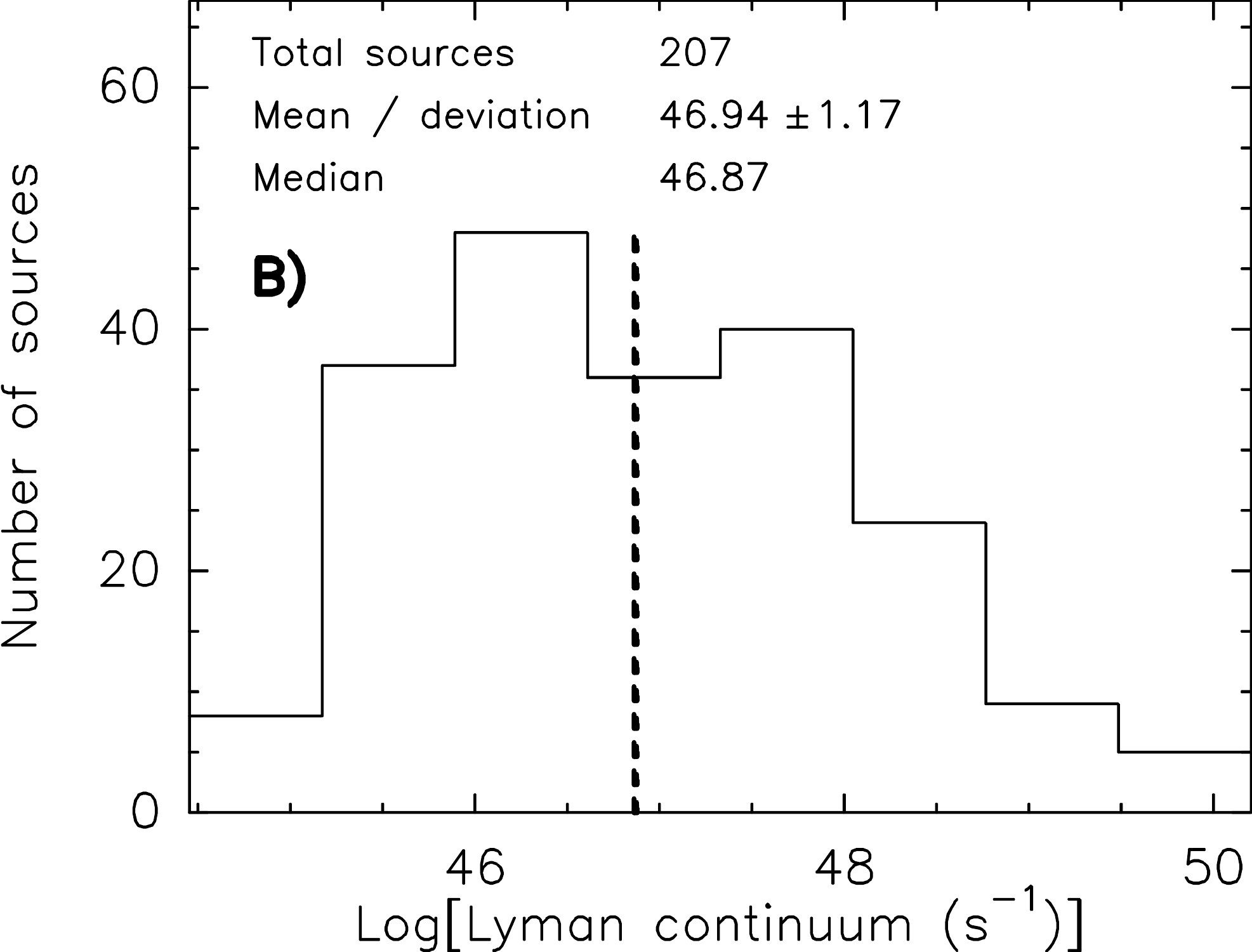, scale=0.27, angle=0} &&
 \epsfig{file=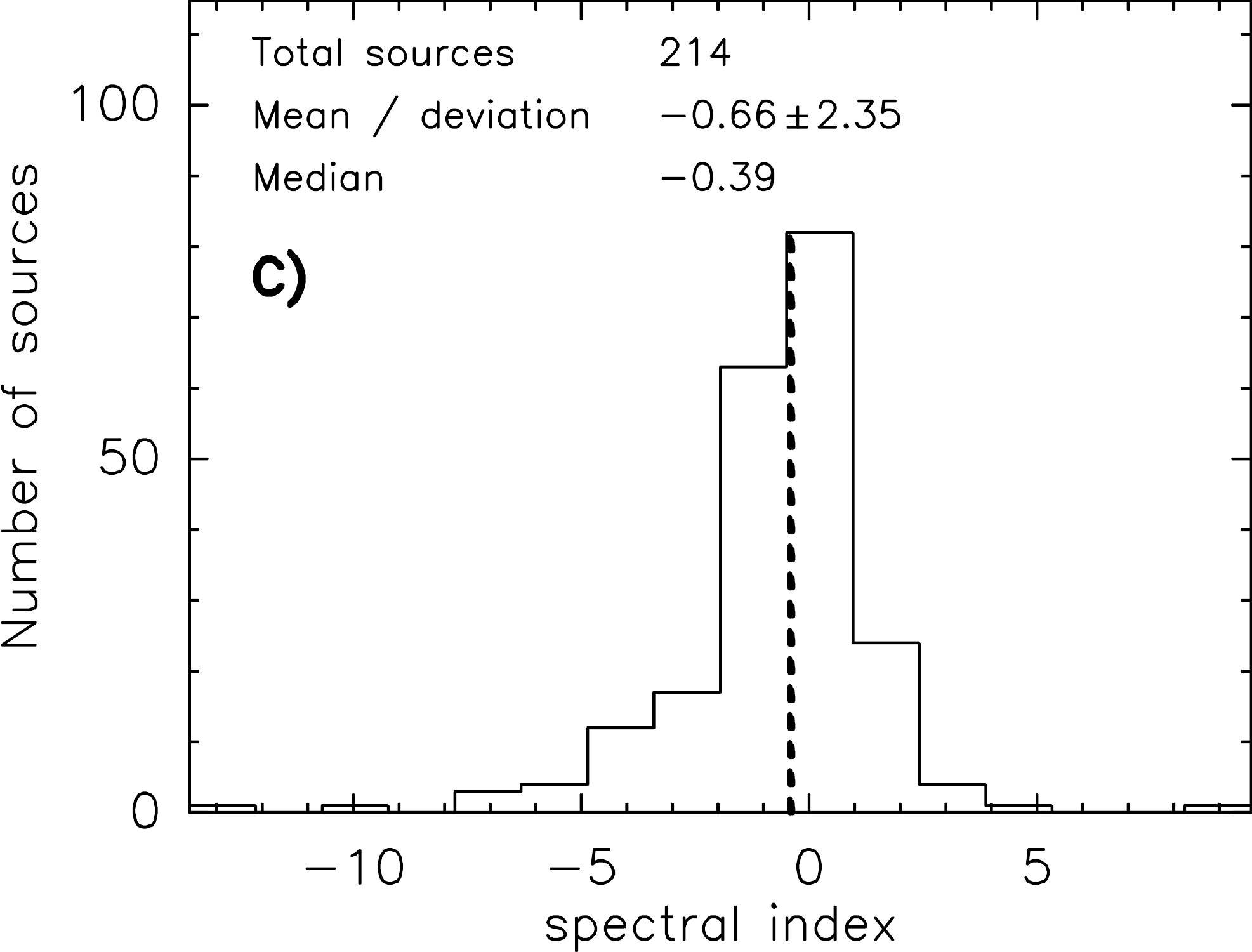, scale=0.27, angle=0} \\
 \noalign{\bigskip}
 \epsfig{file=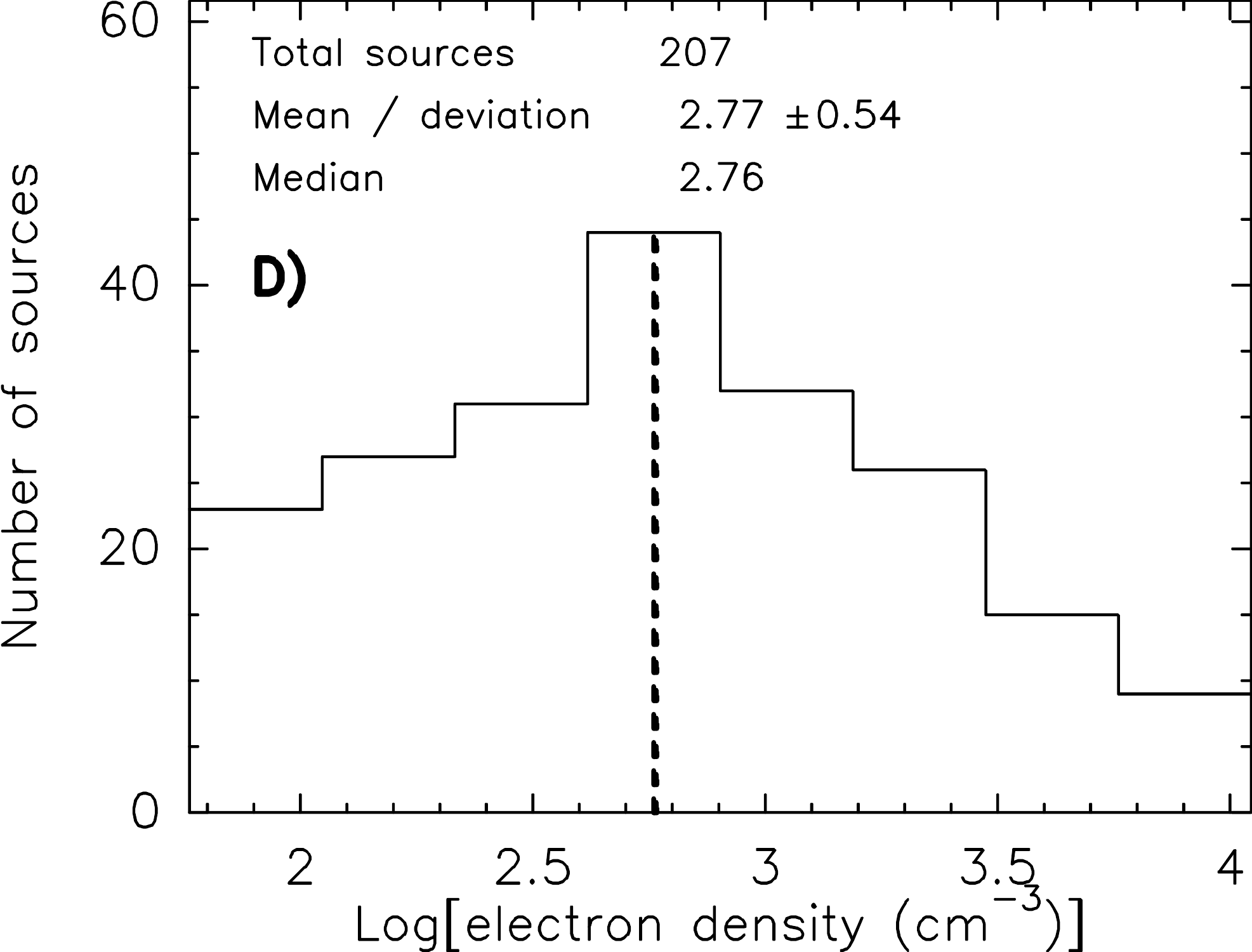, scale=0.27, angle=0} &&
 \epsfig{file=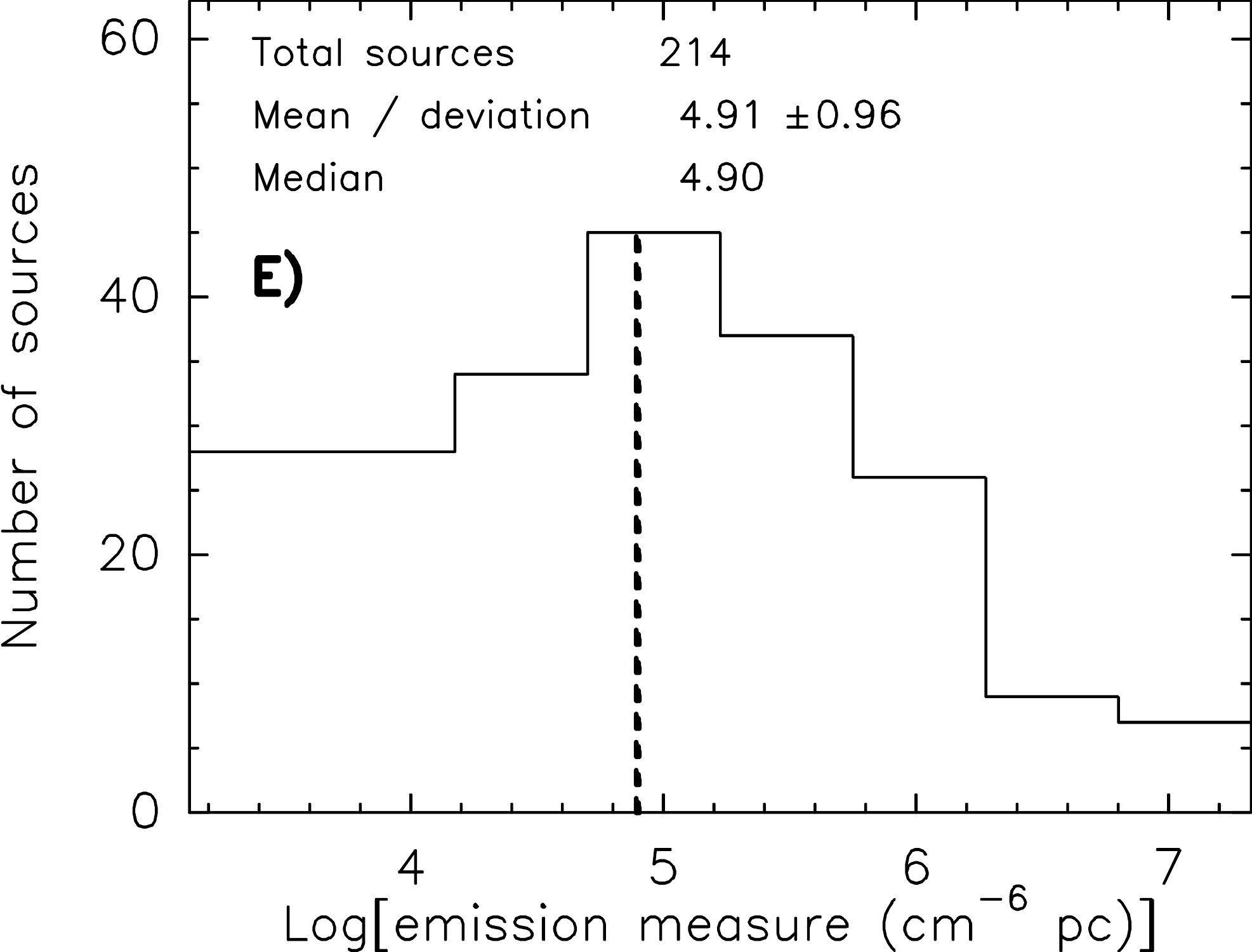, scale=0.27, angle=0} &&
 \epsfig{file=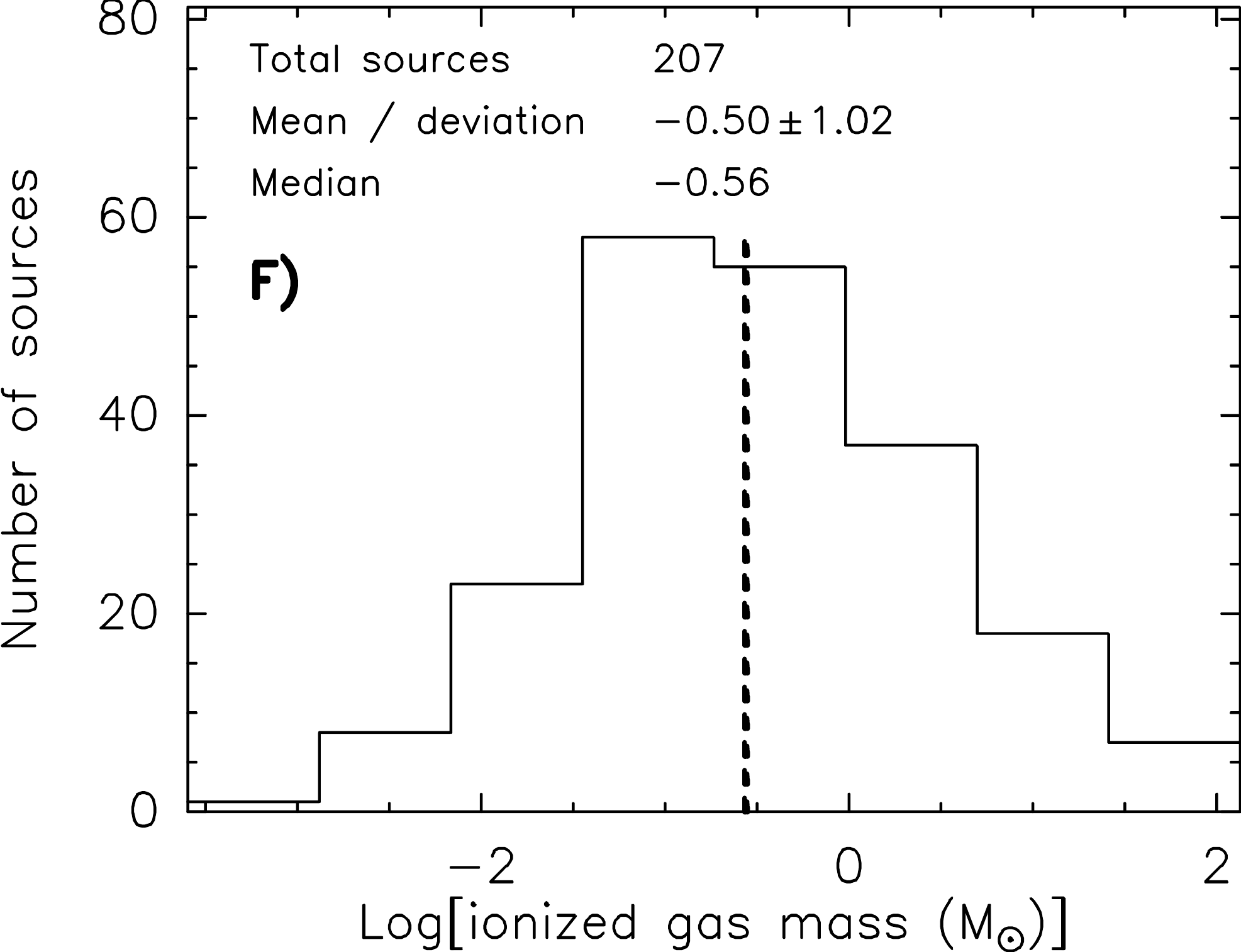, scale=0.27, angle=0} \\
 \noalign{\bigskip}
 \epsfig{file=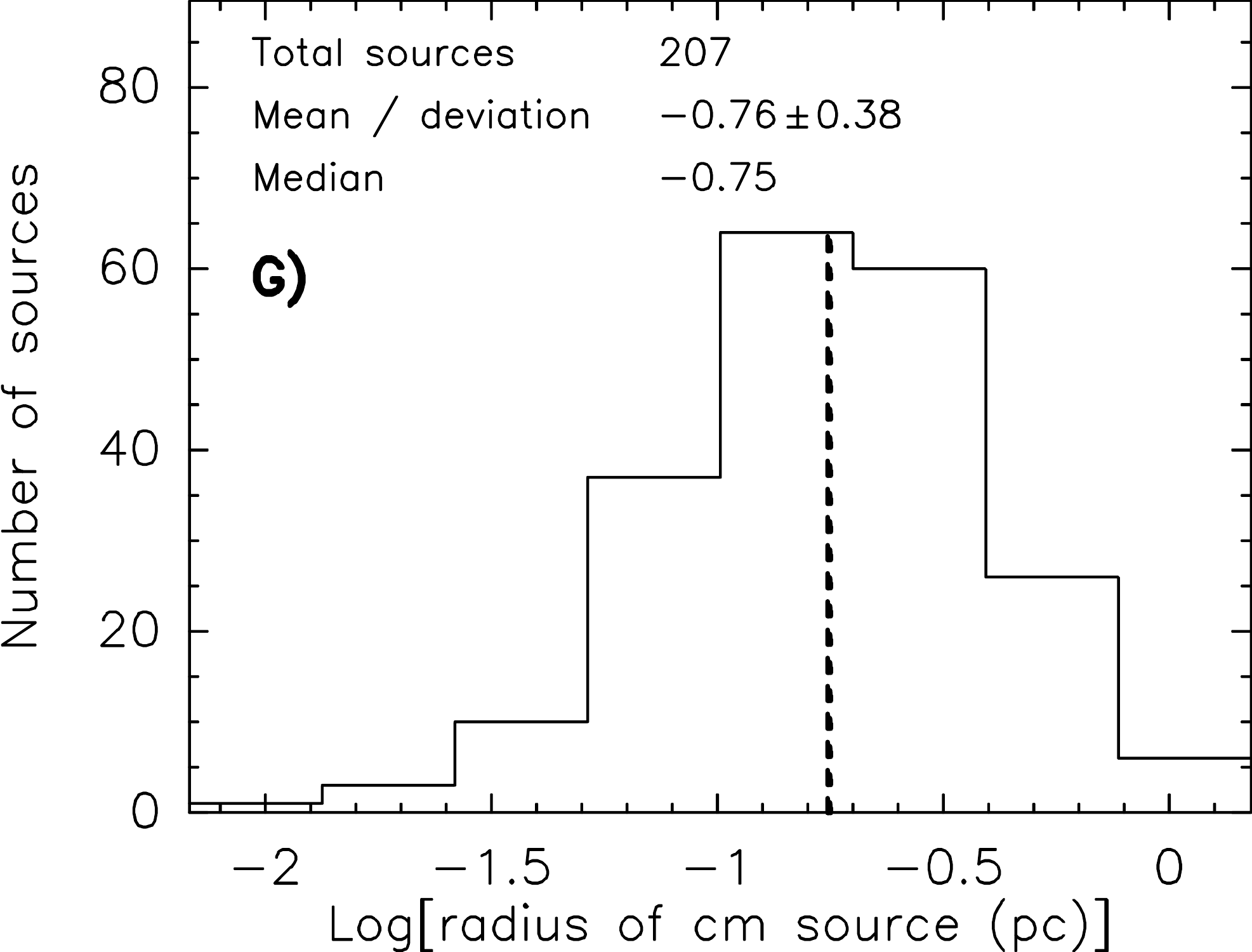, scale=0.27, angle=0} &&
 \epsfig{file=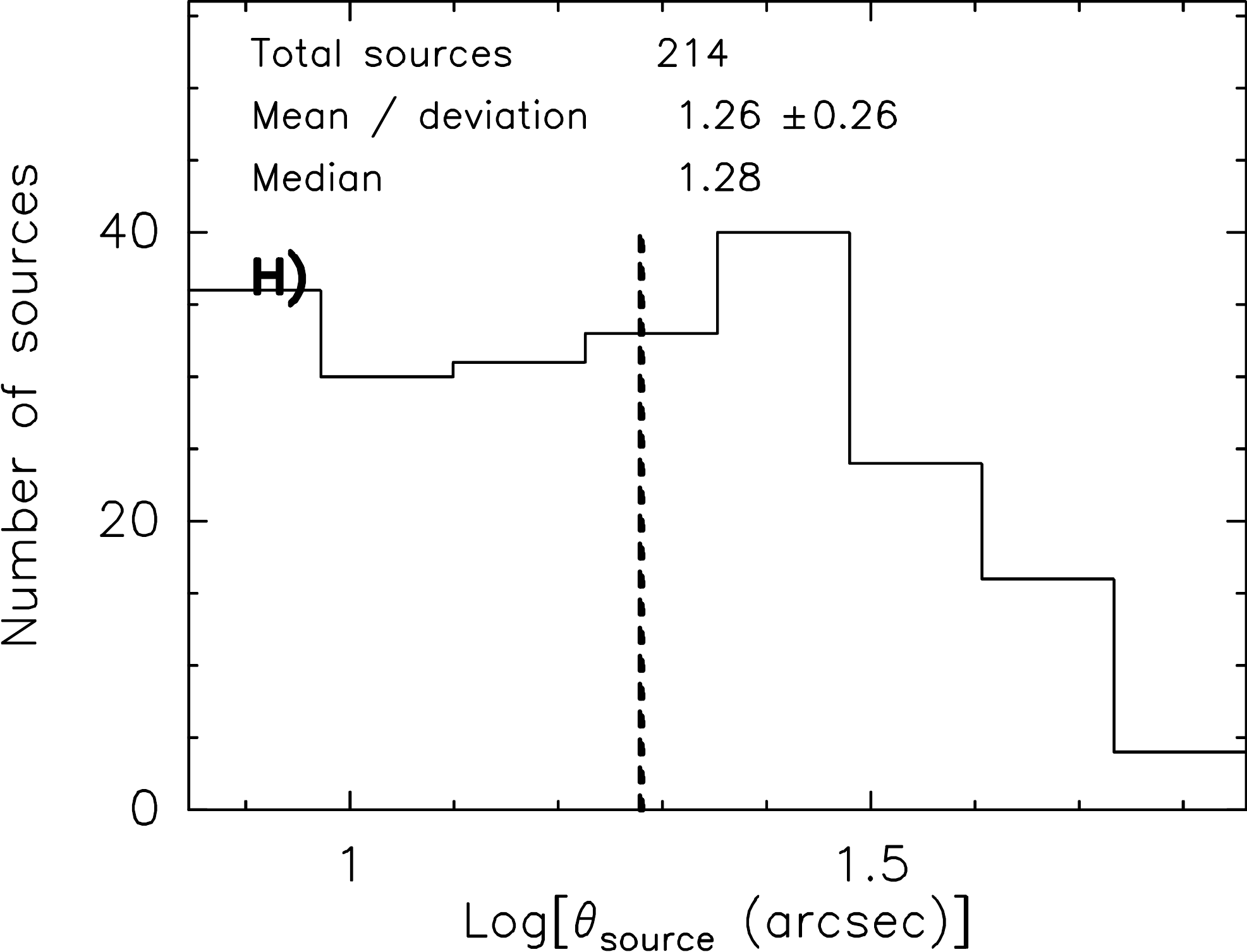, scale=0.27, angle=0} &&
 \epsfig{file=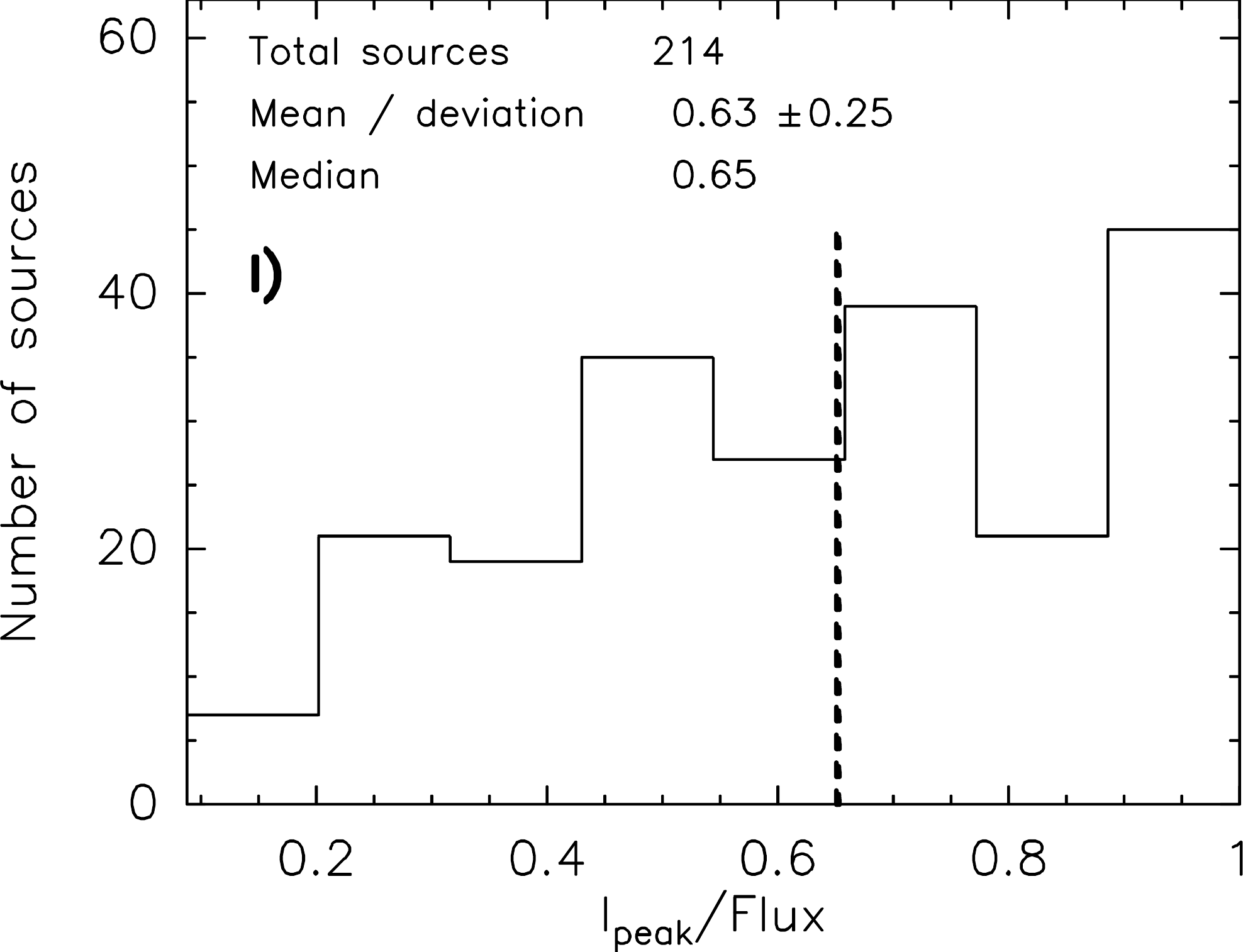, scale=0.27, angle=0} \\
\end{tabular}
\caption{Distributions of {\bf a)} centimeter luminosity; {\bf b)} Lyman continuum; {\bf c)} spectral index; {\bf d)} electron density; {\bf e)} emission measure; {\bf f)} ionized gas mass; {\bf g)} linear radius of the centimeter source; {\bf h)} deconvolved angular diameter, $\theta_{\rm S}$; and {\bf i)} peak intensity to flux density ratio, for the centimeter sources detected with ATCA. The numbers at the top of each panel indicate the total number of sources used in the histogram (in some cases we only can use 207 sources, those with distance determination), and the mean, standard deviation and median values. The vertical dashed line indicates the median value.}
\label{f:cmHisto}
\end{center}
\end{figure*}

\addtocounter{table}{1}  

In Table~\ref{t:results}, we list all the sources detected and, for completeness, the 23 fields with no centimeter emission indicating 5$\sigma$ upper limits. We give the name of the region (Col.~1), the number of each centimeter continuum source detected in a field (Col.~2), the coordinates of the 22.8~GHz detection\footnote{The peak separation in the images at the two frequencies (18.0~GHz and 22.8~GHz) is typically $\lesssim$5\arcsec, smaller than the synthesized beam.} (except where otherwise indicated for which the coordinates of the 18.0~GHz source are reported; Cols.~3 and 4), the peak intensity, integrated flux, observed size and deconvolved size at 18.0~GHz (Cols.~5 to 8) and at 22.8~GHz (Cols.~9 to 12), and the association of a water maser with the centimeter source (Col.~12). The angular diameter, $\theta_\mathrm{S}$, has been calculated assuming the sources are Gaussians from the expression
\begin{displaymath}
\theta_\mathrm{S}=\sqrt{\mathrm{FWHP}^2-\mathrm{HPBW}^2}
\end{displaymath}
with FWHP=2$\sqrt{\mathrm{A}/\pi}$, where $A$ is the area inside the polygon corresponding to the contour at half maximum, and HPBW is the synthesized beam listed in Table~\ref{t:observations}. For unresolved sources, we have considered an upper limit on the angular diameter of 7\arcsec, \ie\ 1/3 of the synthesized beam.

In the top panel of Fig.~\ref{f:cmmaserHisto}, we show the distribution of flux densities for the detected sources. The mean and median values are 47 and 46~mJy at 18.0~GHz, and 40 and 37~mJy at 22.8~GHz. Thus, we are typically measuring fluxes at 22.8~GHz lower than at 18.0~GHz. This can be due either to the flux density decreases with frequency for the detected sources (as expected in optically thin \hii\ regions), or to some part of the centimeter emission is being filtered out at shorter wavelengths (as seen by comparing the images in sources like 10184$-$5748 and 13106$-$6050).

\subsection{Water maser emission\label{s:maserres}}

\addtocounter{table}{1}  

Although the main goal of our observations was the study of the centimeter continuum emission, we also observed the H$_2$O maser line at 22.235~GHz. In Table~\ref{t:masers}, we list the properties of the detected masers: the name of the region, the coordinates, the integrated intensity and the velocity of the peak of the emission. Due to the poor spectral resolution ($\sim$13.5~\kms), the maser emission was typically detected in a single channel. Only in few cases we detected emission in more than one channel: two channels for 16573$-$4214~2, three for 15579$-$5303~0 and 16344$-$4605~0. The integrated intensity was calculated as the sum of the channel intensities multiplied by the corresponding width (13.5~\kms). Out of the 192 observed fields, 78 show water maser emission (detection rate of 41\%), and only four fields show multiple (spatial) components (see Table~\ref{t:masers}). Assuming a typical linewidth of $\approx$0.5~\kms\ for the water masers (\eg\ Gwinn 1994), the dilution factor due to our poor spectral resolution would be $\sim$0.037. Thus, taking into account the mean rms noise level of 18~m\jpb\ per channel, we are only sensitive to strong ($\gtrsim$2~Jy; assuming 5$\sigma$ detections) water maser single components. In the bottom panel of Fig.~\ref{f:cmmaserHisto}, we show the distribution of H$_2$O maser integrated flux densities. In Figures~\ref{f:ATCAmaps} and \ref{f:globalATCAmaps}, the masers are shown as white stars.

\addtocounter{table}{1}  

\begin{table*}
\caption{\label{t:meancm}Median values of main physical parameters of the centimeter sources}
\centering
\begin{tabular}{l l c c c c c c c}
\hline\hline
Parameter
&Units
&all sources
&Low\supa
&High\supa
&Type 1\supb
&Type 2\supb
&Type 3\supb
\\
(1)
&(2)
&(3)
&(4)
&(5)
&(6)
&(7)
&(8)
\\
\hline
number of sources					&-					&214		&75		&74		&15		&99		&55		&		\\
flux$_\mathrm{18.0~GHz}$				&mJy					&$46$	&$37$	&$125$	&$13$	&$107$	&$48$	&		\\
flux$_\mathrm{22.8~GHz}$				&mJy					&$37$	&$33$	&$102$	&$10$	&$120$	&$43$	&		\\
spectral index						&-					&$-0.4$	&$-0.3$	&$-0.3$	&$-0.8$	&$-0.2$	&$-0.2$	&		\\
$L_\mathrm{cm}$						&mJy~kpc$^2$			&$7900$	&$5700$	&$37100$	&$1100$	&$15100$	&$15800$	&		\\
$I_\mathrm{peak}/S_\mathrm{flux}$	&m\jpb/mJy			&$0.65$	&$0.62$	&$0.71$	&$0.52$	&$0.70$	&$0.59$	&		\\
angular size							&arcsec				&$19$	&$22$	&$14$	&$27$	&$15$	&$21$	&		\\
radius								&pc					&$0.18$	&$0.18$	&$0.17$	&$0.36$	&$0.14$	&$0.25$	&		\\
$n_\mathrm{e}$						&$10^3$~cm$^{-3}$	&$0.58$	&$0.46$	&$1.12$	&$0.13$	&$1.23$	&$0.46$	&		\\
$EM$									&10$^6$~cm$^{-6}$~pc	&$0.79$	&$0.60$	&$3.09$	&$0.13$	&$2.82$	&$0.71$	&		\\
$M_\mathrm{i}$						&\mo					&$0.28$	&$0.26$	&$0.46$	&$0.28$	&$0.28$	&$0.60$	&		\\
$N_\mathrm{Ly}$						&$\log$(s$^{-1}$)	&$46.9$	&$46.7$	&$47.6$	&$46.0$	&$47.2$	&$47.2$	&		\\
H$_2$O integrated flux				&mJy					&$400$	&$447$	&$490$	&$355$	&$575$	&$310$	&		\\
\hline
\end{tabular}
\begin{list}{}{}
\item[\supa] Classification (\emph{Low} and \emph{High}) by Palla \et\ (1991).
\item[\supb] Classification (\emph{type~1}, \emph{type~2} and \emph{type~3}) by Molinari \et\ (2008) and this work.
\end{list}
\end{table*}

\section{Analysis\label{s:analysis}}

\subsection{Physical parameters of \hii\ regions\label{s:physpar}}

Assuming that the centimeter continuum emission detected comes from homogeneous optically thin \hii\ regions, we calculated the physical parameters (using the formalism of Mezger \& Henderson 1967 and Rubin 1968) and list them in Table~\ref{t:physpar}. The linear radius of the \hii\ region (Col.~5) was determined from the deconvolved size listed in Table~\ref{t:results}. We calculated the source-averaged brightness temperature ($T_\mathrm{B}$; Col.~6) using
\begin{displaymath}
\label{eq:Tb}
T_\mathrm{B}=\frac{S_\nu~c^2}{2\nu^2~k_\mathrm{B}~\Omega_\mathrm{S}},
\end{displaymath}
where $S_\nu$ is the integrated flux density, $\Omega_\mathrm{S}$ is the solid angle of the source, $c$ the speed of light, $k_\mathrm{B}$ the Boltzmann constant and $\nu$ the observing frequency. The emission measure ($EM$; Col.~8) and the electron density ($n_\mathrm{e}$; Col.~7) were calculated from
\begin{displaymath}
\label{eq:EM}
\bigg[\frac{EM}{\mathrm{cm}^{-6}~\mathrm{pc}}\bigg]\,=\,
1.7\times10^7~\bigg[\frac{S_\nu}{\mathrm{Jy}}\bigg]
\bigg[\frac{\nu}{\mathrm{GHz}}\bigg]^{0.1}
\bigg[\frac{T_\mathrm{e}}{\mathrm{K}}\bigg]^{0.35}
\bigg[\frac{\theta_\mathrm{S}}{\arcsec}\bigg]^{-2},
\end{displaymath}
\begin{displaymath}
\label{eq:ne}
\bigg[\frac{n_\mathrm{e}}{\mathrm{cm}^{-3}}\bigg]\,=\,
2.3\times10^6~\bigg[\frac{S_\nu}{\mathrm{Jy}}\bigg]^{0.5}
\bigg[\frac{\nu}{\mathrm{GHz}}\bigg]^{0.05}
\bigg[\frac{T_\mathrm{e}}{\mathrm{K}}\bigg]^{0.175}
\bigg[\frac{d}{\mathrm{pc}}\bigg]^{-0.5}
\bigg[\frac{\theta_\mathrm{S}}{\arcsec}\bigg]^{-1.5},
\end{displaymath}
where $T_\mathrm{e}$ is the electron temperature assumed to be $10^4$~K, $\theta_\mathrm{S}$ is the angular diameter of the source, and $d$ is the distance. The mass of ionized gas ($M_\mathrm{i}$; Col.~9) was calculated assuming a spherical homogeneous distribution as
\begin{displaymath}
\label{eq:Mi}
\bigg[\frac{M_\mathrm{i}}{M_{\sun}}\bigg]\,=\,
3.5\times10^{-12}~\bigg[\frac{S_\nu}{\mathrm{Jy}}\bigg]^{0.5}
\bigg[\frac{\nu}{\mathrm{GHz}}\bigg]^{0.05}
\bigg[\frac{T_\mathrm{e}}{\mathrm{K}}\bigg]^{0.175}
\bigg[\frac{d}{\mathrm{pc}}\bigg]^{2.5}
\bigg[\frac{\theta_\mathrm{S}}{\arcsec}\bigg]^{1.5}.
\end{displaymath}

The number of Lyman-continuum photons per second ($N_\mathrm{Ly}$; Col.~10; hereafter Lyman continuum for simplicity) was calculated from the flux density and kinematic distance as
\begin{displaymath}
\label{eq:Ni}
\bigg[\frac{N_\mathrm{Ly}}{\mathrm{s}^{-1}}\bigg]\,=\,
8.9\times10^{40}~\bigg[\frac{S_\nu}{\mathrm{Jy}}\bigg]
\bigg[\frac{\nu}{\mathrm{GHz}}\bigg]^{0.1}
\bigg[\frac{T_\mathrm{e}}{10^4\mathrm{K}}\bigg]^{-0.45}
\bigg[\frac{d}{\mathrm{pc}}\bigg]^2.
\end{displaymath}
From the estimated $N_\mathrm{Ly}$, using the tables of Panagia (1973) and Thompson (1984) and assuming that a single ZAMS star is the source of the ionizing photons, we computed the spectral type of the ionizing source (Col.~11). Similar results (with variations of at most one subtype) are obtained when using the more recent calculations of Vacca \et\ (1996), Diaz-Miller \et\ (1998) or Martins \et\ (2005).

Assuming an uncertainty of 10\% on the flux density and angular diameter, one finds that this implies uncertainties of 30\% for the emission measure, 20\% for the electron density, 20\% for the ionized gas mass, and 10\% for the Lyman continuum. Note that for extended sources the measured flux density could be a lower limit should the interferometer be filtering extended emission, and the measured size could be underestimated by up to 50\% if the source is not Gaussian (see Panagia \& Walmsley 1978).

In Column~12 of Table~\ref{t:physpar} we give the spectral index ($\alpha$; $S_\nu\propto\nu^\alpha$) estimated from the 18.0~GHz and 22.8~GHz fluxes. However, these values must be taken with caution due to the large uncertainties involved, mainly because the two frequencies are very close and hence the error on $\alpha$ is large. In addition, the 22.8~GHz images could be more affected than the 18.0~GHz images by filtering of extended structures.

In Fig.~\ref{f:cmHisto}, we show the distribution of the main physical parameters of the centimeter continuum sources. The mean, standard deviation, and median values of each histogram are shown at the top of each panel. In Table~\ref{t:meancm}, we list the median values of these parameters for the whole sample (Col.\ 3), and for the various classes of objects considered in this paper (Cols.\ 4--8; see next Sections). The median diameters ($\sim$0.36~pc), electron densities ($\sim$580~cm$^{-3}$) and emission measures ($\sim$7.9$\times10^4$~cm$^{-6}$~pc) are in agreement with typical values of compact \hii\ regions (Kurtz 2005), with the observed parameters ranging between those of classical and ultracompact \hii\ regions.

\subsection{Lyman continuum\label{s:Lyman}}

In Fig.~\ref{f:LcmLbol} we compare the Lyman continuum, \Nly, obtained from the measured radio flux with the bolometric luminosity, \Lbol, estimated from the IRAS fluxes (Fontani \et\ 2005). We have considered that a centimeter continuum source is associated with the IRAS source if their separation is $\le$60\arcsec\ (corresponding to the HPBW of the IRAS satellite at 25~$\mu$m). We stress, as explained in Sect.~\ref{s:sample}, that for those sources for which it was not possible to discriminate between the near and far kinematic distances, the near estimate was arbitrarily assumed.

The red solid curve in Fig.~\ref{f:LcmLbol} corresponds to the relationship between \Nly\ and \Lbol\ expected for an O-B type star. This has been computed by various authors (\eg\ Panagia 1973; Thompson 1984; Vacca \et\ 1996; Schaerer \& de Koter 1997; D\'iaz-Miller \et\ 1998), resulting in a maximum discrepancy $\la$30\% among different calculations. While most points in Fig.~\ref{f:LcmLbol} fall below the red curve, a significant fraction of the sources happen to lie above it. This is surprising, as the Lyman continuum from a single star is the maximum value obtainable for a given bolometric luminosity, as illustrated by the dashed area in the figure, which corresponds to the expected \Nly\ if \Lbol\ is due to a stellar cluster. This result has been obtained by generating a large collection ($10^6$) of clusters with sizes ranging from 5 to 500000 stars each. The clusters are assembled randomly from a cluster membership distribution which describes the number of clusters ($N_\mathrm{cl}$) composed by a given number of stars ($N_\mathrm{st}$) of the form: $\mathrm{d}N_\mathrm{cl}/\mathrm{d}N_\mathrm{st}\propto N_\mathrm{st}^\alpha$ with $\alpha$=$-$2. Each cluster was populated assuming a randomly sampled Chabrier (2005) initial mass function with stellar masses in the range 0.1--120~\mo. For each cluster we computed the total mass, bolometric luminosity, maximum stellar mass and integrated Lyman continuum. The shaded area in Fig.~\ref{f:LcmLbol} was produced by plotting at each bolometric luminosity the range that includes 90\% of the simulated clusters.

How can one explain the presence of so many objects in the ``forbidden region'' above the red curve in Fig.~\ref{f:LcmLbol}? A priori several explanations are possible and we discuss them in the following:

\begin{figure}[t!]
\begin{center}
\begin{tabular}[b]{c}
 \epsfig{file=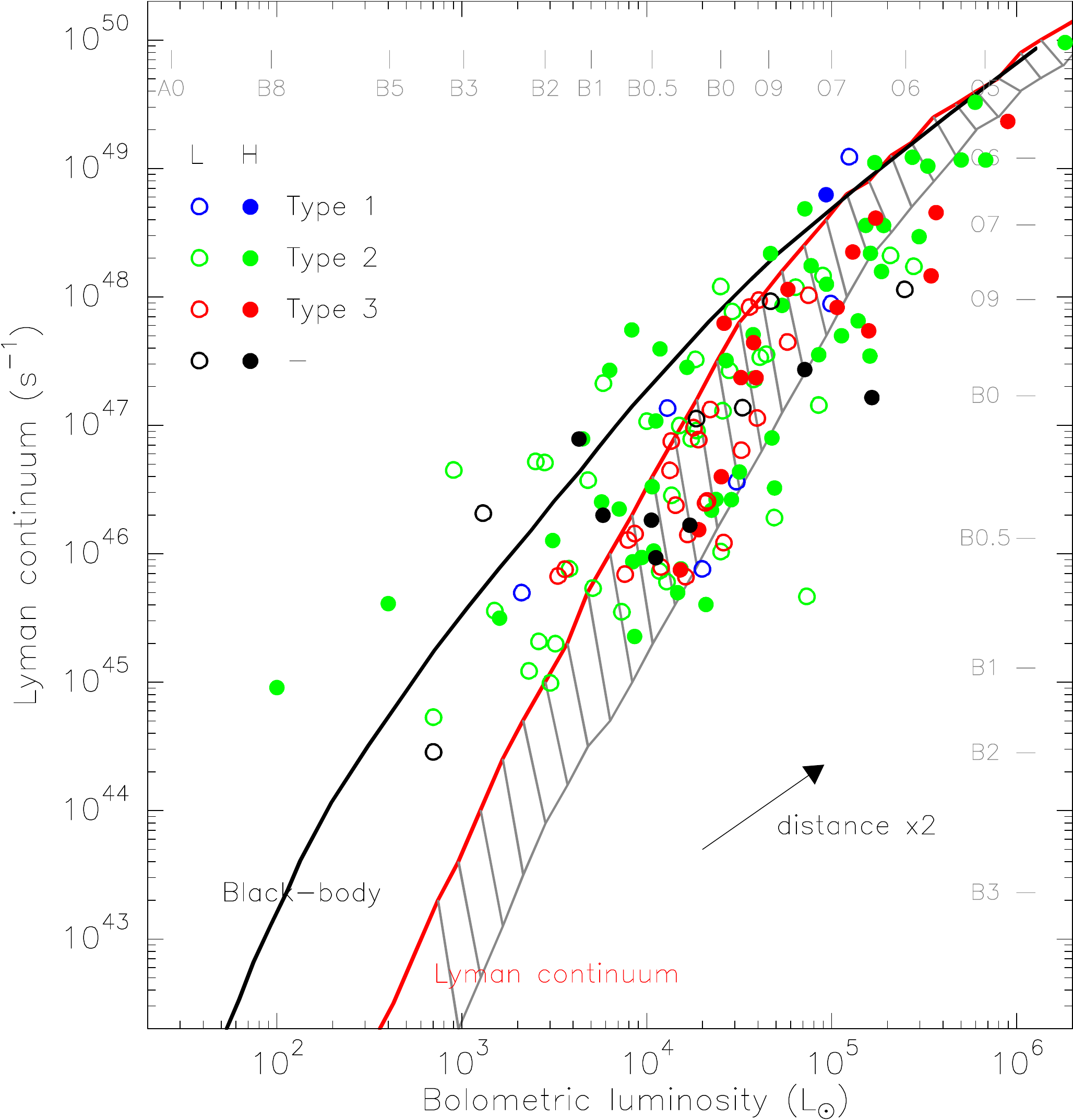, scale=0.38, angle=0} \\
\end{tabular}
\caption{Lyman continuum, $N_\mathrm{Ly}$ (from Table~\ref{t:physpar}), as a function of the bolometric luminosity, $L_\mathrm{bol}$. Open and filled symbols denote \emph{Low} and \emph{High} sources, respectively.  The different colors indicate the \emph{types~1} (blue), \emph{2} (green) and \emph{3} (red), while the black symbols correspond to those centimeter sources not classified in any of the three types (see Sect.~\ref{s:discussion}). The red solid line corresponds to the Lyman continuum of a ZAMS star of a given bolometric luminosity, while the dashed region indicates the expected $N_\mathrm{Ly}$ if $L_\mathrm{bol}$ is due to a cluster (see Sect.~\ref{s:Lyman} for details). The black solid line corresponds to the Lyman continuum expected from a black-body with the same radius and temperature of a ZAMS star. At the top and right axes we list the spectral type corresponding to a given Lyman continuum and a given bolometric luminosity. The arrow indicates how much a point should be shifted if its distance is doubled.}
\label{f:LcmLbol}
\end{center}
\end{figure}

\begin{itemize}

\item The radio flux (from which \Nly\ is computed) could be overestimated. In practice, one cannot think of any reason why this should be the case. On the one hand, if the \hii\ regions were optically thick, density bound, or dusty, the free-free emission would be reduced and the estimated \Nly\ would be a {\it lower} limit. On the other hand, interferometric observations tend to resolve extended emission and rather {\it decrease} the measured flux density (and hence the estimated value of \Nly). Finally, uncertainties in the flux calibration may reach a few 10\%, much less than the observed excess with respect to the single-star curve in Fig.~\ref{f:LcmLbol} (up to almost 3 orders of magnitude).

\item The bolometric luminosity could be underestimated. Since we have derived \Lbol\ from the flux densities reported in the IRAS Point Source Catalogue, this explanation is very unlikely. The IRAS HPBW ranges from 0\farcs5 to 2\arcmin\ and thus encompasses an area even larger than that of the observed \hii\ regions. Consequently, the IRAS fluxes are more likely overestimating the emission from the relevant sources and the estimated \Lbol\ is rather an upper limit.

\item The sources in the ``forbidden region'' could be Planetary Nebulae (PNe) instead of \hii\ regions. In this case the value of \Nly\ would be much greater than for a ZAMS star, for the same \Lbol. We note that most of the objects above the red curve in Fig.~\ref{f:LcmLbol} belong to \emph{type~1} and \emph{2} (see Sect.~\ref{s:discussion}) and are thus associated with molecular clumps, which appears quite unlikely for an evolved star such as those powering PNe.

\item The source of \Nly\ could be other than a ZAMS star. For example, it has been proposed that UV photons from shocks could be ionizing thermal jets (\eg\ Anglada 1996). However, the expected free-free emission at 1.3~cm from such a jet is on the order of 0.1~mJy~kpc$^2$, much less than the measured radio luminosities of the objects in the ``forbidden region'' -- typically 10--$10^6$~mJy~kpc$^2$. Another possibility is that the UV photons are originating from shocks in the accretion flow onto the protostar/disk system. Although intriguing, this hypothesis is yet to be investigated theoretically and thus must be regarded as purely speculative at present.

\item The distances could be wrong. Indeed, a relatively small error on our distance estimates could significantly affect the position of the corresponding points in Fig.~\ref{f:LcmLbol}. The arrow in this figure indicates how much a point should be shifted if its distance is doubled. In fact, for many sources we have arbitrarily assumed the near kinematic distance, while replacing this with the far distance would certainly move the corresponding points to the right of the red curve. However, even if for all sources above the solid red line the far distance were assumed, still $\sim$50\% of them would lie in the ``forbidden region''.

\end{itemize}

In conclusion, we believe that the observed excess of Lyman continuum cannot be trivially explained by any of the above hypotheses. It is worth pointing out, though, that almost all of the sources in the ``forbidden region'' fall very close to or below the Lyman continuum expected from a black-body with the same radius and effective temperature as a ZAMS star (black solid line in Fig.~\ref{f:LcmLbol}). Although not conclusive, this fact suggests that some B-type stars could emit in the UV range much more than predicted by standard models of stellar atmospheres. Given that $\sim$70\% of the \HII\ regions lying above the red curve belong to \emph{types~1} and \emph{2}, one may further speculate that such an excess in \Nly\ occurs preferentially during the earliest evolutionary stages of a high-mass star. Further investigation is needed to confirm our findings and shed light on this intriguing issue, which, to our knowledge, has not been reported previously. Previous interferometric large surveys of \hii\ regions (\eg\ Wood \& Churchwell 1989, Kurtz \et\ 1994) were carried out with angular resolutions $\le$1\arcsec, which could be filtering out a large fraction of the extended centimeter continuum emission that we can image with our $\sim$20\arcsec observations, thus obtaining lower values of \Nly.

\subsection{Source size and electron density relationship\label{s:nesize}}

\begin{figure}[t!]
\begin{center}
\begin{tabular}[b]{c}
 \epsfig{file=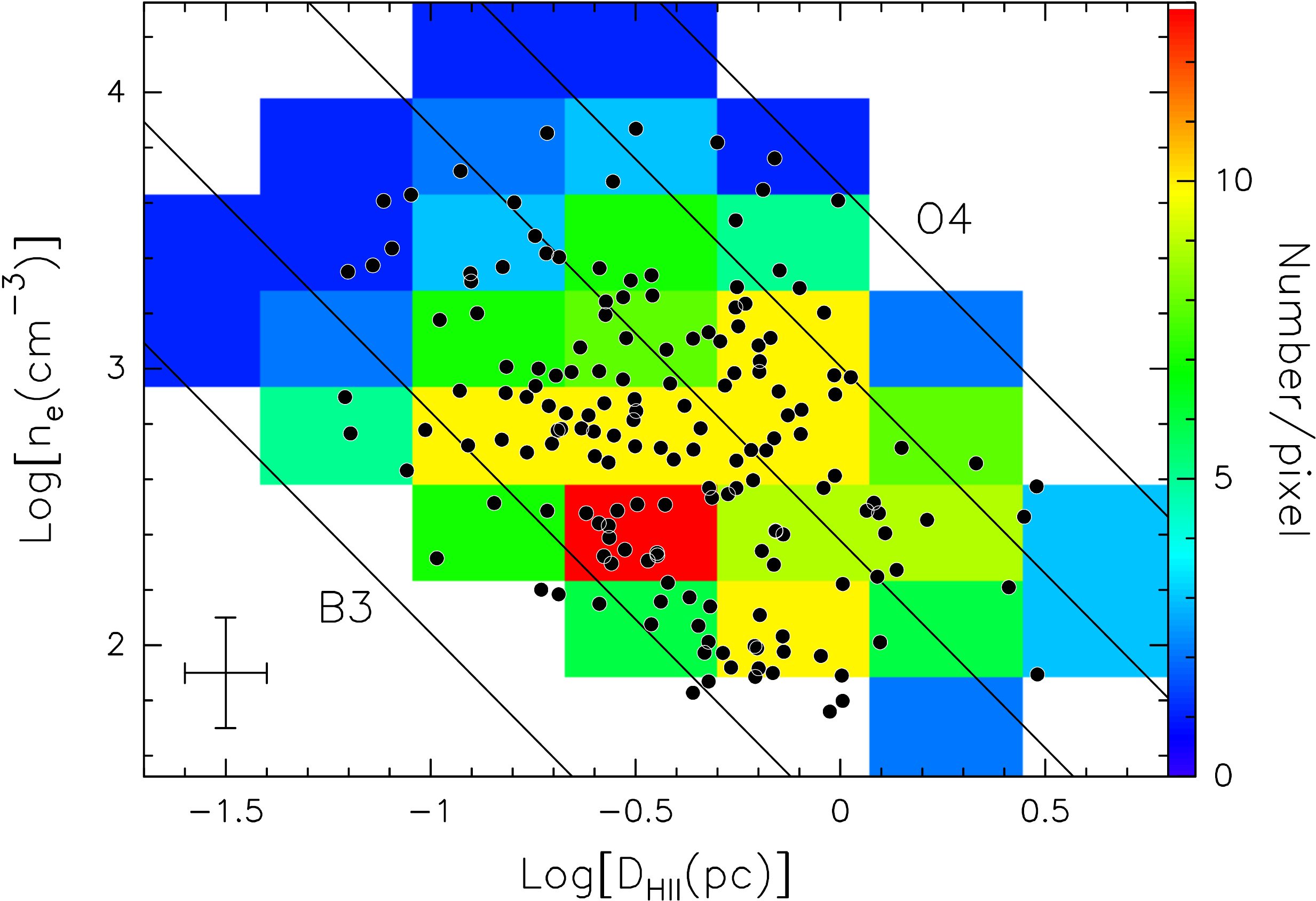, scale=0.35, angle=-90} \\
\end{tabular}
\caption{Electron density of the \hii\ regions detected in our survey versus the corresponding linear diameter. The points denoting the data (listed in Table~\ref{t:physpar}) are overlayed on an image whose intensity (colour scale) gives the number of \hii\ regions expected in each pixel on the basis of the model described in Appendix~\ref{a:model}. The solid lines correspond to ZAMS stars of spectral types B3, B1, B0, O7 and O4. The cross indicates the 20\% error in the electron density and the 10\% error in the source size (see Sect.~\ref{s:physpar}). }
\label{f:nesize}
\end{center}
\end{figure}

It is interesting to investigate the relationship between density and size in our sample of \hii\ regions. This is illustrated in Fig.~\ref{f:nesize}, which shows a plot of the electron density, $n_\mathrm{e}$, versus the diameter of the \hii\ region, $D_\hii$, for all of our objects. Note that for a given Lyman continuum, the relationship between the radius and the electron density of a Str\"omgren \hii\ region is fixed. The solid lines drawn in the figure correspond to such a relationship for ZAMS stars of spectral types ranging from B3 to O4. As expected, all points lie in the region spanned by these spectral types.

The result in Fig.~\ref{f:nesize} is similar to that obtained by other authors (see e.g.\ Garay \& Lizano 1999 and Mart\'{\i}n-Hern\'andez \et\ 2003), who noted that the distribution of the points in the plot appears to be consistent with a steeper relationship than that expected for a simple Str\"omgren sphere ($n_\mathrm{e}\propto D_\hii^{-3/2}$). However, before drawing such a conclusion, one should take into account any possible observational bias that could affect the observed distribution of points in that plot. For example, the largest \hii\ regions may be resolved out by the interferometric observations, while the smallest could fall below our sensitivity limit. To take these and other effects into account, we have attempted a more quantitative approach. This consists in comparing the distribution in Fig.~\ref{f:nesize} with that obtained assuming that O-B type stars are distributed in the Galaxy as established by Mottram \et\ (2011). The details of our procedure are given in Appendix~\ref{a:model}. Here we only stress that with respect to Mottram \et\ we have introduced the additional assumption that the number of \hii\ regions with electron density $\ne$ is given by a function of the type $\fne\propto\ne^\alpha$. In practice, our model depends only on the exponent $\alpha$ and the total number of \hii\ regions in the Galaxy, \NHII. As shown in Appendix~\ref{a:model}, we can fit the observed electron density function $\fne$ and thus obtain the values of the two free parameters, \NHII=15000$^{+2000}_{-1300}$ and $\alpha$=$-0.15\pm0.1$. With this model one can also estimate the expected distribution of points in the plot of Fig.~\ref{f:nesize}. This is shown as a colour scale, indicating the number of detectable \hii\ regions per pixel. Clearly, the expected distribution is far from homogeneous, and consistent with the distribution of the observed points. The agreement between model and data can be better appreciated in Figs.~\ref{f:distra} and~\ref{f:distrb} where the best-fit density ($\fne$) and diameter (\fD) functions are compared to those obtained from our data.

Although the fit is quite satisfactory, a word of caution is in order. Our findings refer to a sample of objects which is not the result of an unbiased search across the Galactic plane, because our observations were performed towards a limited number of targets selected mostly on the basis of their IRAS colours. However, the adopted selection criteria on the IRAS colours (see Sect.~\ref{s:sample}) are expected to identify most Galactic \hii\ regions (and their precursors). It is also worth noting that the number of \hii\ regions estimated by us ($\NHII$=15000) is significantly larger than that obtained by other authors. For instance, Wood \& Churchwell (1989) estimate $\sim$1600 embedded O-type stars in the Galaxy, while from the results of Mottram \et\ (2011) one may obtain $\sim$8200 Galactic massive young stellar objects and compact \hii\ regions. However, the former estimate refers only to stars above $\sim$30000~\lo\ and the latter does not consider extended \hii\ regions, which can be detected with our observations.

\subsection{Background source contamination\label{s:background}}

Most of the sources in our sample are located in the fourth galactic quadrant, covering the range 254\degr<$l$<360\degr, with a few at $l$$\approx$10\degr. Regarding the galactic latitudes, our sources are confined to the galactic disk with $|b|$<3\degr, with most of them (85\%) restricted to $|b|$<1\degr. Note that most of these sources are included in the range of galactic coordinates covered by recent infrared and submillimeter unbiased surveys (\eg\ GLIMPSE and MIPSGAL with \emph{Spitzer}; Hi-GAL with \emph{Herschel}; ATLASGAL with APEX) that can be used to construct the spectral energy distributions from infrared to millimeter wavelengths. This will be the subject of a forthcoming paper (S\'anchez-Monge \et, in prep). The location of our sources within the galactic disk is consistent with them being associated with star-forming regions, with the centimeter continuum emission coming from \hii\ regions (as shown in the previous section). However, we cannot exclude the possibility that some of the radio continuum sources are in fact extragalactic sources. Following the Appendix of Anglada \et\ (1998), and considering our ATCA observations at a frequency of 20.4~GHz, we can estimate the number of expected background sources as
\begin{displaymath}
<N>\,=\,0.041
\bigg[1-\exp^{-0.109~\theta_\mathrm{FOV}^2}\bigg]
\bigg(\frac{S_\mathrm{0}}{\mathrm{mJy}}\bigg)^{-0.75},
\end{displaymath}
with $\theta_\mathrm{FOV}$ the field of view in arcmin, and $S_\mathrm{0}$ the detectable flux density threshold at the center of the field that we will consider to be 5$\sigma$ ($\sim$2~mJy). For a field of view of 2$\farcm$5, and considering that we have 192 distinct fields, we expect to detect $<\!N\!>\simeq3$ background sources. Since we have detected a total of 214 centimeter continuum sources, only $\sim$2\% of them could be in fact extragalactic sources.

There are 8 sources with large negative spectral indices (<$-5$) that could be extragalactic non-thermal sources: 10184$-$5748~0, 10317$-$5936~0, 13106$-$6050~0, 13592$-$6153~0, 14214$-$6017~0, 15015$-$5720~0, 16363$-$4645~0, 17242$-$3513 (see Table~\ref{t:physpar}). However, visual inspection of the radio continuum maps reveals that these sources are associated with strong millimeter condensations or infrared sources, and some of them are coincident with or close to H$_2$O masers, suggesting a galactic origin. Also, the spectral indices are too negative even for synchrotron emission from jets found in star-forming regions (\eg\ Reid \et\ 1995, Carrasco-Gonz\'alez \et\ 2010). Therefore, such spectral indices are probably the result of filtering out more emission at 22.8~GHz than at 18.0~GHz. Thus, these regions are probably extended and more evolved \hii\ regions, rather than extragalactic sources. In fact, it is interesting to note that the 8 sources with large negative spectral indices correspond to sources located in regions with multiple centimeter sources, suggesting that confusion in combination with the poor \emph{uv}-coverage of our snapshot observations could complicate the estimate of the flux, eventually resulting in unreliable spectral indices.

\begin{figure*}[t!]
\begin{center}
\begin{tabular}[b]{c c c c c c c}
 \epsfig{file=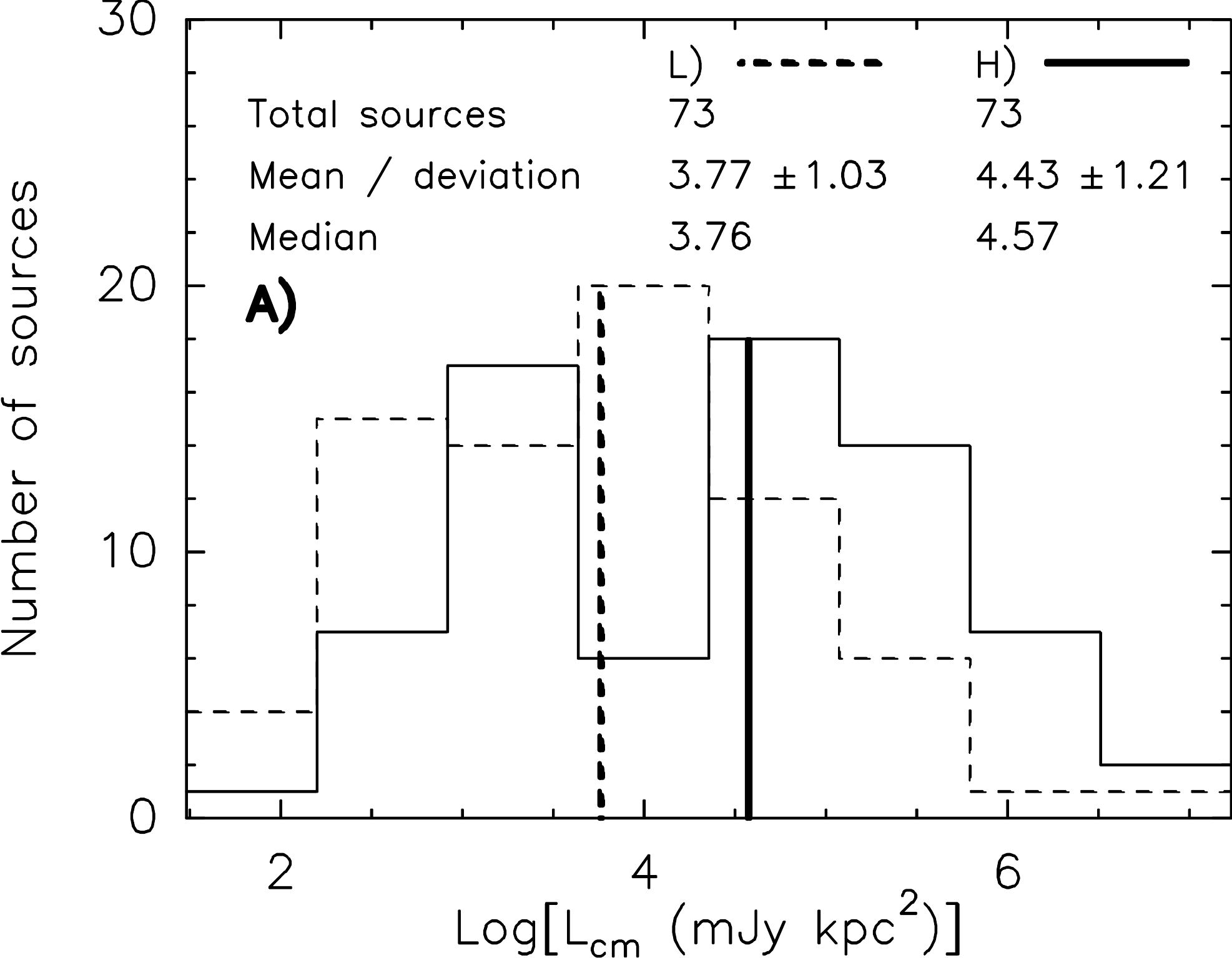, scale=0.27, angle=0} &&
 \epsfig{file=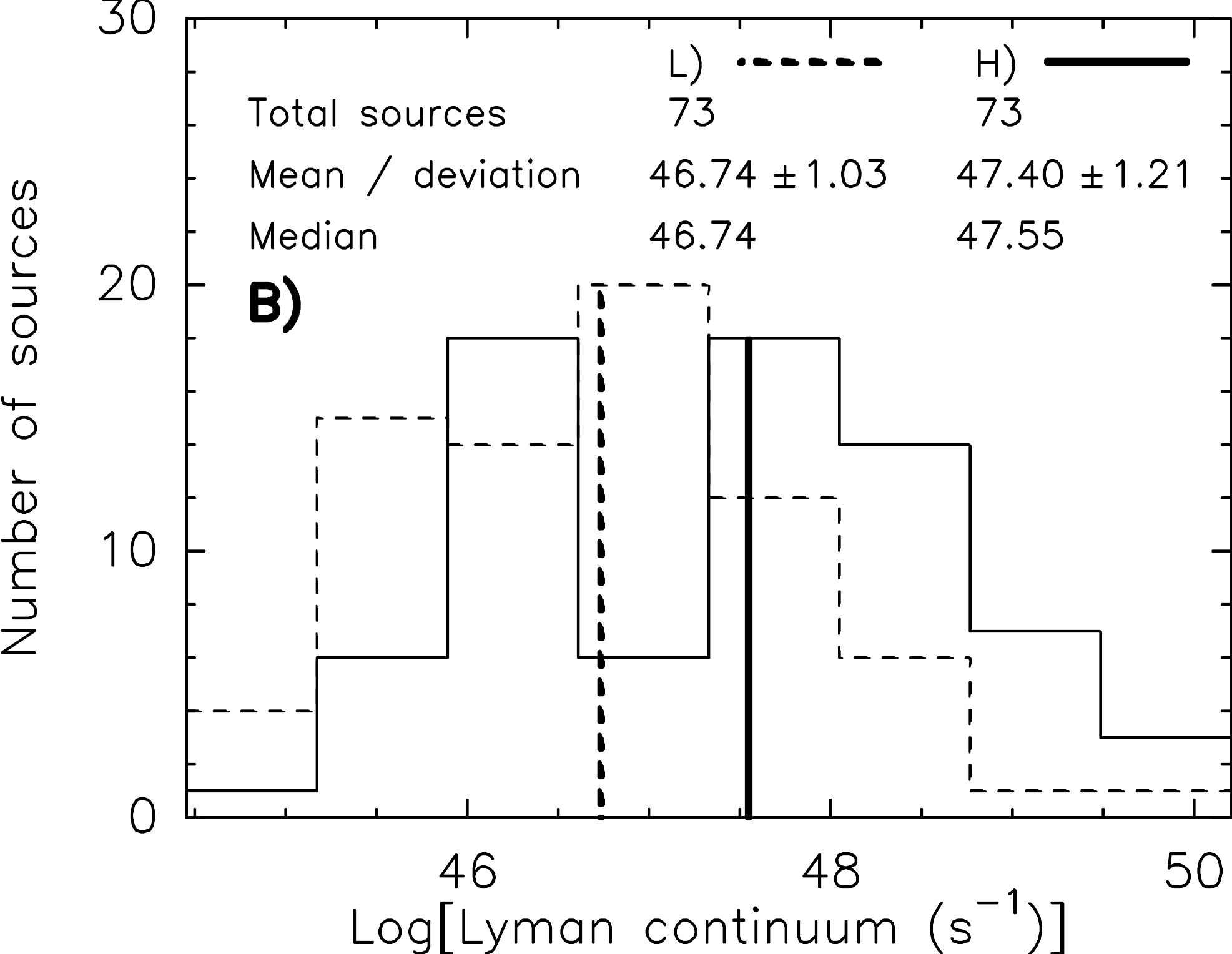, scale=0.27, angle=0} &&
 \epsfig{file=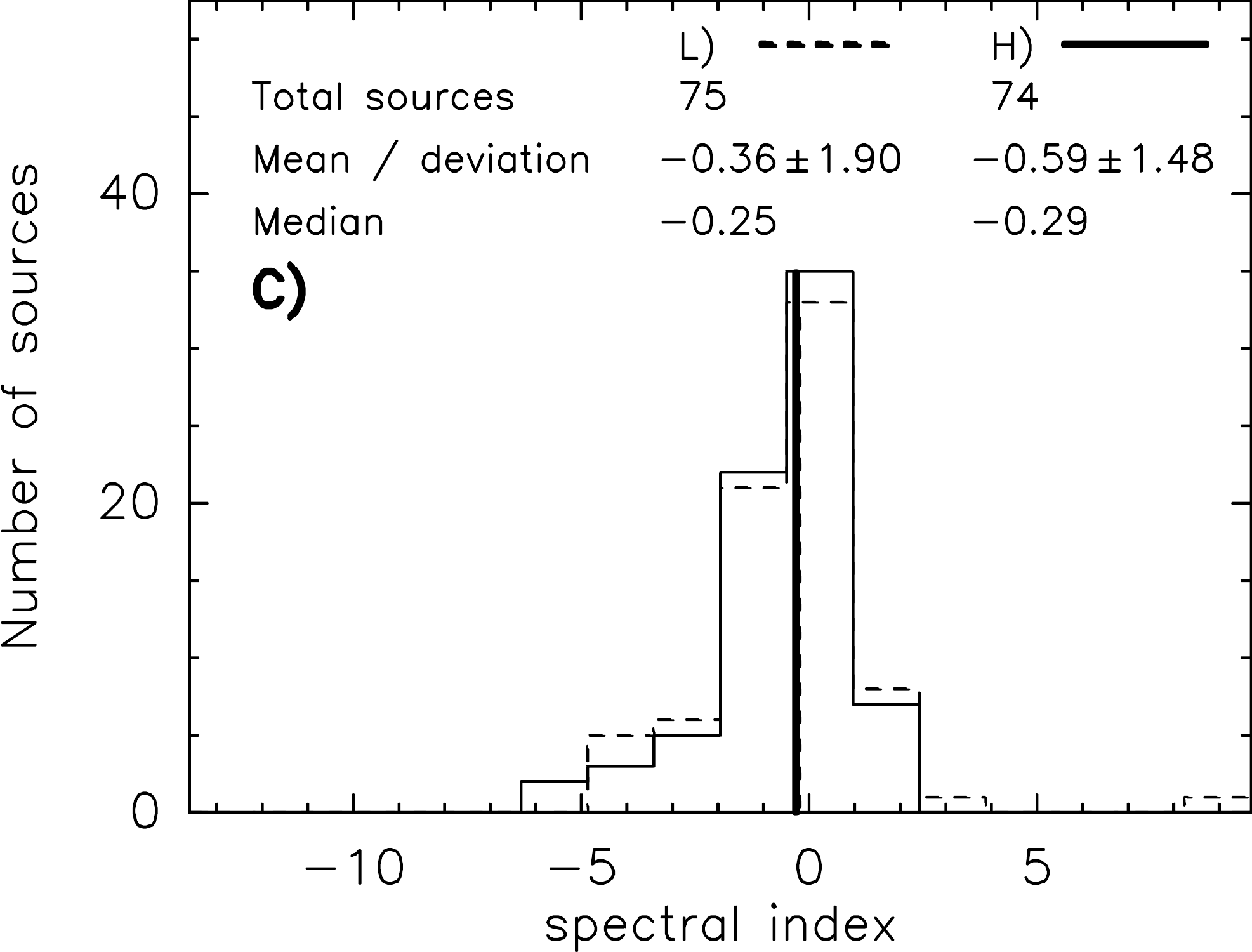, scale=0.27, angle=0} \\
 \noalign{\bigskip}
 \epsfig{file=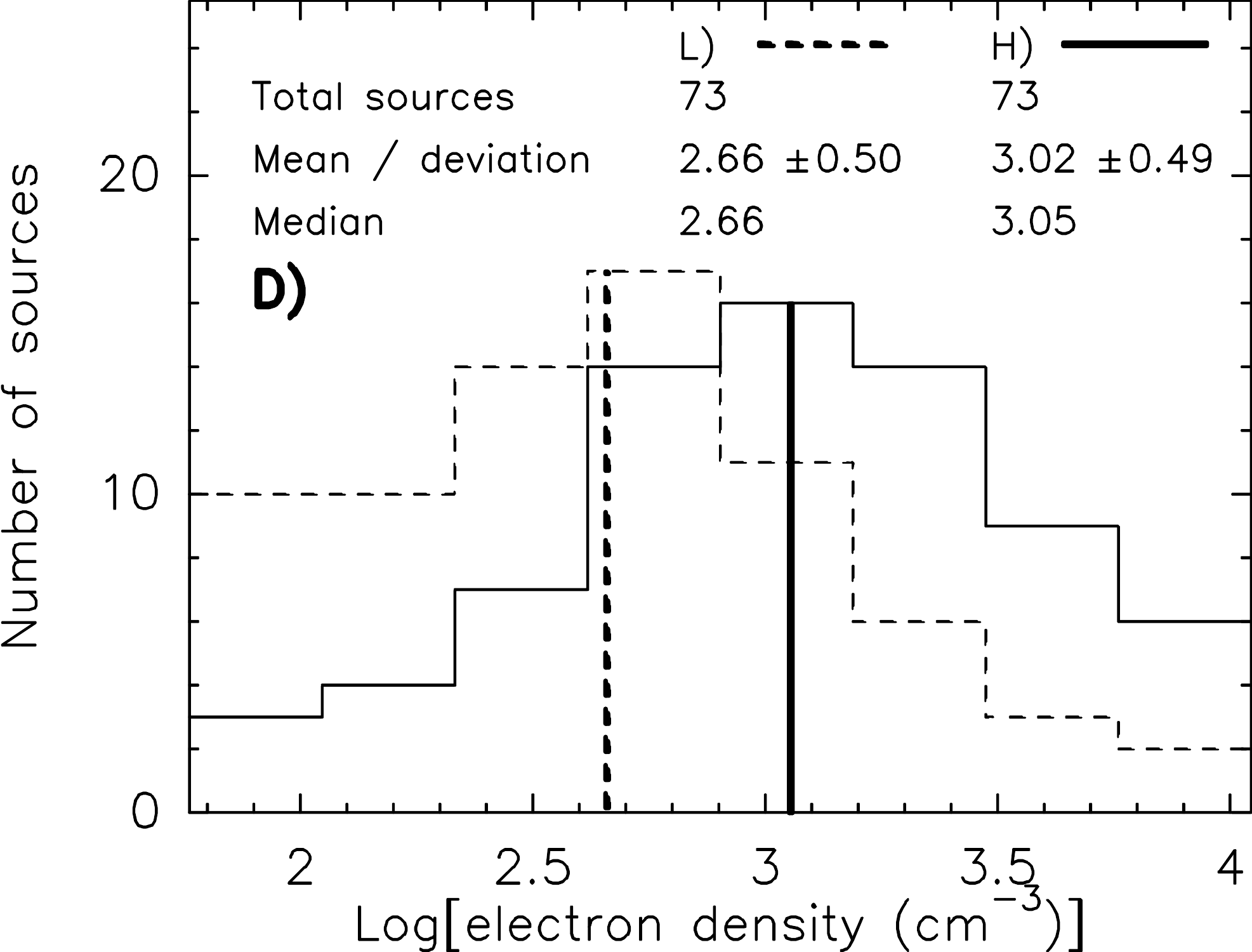, scale=0.27, angle=0} &&
 \epsfig{file=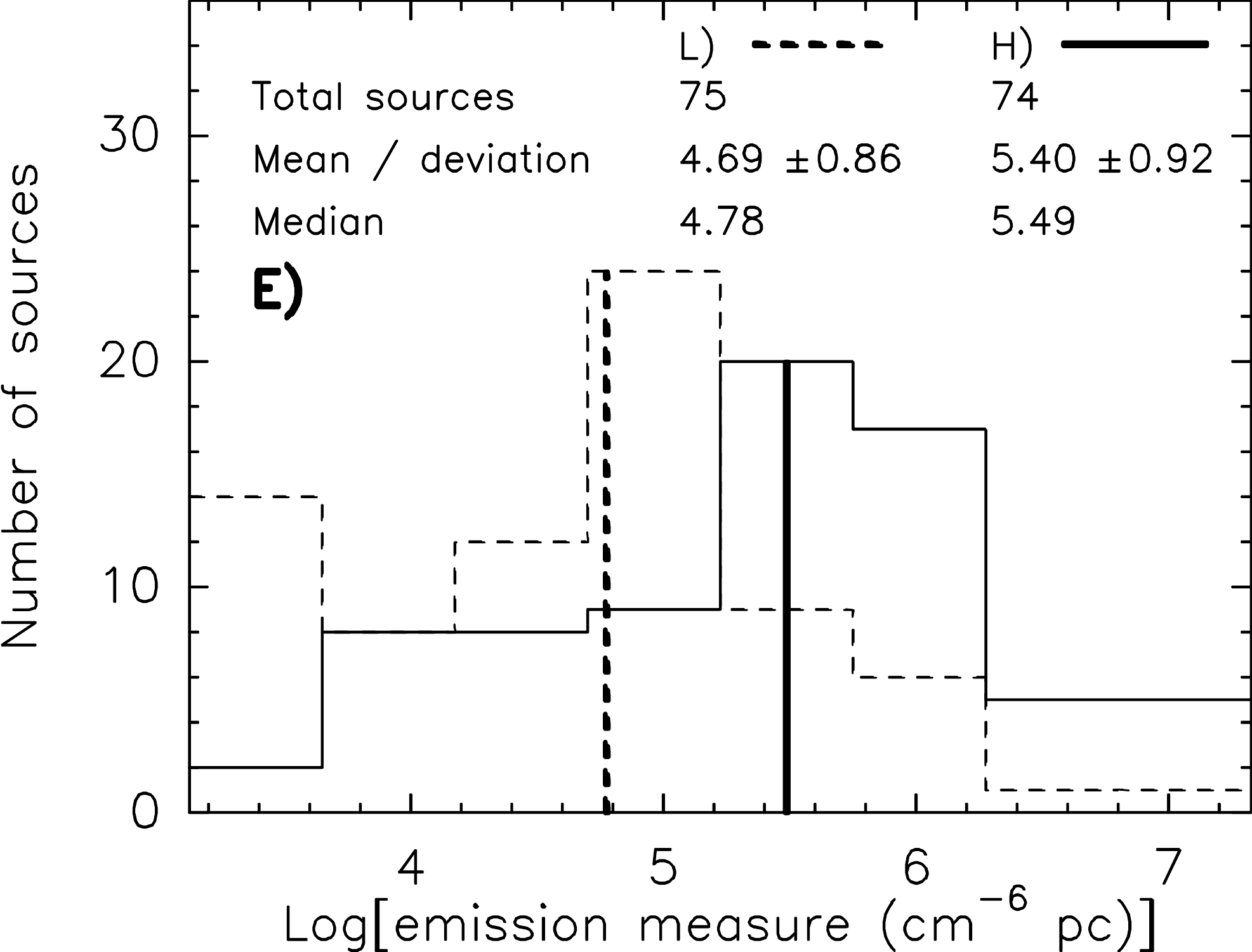, scale=0.27, angle=0} &&
 \epsfig{file=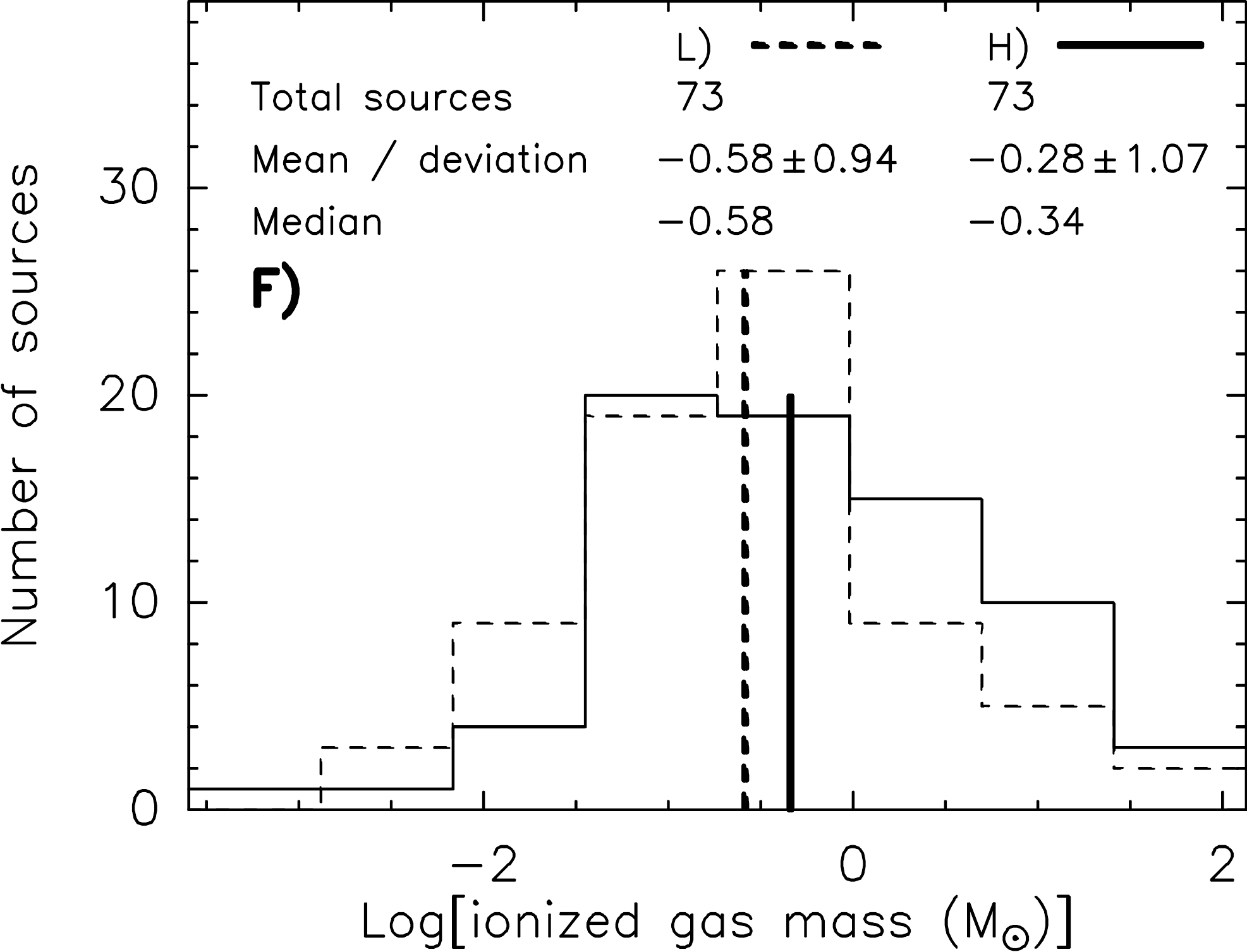, scale=0.27, angle=0} \\
 \noalign{\bigskip}
 \epsfig{file=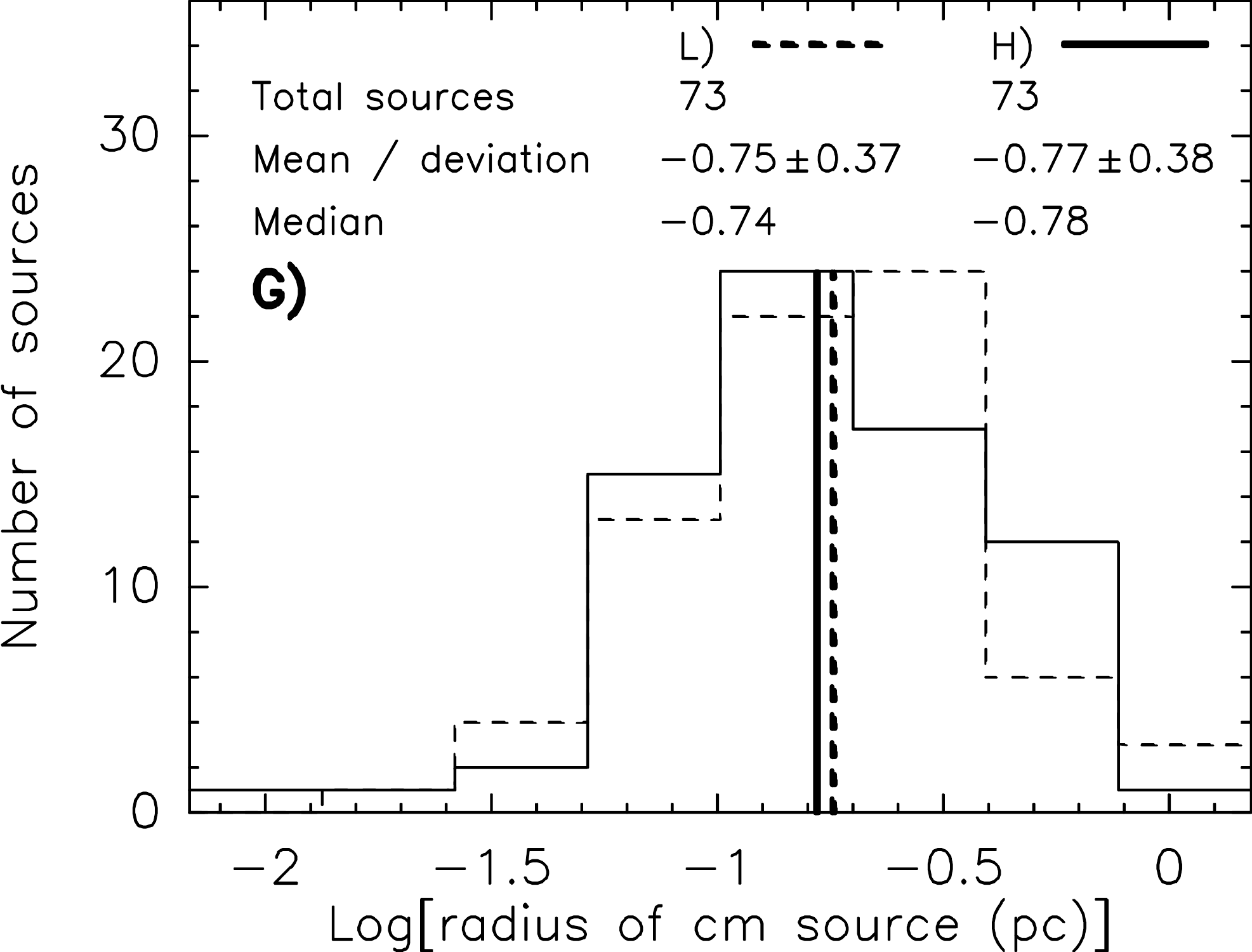, scale=0.27, angle=0} &&
 \epsfig{file=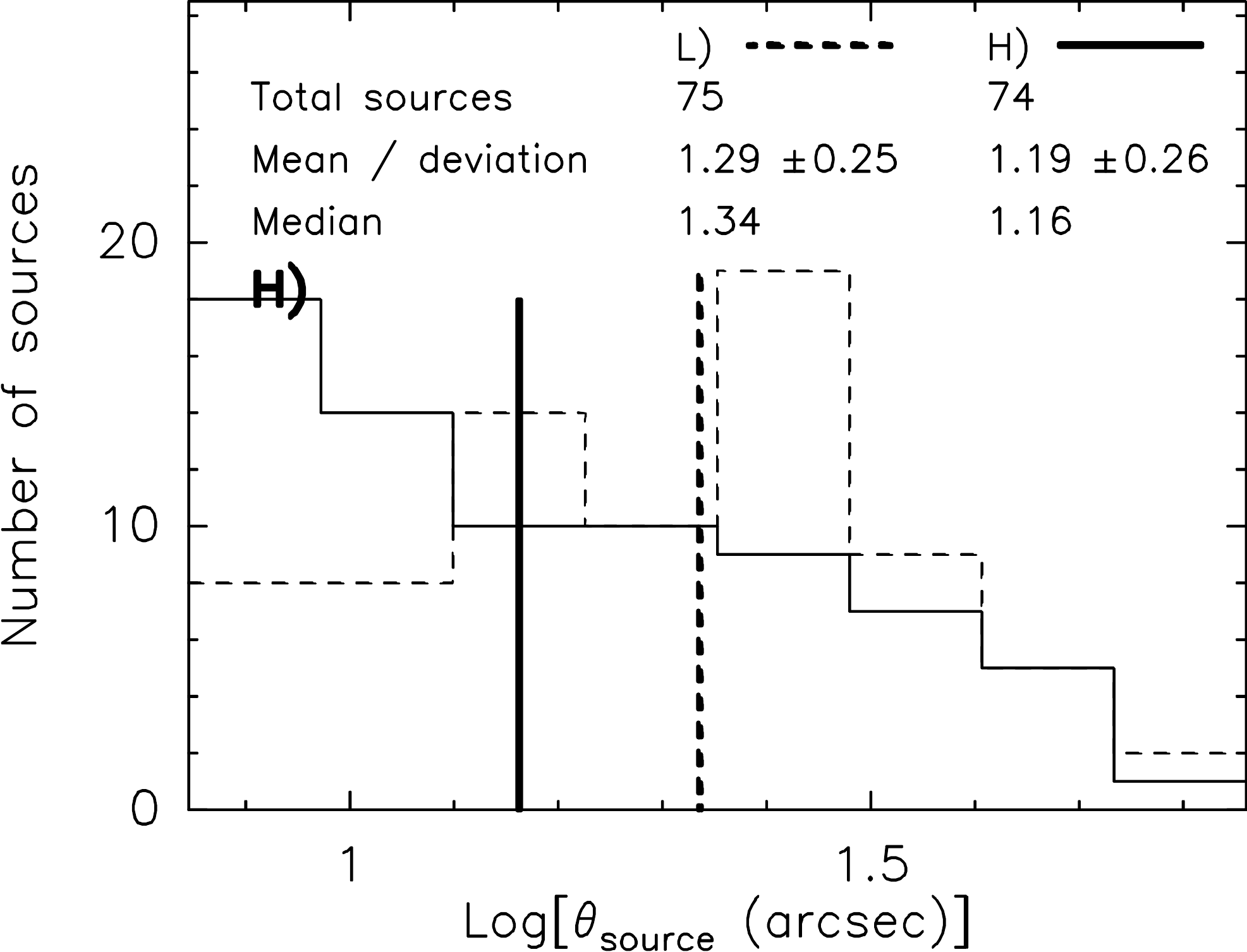, scale=0.27, angle=0} &&
 \epsfig{file=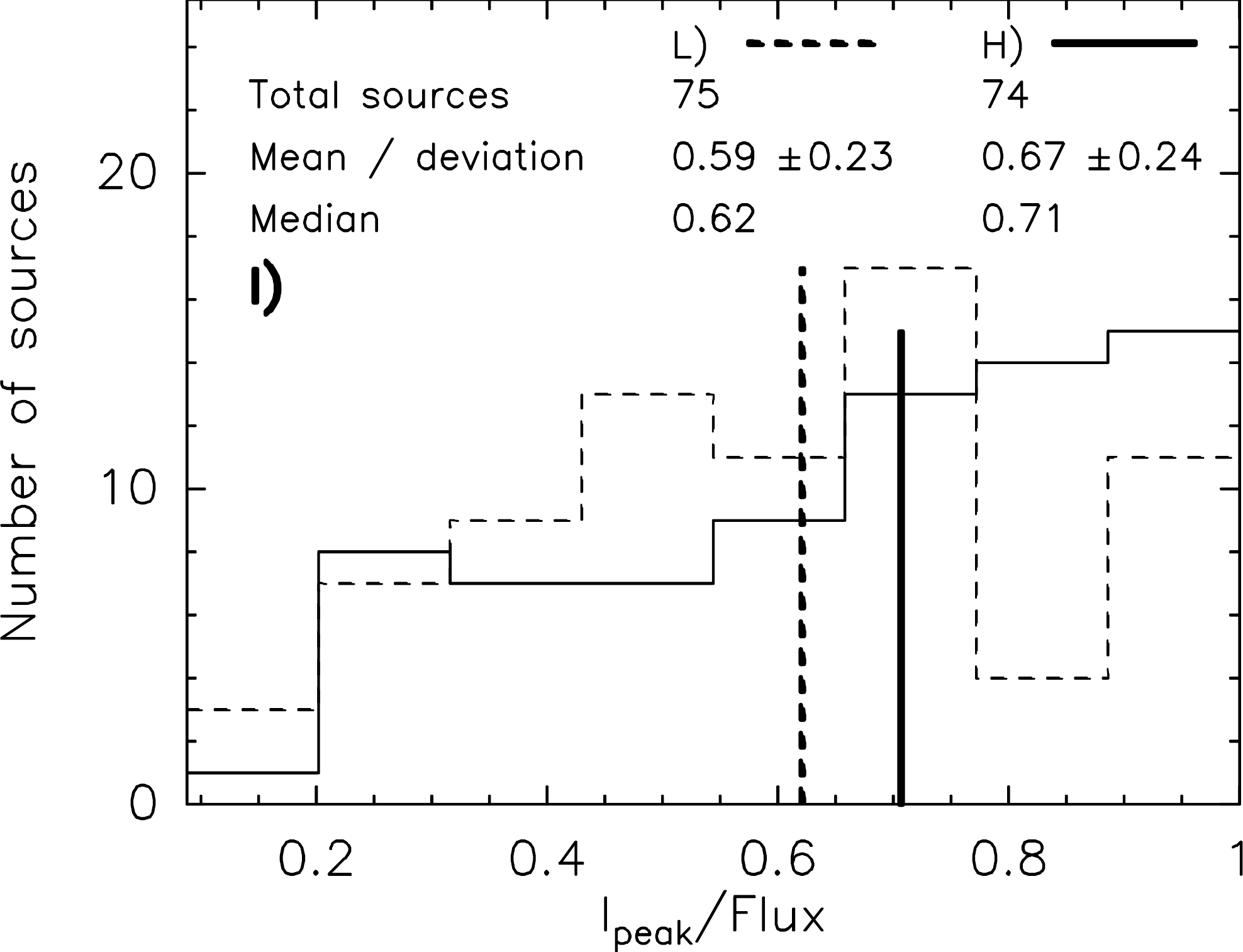, scale=0.27, angle=0} \\
\end{tabular}
\caption{Distributions of {\bf a)} centimeter luminosity; {\bf b)} Lyman continuum; {\bf c)} spectral index; {\bf d)} electron density; {\bf e)} emission measure; {\bf f)} ionized gas mass; {\bf g)} linear radius of the centimeter source; {\bf h)} deconvolved angular diameter, $\theta_{\rm S}$; and {\bf i)} peak intensity to flux density ratio, for the centimeter sources associated with \emph{Low} (dashed line) and \emph{High} (solid line) regions. The numbers at the top of each panel are as in Fig.~\ref{f:cmHisto}. The vertical thick lines indicate the median values.}
\label{f:HLHisto}
\end{center}
\end{figure*}

\subsection{Low versus High: centimeter continuum properties\label{s:analysisHL}}

The two groups (\emph{High} and \emph{Low}) defined by Palla \et\ (1991) to classify the IRAS sources associated with massive star-forming regions, have been proposed to include sources in different evolutionary stages, with the \emph{High} sources being more evolved and more likely associated with \uchii\ regions. Several works have studied the molecular and dust content in these two groups of sources. Brand \et\ (2001) compared the molecular properties of two samples containing \emph{Low} and \emph{High} sources, and reported that molecular clumps associated with \emph{Low} sources have larger sizes, are less massive, cooler and less turbulent than \emph{High} sources. More recently, Fontani \et\ (2005) observed different transitions of CS and C$^{17}$O tracing the dense gas associated with a larger (130) sample of \emph{Low} sources. They found that the temperatures and linewidths estimated for the \emph{Low} sources were similar to those found toward different samples of \emph{High} sources (Bronfman \et\ 1989; Sridharan \et\ 2002; Beuther \et\ 2002), with only slightly narrower lines for the \emph{Low} sources, suggesting that the gas could be more perturbed in \emph{High} sources, due to the presence of \hii\ regions. Beltr\'an \et\ (2006) studied the dust properties by observing the 1.2~mm continuum emission toward a large (235) sample of \emph{Low} and \emph{High} sources. The sizes of the clumps, masses, surface densities and H$_2$ volume densities were similar between the two groups of sources, with only small variations found in the size and mass (slightly larger for \emph{High} sources). Thus, it seems that there are no clear differences in the molecular and dust properties between \emph{High} and \emph{Low} sources.  

In this section, we compare the properties of the centimeter emission (likely tracing \hii\ regions) associated with the two different groups of sources. In our sample --- 79 IRAS \emph{High} sources and 81 IRAS \emph{Low} sources --- the detection rate of centimeter continuum emission is 94\% and 93\% for \emph{High} and \emph{Low} sources, respectively. Thus, there are no differences between both groups regarding the presence of \hii\ regions. Is there any difference in the physical properties of the \hii\ regions associated with the \emph{High} and \emph{Low} sources? In Fig.~\ref{f:HLHisto}, we show the distribution of the main physical parameters of the centimeter continuum sources for both groups. The physical properties of the \hii\ regions in both groups seem to be similar, with values of the electron density, emission measure, flux density and Lyman continuum slightly higher for \emph{High} sources (see also the median values listed in the Columns~4 and 5 of Table~\ref{t:meancm}). Moreover, the plot of the Lyman continuum as a function of the bolometric luminosity (see Fig.~\ref{f:LcmLbol}) does not show large differences between the distribution of \emph{High} (solid symbols) and that of \emph{Low} (open symbols) sources. It is worth noting that our detection rates (>90\%) seem to contradict those obtained by Molinari \et\ (1998a): 43\% and 24\% for \emph{High} and \emph{Low} sources, respectively. However, their higher angular resolution observations ($\sim$10 times better) could be filtering out the extended \hii\ regions (still observable in our ATCA observations), which results in a decrease of their detection rates. It is worth noting that Molinari \et\ (1996) suggest that \emph{Low} sources could be contaminated by evolved (and more extended) \hii\ regions. Regarding the water maser association, Palla \et\ (1991) find that the detection rate for \emph{High} sources (26\%) is almost 3 times larger than that of \emph{Low} sources (9\%). In our sample, we obtain a water maser detection rate of 46\% and 33\% for the \emph{High} and \emph{Low} sources, respectively. A direct comparison of the detection rates between the two works is difficult since the observing conditions (sensitivity and spectral resolution) differ considerably between our observations and Palla \et\ (1991).

In summary, it seems that the presence of \hii\ regions or water masers, as well as the properties of the millimeter and dense gas emission, are similar for both groups, suggesting that there may exist more effective criteria to make an evolutionary classification rather than the infrared colors of \emph{Low} and \emph{High} sources. This will be discussed in the next section.

\begin{figure}[th!]
\begin{center}
\begin{tabular}[b]{c}
 \epsfig{file=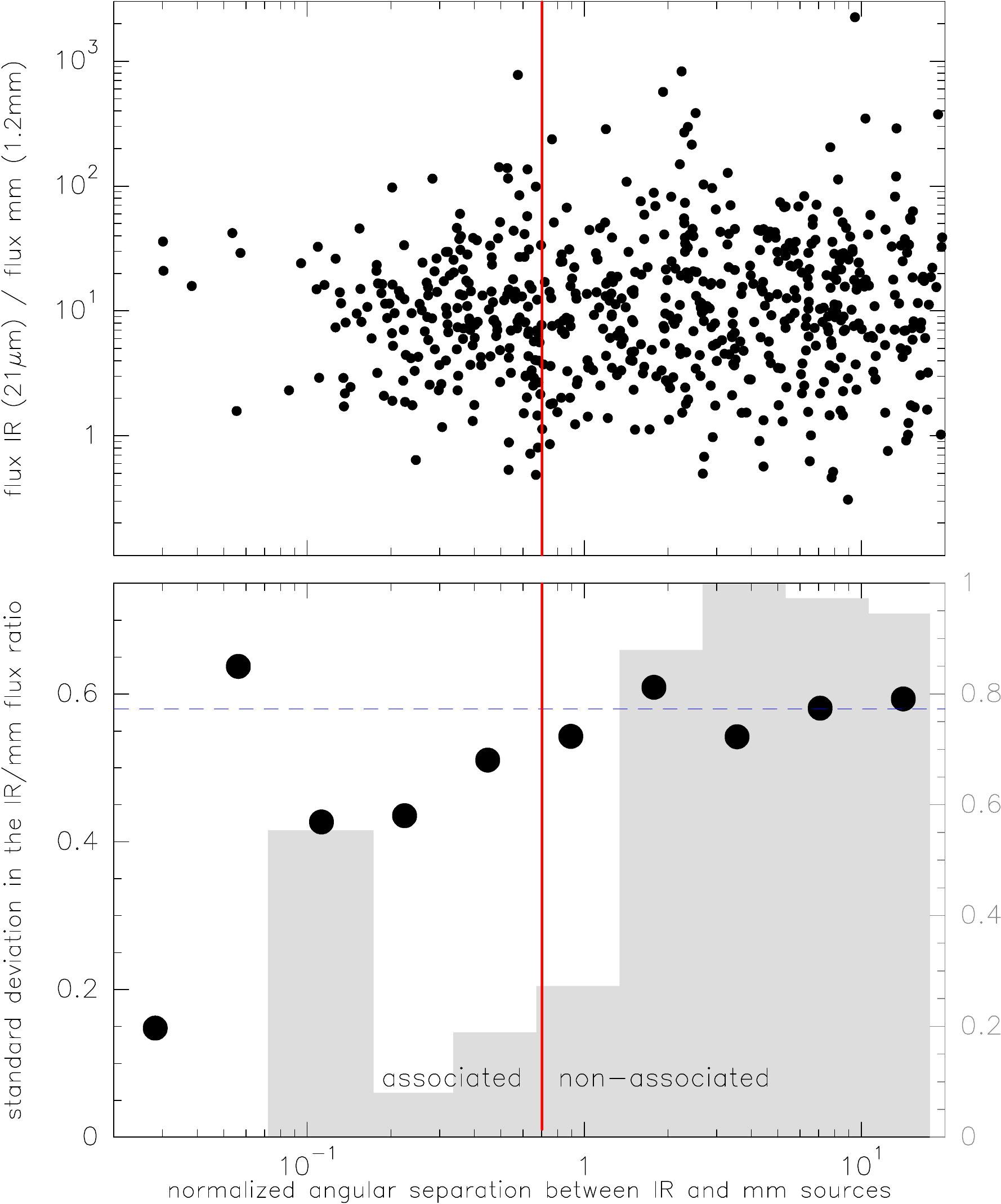, scale=0.49, angle=0} \\
\end{tabular}
\caption{{\bf Top}: 21~$\mu$m-to-1.2~mm flux ratio versus the normalized angular separation between millimeter and infrared sources (see Sect.~\ref{s:association} for details). The flux at 21~$\mu$m flux has been obtained from the MSX catalogue, and the flux at 1.2~mm has been obtained from Beltr\'an \et\ (2006). {\bf Bottom}: standard deviation in the 21~$\mu$m-to-1.2~mm flux ratio versus the normalized angular separation (black dots). The dashed horizontal blue line correspond to the standard deviation ($\simeq$0.57) for data with $\Delta_{\rm n}$>0.7. The grey histogram indicates the probability that the distribution of data within one bin is the same as the global distribution with $\Delta_{\rm n}$>1, using the Kolmogorov-Smirnov statistical test. The red vertical line indicates the normalized angular separation that differentiates between millimeter and infrared associated and non-associated sources (see Section~\ref{s:association}).}
\label{f:IRmmSeps}
\end{center}
\end{figure}
\begin{figure}[th!]
\begin{center}
\begin{tabular}[b]{c}
 \epsfig{file=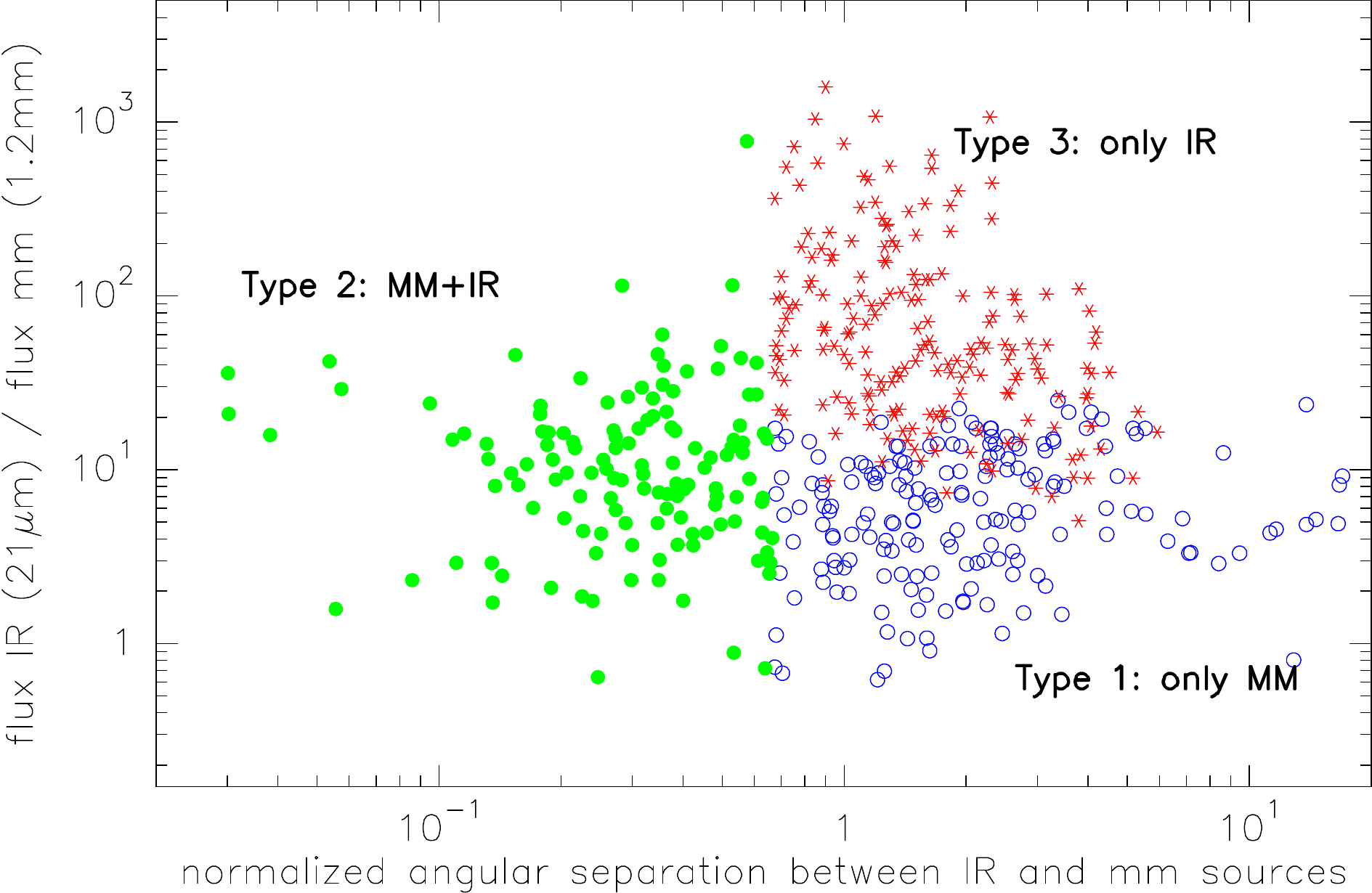, scale=0.44, angle=0} \\
\end{tabular}
\caption{21~$\mu$m-to-1.2~mm flux ratio versus the normalized angular separation between millimeter and infrared sources. Blue open circles correspond to \emph{type~1} sources (only millimeter emission), green filled circles correspond to \emph{type~2} sources (association between millimeter and infrared emission); and red stars correspond to \emph{type~3} sources (only infrared emission). Blue symbols (millimeter-only sources) are upper limits in which we have considered an upper limit of 4 times the rms noise in the MSX image. Red symbols (infrared-only sources) are lower limits in which we have considered an upper limit of 4 times the rms noise in the SEST image.}
\label{f:groups}
\end{center}
\end{figure}
\begin{figure}[th!]
\begin{center}
\begin{tabular}[b]{c}
 \epsfig{file=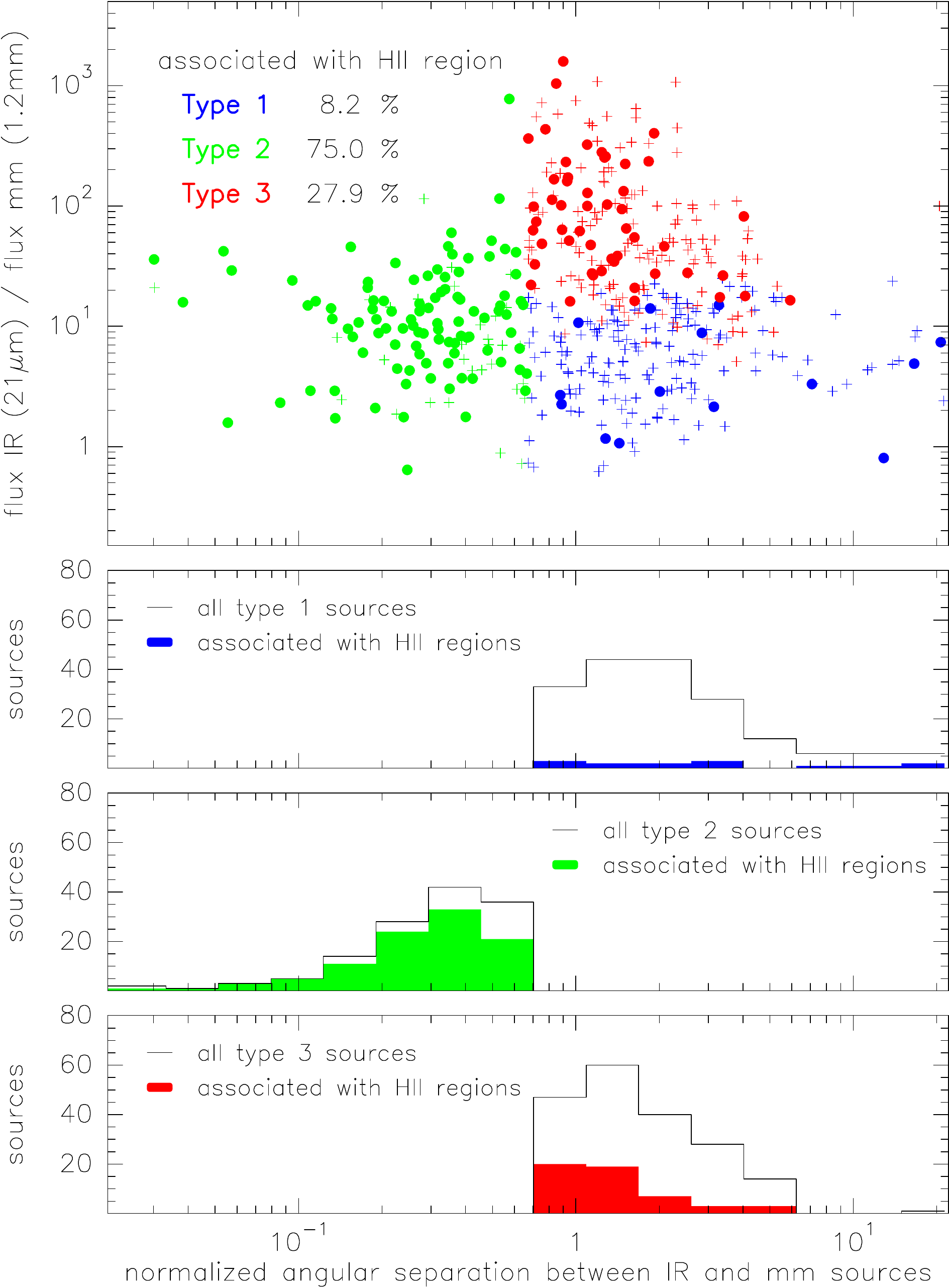, scale=0.44, angle=0} \\
\end{tabular}
\caption{Associations with \hii\ regions. {\bf Top}: 21~$\mu$m-to-1.2~mm flux ratio versus the normalized angular separation between millimeter and infrared sources as in Fig.~\ref{f:groups}. Blue symbols correspond to \emph{type~1} sources (only millimeter emission), green symbols correspond to \emph{type~2} sources (association between millimeter and infrared emission); and red symbols correspond to \emph{type~3} sources (only infrared emission). Colored filled dots show those objects associated with \hii\ regions detected with ATCA (see Table~\ref{t:results}). The numbers at the top of the panel indicate the percentage of association of \hii\ regions in each of three groups. {\bf Bottom}: Histograms of the normalized angular separation for the different groups. The solid black lines correspond to all the sources of each group, and the colored filled histograms correspond those objects associated with \hii\ regions.}
\label{f:cmassociations}
\end{center}
\end{figure}
\begin{figure}[th!]
\begin{center}
\begin{tabular}[b]{c}
 \epsfig{file=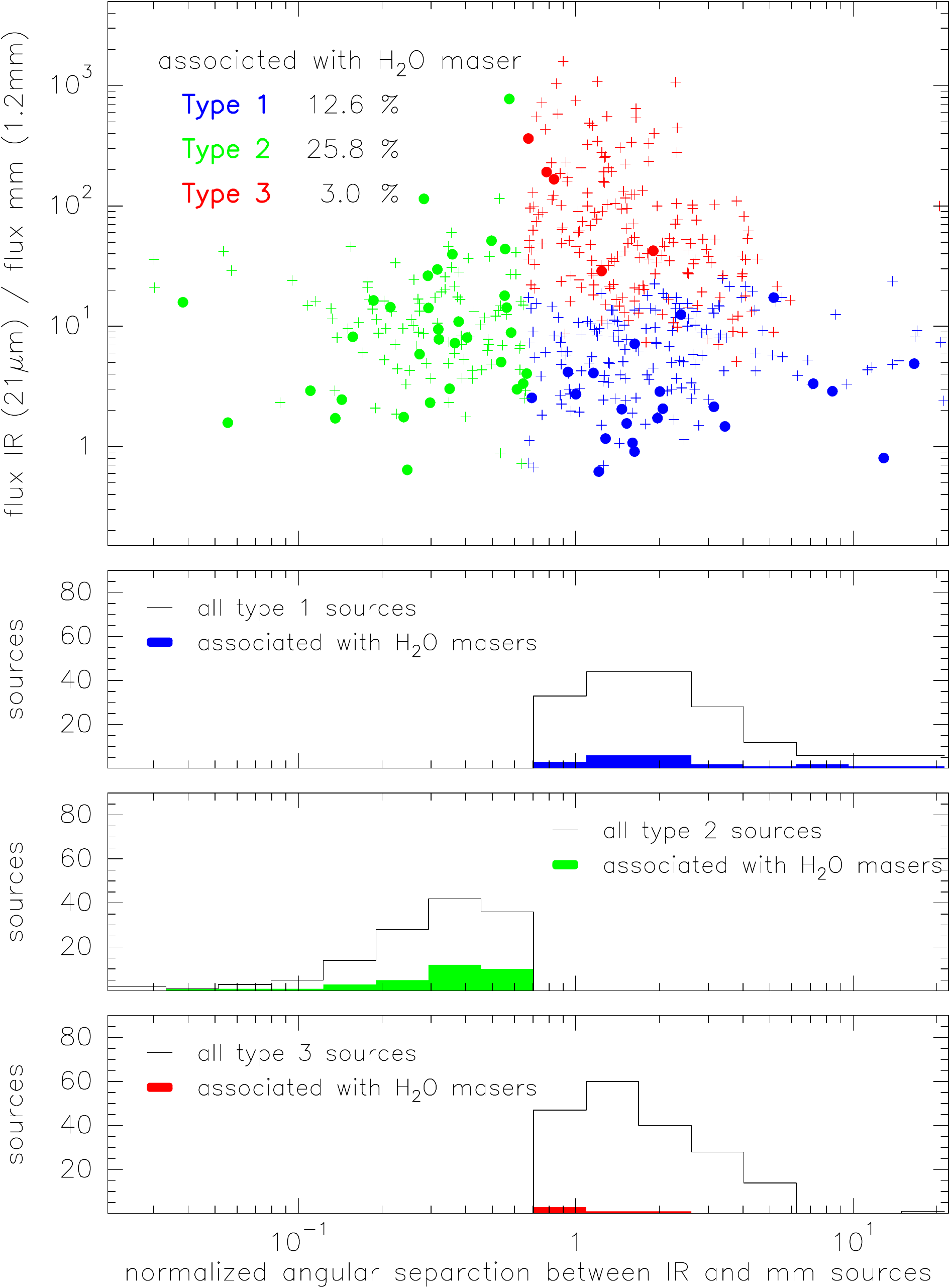, scale=0.44, angle=0} \\
\end{tabular}
\caption{Associations with H$_2$O masers. {\bf Top}: 21~$\mu$m-to-1.2~mm flux ratio versus the normalized angular separation between millimeter and infrared sources as in Fig.~\ref{f:cmassociations}. Colored filled dots show those objects associated with H$_2$O masers detected with ATCA (see Table~\ref{t:masers}). The numbers at the top of the panel indicate the percentage of association of H$_2$O masers in each of three groups. {\bf Bottom}: Histograms of the normalized angular separation for the different groups. The solid black lines correspond to all the sources of each group, and the colored filled histograms correspond those objects associated with H$_2$O masers.}
\label{f:h2oassociations}
\end{center}
\end{figure}

\section{Evolutionary sequence\label{s:discussion}}

\subsection{Millimeter and infrared counterparts\label{s:association}}

In this section, we discuss the association between millimeter and infrared sources, with the aim of establishing a classification of the objects based on their emission in the two wavelength regimes. For the millimeter sources, we used the sample of 210 regions, containing 667 millimeter clumps, studied by Beltr\'an \et\ (2006) with the SEST telescope at 1.2~mm. For the infrared sources, we used the MSX Point Source Catalogue (Price \et\ 1999), providing information between 8~$\mu$m and 21~$\mu$m (mid-IR) with an angular resolution (20\arcsec) similar to that of the SEST (24\arcsec) and ATCA (20--25\arcsec) maps . We considered only the MSX sources detected at 21~$\mu$m satisfying one of two additional criteria: (a) quality factor $\ge$3 in the E-band (21.3~$\mu$m), \ie\ well detected; (b) quality factor 1 or 2 in the E-band, and quality factor $\ge$3 in at least one of the other bands (A: 8.3~$\mu$m, C: 12.1~$\mu$m, D: 14.7~$\mu$m). We used the MSX data, instead of other more sensitive and recent surveys (\eg\ MIPSGAL at 24~$\mu$m) for four reasons: the existence of a point source catalogue for the MSX data, not yet available for other surveys; the saturation of many regions in our sample in the MIPSGAL images; the similar resolution with our millimeter and centimeter images; and consistency with the method used by Molinari \et\ (2008; see next sections). 

In order to determine the association between millimeter and infrared sources, we performed the following analysis. We calculated the angular separation between millimeter and infrared sources, and searched for the closest companion of each millimeter source. In the top panel of Fig.~\ref{f:IRmmSeps}, we plot the 21~$\mu$m-to-1.2~mm flux ratio (hereafter $S_{21\mu{\rm m}}/S_{1.2{\rm mm}}$) versus the angular separation for each pair of millimeter and infrared sources (hereafter $\Delta_{\rm n}$). Note that we use a normalized angular separation (to get rid of the dependence on distance), computed by dividing the angular separation by the sum of the angular radii of the millimeter and infrared sources. The angular radius is equal to half the angular diameter reported in Beltr\'an \et\ (2006) for the millimeter sources, and is assumed to be half the 20\arcsec\ HPBW for the MSX sources. Our expectation is that $S_{21\mu{\rm m}}/S_{1.2{\rm mm}}$ could take any value if the millimeter and infrared fluxes come from two different (unrelated) objects, whereas it should span a narrow range of values if the two fluxes belong to the same object (as young stellar objects have roughly similar spectral energy distributions). We expect that the probability that the MSX source is the real counterpart of the mm source decreases with such a separation.  This expectation is confirmed by the fact that the flux ratio spans a larger range of values for increasing separation, thus confirming that the MSX source is really associated with the mm source only for sufficiently small $\Delta_{\rm n}$. This can be seen in the bottom panel of Fig.~\ref{f:IRmmSeps} (black dots), where the dispersion $\sigma_{\rm r}$ of $S_{21\mu{\rm m}}/S_{1.2{\rm mm}}$ appears to decrease below $\Delta_{\rm n}\simeq0.7$, marked by the vertical line in the figure. This normalized separation corresponds approximately to an angular separation between the millimeter and infrared sources of $\sim$15\arcsec\ (taking into account the typical sizes of the millimeter sources: $\sim$20--40\arcsec; Beltr\'an \et\ 2006). For a range of distances between 1 and 7~kpc, the resulting spatial separation is $\sim$0.07--0.5~pc.

In addition, we used the Kolmogorov-Smirnov statistical test to compare the $S_{21\mu{\rm m}}/S_{1.2{\rm mm}}$ distribution in different bins with the global distribution containing data with $\Delta_{\rm n}>1$. The result is shown by the grey histogram in the bottom panel of Fig.~\ref{f:IRmmSeps}, and demonstrates that the distributions for almost all bins with $\Delta_{\rm n}\lesssim1$ are statistically different with respect to the global distribution: the test shows that the probability of the distribution in each bin being the same of the global distribution becomes very low ($P\lesssim0.3$) for $\Delta_{\rm n}\lesssim1$. Furthermore, the comparison of the distributions with $\Delta_{\rm n}<1$ and $\Delta_{\rm n}>1$ results in a minimum probability of $P\simeq0.05$. The results of the Kolmogorov-Smirnov statistical tests, together with the decrease in the dispersion for $\Delta_{\rm n}\lesssim0.7$, are strong evidence of a different behavior between IR-mm associations with respect to the normalized angular separation. We thus conclude that only an IR counterpart satisfying the condition $\Delta_{\rm n}<0.7$ (as a conservative value from the range 0.7--1, corresponding to a spatial separation of 0.07--0.5~pc; see above) is physically associated with the corresponding mm source.

\subsection{Source classification}

The analysis presented in the previous section allows us to identify three types of YSOs: millimeter sources associated with infrared counterparts, millimeter sources not associated with an infrared source, and infrared sources lying relatively close to millimeter sources (and therefore likely associated with star forming regions) but not associated with them. Note that while the millimeter continuum emission probably comes from dust envelopes around YSOs, we cannot be sure that the same is true for the infrared emission, which could be produced by more evolved objects or background/foreground stars. With the aim of selecting only YSO candidates we applied the following restriction to our catalogue of infrared-only sources: $F_\mathrm{21~{\mu}m}$$\ge$$F_\mathrm{15~{\mu}m}$ and $F_\mathrm{21~{\mu}m}$$\ge$$F_\mathrm{12~{\mu}m}$ in the MSX bands, to ensure an increasing spectral energy distribution at infrared wavelengths typical of YSOs\footnote{These selection criteria based on the MSX fluxes are similar to those used for the Red MSX Source (RMS) survey (\eg\ Hoare \et\ 2004; Urquhart \et\ 2008): $F_\mathrm{21~{\mu}m}$$\ge$2.7~Jy, $F_\mathrm{21~{\mu}m}$$\ge$$2\,F_\mathrm{8~{\mu}m}$, and $F_\mathrm{21~{\mu}m}$$\ge$$F_\mathrm{15~{\mu}m}$$\ge$$F_\mathrm{8~{\mu}m}$}. In Fig.~\ref{f:groups} we plot the ratio $S_{21\mu{\rm m}}/S_{1.2{\rm mm}}$ versus the normalized angular separation, taking into account the previous restriction, and considering the sources within the ATCA fields. For those millimeter sources with no infrared counterpart we assumed a flux at 21~$\mu$m equal to an upper limit 4 times the rms noise level of the MSX map, and, similarly, for sources only detected at infrared wavelengths we assumed the flux at 1.2~mm to be an upper limit equal to 4 times the rms noise level of the millimeter map. The resulting plot shows the three different groups of sources: \emph{type~1} only millimeter continuum sources (blue circles), \emph{type~2} millimeter continuum sources associated with an infrared counterpart (green dots), and \emph{type~3} infrared sources located far from a millimeter source (red stars).

\begin{figure*}[t!]
\begin{center}
\begin{tabular}[b]{c c c}
 \epsfig{file=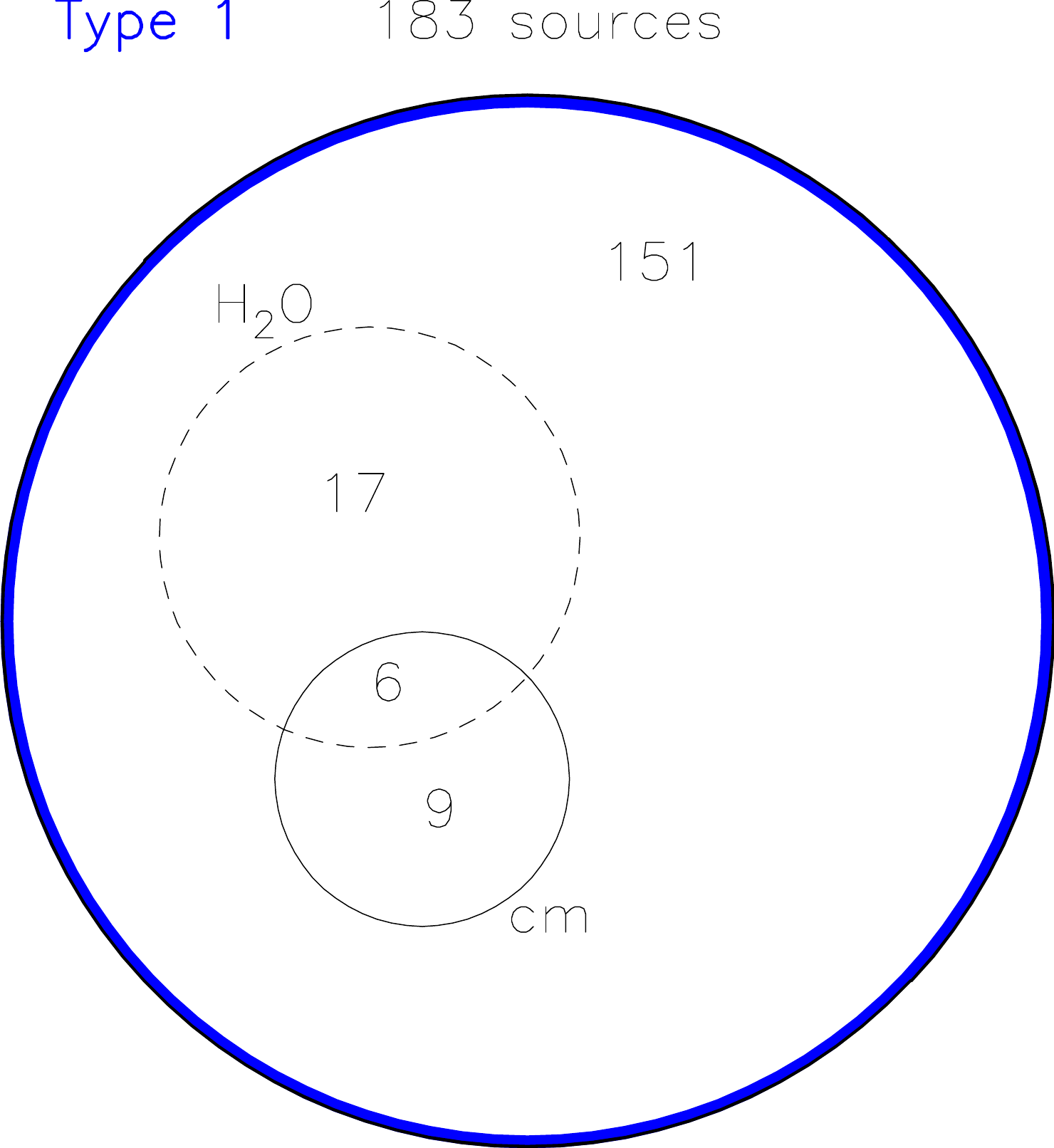, scale=0.37, angle=0} &
 \epsfig{file=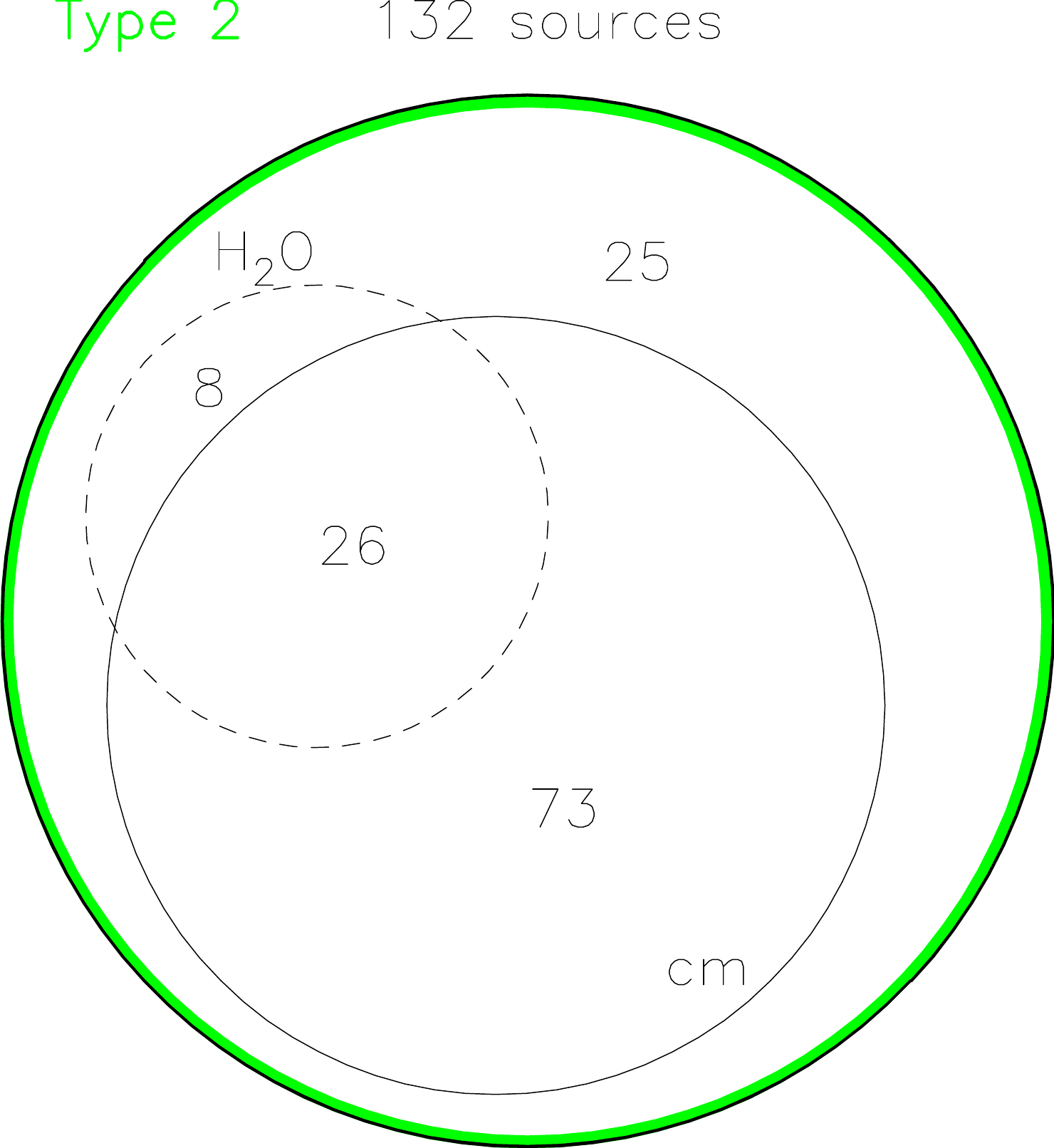, scale=0.37, angle=0} &
 \epsfig{file=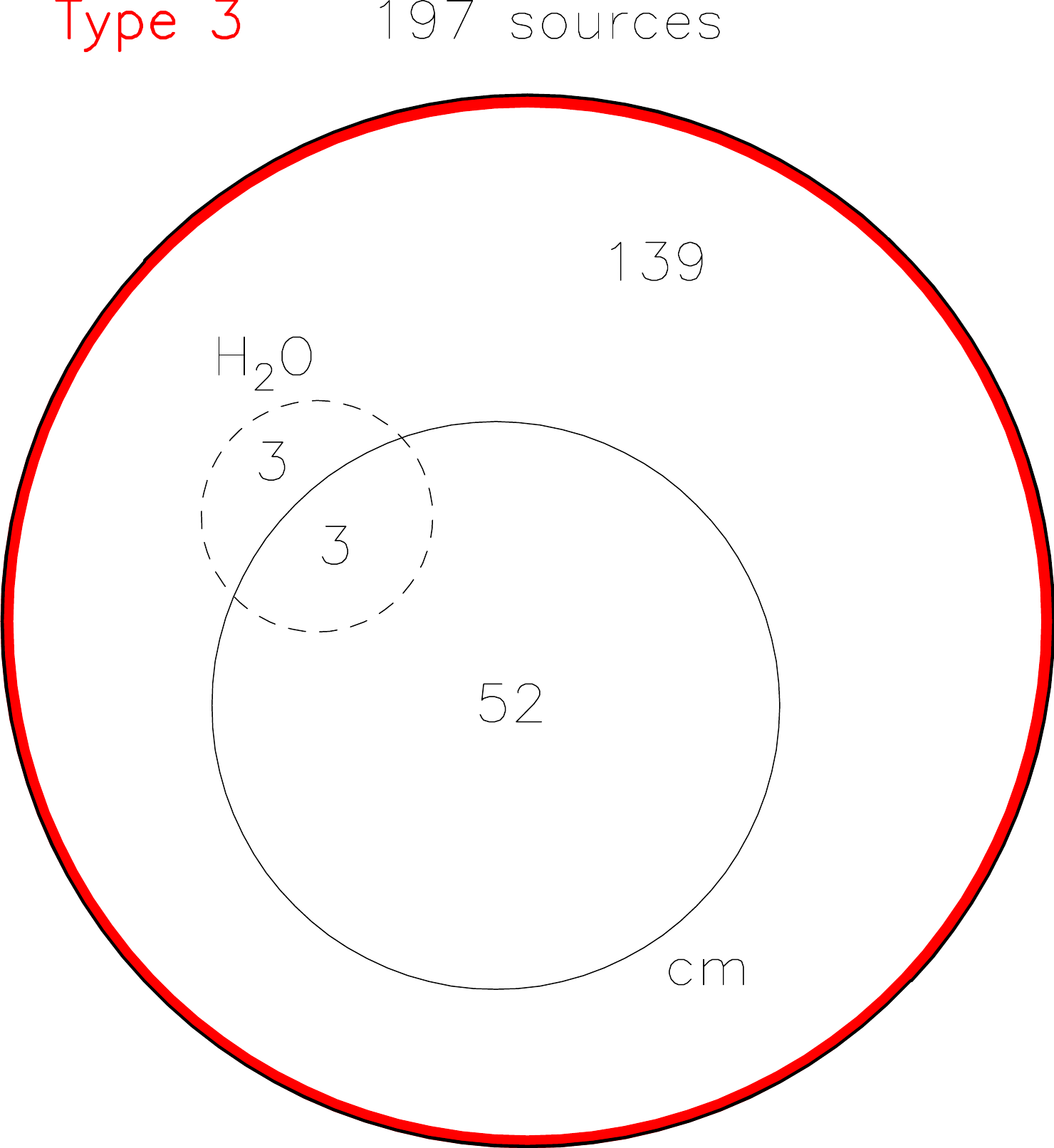, scale=0.37, angle=0} \\
\end{tabular}
\caption{Sketch showing the number of sources in each of the three types of sources. We indicate the number of sources associated with only \hii\ regions, only water maser emission, and both \hii\ regions and water masers.}
\label{f:sketch}
\end{center}
\end{figure*}

These three groups of sources correspond to the classification proposed by Molinari \et\ (2008) from a study of the infrared and millimeter properties of a sample of 42 regions. Molinari \et\ (2008) checked by eye the infrared and millimeter emission, and by fitting different spectral energy distributions, identified associations between infrared and millimeter sources. A simple model allowed the authors to establish an evolutionary sequence, in order of increasing age going from \emph{type~1} to \emph{type~3} objects. In the sample studied by Molinari \et\ (2008) there are 33 millimeter sources taken from Beltr\'an \et\ (2006) and thus included in our analysis. Taking into account our method for determining associations between infrared and millimeter sources, and comparing it with the results of Molinari \et\ (2008), we conclude that $\sim$80\% of the 33 sources agree with respect to the source classification in type 1, 2 or 3. Thus, our more automatic classification method, used for the classification in different evolutionary stages of our sources, agrees with the Molinari \et\ method, and will be used for further analysis. In Tables~\ref{t:type1} to \ref{t:type3}, we list the millimeter and/or infrared sources classified in the three evolutionary stages.

\addtocounter{table}{1}  
\addtocounter{table}{1}  
\addtocounter{table}{1}  

\begin{table}[t!]
\caption{\label{t:sketch}Number (and percentage) of sources associated with \hii\ regions and water masers for types 1, 2 and 3.}
\centering
\begin{tabular}{l | r l | r l | r l |}
\hline\hline

&\multicolumn{2}{c}{\emph{type~1}}
&\multicolumn{2}{c}{\emph{type~2}}
&\multicolumn{2}{c}{\emph{type~3}}
\\
\hline
\hii\ region			&\phn15	&\phnn8\%	&\phn99	&\phn75\%	&\phn55	&\phn28\%	\\
H$_2$O maser			&\phn23	&\phn13\%	&\phn34	&\phn26\%	&\phnn6	&\phnn3\%	\\
all					&183		&100\%		&132		&100\%		&197		&100\%		\\
\hline
\end{tabular}
\end{table}

At this point, we can compare the evolutionary classification derived from the IR--mm analysis with the \emph{High}--\emph{Low} classification obtained from the IRAS colors. If we consider the millimeter (SEST) and infrared (MSX) sources located within an IRAS beam ($\sim$2\arcmin), we find that an IRAS source with \emph{Low} colors contains 42\%, 27\% and 31\% of \emph{type~1}, \emph{2} and \emph{3} sources, respectively; while within the beam of a \emph{High} IRAS source we can find 23\%, 37\% and 40\% of \emph{type~1}, \emph{2} and \emph{3} sources, respectively. This result agrees with the analysis done by Molinari \et\ (2008) in a much smaller sample, and suggests that less evolved sources (\emph{type~1}) are typically found closer to a \emph{Low} IRAS source than more evolved sources (\emph{type~2} and \emph{3}). However, evolved objects can also be found associated with \emph{Low} IRAS sources, as shown by distinct studies (\eg\ Molinari \et\ 1998a, 2000), and as expected by the large detection rate of \hii\ regions found associated with \emph{Low} sources (see Section~\ref{s:analysisHL}).

\begin{figure}[t!]
\begin{center}
\begin{tabular}[b]{c}
 \epsfig{file=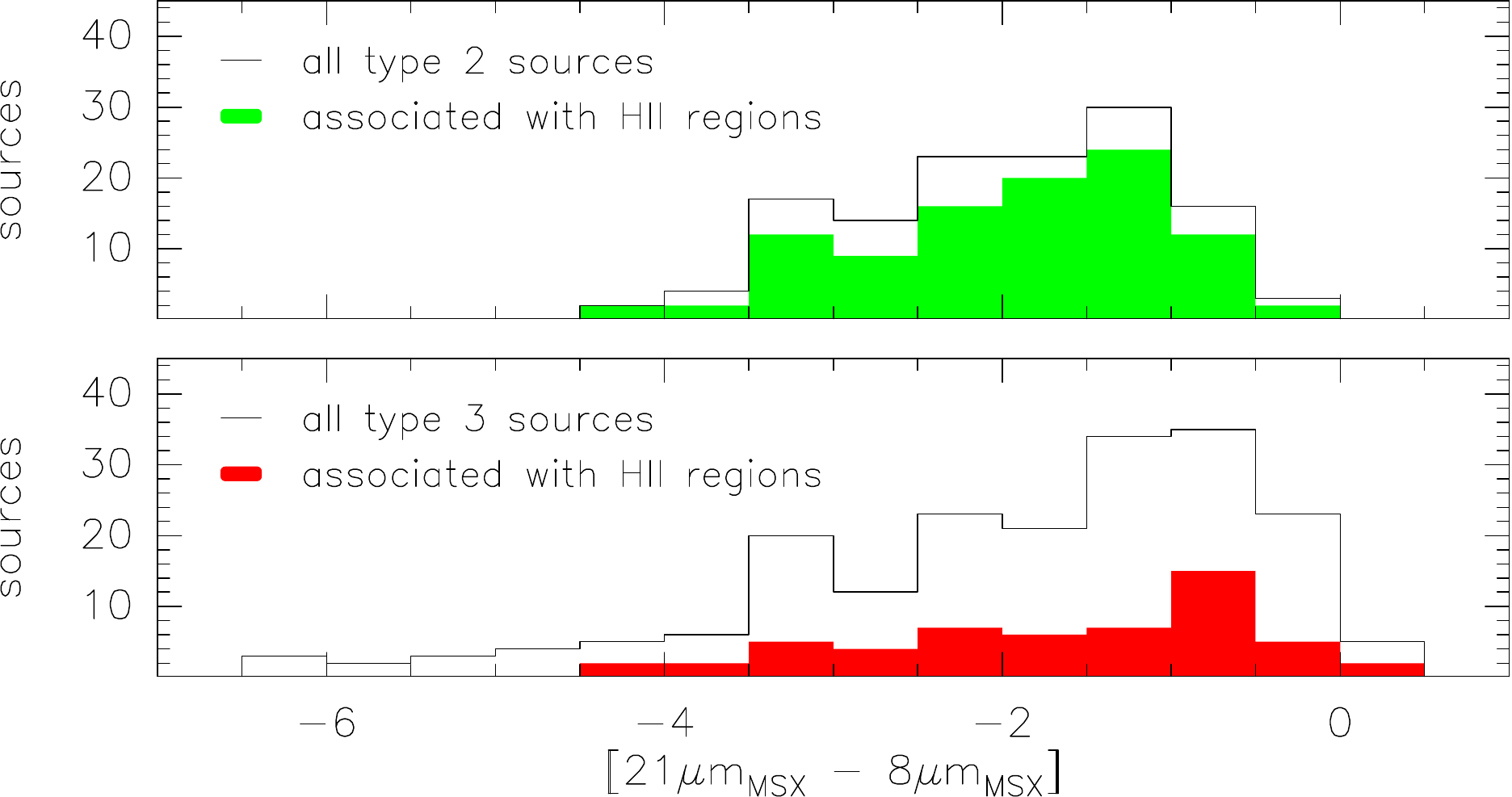, scale=0.43, angle=0} \\
 \noalign{\bigskip}
 \epsfig{file=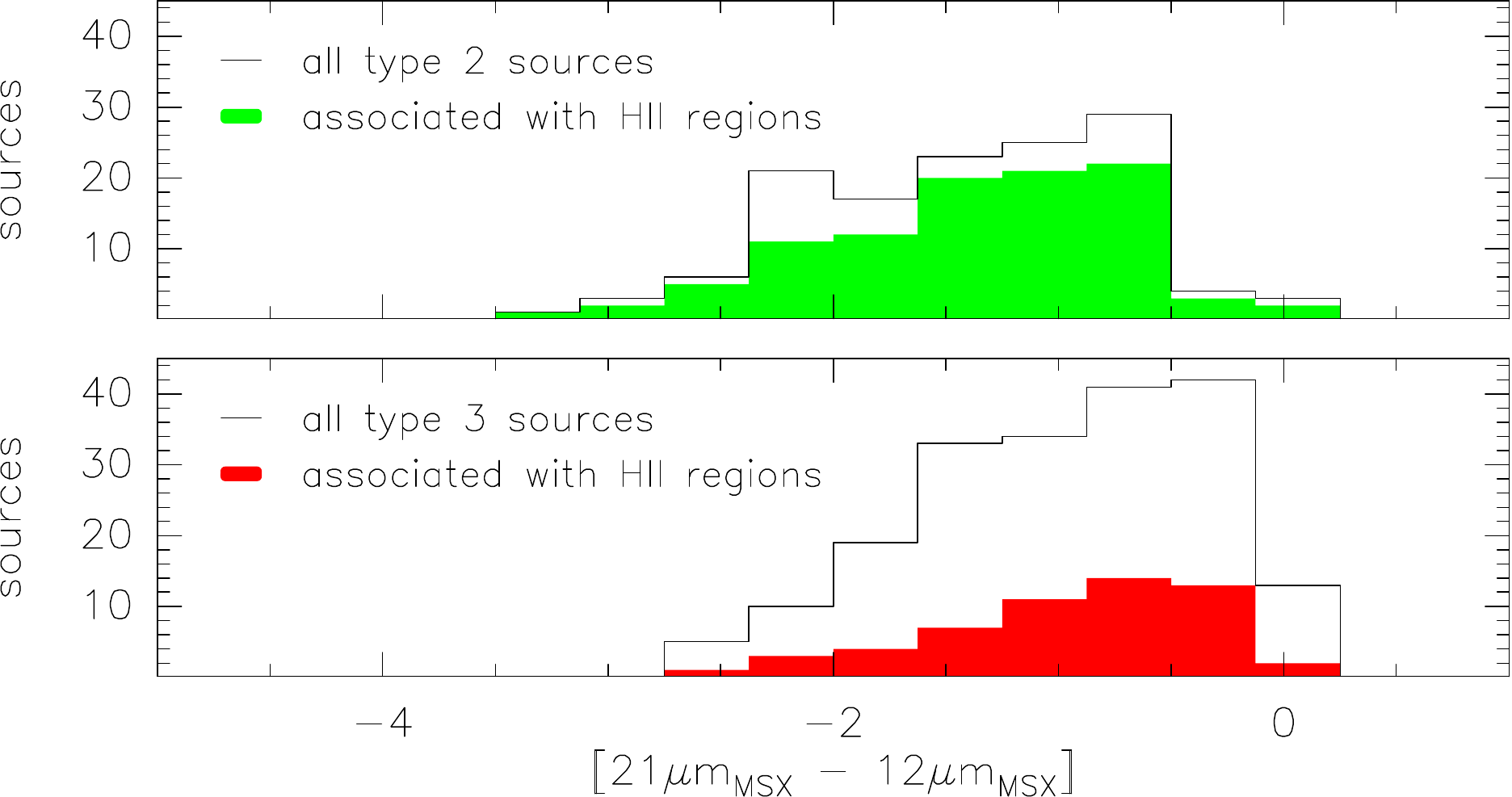, scale=0.43, angle=0} \\
\end{tabular}
\caption{Distributions of the MSX colors [21~$\mu$m--8~$\mu$m] (two top panels) and [21~$\mu$m--12~$\mu$m] (two bottom panels) for \emph{type~2} (green histograms) and \emph{type~3} (red histograms) sources. The solid black lines show the total number of sources of each group, and the colored filled histograms show those sources associated with \hii\ regions.}
\label{f:IRcolours}
\end{center}
\end{figure}

\begin{figure*}[t!]
\begin{center}
\begin{tabular}[b]{c c c c c c c}
 \epsfig{file=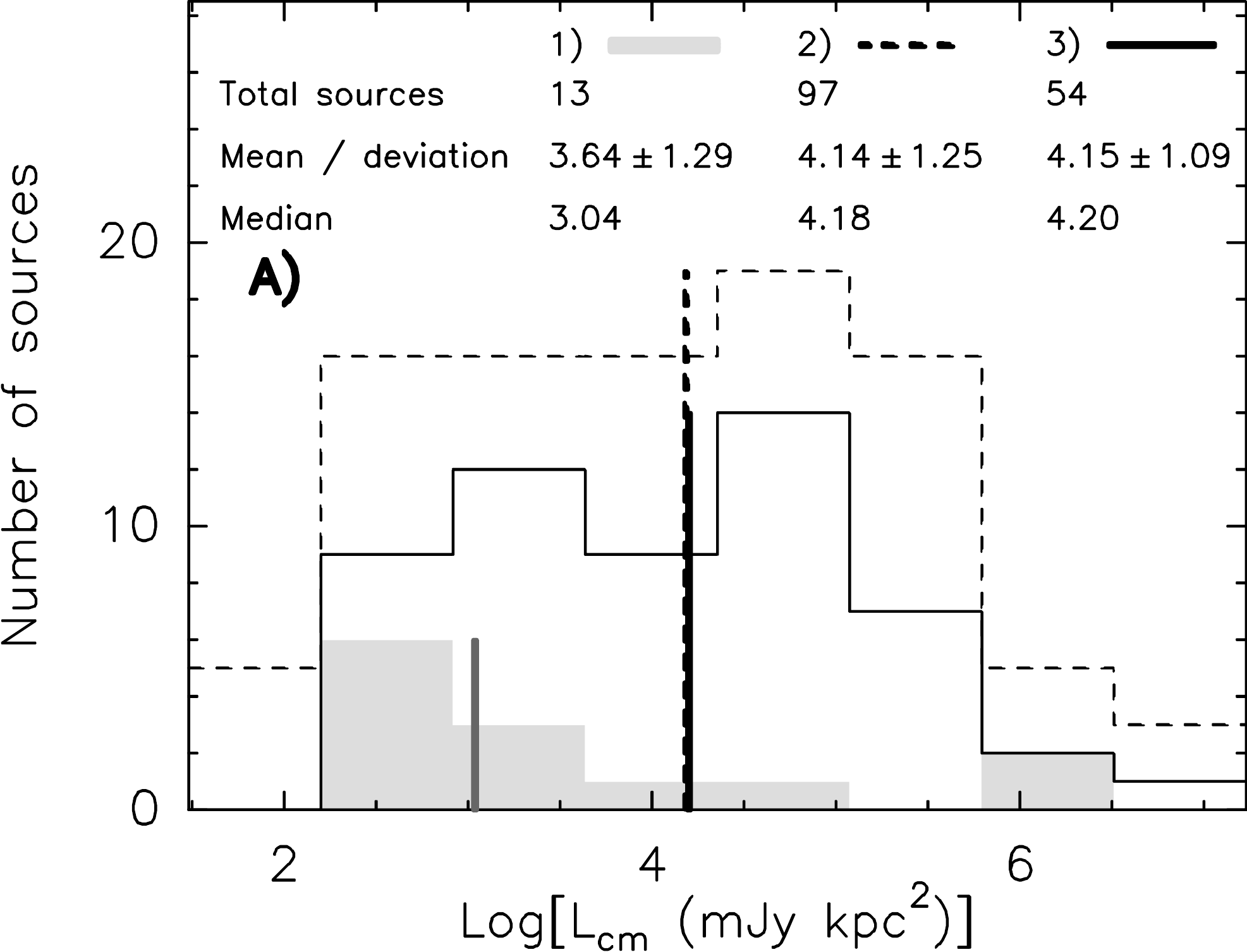, scale=0.27, angle=0} &&
 \epsfig{file=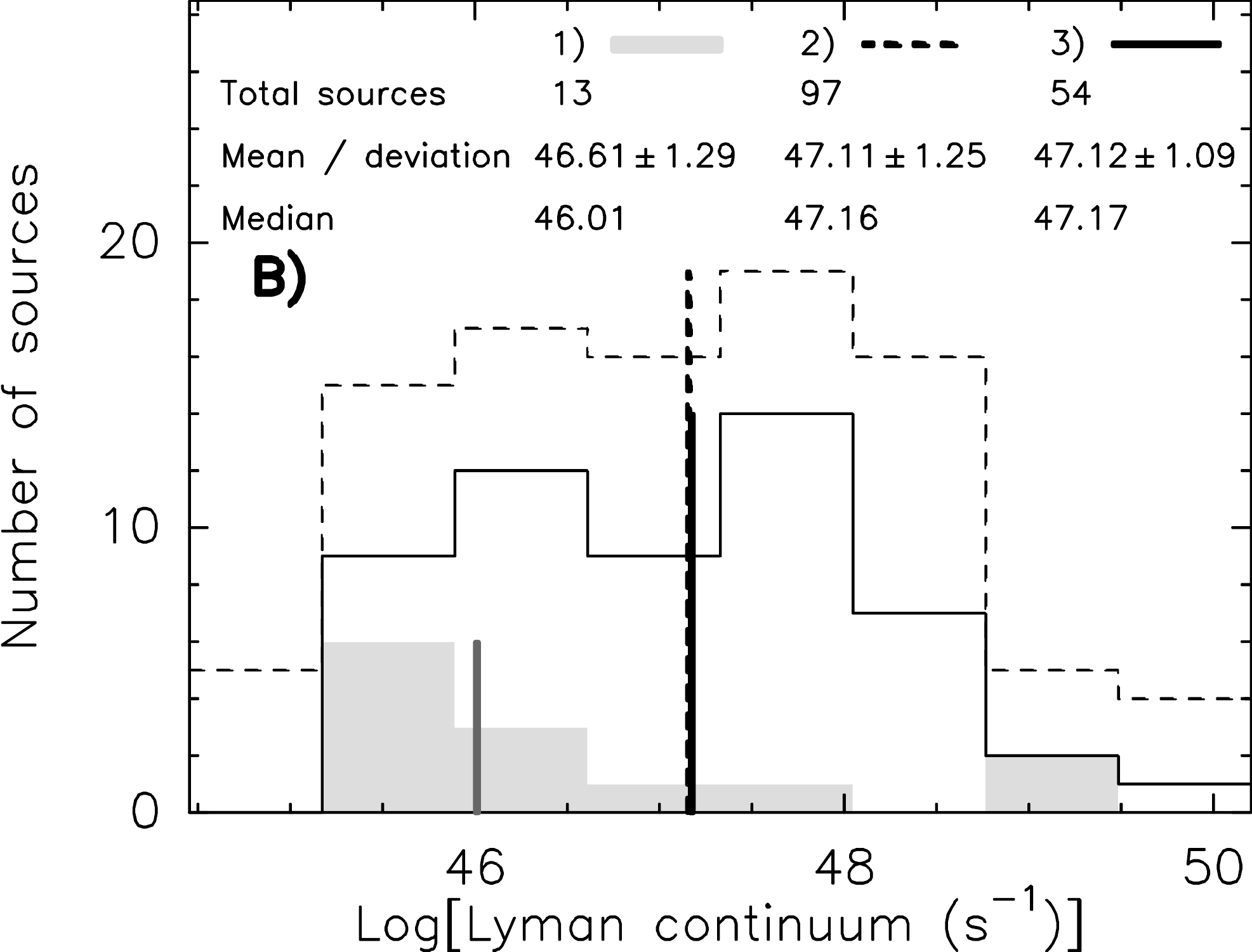, scale=0.27, angle=0} &&
 \epsfig{file=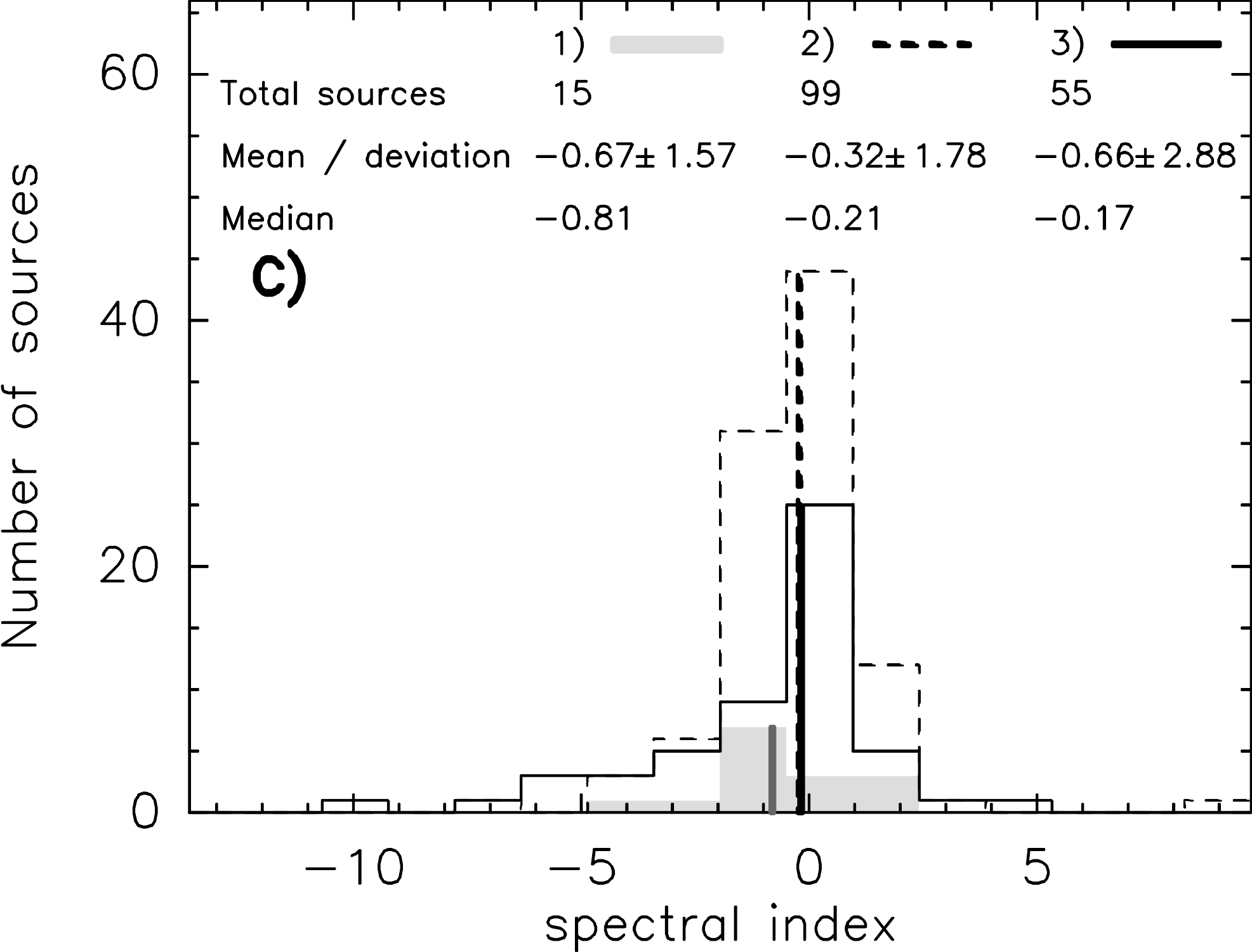, scale=0.27, angle=0} \\
 \noalign{\bigskip}
 \epsfig{file=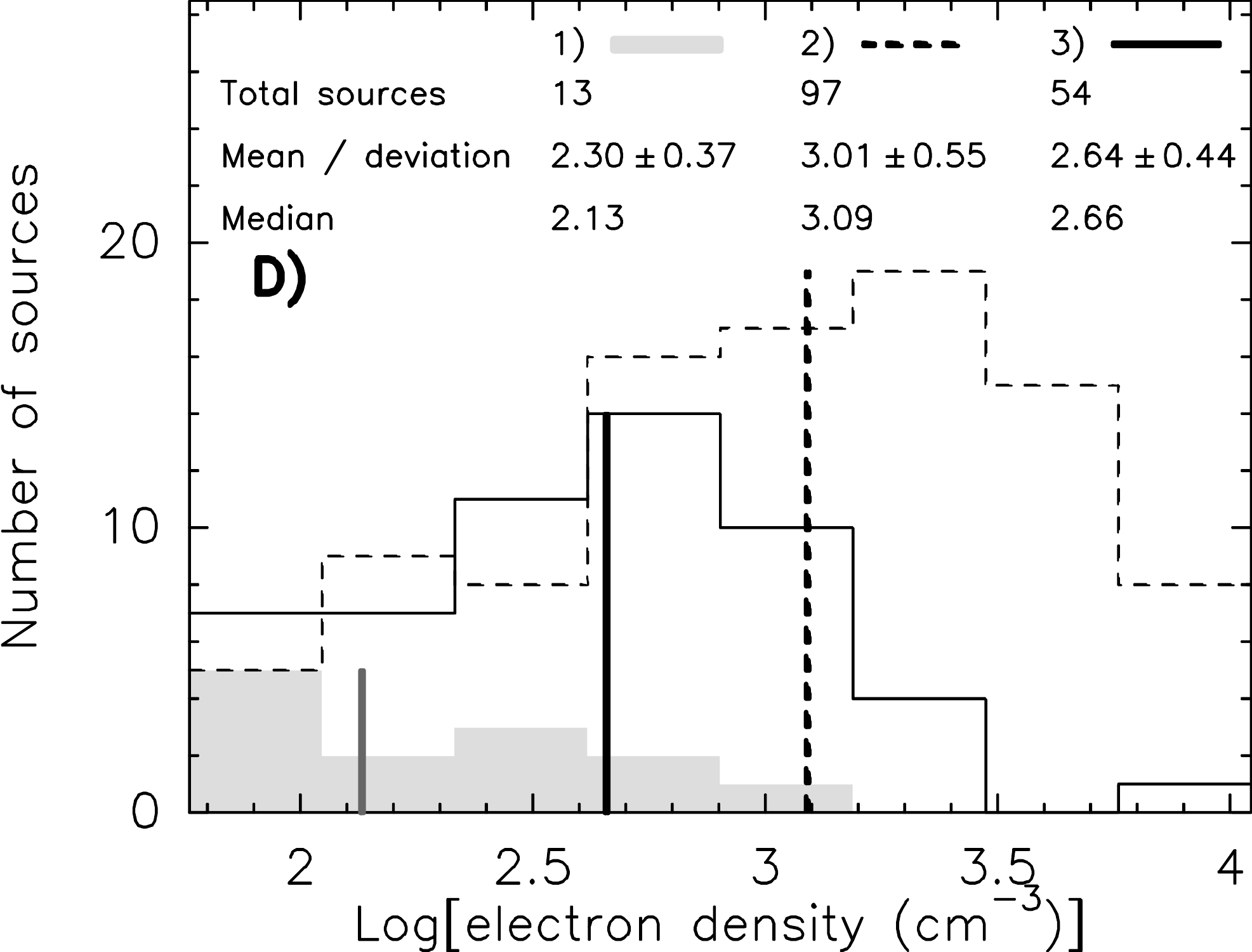, scale=0.27, angle=0} &&
 \epsfig{file=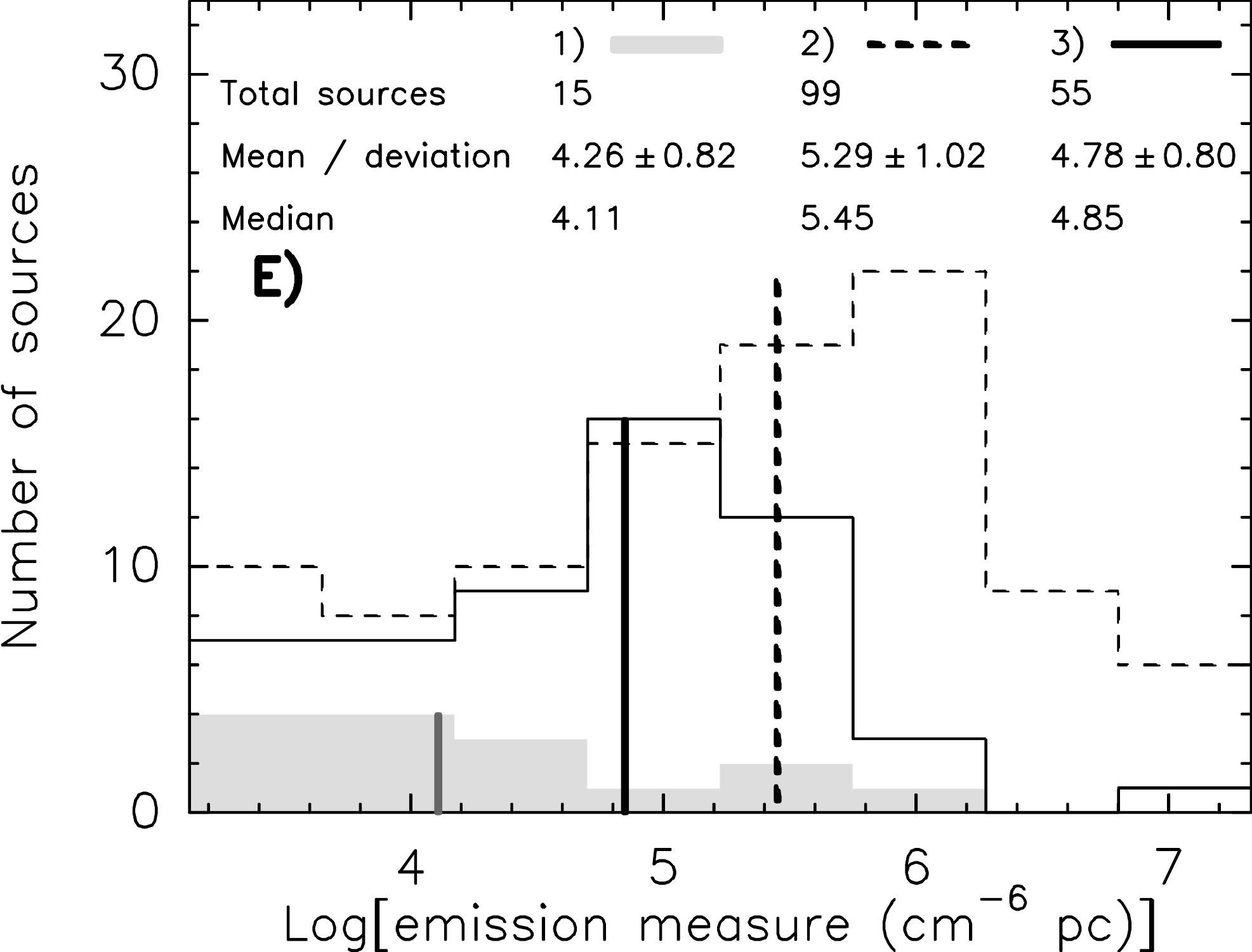, scale=0.27, angle=0} &&
 \epsfig{file=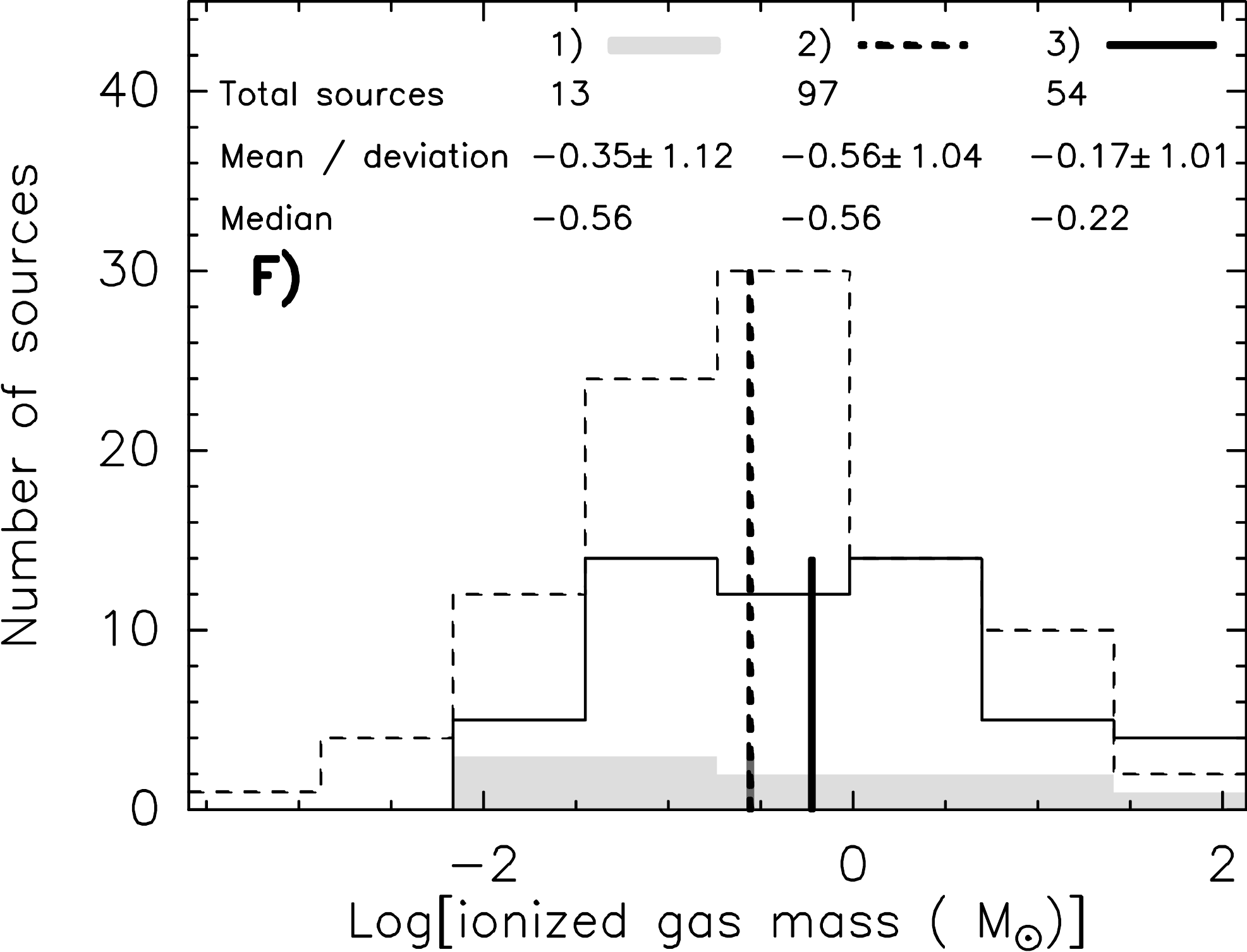, scale=0.27, angle=0} \\
 \noalign{\bigskip}
 \epsfig{file=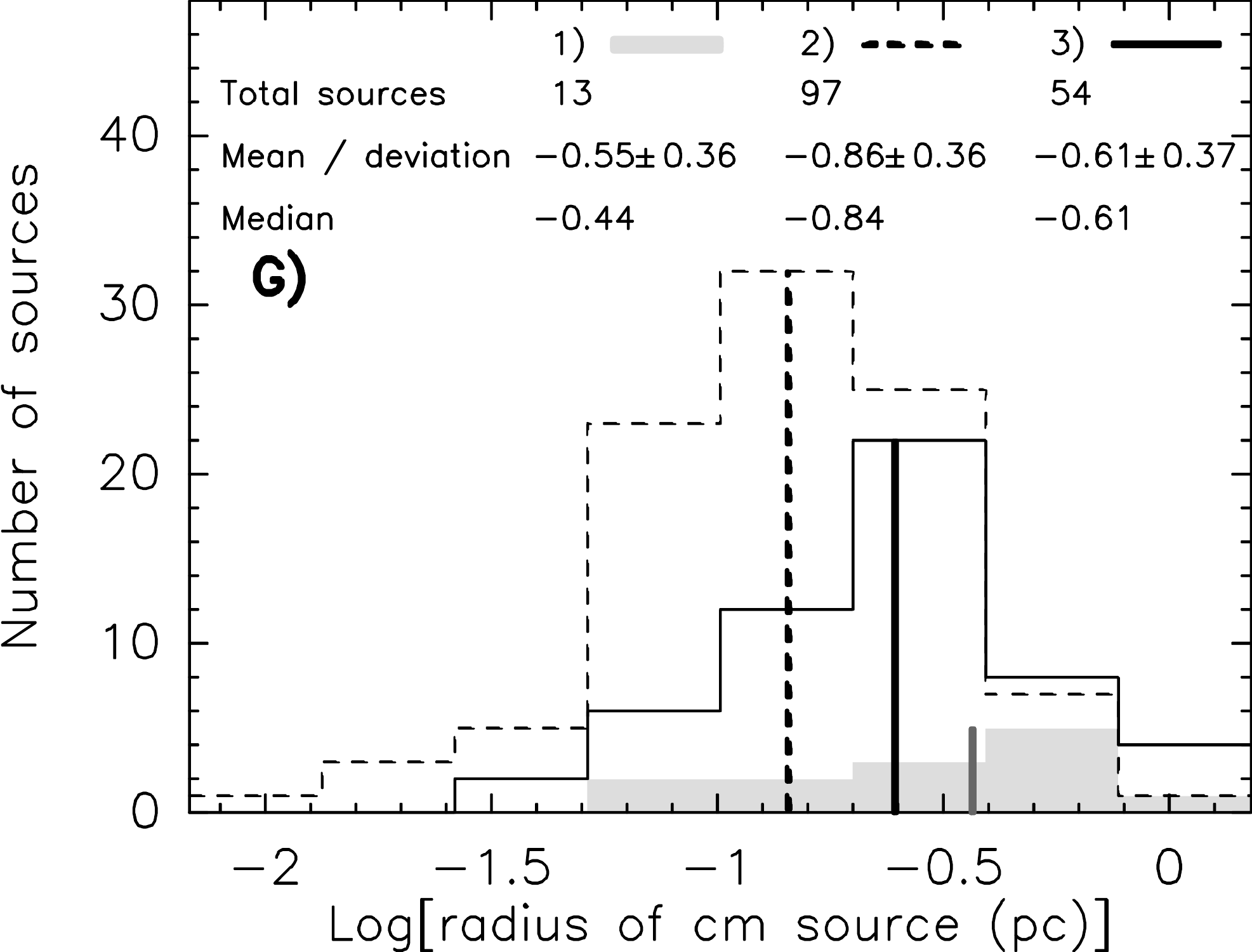, scale=0.27, angle=0} &&
 \epsfig{file=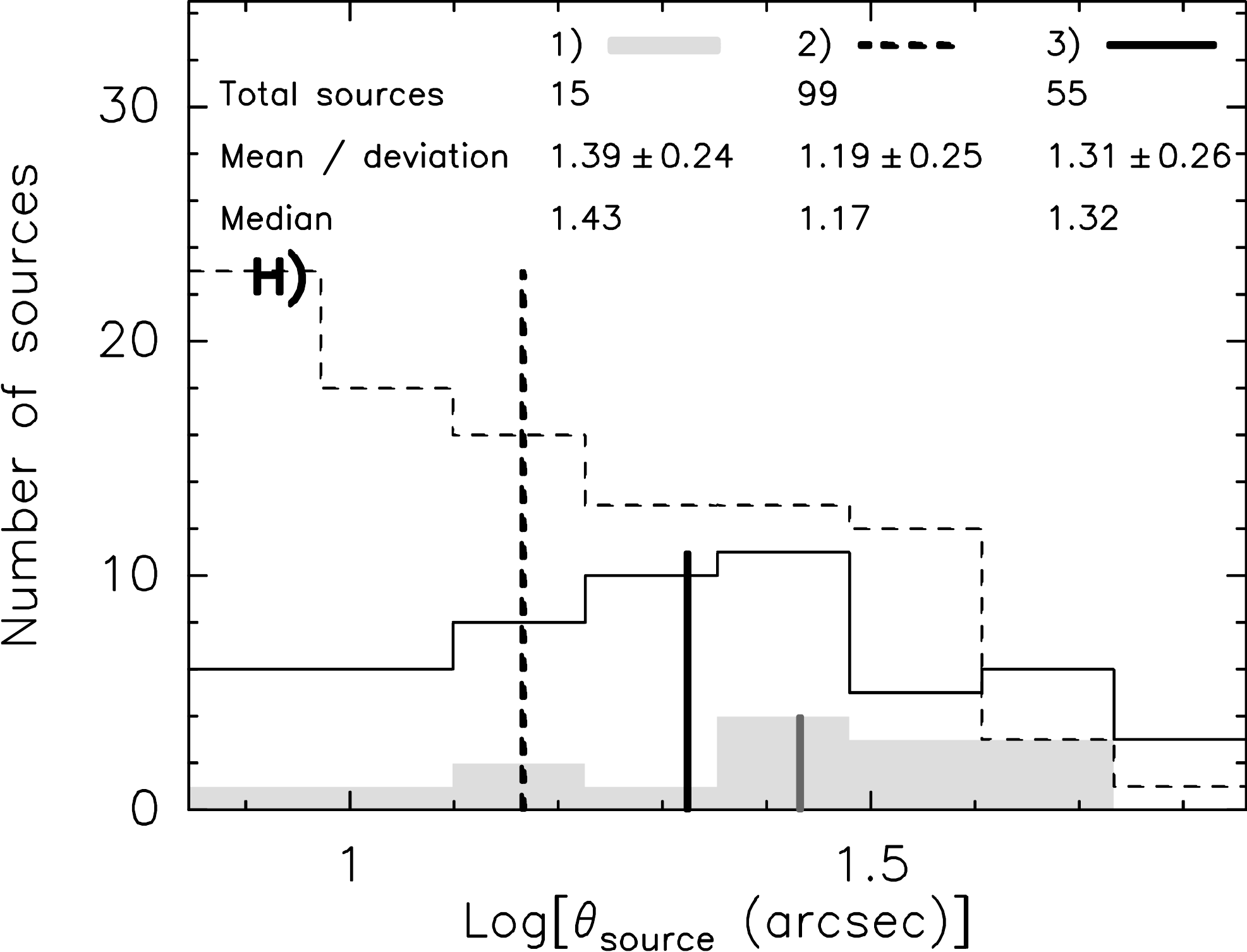, scale=0.27, angle=0} &&
 \epsfig{file=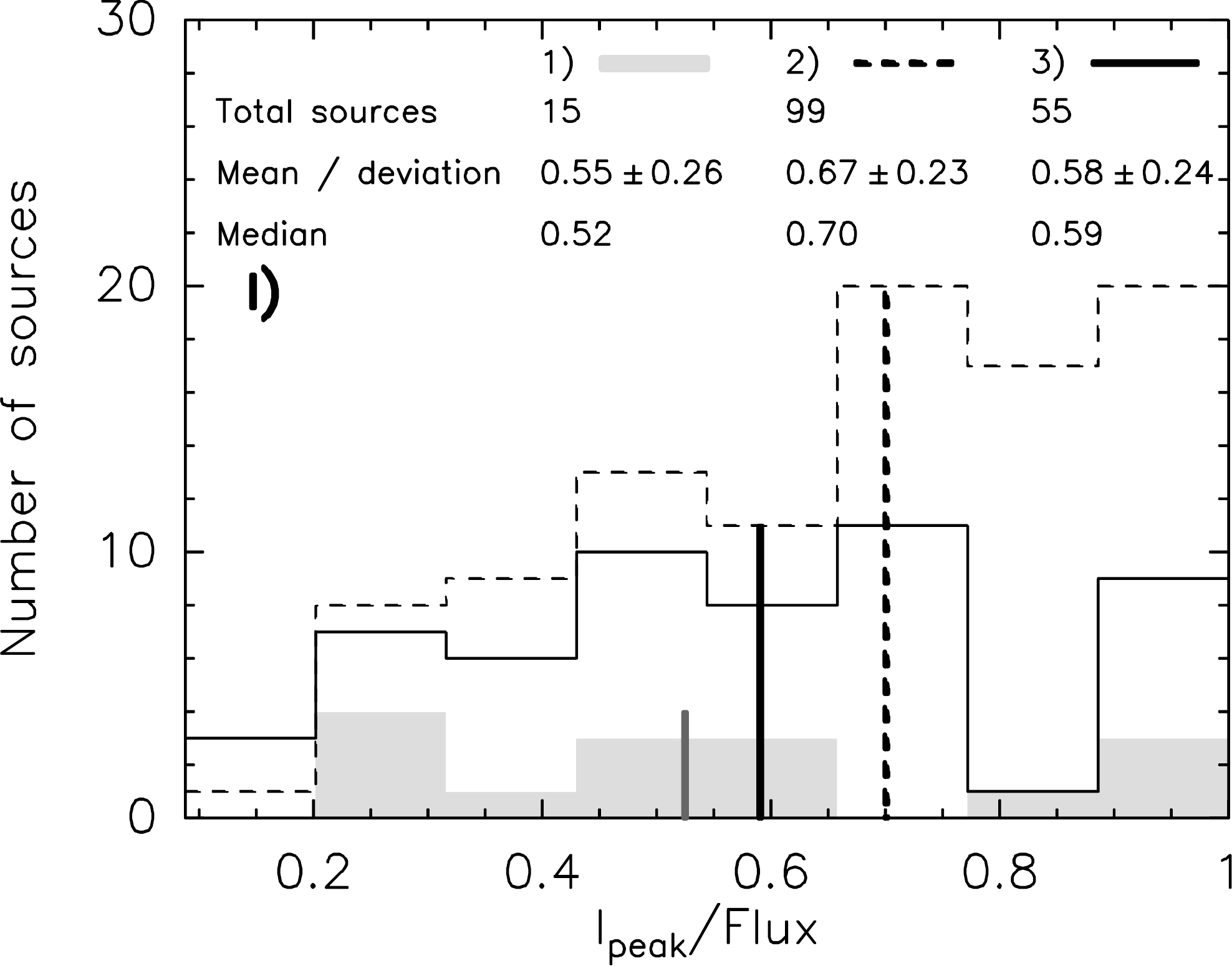, scale=0.27, angle=0} \\
\end{tabular}
\caption{Distributions of {\bf a)} centimeter luminosity; {\bf b)} Lyman continuum; {\bf c)} spectral index; {\bf d)} electron density; {\bf e)} emission measure; {\bf f)} ionized gas mass; {\bf g)} linear radius of the centimeter source; {\bf h)} deconvolved angular diameter, $\theta_{\rm S}$; and {\bf i)} peak intensity to flux density ratio, for the centimeter sources in the evolutionary stages \emph{type~1} (grey filled), \emph{type~2} (dashed line) and \emph{type~3} (solid line). The numbers at the top of each panel are as in Fig.~\ref{f:cmHisto}. The vertical thick lines indicate the median values.}
\label{f:cmtypeHisto}
\end{center}
\end{figure*}

\subsection{\hii\ regions and H$_2$O masers}

In this section, we investigate the presence of \hii\ regions and water masers in the three evolutionary stages (see Figures~\ref{f:cmassociations} and \ref{f:h2oassociations}). We use the same criteria to establish if an \hii\ region or a water maser is associated with a millimeter or infrared source, \ie\ the normalized distance must be <0.7, using the size listed in Table~\ref{t:results} for the \hii\ regions, and the HPBW at 22~GHz for the water maser (the water maser emission is unresolved in our observations). In Fig.~\ref{f:sketch}, we show a simple sketch indicating the associations with \hii\ regions and water masers for the three source types, and in Table~\ref{t:sketch}, we summarize the number of sources and percentages of associations with \hii\ regions and water masers in the \emph{type~1}, \emph{2} and \emph{3} sources.

Regarding the \hii\ regions, most of the centimeter continuum sources appear associated with \emph{type~2} and \emph{3} objects (corresponding to 72\% of all detected sources), and only a low percentage (7\%) are associated with \emph{type~1} objects. The remaining 21\% corresponds to \hii\ regions not associated with millimeter nor infrared sources. These results indicate on the one hand that for the first evolutionary stage (\emph{type~1}) the \hii\ region has not developed yet (only an 8\% of \emph{type~1} sources are associated with \hii\ regions; see Fig.~\ref{f:cmassociations}), and that the massive protostar begins to ionize the surrounding gas when the millimeter core is detectable at infrared wavelengths (75\% of \emph{type~2} sources are associated with \hii\ regions). On the other hand, in the last stages when the infrared source dominates the emission and most of the dusty clump has been destroyed (\ie\ \emph{type~3}), one would expect that the association of \hii\ regions with infrared sources was higher than for \emph{type~2} objects. In contrast, we observe a decrease in the percentage of associations (from 75\% down to 28\%). We investigated if there was a contamination of infrared sources not associated with star formation in our sample of \emph{type~3} objects. From MSX color-color diagrams we do not detect a differentiation between infrared sources associated with \hii\ regions and infrared sources not associated. In Fig.~\ref{f:IRcolours}, we show two MSX-color histograms for \emph{type~2} and \emph{type~3} objects. We can see that for each type the distribution of sources associated with \hii\ regions is similar to the total distribution, and thus we conclude that no significant contamination of not associated sources should affect the \emph{type~3} sample.

In Fig.~\ref{f:cmtypeHisto}, we show the distribution of the physical parameters of the \hii\ regions associated with the three types of objects. It seems that the continuum flux and Lyman continuum is greater in more evolved objects (\emph{type~2} and \emph{3}) than in the regions associated with only millimeter emission (see last three columns of Table~\ref{t:meancm}). Focusing on the two more evolved types, we can see that the electron density and the emission measure decrease from \emph{type~2} to \emph{type~3}, while the size increases, as expected if the \hii\ regions associated with type~3 objects are more evolved. Taking this into account, the relatively low percentage of association for \emph{type~3} objects could be understood in terms of the properties of their \hii\ regions: if the \hii\ region is larger and has a lower density, our snapshot observations would be inadequate to detect these \hii\ regions, and thus the percentage of associations would appear reduced in comparison with \emph{type~2} objects, the latter being associated with more compact and dense \hii\ regions, more easily detectable in our interferometric observations. The small number of \hii\ regions associated with \emph{type~1} objects, makes it difficult to derive typical values for these \hii\ regions. However, it is worth noting that the sizes of these centimeter continuum sources are larger than those of \emph{type~2} and \emph{3} objects, contrary to what is expected if the \hii\ regions of \emph{type~1} objects are less evolved. Similarly, low values of the density and emission measure, due to the large sizes, are also unexpected for less evolved \hii\ regions. An inspection of the maps for these sources reveals that most of these centimeter continuum sources are faint, which results in a bad estimation of the size, and in other cases a faint MSX infrared emission appears at the position of the \hii\ region, suggesting that there could be some contamination of \emph{type~2} objects. This would imply, that the detection rate of \hii\ regions in the \emph{type~1} group would be lower, as expected if these sources are younger.

\begin{figure}[t!]
\begin{center}
\begin{tabular}[b]{c}
 \epsfig{file=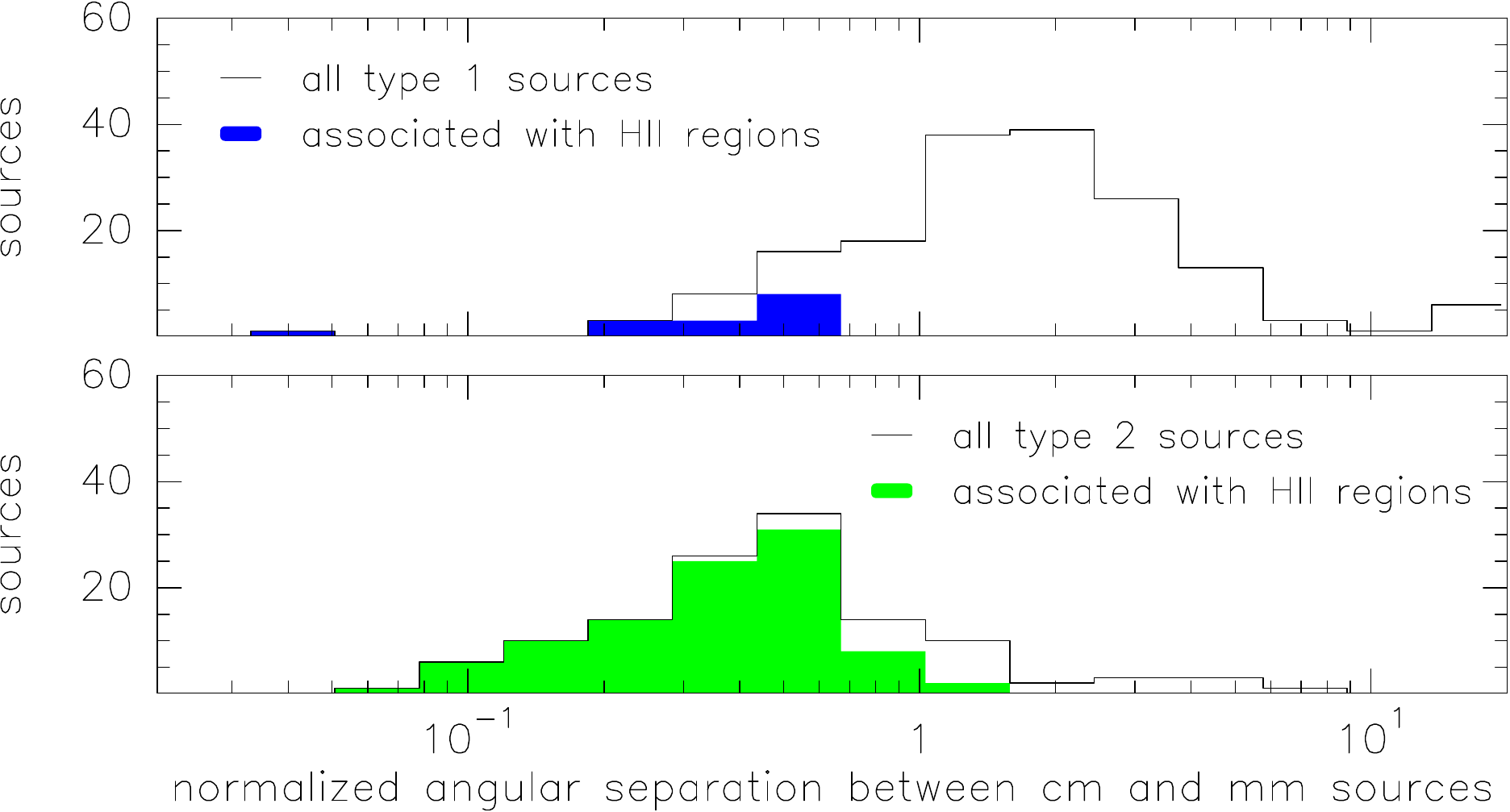, scale=0.43, angle=0} \\
\end{tabular}
\caption{Histograms of the normalized angular separation between the millimeter continuum source and the nearest \hii\ region for \emph{type~1} (top panel) and \emph{type~2} (bottom panel) sources.}
\label{f:cmmmSeps}
\end{center}
\end{figure}

In Fig.~\ref{f:cmmmSeps}, we report the distribution of normalized distances between the millimeter source and the nearest centimeter continuum source. The distribution for \emph{type~2} objects peaks at smaller normalized distances than for millimeter-only \emph{type~1} objects. In massive YSOs the association with centimeter continuum emission typically indicates the developing of an \hii\ region, and thus, objects that have reached the ZAMS. The results shown in Fig.~\ref{f:cmmmSeps} agree with the tentative result shown by Molinari \et\ (2008) in a smaller sample.

Regarding the water maser emission (see Fig.~\ref{f:h2oassociations}), we find that most of the water masers are associated with the earliest stages (67\% for \emph{type~1} and \emph{2}), and only few of them (7\%) are associated with \emph{type~3} objects. The evolution might begin with millimeter-only objects (\emph{type~1}) with only some of them associated with water maser emission. As the object evolves and becomes visible in the infrared (\emph{type~2}), the number of masers associated with these objects increases by a factor of 2, most of them being associated with \hii\ regions. Finally, as the object evolves and becomes less embedded, the association with masers also decreases.

The sketch shown in Fig.~\ref{f:sketch} summarizes the results found in this work: while the \hii\ region phase is dominant in the \emph{type~2} and \emph{type~3} objects, the water masers are most likely associated with the first stages. These results confirm the evolutionary classification proposed by Molinari \et\ (2008), in which \emph{type~1} sources would be high-mass protostars embedded in dust clumps with maser emission but still not developing an \hii\ region, \emph{type~2} sources would correspond to ZAMS OB stars with developed \hii\ regions, but still embedded in dust condensations and associated with maser emission, and \emph{type~3} sources would be more evolved ZAMS OB stars surrounded only by remnants of their parental clouds and with more extended and less dense \hii\ regions. This scenario is consistent with the evolutionary schemes presented by Ellingsen \et\ (2007) and Breen \et\ (2010), in which water masers appear first in the evolution of a massive protostar, coexist with the \hii\ regions, and disappear while the \hii\ regions are still detectable. Similarly, the evolutionary stages proposed by Molinari \et\ (2008) and this work compare well with the evolutionary classifications described by Beuther \et\ (2007) or Zinnecker \& Yorke (2007): \emph{type~1} sources would correspond to High-Mass Starless Cores (HMSCs) or young High-Mass Protostellar Objects (HMPOs) likely associated with infrared dark clouds (IRDCs) and hot molecular cores (HMCs), \emph{type~2} sources would be mainly evolved HMPOs or final stars associated with ultracompact \hii\ regions, and \emph{type~3} objects would be final stars associated with compact or classical \hii\ regions.

\section{Summary\label{s:conclusions}}

We have carried out ATCA observations of the H$_2$O maser line and radio continuum emission at 18.0~GHz and 22.8~GHz, toward a sample of 192 southern ($\delta$<$-$30\degr) massive star forming regions containing several clumps already imaged at 1.2~mm (Beltr\'an \et\ 2006). The sample consisted of 160 fields centered on an IRAS source plus 32 fields centered on a millimeter clump located $>$150\arcsec\ from an IRAS source. In total we observed 79 \emph{High} IRAS sources and 81 \emph{Low} IRAS sources (according to the criteria of Palla \et\ 1991). The main findings obtained in this work are:

\begin{itemize}

\item We detected centimeter continuum emission in 169 out of 192 fields, corresponding to a detection rate of 88\%. In total, 12\% of the fields do not show centimeter continuum emission (up to a level of $\sim$2~mJy), in 68\% we find a single component (with an angular resolution of $\sim$20\arcsec), and in 20\% we find multiple components. 

\item For the water maser emission, we have detected 85 distinct components in 78 fields, corresponding to a detection rate of 41\%. Due to the poor spectral resolution (12.5~\kms), we are probably detecting only relatively strong ($\gtrsim$1~Jy) water masers.

\item Assuming that the centimeter continuum emission comes from optically thin \hii\ regions, we have derived the physical parameters obtaining values in agreement with compact \hii\ regions (diameters $\sim$0.36~pc, electron densities $\sim$580~cm$^{-3}$, emission measures $\sim$7.9$\times10^5$~cm$^{-6}$~pc). The derived number of Lyman continuum photons spans a range between $10^{44}$~s$^{-1}$ and $10^{50}$~s$^{-1}$, corresponding to spectral types B5 to O5.

\item No large differences are found when studying the association or physical parameters of the \hii\ regions in the \emph{High} and \emph{Low} groups of sources. We found an \hii\ region detection rate of 94\% and 93\% for \emph{High} and \emph{Low} sources, respectively, with the electron density and emission measure slightly higher for \emph{High} sources. For the water maser emission, we found a detection rate of 46\% and 33\% for \emph{High} and \emph{Low} sources. These detection rates differ from previous results (\eg\ Palla \et\ 1991; Molinari \et\ 1998a) probably due to the different observing conditions (sensitivity, angular resolution and spectral resolution).

\item We have compared the Lyman continuum obtained from the measured radio flux with the bolometric luminosity estimated from the IRAS fluxes. Several sources (preferentially associated with the earliest evolutionary stages, and with B-type stars) show an excess of Lyman continuum that cannot be trivially explained. If the distances, which are the main source of uncertainty in this comparison, are confirmed (or more accurately estimated), and the excess of Lyman continuum is still appreciable, we should consider that B-stars could emit in the UV range much more than predicted by standard models of stellar atmospheres or that there exists another source of Lyman continuum in addition to the ZAMS star.

\item We have investigated the relation of the electron density and size of the \hii\ regions, and compared our distribution with that obtained assuming that O-B type stars are distributed in the Galaxy as proposed by Mottram \et\ (2011). From this analysis, we estimate the number of compact and extended \hii\ regions in the Galaxy to be $\sim$15000.

\end{itemize} 

We have used the SEST millimeter survey by Beltr\'an \et\ (2006) and the MSX Point Source Catalogue (Price \et\ 1999), to investigate the presence and association of millimeter and infrared sources, with the aim of establishing an evolutionary classification. Following the work of Molinari \et\ (2008), we have established three different groups of sources, in order of increasing age: (\emph{type~1}) only millimeter continuum sources, (\emph{type~2}) millimeter continuum sources associated with an infrared counterpart, and (\emph{type~3}) infrared sources without millimeter continuum emission. The main findings for the \hii\ region and water maser properties in this evolutionary classification are:

\begin{itemize}

\item \hii\ regions are mainly associated (72\%) with \emph{type~2} and \emph{3} sources, while only 7\% are associated with \emph{type~1} objects. Water masers are mainly associated with \emph{type~1} and \emph{2} objects: 67\% compared to the 7\% association with \emph{type~3} sources.

\item \hii\ regions associated with \emph{type~3} sources are larger in size and less dense than those associated with \emph{type~2} sources, as expected if the \hii\ region expands as it evolves.

\item The detailed analysis of each group results in an evolutionary trend for the association of \hii\ regions (8\% for \emph{type~1}, 75\% for \emph{type~2}, and 28\% for \emph{type~3}) and water masers (13\% for \emph{type~1}, 26\% for \emph{type~2}, and 3\% for \emph{type~3}). This scenario is consistent with the evolutionary schemes presented by Ellingsen \et\ (2007) and Breen \et\ (2010), in which water masers appear first in the evolution of a massive protostar, coexist with the \hii\ regions, and disappear while the \hii\ regions are still detectable. Our results of \hii\ regions and H$_2$O maser associations with different evolutionary types, confirm the evolutionary classification proposed by Molinari \et\ (2008), which compares well with the evolutionary classifications described by Beuther \et\ (2007) and Zinnecker \& Yorke (2007). Thus, \emph{Type~1} sources, associated with the earliest stages in our evolutionary scheme, appear to be good candidates to discriminate between the different theoretical models of massive star formation.

\end{itemize}

\begin{acknowledgements}
We thanks the anonymous referee for his/her comments. The figures of this paper have been done with the software package Greg of GILDAS (http://www.iram.fr/IRAMFR/GILDAS).
\end{acknowledgements}


\begin{appendix}
\section{Model description\label{a:model}}

\begin{figure}[t!]
\begin{center}
\begin{tabular}[b]{c}
 \epsfig{file=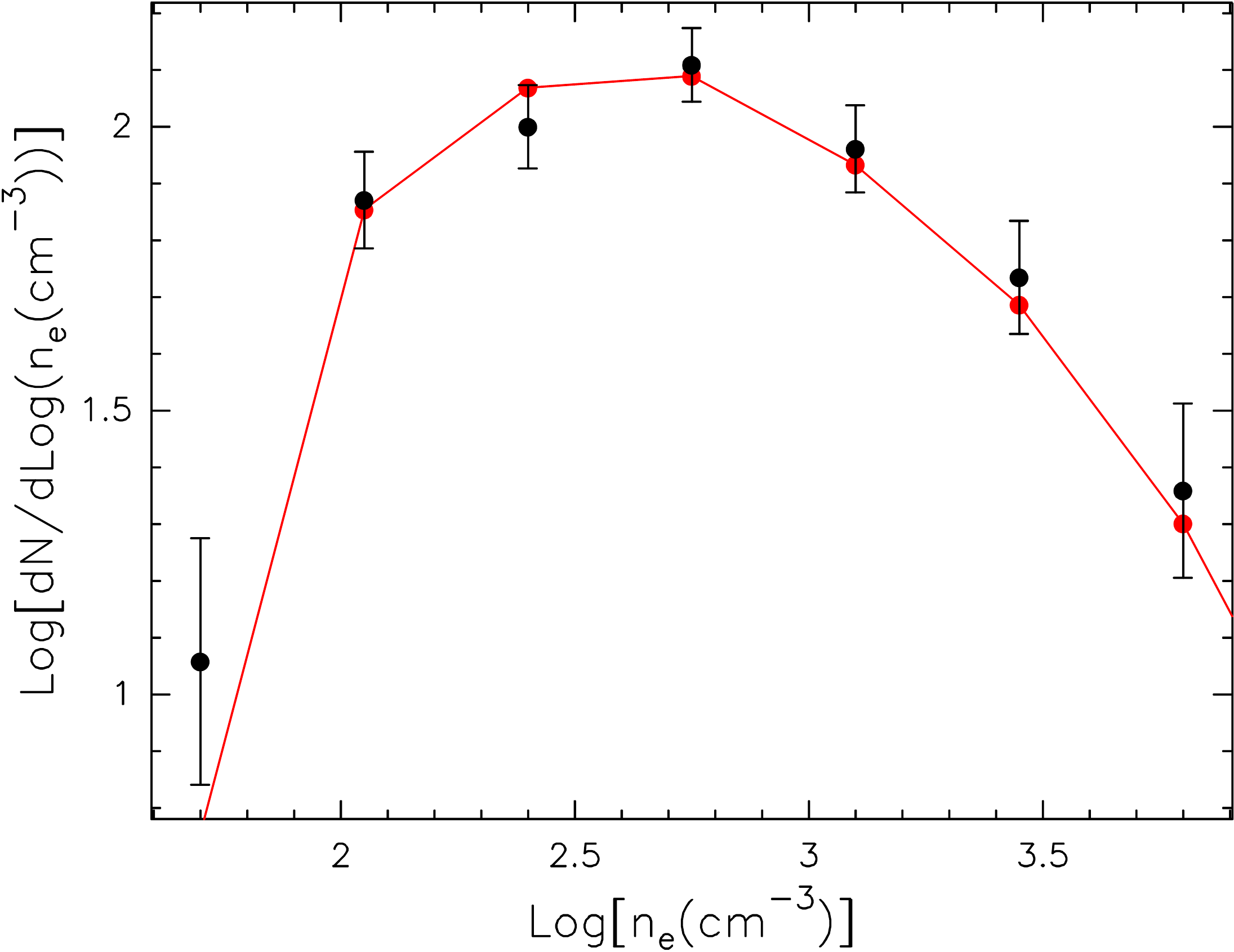, scale=0.35, angle=-90} \\
\end{tabular}
\caption{Electron density function obtained from the data (points with error bars) and best fit from the model (solid line) for $\alpha$=$-0.15$ and \NHII=15000.}
\label{f:distra}
\end{center}
\end{figure}
\begin{figure}[t!]
\begin{center}
\begin{tabular}[b]{c}
 \epsfig{file=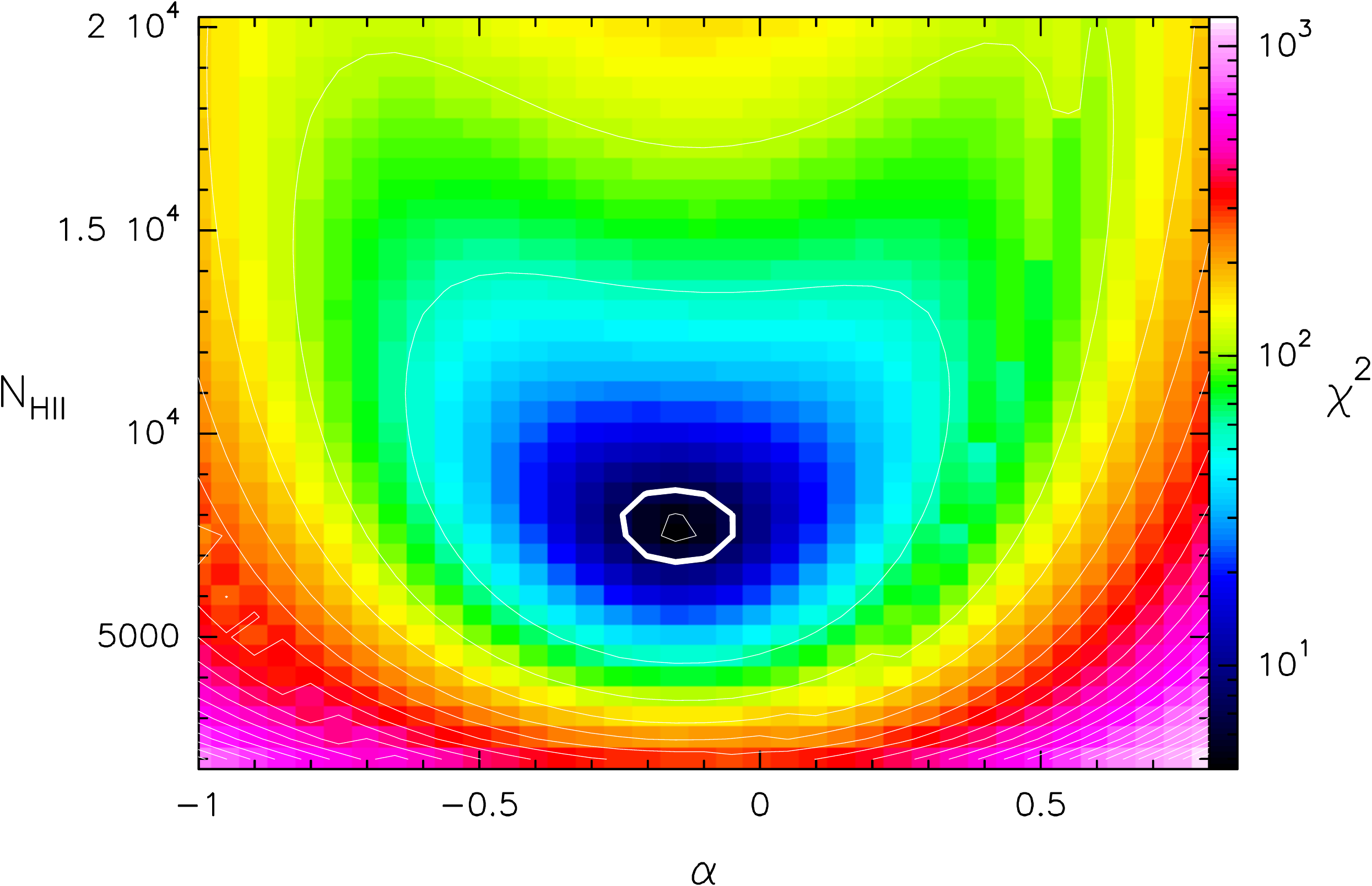, scale=0.35, angle=-90} \\
\end{tabular}
\caption{$\chi^2$ obtained by comparing the \ne\ function from the model to that obtained from the data (see Fig.~\ref{f:distra}), as a function of the two free parameters, $\alpha$ and \NHII. The minimum value (4.64) is obtained for $\alpha$=$-0.15$ and \NHII=15000. The thick contour corresponds to $\chi^2$ exceeding the minimum value by 2.3, as recommended by Lampton \et\ (1976) to estimate the $1\sigma$ confidence level.}
\label{f:chiq}
\end{center}
\end{figure}
\begin{figure}[t!]
\begin{center}
\begin{tabular}[b]{c}
 \epsfig{file=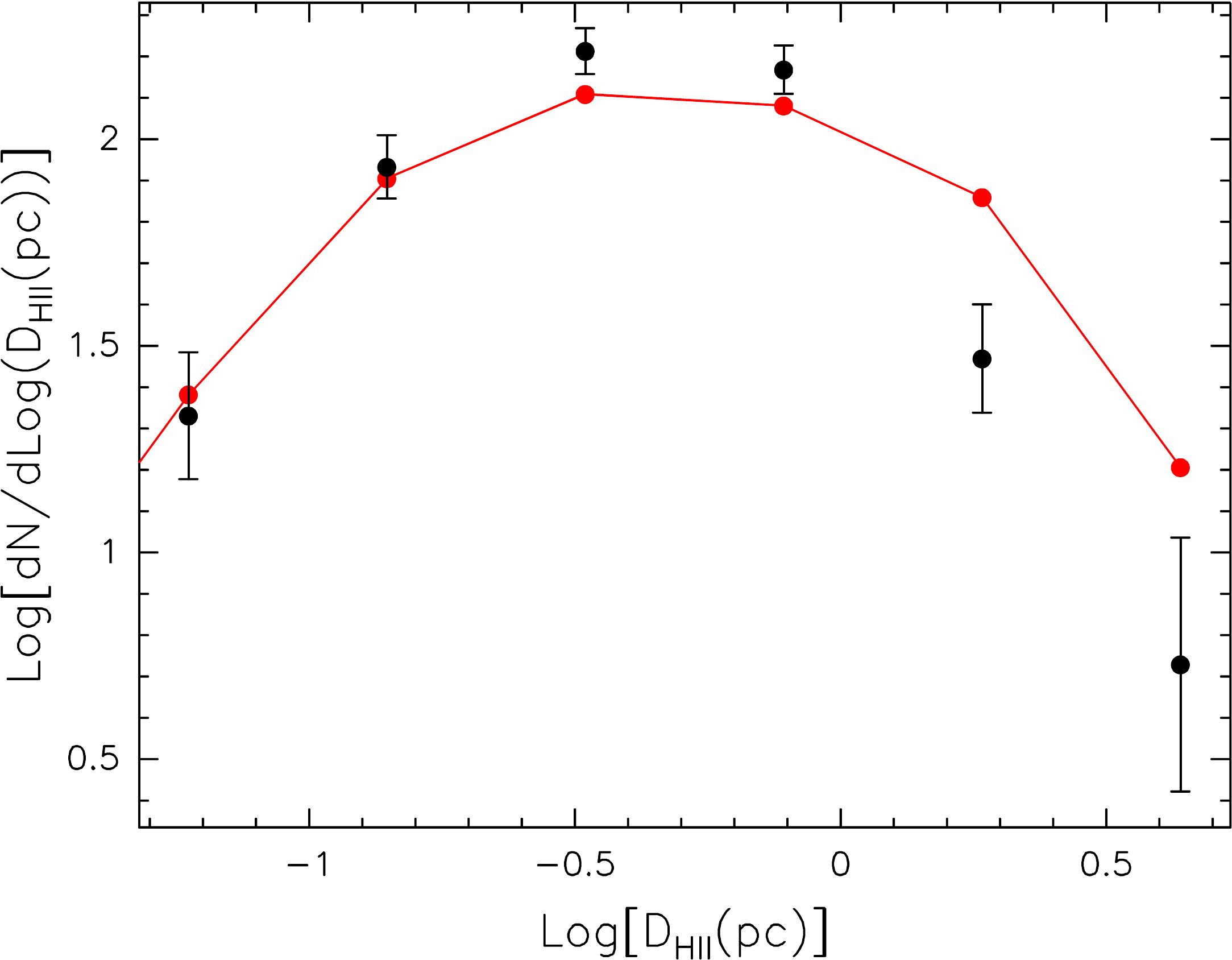, scale=0.35, angle=-90} \\
\end{tabular}
\caption{\hii\ region diameter function obtained from the data (points with error bars) and from the same model fit (solid line) as in Fig.~\ref{f:distra}.}
\label{f:distrb}
\end{center}
\end{figure}

The purpose of this appendix is to give an estimate of the number of \hii\ regions that can be detected in our survey. This obviously depends on the physical parameters of the \hii\ regions, their distances, and the sensitivity of our observations. We make the following assumptions:

\begin{itemize}
\item the \hii\ regions are spatially distributed across the Galaxy according
 to the equation
\begin{eqnarray*}
\fvol & = & \frac{\exp\left[-\frac{|z|}{\zd}\right]}{2\zd} \,
            \frac{\exp\left[-\left(\frac{R-\Ro}{\Rd}\right)^2\right] -
	        \exp\left[-\left(\frac{R-\Ro}{\Rh}\right)^2\right]}
	       {\pi\left[\Rd\left(\Rd+\sqrt{\pi}\Ro\right) -
		    \Rh\left(\Rh+\sqrt{\pi}\Ro\right)\right]} \nonumber \\
      &   & \Leftrightarrow\ R\ge\Ro \\
\fvol & = & 0  ~~\Leftrightarrow\ ~~R<\Ro
\end{eqnarray*}
which is the same as Eq.~(1) of Mottram \et\ (2011), normalized in such a way that $\int_{-\infty}^{+\infty}\,\d z \, \int_{\Ro}^{+\infty} \fvol 2\pi R\,\d R = 1$; here $R$ is the galactocentric distance, $z$ the height on the Galactic plane, $\Ro$=2.2~kpc, $\Rd$=6.99~kpc, $\Rh$=1.71~kpc, and $\zd$=0.039~kpc (from Mottram \et\ 2011);

\item the \hii\ regions are homogeneous Str\"omgren spheres ionized by stars with luminosities in the range $10^3$--$10^6$~\lo;

\item the normalized luminosity function ($\int_{\Log L_1}^{\Log L_2} \flum \d \Log L = 1$) of the ionizing star(s) is described by the power law
\begin{displaymath}
\flum = \frac{\beta\,\ln10\,L^\beta}{L_2^\beta-L_1^\beta}
\end{displaymath}
with $L_1$=$10^3$~\lo, $L_2$=$10^6$~\lo, and $\beta$=$-0.9$, as found by Mottram \et\ (2011);

\item the normalized electron density distribution ($\int_{\Log {n_{\rm e}}_1}^{\Log {n_{\rm e}}_2} \fne \d \Log \ne = 1$) is also described by a power law, \ie\
\begin{eqnarray*}
\fne & = & \frac{\alpha\,\ln10\,\ne^\alpha}{\ne_2^\alpha-\ne_1^\alpha} ~~\Leftrightarrow\  \alpha\neq 0 \\
\fne & = & \frac{\ln10}{\ln\left(\ne_2/\ne_1\right)} ~~\Leftrightarrow\  \alpha = 0
\end{eqnarray*}
with $\ne_1$=30~cm$^{-3}$ and $\ne_2$=$2\times10^4$~cm$^{-3}$ to span the range of $\ne$ in Fig.~\ref{f:nesize}.
\end{itemize}
%

Under these assumptions, the number of Galactic \hii\ regions per unit volume, dex in luminosity, and dex in electron density is given by the expression
\begin{equation}\label{e:nel}
 {\cal N}(R,z,\Log L, \Log \ne) = \NHII\ \, \fvol \, \flum \, \fne,
\end{equation}
where \NHII\ is the total number of \hii\ regions. There are only two free parameters in this model: the index $\alpha$ and \NHII. In order to find the corresponding values, we can fit the electron density function obtained from our data and represented by the points with error bars in Fig.~\ref{f:distra}. The number of \hii\ regions in each $\ne$ bin of this figure can be calculated from the model in the following way. The volume of the Galaxy is divided in a suitable number of cells in cylindrical coordinates and for each of these, Equation~(\ref{e:nel}) is used to calculate the number of \hii\ regions contained in that cell, spanning all luminosities in the range $L_1$--$L_2$, and with density falling in the chosen bin. In this process, only cells falling in the region surveyed by us (\ie\ those with 254<$l$<360$\degr$ and $|b|$<10$\degr$) were counted. Moreover, for each $L$ the corresponding Lyman continuum, \Nly, was computed assuming the relationship for a cluster rather than that for a single star (see Section~\ref{s:physpar}), as it seems unlikely that \hii\ regions with linear diameters up to 3~pc may be ionized by only one star. From $\ne$ and \Nly, one can calculate the diameter of the Str\"omgren \hii\ region, $D_\hii$, and the corresponding angular diameter and radio flux measured in the instrumental beam.

In order to take into account the bias due to the source selection criteria, as well as those introduced by the limited sensitivity of the observations, we took into account only the sources that satisfy the following requirements:
\begin{itemize}
\item $L/(4\pi d^2)$>1.6~\lo~kpc$^{-2}$, the minimum bolometric flux of the IRAS sources selected by Palla \et\ (1991) and Fontani \et\ (2005);
\item radio flux above 0.9~mJy~beam$^{-1}$, the mean $3\sigma$ sensitivity of our maps;
\item angular diameter below 60\arcsec, the maximum size imaged in our interferometric observations.
\end{itemize}
Finally, to construct the functions in Fig.~\ref{f:distra}, we rejected all objects (both in the model and the data) with diameter $\le$7$\arcsec$, as these are basically unresolved and the corresponding value of $\ne$ in Table~\ref{t:physpar} is a lower limit.

The best fit to the electron density function is compared to the data in Fig.~\ref{f:distra}. Such a fit has been obtained with $\chi^2$ minimization by varying the two free parameters of the models over a sufficiently large range of values. This is illustrated in Fig.~\ref{f:chiq}, where the $\chi^2$ is plotted as a function of $\alpha$ and \NHII. The thick contour corresponds to the minimum $\chi^2$ value (4.64) plus 2.3, which gives the $1\sigma$ confidence level according to Table~1 of Lampton \et\ (1976). We conclude that the best fit is obtained for \NHII=15000$^{+2000}_{-1300}$ and $\alpha$=$-0.15\pm0.1$

As previously explained, for given $\ne$ and \Nly, the \hii\ region size is univocally determined. This means that knowledge of the density function $\fne$ provides us also with the analogous function for the linear diameter, \fD. The latter is shown in Fig.~\ref{f:distrb} and compared to the observed distribution. Clearly the two are quite consistent for all diameters, with some discrepancy for the largest $D_\hii$, where the model appears to overestimate the measured value. However, the largest \hii\ regions are also the most difficult to image with an interferometer, and this may explain such a small discrepancy.

\end{appendix}

\clearpage
\longtab{1}{
\begin{scriptsize}
\begin{longtable}{l c c c c c c c c c c c c c l}
\caption{\label{t:observations} Target fields observed by ATCA at 18.0~GHz and 22.8~GHz}\\
\hline\hline
&&&&&&
&\multicolumn{3}{c}{18.0~GHz}
&
&\multicolumn{3}{c}{22.8~GHz}
\\
\cline{8-10}\cline{12-14}\noalign{\smallskip}
&
&Dist.\tablefootmark{c}
&$\alpha({\rm J2000})$
&$\delta({\rm J2000})$
&$l$
&$b$
&Beam
&P.A.
&Rms
&
&Beam
&P.A.
&Rms
\\
Region\tablefootmark{a}
&Type\tablefootmark{b}
&(kpc)
&(~\raun~)
&(~\deun~)
&($\degr$)
&($\degr$)
&($\arcsec\times\arcsec$)
&($\degr$)
&(\jpb)
&
&($\arcsec\times\arcsec$)
&($\degr$)
&(\jpb)
&Date\tablefootmark{d}
\\
\hline
\endfirsthead
\caption{continued.}\\
\hline\hline
&&&&&&
&\multicolumn{3}{c}{18.0~GHz}
&
&\multicolumn{3}{c}{22.8~GHz}
\\
\cline{8-10}\cline{12-14}\noalign{\smallskip}
&
&Dist.\tablefootmark{c}
&$\alpha({\rm J2000})$
&$\delta({\rm J2000})$
&$l$
&$b$
&Beam
&P.A.
&Rms
&
&Beam
&P.A.
&Rms
\\
Region\tablefootmark{a}
&Type\tablefootmark{b}
&(kpc)
&(~\raun~)
&(~\deun~)
&($\degr$)
&($\degr$)
&($\arcsec\times\arcsec$)
&($\degr$)
&(\jpb)
&
&($\arcsec\times\arcsec$)
&($\degr$)
&(\jpb)
&Date\tablefootmark{d}
\\
\hline
\endhead
\hline
\endfoot
08140$-$3559~0	&H  &\phn3.8\tabz	&08 15 59.00  &$-$36 08 18.0  &254.05 &$-$0.56    &$29.1\times18.7$  &$-$74 &$2.8\times10^{-4}$     &&$21.9\times14.8$  &$-$73  &$2.5\times10^{-4}$       &25\texttt{J} \\
08211$-$4158~0	&L  &\phn1.7\tabz	&08 22 52.30  &$-$42 07 57.0  &259.76 &$-$2.84    &$26.8\times19.2$  &$-$76 &$2.0\times10^{-4}$     &&$20.8\times15.3$  &$-$75  &$1.0\times10^{-4}$       &09\texttt{O} \\
08438$-$4340~0	&H  &\phn1.3\tabz	&08 45 36.00  &$-$43 51 01.0  &263.62 &$-$0.53    &$26.9\times19.4$  &$-$79 &$2.5\times10^{-3}$     &&$20.4\times18.6$  &$+$86  &$1.5\times10^{-3}$       &09\texttt{O} \\
08470$-$4243~0	&H  &\phn2.2\tabz	&08 48 47.90  &$-$42 54 22.0  &263.25 &$+$0.51    &$26.6\times19.4$  &$-$76 &$1.7\times10^{-4}$     &&$20.3\times15.4$  &$-$77  &$2.5\times10^{-4}$       &25\texttt{J} \\
08477$-$4359~0	&L  &\phn1.8\tabz	&08 49 32.90  &$-$44 10 47.0  &264.32 &$-$0.18    &$26.4\times19.3$  &$-$77 &$2.0\times10^{-4}$     &&$20.2\times15.3$  &$-$75  &$1.0\times10^{-4}$       &09\texttt{O} \\
08477$-$4359~1	&L  &\phn1.8\tabz	&08 49 35.13  &$-$44 11 59.0  &264.34 &$-$0.19    &$26.6\times19.3$  &$-$76 &$2.7\times10^{-4}$     &&$20.1\times15.5$  &$-$76  &$2.4\times10^{-4}$	    &25\texttt{J} \\
08563$-$4225~0	&L  &\phn1.7\tabz	&08 58 12.50  &$-$42 37 34.0  &264.14 &$+$2.02    &$27.3\times19.2$  &$-$76 &$5.0\times10^{-5}$     &&$20.2\times18.0$  &$+$89  &$7.0\times10^{-5}$	    &09\texttt{O} \\
08589$-$4714~0	&H  &\phn1.5\tabz	&09 00 40.50  &$-$47 25 55.0  &268.06 &$-$0.80    &$26.2\times20.1$  &$-$76 &$2.0\times10^{-4}$     &&$20.2\times18.1$  &$-$83  &$8.0\times10^{-5}$       &09\texttt{O} \\
\end{longtable}
\tablefoot{
\tablefoottext{a}{First number correspond to the IRAS source name. The second number if equal to 0: phase center corresponding to the IRAS source; if different of 0: phase center corresponding to the millimeter clump detected by Beltr\'an \et\ (2006), the number corresponds to the number of the millimeter clump.}
\tablefoottext{b}{H (High) or L (Low) according to the classification of Palla \et\ (1991).}
\tablefoottext{c}{Kinematic distance from Beltr\'an \et\ (2006). The near/far ambiguity has been solved where indicated, if not, the near distance has been adopted.}
\tablefoottext{d}{Day of observation. 25\texttt{J}: 25 July 2009, 26\texttt{J}: 26 July 2009, 09\texttt{O}: 9 October 2009, and 10\texttt{O}: 10 October 2009.}
\tablefoottext{e}{Distance not available in Beltr\'an \et\ (2006), obtained from different works: 14394$-$600, 17040$-$3959 and 17352$-$3153 from Green \& McClure-Griffiths (2011); 15506$-$5325 and 15507$-$5347 from Grave \& Kumar 2009; 17425$-$3017 from Molinari \et\ (2008).}
\tablefoottext{f}{Near/far ambiguity solved. Refs: Green \& McClure-Griffiths (2011); Urquhart \et\ (2012).}
\tablefoottext{g}{Also observed on 10 October 2009.}
}
\end{scriptsize}
}

\longtab{3}{
\begin{scriptsize}
\begin{longtable}{l c c c c c c c c c c c c c c}
\caption{\label{t:results} ATCA centimeter continuum sources at 18.0~GHz and 22.8~GHz}\\
\hline\hline
&
&
&
&\multicolumn{4}{c}{18.0~GHz}
&
&\multicolumn{4}{c}{22.8~GHz}
&
\\
\cline{5-8}\cline{10-13}\noalign{\smallskip}
&
&$\alpha_\mathrm{J2000}$
&$\delta_\mathrm{J2000}$
&I$_\mathrm{peak}$
&Flux\tabc
&FWHP\tabd
&Deconv\tabd
&
&I$_\mathrm{peak}$\tabc
&Flux\tabc
&FWHP\tabd
&Deconv\tabd
&\water
\\
Region\taba
&\#\tabb
&(~\raun~)
&(~\deun~)
&(Jy~beam$^{-1}$)
&(Jy)
&(arcsec)
&(arcsec)
&
&(Jy~beam$^{-1}$)
&(Jy)
&(arcsec)
&(arcsec)
&maser\tabe
\\
\hline
\endfirsthead
\caption{continued.}\\
\hline\hline
&
&
&
&\multicolumn{4}{c}{18.0~GHz}
&
&\multicolumn{4}{c}{22.8~GHz}
&
\\
\cline{5-8}\cline{10-13}\noalign{\smallskip}
&
&$\alpha_\mathrm{J2000}$
&$\delta_\mathrm{J2000}$
&I$_\mathrm{peak}$\tabc
&Flux\tabc
&FWHP\tabd
&Deconv\tabd
&
&I$_\mathrm{peak}$
&Flux\tabc
&FWHP\tabd
&Deconv\tabd
&\water
\\
Region\taba
&\#\tabb
&(~\raun~)
&(~\deun~)
&(Jy~beam$^{-1}$)
&(Jy)
&(arcsec)
&(arcsec)
&
&(Jy~beam$^{-1}$)
&(Jy)
&(arcsec)
&(arcsec)
&maser\tabe
\\
\hline
\endhead
\hline
\endfoot
08140$-$3559~0\tabz &1  &08 16 00.4\tabz  &$-$36 08 27.8\tabz  &\phb0.0090  &0.0107  &25.9  &\phb11.1	&&\phb0.0065  &0.0107  &23.8  &\phb15.5    &$-$   \\
08211$-$4158~0\tabz &1  &08 22 52.2\tabz  &$-$42 07 57.4\tabz  &\phb0.0021  &0.0029  &32.1  &\phb22.6	&&\phb0.0021  &0.0033  &23.5  &\phb15.2    &$-$   \\
                    &2  &08 22 52.7\tabz  &$-$42 07 32.2\tabz  &$<$0.0010   &	     &      &		&&\phb0.0021  &0.0021  &20.8  &\phb10.6    &$-$   \\
08438$-$4340~0\tabz &1  &08 45 34.5\tabz  &$-$43 50 31.2\tabz  &\phb0.0728  &0.1407  &41.0  &\phb34.0	&&\phb0.0336  &0.0713  &32.2  &\phb25.6    &$-$   \\
                    &2  &08 45 36.9\tabz  &$-$43 51 20.3\tabz  &\phb0.0871  &0.3917  &54.7  &\phb49.7	&&\phb0.0729  &0.1829  &32.1  &\phb25.5    &$-$   \\
                    &3  &08 45 37.3\tabz  &$-$43 50 36.0\tabz  &\phb0.0441  &0.0441  &21.7  &\phn$<$7.0 &&\phb0.0482  &0.0774  &25.7  &\phb16.7    &$-$   \\
08470$-$4243~0\tabz &1  &08 48 47.9\tabz  &$-$42 54 16.7\tabz  &\phb0.0072  &0.0131  &32.0  &\phb22.5	&&\phb0.0082  &0.0148  &23.2  &\phb14.9    &Y   \\
08477$-$4359~0\tabz &1  &08 49 32.8\tabz  &$-$44 10 46.3\tabz  &\phb0.0102  &0.0106  &23.2  &\phn$<$7.0 &&\phb0.0093  &0.0174  &22.2  &\phb13.5    &$-$   \\
\end{longtable}
\tablefoot{
\tablefoottext{a}{See Table~\ref{t:observations}.}
\tablefoottext{b}{ATCA source number (from this work).}
\tablefoottext{c}{Primary beam corrected.} 
\tablefoottext{d}{FWHP: full width half power, and Deconv: deconvolved size (in arcsec).} 
\tablefoottext{e}{Y: water maser associated with the centimeter source.}
\tablefoottext{f}{The source seen in the map is that associated with 08477$-$4359~0.}
\tablefoottext{g}{Coordinates from the 18.0~GHz map.}
\tablefoottext{h}{The source seen in the map is that associated with 15278$-$5620~0.}
\tablefoottext{i}{The source seen in the map is that associated with 15278$-$5620~0.}
\tablefoottext{j}{The source seen in the map is that associated with 15479$-$54190.}
}
\end{scriptsize}
}

\longtab{4}{
\begin{scriptsize}
\begin{longtable}{r l c c c c}
\caption{\label{t:masers} Regions with water maser emission}\\
\hline\hline
&
&$\alpha_\mathrm{J2000}$
&$\delta_\mathrm{J2000}$
&$\int S_\nu~\mathrm{d}v$\tablefootmark{a}
&$v_\mathrm{peak}$
\\
\#
&Region
&(~\raun~)
&(~\deun~)
&(Jy~\kms)
&(\kms)
\\
\hline
\endfirsthead
\caption{continued.}\\
\hline\hline
&
&$\alpha_\mathrm{J2000}$
&$\delta_\mathrm{J2000}$
&$\int S_\nu~\mathrm{d}v$ \tablefootmark{a}
&$v_\mathrm{peak}$
\\
\#
&Region
&(~\raun~)
&(~\deun~)
&(Jy~\kms)
&(\kms)
\\
\hline
\endhead
\hline
\endfoot
1	&08438$-$4340~0   &08 45 33.718   &$-$43 52 05.85   &$1.5891\times10^{-1}$   &\phnn$+9$  \\
2	&08470$-$4243~0   &08 48 47.887   &$-$42 54 30.16   &$2.3593\times10^{-1}$   &\phn$+17$  \\
3	&08563$-$4225~0   &08 58 11.630   &$-$42 37 29.95   &$1.2415\times10^{+0}$   &\phnn$+9$  \\
4	&08589$-$4714~0   &09 00 40.898   &$-$47 26 04.88   &$2.1442\times10^{+1}$   &\phnn$+7$  \\
5	&09131$-$4723~0   &09 14 51.711   &$-$47 36 49.99   &$9.3873\times10^{-1}$   &\phnn$+7$  \\
6	&09166$-$4813~1   &09 17 54.507   &$-$48 27 23.35   &$8.1421\times10^{-2}$   &\phnn$+7$  \\
7	&09578$-$5649~0   &09 59 31.319   &$-$57 04 42.52   &$9.6140\times10^{-2}$   &\phnn$+3$  \\
8	&10019$-$5712~0   &10 03 40.288   &$-$57 26 35.68   &$6.0604\times10^{+0}$   &\phn$-11$  \\
9	&10088$-$5730~2   &10 10 46.616   &$-$57 43 56.77   &$1.2076\times10^{-1}$   &\phn$-11$  \\
10	&10095$-$5843~0   &10 11 14.465   &$-$58 58 11.68   &$1.1166\times10^{-1}$   &\phn$-11$  \\
11	&10276$-$5711~0   &10 29 31.326   &$-$57 26 49.26   &$2.3131\times10^{-1}$   &\phn$+15$  \\
12	&10295$-$5746~0   &10 31 29.501   &$-$58 02 17.06   &$3.3319\times10^{+2}$   &\phn$+15$  \\
13	&10317$-$5936~0   &10 33 36.712   &$-$59 51 32.09   &$1.7312\times10^{-1}$   &\phn$+29$  \\
14	&10337$-$5710~0   &10 35 40.044   &$-$57 26 05.82   &$3.9008\times10^{-1}$   &\phn$-11$  \\
15	&10501$-$5556~0   &10 52 12.860   &$-$56 12 41.01   &$6.5700\times10^{-2}$   &\phn$-11$  \\
16	&10545$-$6244~1   &10 57 33.437   &$-$62 59 02.56   &$2.4358\times10^{+1}$   &\phnn$+1$  \\
17	&10555$-$6242~0   &10 57 33.887   &$-$62 59 01.95   &$1.8009\times10^{+1}$   &\phnn$+0$  \\
18	&10559$-$5914~0   &10 58 00.455   &$-$59 30 40.67   &$7.8435\times10^{-2}$   &\phn$-66$  \\
19	&11380$-$6311~0   &11 40 29.273   &$-$63 27 54.88   &$3.8797\times10^{-1}$   &\phnn$-0$  \\
20	&12102$-$6133~1   &12 12 56.019   &$-$61 50 08.00   &$1.6179\times10^{-1}$   &\phn$-27$  \\
21	&11265$-$6158~0   &11 28 50.655   &$-$62 14 59.88   &$4.6810\times10^{-1}$   &\phn$-26$  \\
22	&12268$-$6156~0   &12 29 44.423   &$-$62 13 17.87   &$7.0320\times10^{-2}$   &\phn$-28$  \\
23	&12272$-$6240~0   &12 30 03.434   &$-$62 56 51.22   &$6.0996\times10^{+1}$   &\phn$+12$  \\
24	&11404$-$6215~0   &11 42 49.414   &$-$62 32 40.87   &$1.5767\times10^{-1}$   &\phn$+54$  \\
25	&13333$-$6234~0   &13 36 43.564   &$-$62 49 45.14   &$6.5621\times10^{-1}$   &\phnn$-2$  \\
26	&13395$-$6153~0   &13 43 01.604   &$-$62 08 53.06   &$9.9340\times10^{-1}$   &\phn$-63$  \\
27	&13438$-$6203~0   &13 47 23.748   &$-$62 18 10.40   &$8.5210\times10^{-1}$   &\phn$-50$  \\
28	&13534$-$6152~0   &13 57 05.959   &$-$62 07 52.40   &$5.9344\times10^{-1}$   &\phn$+52$  \\
29	&13560$-$6133~2   &13 59 30.762   &$-$61 48 40.12   &$4.7719\times10^{-1}$   &\phn$-50$  \\
30	&13585$-$6133~0   &14 02 27.156   &$-$61 45 13.05   &$3.5356\times10^{-1}$   &\phn$+31$  \\
31	&14000$-$6104~0   &14 03 35.149   &$-$61 18 26.87   &$1.6244\times10^{-1}$   &\phn$-63$  \\
32	&14201$-$6044~1   &14 24 30.603   &$-$60 56 36.28   &$4.3214\times10^{-1}$   &\phn$-17$  \\
33	&14214$-$6017~0   &14 25 14.179   &$-$60 32 21.79   &$1.1790\times10^{+0}$   &\phn$-62$  \\
34	&14394$-$6004~0   &14 43 11.057   &$-$60 17 16.35   &$1.9216\times10^{+0}$   &\phnn$+6$  \\
35	&14557$-$5849~0   &14 59 32.635   &$-$59 00 33.53   &$5.1293\times10^{+1}$   &$-138$     \\
36	&15061$-$5806~0   &15 10 00.528   &$-$58 17 30.35   &$6.0789\times10^{-1}$   &\phnn$-4$  \\
37	&15219$-$5658~0   &15 25 47.784   &$-$57 09 17.94   &$5.9701\times10^{-2}$   &\phn$-18$  \\
38	&15278$-$5620~0   &15 31 45.509   &$-$56 30 50.73   &$1.4787\times10^{+0}$   &\phn$-45$  \\
	&15278$-$5620~2\tablefootmark{b}
	                  &\ldots         &\ldots           &\ldots                  &\ldots     \\
39	&15454$-$5335~0   &15 49 19.409   &$-$53 45 18.64   &$4.2513\times10^{+0}$   &\phn$-86$  \\
40	&15454$-$5507~0   &15 49 18.628   &$-$55 16 53.84   &$2.6990\times10^{-1}$   &\phn$-72$  \\
41	&                 &15 49 17.947   &$-$55 16 51.61   &$1.4936\times10^{-1}$   &$-113$     \\
42	&15470$-$5419~1   &15 51 29.544   &$-$54 31 23.06   &$1.3684\times10^{-1}$   &\phn$-59$  \\
43	&15470$-$5419~3   &15 51 01.338   &$-$54 26 37.45   &$8.5271\times10^{-1}$   &\phn$-59$  \\
44	&15519$-$5430~0   &15 55 46.767   &$-$54 39 17.64   &$2.2018\times10^{-1}$   &\phn$-18$  \\
45	&15530$-$5231~0   &15 56 51.367   &$-$52 40 24.87   &$8.6716\times10^{-1}$   &\phn$-99$  \\
46	&15557$-$5215~0   &15 59 36.170   &$-$52 22 55.17   &$1.1532\times10^{+0}$   &\phn$-99$  \\
47	&15579$-$5303~0   &16 01 46.569   &$-$53 11 42.09   &$1.3761\times10^{+1}$   &\phn$-45$  \\
48	&16061$-$5048~1   &16 10 06.427   &$-$50 50 28.61   &$3.3032\times10^{+0}$   &\phn$-71$  \\
49	&16061$-$5048~4   &16 10 05.966   &$-$50 57 11.58   &$1.2589\times10^{-1}$   &\phn$-31$  \\
	&16061$-$5048~5\tablefootmark{c}
	                  &\ldots         &\ldots           &\ldots                  &\ldots     \\
50	&16069$-$4858~0   &16 10 39.103   &$-$49 05 56.75   &$9.5290\times10^{-2}$   &\phn$-17$  \\
51	&16093$-$5015~0   &16 13 01.495   &$-$50 22 36.76   &$2.0062\times10^{-1}$   &\phn$-31$  \\
52	&16112$-$4943~0   &16 15 05.703   &$-$49 50 26.96   &$5.3459\times10^{-1}$   &\phn$-44$  \\
53	&                 &16 14 59.418   &$-$49 50 41.30   &$3.2185\times10^{-1}$   &\phn$-44$  \\
54	&                 &16 14 06.703   &$-$49 46 09.56   &$1.9700\times10^{-1}$   &\phn$-44$  \\
55	&16153$-$5016~1   &16 18 56.682   &$-$50 23 51.12   &$2.7601\times10^{-1}$   &\phn$-45$  \\
56	&16204$-$4916~0   &16 24 15.247   &$-$49 22 24.62   &$3.0337\times10^{-1}$   &\phn$-84$  \\
57	&16218$-$4931~0   &16 25 38.389   &$-$49 38 13.02   &$3.2916\times10^{-1}$   &\phn$-99$  \\
58	&16252$-$4853~0   &16 29 00.539   &$-$48 58 57.91   &$1.1972\times10^{-1}$   &\phn$-56$  \\
59	&16254$-$4844~1   &16 29 01.156   &$-$48 50 32.09   &$2.3746\times10^{-1}$   &\phn$-43$  \\
60	&16344$-$4605~0   &16 38 09.058   &$-$46 11 02.29   &$3.8671\times10^{+1}$   &\phn$-58$  \\
61	&16363$-$4645~0   &16 39 59.623   &$-$46 50 20.40   &$3.3537\times10^{-1}$   &\phn$-69$  \\
62	&16435$-$4515~0   &16 47 02.887   &$-$45 21 27.95   &$5.0995\times10^{-1}$   &\phn$-55$  \\
63	&16464$-$4359~0   &16 50 01.312   &$-$44 05 03.35   &$2.6794\times10^{-1}$   &\phn$-81$  \\
64	&16482$-$4443~0   &16 51 46.752   &$-$44 48 54.80   &$1.0681\times10^{-1}$   &\phn$-27$  \\
65	&16573$-$4214~2   &17 00 32.633   &$-$42 25 08.71   &$8.6345\times10^{+0}$   &\phn$-26$  \\
66	&17033$-$4035~0   &17 06 44.806   &$-$40 39 29.09   &$3.1869\times10^{-1}$   &\phn$-39$  \\
67	&17040$-$3959~0   &17 07 33.739   &$-$40 03 08.75   &$1.7729\times10^{-1}$   &\phnn$-4$  \\
68	&17082$-$4114~0   &17 11 45.220   &$-$41 17 49.42   &$2.6972\times10^{-1}$   &\phn$-45$  \\
69	&17118$-$3909~0   &17 15 20.373   &$-$39 13 41.13   &$2.0184\times10^{-1}$   &$-132$     \\
70	&17149$-$3916~0   &17 18 23.467   &$-$39 18 40.29   &$5.8627\times10^{-1}$   &\phn$+57$  \\
71	&17184$-$3638~0   &17 21 50.382   &$-$36 41 13.95   &$8.2766\times10^{-1}$   &\phn$+10$  \\
72	&17195$-$3811~0   &17 23 00.277   &$-$38 14 58.11   &$8.4816\times10^{-1}$   &\phn$-18$  \\
73	&17195$-$3811~2   &17 23 00.399   &$-$38 14 55.42   &$1.0467\times10^{+0}$   &\phn$-50$  \\
74	&                 &17 23 01.191   &$-$38 13 52.39   &$2.5878\times10^{-1}$   &\phn$-37$  \\
75	&                 &17 22 57.048   &$-$38 14 37.09   &$2.0529\times10^{-1}$   &\phn$-24$  \\
76	&                 &17 23 00.216   &$-$38 15 00.55   &$3.4993\times10^{-1}$   &\phn$-10$  \\
77	&17200$-$3658~0   &17 23 28.498   &$-$37 01 48.34   &$3.1568\times10^{+0}$   &\phn$-17$  \\
78	&17221$-$3619~1   &17 25 32.111   &$-$36 22 05.75   &$1.4298\times10^{-1}$   &\phn$-17$  \\
79	&17230$-$3531~0   &17 26 26.572   &$-$35 33 44.14   &$1.8233\times10^{-1}$   &$-125$     \\
80	&                 &17 26 27.454   &$-$35 34 42.03   &$9.5942\times10^{-2}$   &\phn$-84$  \\
81	&17249$-$3501~0   &17 28 17.809   &$-$35 04 12.48   &$2.4608\times10^{-1}$   &\phn$-57$  \\
82	&17279$-$3350~0   &17 31 15.804   &$-$33 51 51.72   &$5.8603\times10^{-1}$   &\phn$-30$  \\
83	&17377$-$3109~0   &17 40 57.067   &$-$31 11 00.35   &$2.0537\times10^{+1}$   &\phnn$-3$  \\
84	&18018$-$2426~9   &18 04 53.462   &$-$24 26 49.52   &$5.4888\times10^{-1}$   &\phn$+11$  \\
85	&18144$-$1723~0   &18 17 23.980   &$-$17 22 14.08   &$5.2644\times10^{+0}$   &\phn$+40$  \\
\end{longtable}
\tablefoot{
\tablefoottext{a}{Emission integrated over all the velocity range. In most of the cases the water maser emission is only detected in one channel of 13.5~\kms.}
\tablefoottext{b}{The maser emission seen in the map is that also detected in 15278$-$5620~0.}
\tablefoottext{c}{The maser emission seen in the map is that also detected in 16061$-$5048~1.}
}
\end{scriptsize}
}

\longtab{5}{
\begin{scriptsize}
\begin{longtable}{l c c c c c c c c c c c}
\caption{\label{t:physpar} Physical parameters of optically thin \hii\ regions, and spectral indices}\\
\hline\hline
&
&Freq.
&Dist.
&radius
&$T_\mathrm{B}$
&$n_\mathrm{e}$
&EM
&$M_\mathrm{i}$
&$\log\Nly$
&Spectral
&Spectral
\\
Region
&\#
&(GHz)
&(kpc)
&(pc)
&(K)
&(cm$^{-3}$)
&(pc~cm$^{-6}$)
&(\mo)
&(s$^{-1}$)
&type
&index
\\
\hline
\endfirsthead
\caption{continued.}\\
\hline\hline
&
&Freq.
&Dist.
&radius
&$T_\mathrm{B}$
&$n_\mathrm{e}$
&EM
&$M_\mathrm{i}$
&$\log\Nly$
&Spectral
&Spectral
\\
Region
&\#
&(GHz)
&(kpc)
&(pc)
&(K)
&(cm$^{-3}$)
&(pc~cm$^{-6}$)
&(\mo)
&(s$^{-1}$)
&type
&index
\\
\hline
\endhead
\hline
\endfoot
08140$-$3559~0  &1  &18.0  &\phn3.8  &\phb0.102  &\phb$3.28\times10^{-1}$  &\phb$6.00\times10^{+2}$  &\phb$4.91\times10^{+4}$  &\phb$6.64\times10^{-2}$  &46.26  &B0.5       &\phn$-$0.0  \\
08211$-$4158~0  &1  &18.0  &\phn1.7  &\phb0.093  &\phb$2.11\times10^{-2}$  &\phb$1.59\times10^{+2}$  &\phb$3.16\times10^{+3}$  &\phb$1.34\times10^{-2}$  &44.99  &B2         &\phn$+$0.5  \\
                &2  &22.8  &\phn1.7  &\phb0.044  &\phb$4.39\times10^{-2}$  &\phb$4.30\times10^{+2}$  &\phb$1.08\times10^{+4}$  &\phb$3.74\times10^{-3}$  &44.87  &B2         &\phn$+$3.2  \\
08438$-$4340~0  &1  &18.0  &\phn1.3  &\phb0.107  &\phb$4.58\times10^{-1}$  &\phb$6.92\times10^{+2}$  &\phb$6.85\times10^{+4}$  &\phb$8.82\times10^{-2}$  &46.45  &B0.5       &\phn$-$2.9  \\
                &2  &18.0  &\phn1.3  &\phb0.157  &\phb$5.98\times10^{-1}$  &\phb$6.54\times10^{+2}$  &\phb$8.94\times10^{+4}$  &\phb$2.60\times10^{-1}$  &46.89  &B0.5       &\phn$-$3.2  \\
                &3  &22.8  &\phn1.3  &\phb0.053  &\phb$6.50\times10^{-1}$  &\phb$1.51\times10^{+3}$  &\phb$1.60\times10^{+5}$  &\phb$2.29\times10^{-2}$  &46.20  &B0.5       &\phn$+$2.4  \\
08470$-$4243~0  &1  &18.0  &\phn2.2  &\phb0.120  &\phb$9.72\times10^{-2}$  &\phb$3.01\times10^{+2}$  &\phb$1.45\times10^{+4}$  &\phb$5.40\times10^{-2}$  &45.87  &B0.5       &\phn$+$0.5  \\
08477$-$4359~0  &1  &22.8  &\phn1.8  &\phb0.059  &\phb$2.23\times10^{-1}$  &\phb$8.35\times10^{+2}$  &\phb$5.49\times10^{+4}$  &\phb$1.78\times10^{-2}$  &45.83  &B0.5       &\phn$+$2.1  \\
\end{longtable}
\end{scriptsize}
}

\longtab{7}{
\begin{scriptsize}
\begin{longtable}{l c c c c l c c l c c c}
\caption{\label{t:type1} Sources classified as \emph{type~1} objects: millimeter-only sources}\\
\hline\hline
\multicolumn{4}{c}{millimeter\taba}
&
&\multicolumn{2}{c}{centimeter\tabb}
&
&\multicolumn{2}{c}{H$_2$O maser\tabc}
\\\cline{1-4}\cline{6-7}\cline{9-10}
Region
&\#
&$\alpha_\mathrm{J2000}$
&$\delta_\mathrm{J2000}$
&
&Region
&\#
&
&Region
&\#
\\
\hline
\endfirsthead
\caption{continued.}\\
\hline\hline
\multicolumn{4}{c}{millimeter\taba}
&
&\multicolumn{2}{c}{centimeter\tabb}
&
&\multicolumn{2}{c}{H$_2$O maser\tabc}
\\\cline{1-4}\cline{6-7}\cline{9-10}
Region
&\#
&$\alpha_\mathrm{J2000}$
&$\delta_\mathrm{J2000}$
&
&Region
&\#
&
&Region
&\#
\\
\hline
\endhead
\hline
\endfoot
08438$-$4340  &\phn2  &131.382  &$-$43.839  &&                &   &&                &    \\
08438$-$4340  &\phn1  &131.394  &$-$43.868  &&                &   &&08438$-$4340~0  &\phn1  \\
08470$-$4243  &\phn1  &132.197  &$-$42.911  &&                &   &&                &    \\
08477$-$4359  &\phn3  &132.378  &$-$44.191  &&                &   &&                &    \\
08477$-$4359  &\phn2  &132.378  &$-$44.184  &&                &   &&                &    \\
08477$-$4359  &\phn1  &132.396  &$-$44.200  &&                &   &&                &    \\
08563$-$4225  &\phn4  &134.540  &$-$42.644  &&                &   &&                &    \\
09014$-$4736  &\phn2  &135.784  &$-$47.812  &&                &   &&                &    \\
\end{longtable}
\tablefoot{
\tablefoottext{a}{Millimeter source: Region indicates the IRAS name, \# indicates the millimeter clump from Beltr\'an \et\ (2006), $\alpha_\mathrm{J2000}$ and $\delta_\mathrm{J2000}$ are the right ascension (in degrees) and the declination (in degrees).}
\tablefoottext{b}{Centimeter source: Region indicates the name of the ATCA field (see Table~\ref{t:observations}), \# indicates the number of the ATCA source detected in the observed field (see Table~\ref{t:results}).}
\tablefoottext{c}{H$_2$O maser spot: Region indicates the name of the ATCA field (see Table~\ref{t:observations}), \# indicates the number of the water maser component as listed in Table~\ref{t:masers}.}
}
\end{scriptsize}
}

\longtab{8}{
\begin{scriptsize}
\begin{longtable}{l c c c c l c c c l c c l c c c}
\caption{\label{t:type2} Sources classified as \emph{type~2} objects: millimeter and infrared sources}\\
\hline\hline
\multicolumn{4}{c}{millimeter\taba}
&
&\multicolumn{3}{c}{infrared\tabb}
&
&\multicolumn{2}{c}{centimeter\tabc}
&
&\multicolumn{2}{c}{H$_2$O maser\tabd}
\\\cline{1-4}\cline{6-8}\cline{10-11}\cline{13-14}
Region
&\#
&$\alpha_\mathrm{J2000}$
&$\delta_\mathrm{J2000}$
&
&MSX name
&$\alpha_\mathrm{J2000}$
&$\delta_\mathrm{J2000}$
&
&Region
&\#
&
&Region
&\#
\\
\hline
\endfirsthead
\caption{continued.}\\
\hline\hline
\multicolumn{4}{c}{millimeter\taba}
&
&\multicolumn{3}{c}{infrared\tabb}
&
&\multicolumn{2}{c}{centimeter\tabc}
&
&\multicolumn{2}{c}{H$_2$O maser\tabd}
\\\cline{1-4}\cline{6-8}\cline{10-11}\cline{13-14}
Region
&\#
&$\alpha_\mathrm{J2000}$
&$\delta_\mathrm{J2000}$
&
&MSX name
&$\alpha_\mathrm{J2000}$
&$\delta_\mathrm{J2000}$
&
&Region
&\#
&
&Region
&\#
\\
\hline
\endhead
\hline
\endfoot
08140$-$3559  &\phn1  &123.985  &$-$36.134  &&G254.0491$-$00.5615  &123.988  &$-$36.136  &&                &   &&                &    \\
08211$-$4158  &\phn1  &125.718  &$-$42.135  &&G259.7592$-$02.8378  &125.718  &$-$42.132  &&08211$-$4158~0  &1  &&                &    \\
08563$-$4225  &\phn1  &134.549  &$-$42.628  &&G264.1444$+$02.0190  &134.553  &$-$42.627  &&08563$-$4225~0  &1  &&08563$-$4225~0  &\phn3  \\
08589$-$4714  &\phn1  &135.165  &$-$47.436  &&G268.0594$-$00.8040  &135.171  &$-$47.434  &&                &   &&08589$-$4714~0  &\phn4  \\
09014$-$4736  &\phn1  &135.764  &$-$47.808  &&G268.6056$-$00.7476  &135.770  &$-$47.806  &&                &   &&                &    \\
09026$-$4842  &\phn3  &136.096  &$-$48.910  &&G269.5703$-$01.3202  &136.094  &$-$48.907  &&09026$-$4842~0  &1  &&                &    \\
09131$-$4723  &\phn1  &138.721  &$-$47.597  &&G269.8017$+$00.8510  &138.721  &$-$47.599  &&09131$-$4723~0  &1  &&                &    \\
09209$-$5143  &\phn3  &140.651  &$-$51.944  &&G273.7699$-$01.3089  &140.650  &$-$51.943  &&09209$-$5143~0  &1  &&                &    \\
09566$-$5607  &\phn1  &149.593  &$-$56.371  &&G280.6208$-$01.1879  &149.597  &$-$56.371  &&09566$-$5607~0  &1  &&                &    \\
09578$-$5649  &\phn1  &149.883  &$-$57.063  &&G281.1641$-$01.6418  &149.879  &$-$57.061  &&09578$-$5649~0  &1  &&                &    \\
10019$-$5712  &\phn1  &150.919  &$-$57.446  &&G281.8449$-$01.6094  &150.921  &$-$57.444  &&10019$-$5712~0  &1  &&10019$-$5712~0  &\phn8  \\
10038$-$5705  &\phn1  &151.379  &$-$57.332  &&G281.9780$-$01.3712  &151.381  &$-$57.331  &&                &   &&                &    \\
10095$-$5843  &\phn1  &152.812  &$-$58.973  &&G283.5475$-$02.2655  &152.815  &$-$58.971  &&10095$-$5843~0  &1  &&10095$-$5843~0  &10  \\
10123$-$5727  &\phn3  &153.574  &$-$57.690  &&G283.1465$-$00.9828  &153.575  &$-$57.688  &&                &   &&                &    \\
10123$-$5727  &\phn1  &153.533  &$-$57.697  &&G283.1277$-$01.0041  &153.524  &$-$57.695  &&10123$-$5727~0  &1  &&                &    \\
10184$-$5748  &\phn1  &155.065  &$-$58.067  &&G284.0155$-$00.8579  &155.065  &$-$58.066  &&10184$-$5748~0  &2  &&                &    \\
10276$-$5711  &\phn1  &157.388  &$-$57.447  &&G284.7336$+$00.3265  &157.385  &$-$57.443  &&10276$-$5711~0  &1  &&                &    \\
10286$-$5838  &\phn1  &157.636  &$-$58.898  &&G285.5970$-$00.8508  &157.639  &$-$58.897  &&10286$-$5838~0  &2  &&                &    \\
10295$-$5746  &\phn1  &157.872  &$-$58.040  &&G285.2611$-$00.0492  &157.871  &$-$58.037  &&10295$-$5746~0  &1  &&10295$-$5746~0  &12  \\
10320$-$5928  &\phn1  &158.485  &$-$59.734  &&G286.3938$-$01.3514  &158.485  &$-$59.733  &&10320$-$5928~0  &1  &&                &    \\
10337$-$5710  &\phn1  &158.915  &$-$57.438  &&G285.4410$+$00.7510  &158.918  &$-$57.436  &&10337$-$5710~0  &1  &&10337$-$5710~0  &14  \\
10439$-$5941  &\phn1  &161.479  &$-$59.949  &&G287.8134$-$00.8140  &161.476  &$-$59.949  &&10439$-$5941~0  &1  &&                &    \\
10501$-$5556  &\phn1  &163.046  &$-$56.212  &&G286.8574$+$02.8949  &163.044  &$-$56.209  &&10501$-$5556~0  &1  &&10501$-$5556~0  &15  \\
10555$-$6242  &\phn1  &164.389  &$-$62.986  &&G290.4086$-$02.9130  &164.389  &$-$62.982  &&10545$-$6244~1  &1  &&10545$-$6244~1  &16  \\
10559$-$5914  &\phn1  &164.462  &$-$59.493  &&G288.9606$+$00.2645  &164.464  &$-$59.490  &&                &   &&                &    \\
10589$-$6034  &\phn1  &165.245  &$-$60.841  &&G289.8794$-$00.7979  &165.249  &$-$60.840  &&10589$-$6034~0  &1  &&                &    \\
11079$-$6101  &\phn1  &167.501  &$-$61.309  &&G291.0670$-$00.7941  &167.506  &$-$61.308  &&11079$-$6101~0  &1  &&                &    \\
11220$-$6147  &\phn1  &171.066  &$-$62.068  &&G292.9183$-$00.9020  &171.073  &$-$62.068  &&11220$-$6147~0  &1  &&                &    \\
11265$-$6158  &\phn1  &172.212  &$-$62.252  &&G293.4823$-$00.9047  &172.214  &$-$62.252  &&11265$-$6158~0  &2  &&11265$-$6158~0  &21  \\
12063$-$6259  &\phn1  &182.250  &$-$63.268  &&G298.1829$-$00.7860  &182.255  &$-$63.266  &&12063$-$6259~0  &1  &&                &    \\
12127$-$6244  &\phn5  &183.864  &$-$63.040  &&G298.8691$-$00.4515  &183.869  &$-$63.039  &&                &   &&                &    \\
12127$-$6244  &\phn1  &183.845  &$-$63.027  &&G298.8591$-$00.4372  &183.852  &$-$63.024  &&12127$-$6244~0  &1  &&                &    \\
12132$-$6211  &\phn2  &183.990  &$-$62.468  &&G298.8449$+$00.1221  &183.990  &$-$62.468  &&12132$-$6211~0  &2  &&                &    \\
12268$-$6156  &\phn1  &187.425  &$-$62.221  &&G300.4013$+$00.5458  &187.424  &$-$62.219  &&12268$-$6156~0  &1  &&                &    \\
12272$-$6240  &\phn1  &187.517  &$-$62.949  &&G300.5035$-$00.1804  &187.513  &$-$62.951  &&                &   &&12272$-$6240~0  &23  \\
12300$-$6119  &\phn1  &188.208  &$-$61.593  &&G300.7221$+$01.2007  &188.212  &$-$61.591  &&                &   &&                &    \\
13023$-$6213  &\phn1  &196.383  &$-$62.499  &&G304.5572$+$00.3279  &196.381  &$-$62.499  &&13023$-$6213~0  &1  &&                &    \\
13024$-$6158  &\phn1  &196.411  &$-$62.246  &&G304.5848$+$00.5804  &196.411  &$-$62.246  &&13024$-$6158~0  &1  &&                &    \\
13039$-$6331  &\phn1  &196.784  &$-$63.787  &&G304.6668$-$00.9654  &196.789  &$-$63.784  &&                &   &&                &    \\
13054$-$6159  &\phn2  &197.160  &$-$62.259  &&G304.9323$+$00.5462  &197.160  &$-$62.259  &&                &   &&                &    \\
13106$-$6050  &\phn2  &198.460  &$-$61.110  &&G305.6327$+$01.6467  &198.454  &$-$61.108  &&13106$-$6050~0  &1  &&                &    \\
13107$-$6208  &\phn1  &198.496  &$-$62.416  &&G305.5393$+$00.3394  &198.498  &$-$62.418  &&                &   &&                &    \\
13333$-$6234  &\phn2  &204.190  &$-$62.831  &&G308.0813$-$00.4131  &204.195  &$-$62.831  &&13333$-$6234~0  &2  &&13333$-$6234~0  &25  \\
13333$-$6234  &\phn1  &204.142  &$-$62.818  &&G308.0568$-$00.3960  &204.136  &$-$62.818  &&13333$-$6234~0  &1  &&                &    \\
13384$-$6152  &\phn2  &205.757  &$-$62.151  &&G308.9176$+$00.1231  &205.757  &$-$62.148  &&13395$-$6153~0  &2  &&13395$-$6153~0  &26  \\
13395$-$6153  &\phn1  &205.757  &$-$62.145  &&G308.9176$+$00.1231  &205.757  &$-$62.148  &&                &   &&13395$-$6153~0  &26  \\
13481$-$6124  &\phn2  &207.907  &$-$61.652  &&G310.0135$+$00.3892  &207.908  &$-$61.652  &&13481$-$6124~0  &1  &&                &    \\
13534$-$6152  &\phn1  &209.257  &$-$62.117  &&G310.5163$-$00.2179  &209.254  &$-$62.120  &&13534$-$6152~0  &1  &&                &    \\
13563$-$6109  &\phn1  &209.991  &$-$61.409  &&G311.0341$+$00.3791  &209.988  &$-$61.410  &&13563$-$6109~0  &1  &&                &    \\
13585$-$6133  &\phn1  &210.533  &$-$61.808  &&G311.1794$-$00.0720  &210.535  &$-$61.806  &&13585$-$6133~0  &1  &&                &    \\
13590$-$6051  &\phn2  &210.654  &$-$61.097  &&G311.4255$+$00.5964  &210.652  &$-$61.096  &&13590$-$6051~0  &1  &&                &    \\
13592$-$6153  &\phn1  &210.720  &$-$62.124  &&G311.1771$-$00.3998  &210.720  &$-$62.123  &&13592$-$6153~0  &1  &&                &    \\
14000$-$6104  &\phn3  &210.903  &$-$61.306  &&G311.4824$+$00.3602  &210.899  &$-$61.308  &&14000$-$6104~0  &1  &&14000$-$6104~0  &31  \\
14039$-$6113  &\phn1  &211.901  &$-$61.455  &&G311.8998$+$00.0812  &211.900  &$-$61.457  &&14039$-$6113~0  &1  &&                &    \\
14057$-$6032  &\phn1  &212.352  &$-$60.784  &&G312.3070$+$00.6613  &212.353  &$-$60.783  &&14057$-$6032~0  &1  &&                &    \\
14090$-$6132  &\phn1  &213.181  &$-$61.788  &&G312.3825$-$00.4143  &213.182  &$-$61.785  &&14090$-$6132~0  &2  &&                &    \\
14166$-$6118  &\phn1  &215.081  &$-$61.531  &&G313.3153$-$00.4640  &215.075  &$-$61.531  &&14166$-$6118~0  &1  &&                &    \\
14183$-$6050  &\phn1  &215.645  &$-$61.143  &&G313.7051$-$00.1895  &215.644  &$-$61.141  &&                &   &&                &    \\
14188$-$6054  &\phn1  &215.647  &$-$61.143  &&G313.7051$-$00.1895  &215.644  &$-$61.141  &&                &   &&                &    \\
14214$-$6017  &\phn1  &216.306  &$-$60.528  &&G314.2204$+$00.2726  &216.304  &$-$60.527  &&                &   &&                &    \\
14425$-$6023  &\phn1  &221.602  &$-$60.598  &&G316.5871$-$00.8086  &221.597  &$-$60.596  &&14425$-$6023~0  &1  &&                &    \\
14557$-$5849  &\phn2  &224.895  &$-$59.025  &&G318.7748$-$00.1513  &224.894  &$-$59.024  &&14557$-$5849~0  &1  &&                &    \\
15061$-$5806  &\phn1  &227.501  &$-$58.292  &&G320.3150$-$00.1770  &227.501  &$-$58.293  &&15061$-$5806~0  &1  &&15061$-$5806~0  &36  \\
15068$-$5733  &\phn1  &227.684  &$-$57.748  &&G320.6747$+$00.2452  &227.682  &$-$57.746  &&15068$-$5733~0  &1  &&                &    \\
15100$-$5613  &\phn2  &228.432  &$-$56.426  &&G321.7020$+$01.1747  &228.427  &$-$56.422  &&                &   &&                &    \\
15100$-$5613  &\phn1  &228.452  &$-$56.413  &&G321.7209$+$01.1711  &228.460  &$-$56.415  &&15100$-$5613~0  &1  &&                &    \\
15178$-$5641  &\phn3  &230.443  &$-$56.878  &&G322.3954$+$00.2031  &230.434  &$-$56.878  &&15178$-$5641~0  &1  &&                &    \\
15178$-$5641  &\phn1  &230.423  &$-$56.876  &&G322.3946$+$00.2093  &230.426  &$-$56.873  &&                &   &&                &    \\
15219$-$5658  &\phn1  &231.449  &$-$57.153  &&G322.7056$-$00.3283  &231.446  &$-$57.152  &&15219$-$5658~0  &1  &&15219$-$5658~0  &37  \\
15239$-$5538  &\phn1  &231.975  &$-$55.814  &&G323.6920$+$00.6228  &231.966  &$-$55.812  &&                &   &&                &    \\
15246$-$5612  &\phn1  &232.132  &$-$56.386  &&G323.4468$+$00.0968  &232.133  &$-$56.385  &&15246$-$5612~0  &1  &&                &    \\
15278$-$5620  &\phn1  &232.938  &$-$56.513  &&G323.7399$-$00.2617  &232.938  &$-$56.513  &&                &   &&15278$-$5620~0  &38  \\
15347$-$5518  &\phn1  &234.642  &$-$55.469  &&G325.1240$+$00.0312  &234.643  &$-$55.468  &&15347$-$5518~0  &1  &&                &    \\
15454$-$5335  &\phn1  &237.332  &$-$53.754  &&G327.4014$+$00.4454  &237.329  &$-$53.754  &&15454$-$5335~0  &1  &&15454$-$5335~0  &39  \\
15470$-$5419  &\phn2  &237.726  &$-$54.484  &&G327.1309$-$00.2655  &237.728  &$-$54.477  &&15470$-$5419~0  &1  &&                &    \\
15507$-$5359  &\phn2  &238.660  &$-$54.138  &&G327.7726$-$00.3426  &238.663  &$-$54.131  &&                &   &&                &    \\
15519$-$5430  &\phn1  &238.948  &$-$54.654  &&G327.5646$-$00.8489  &238.945  &$-$54.653  &&15519$-$5430~0  &1  &&15519$-$5430~0  &44  \\
15530$-$5231  &\phn1  &239.213  &$-$52.674  &&G328.9580$+$00.5671  &239.211  &$-$52.673  &&15530$-$5231~0  &2  &&15530$-$5231~0  &45  \\
15557$-$5337  &\phn1  &239.902  &$-$53.766  &&G328.5669$-$00.5327  &239.903  &$-$53.764  &&15557$-$5337~0  &1  &&                &    \\
15579$-$5347  &\phn1  &240.469  &$-$53.941  &&G328.7050$-$00.8841  &240.470  &$-$53.939  &&15579$-$5347~0  &1  &&                &    \\
15583$-$5314  &\phn1  &240.546  &$-$53.376  &&G329.1087$-$00.4908  &240.547  &$-$53.378  &&15583$-$5314~0  &1  &&                &    \\
16056$-$5125  &\phn2  &242.348  &$-$51.552  &&G331.1465$+$00.1343  &242.352  &$-$51.552  &&16056$-$5125~0  &1  &&                &    \\
16069$-$4858  &\phn1  &242.663  &$-$49.102  &&G332.9565$+$01.8035  &242.661  &$-$49.098  &&                &   &&16069$-$4858~0  &50  \\
16085$-$5138  &\phn6  &243.081  &$-$51.766  &&G331.3312$-$00.3316  &243.082  &$-$51.768  &&16085$-$5138~0  &1  &&                &    \\
16085$-$5138  &\phn2  &243.110  &$-$51.771  &&G331.3402$-$00.3444  &243.107  &$-$51.771  &&                &   &&                &    \\
16085$-$5138  &\phn1  &243.207  &$-$51.729  &&G331.4181$-$00.3546  &243.209  &$-$51.725  &&16093$-$5128~2  &1  &&                &    \\
16093$-$5128  &\phn4  &243.295  &$-$51.607  &&G331.5383$-$00.3072  &243.298  &$-$51.608  &&16093$-$5128~0  &1  &&                &    \\
16093$-$5128  &\phn1  &243.206  &$-$51.725  &&G331.4181$-$00.3546  &243.209  &$-$51.725  &&16093$-$5128~2  &1  &&                &    \\
16112$-$4943  &\phn3  &243.752  &$-$49.846  &&G332.9639$+$00.7718  &243.758  &$-$49.844  &&16112$-$4943~0  &1  &&16112$-$4943~0  &53  \\
16148$-$5011  &\phn2  &244.661  &$-$50.317  &&G333.0494$+$00.0324  &244.655  &$-$50.316  &&16148$-$5011~0  &1  &&                &    \\
16148$-$5011  &\phn1  &244.737  &$-$50.399  &&G333.0299$-$00.0645  &244.739  &$-$50.399  &&                &   &&16153$-$5016~1  &55  \\
16153$-$5016  &\phn1  &244.736  &$-$50.399  &&G333.0299$-$00.0645  &244.739  &$-$50.399  &&16153$-$5016~1  &1  &&                &    \\
16164$-$4929  &\phn2  &245.078  &$-$49.582  &&G333.7595$+$00.3624  &245.082  &$-$49.583  &&16164$-$4929~3  &1  &&                &    \\
16164$-$4929  &\phn1  &245.040  &$-$49.606  &&G333.7261$+$00.3678  &245.039  &$-$49.603  &&16164$-$4929~0  &1  &&                &    \\
16164$-$4837  &\phn1  &245.054  &$-$48.754  &&G334.3321$+$00.9645  &245.056  &$-$48.752  &&16164$-$4837~0  &1  &&                &    \\
16194$-$4934  &\phn2  &245.822  &$-$49.685  &&G334.0256$-$00.0502  &245.822  &$-$49.687  &&16194$-$4934~0  &1  &&                &    \\
16204$-$4916  &\phn1  &246.058  &$-$49.395  &&G334.3417$+$00.0455  &246.060  &$-$49.395  &&16204$-$4916~0  &1  &&                &    \\
16218$-$4931  &\phn1  &246.408  &$-$49.639  &&G334.3272$-$00.2884  &246.412  &$-$49.639  &&16218$-$4931~0  &1  &&16218$-$4931~0  &57  \\
16252$-$4853  &\phn2  &247.263  &$-$48.999  &&G335.1734$-$00.2428  &247.262  &$-$48.998  &&16252$-$4853~0  &1  &&                &    \\
16344$-$4605  &\phn2  &249.552  &$-$46.206  &&G338.2717$+$00.5211  &249.553  &$-$46.205  &&                &   &&                &    \\
16344$-$4605  &\phn1  &249.539  &$-$46.184  &&G338.2801$+$00.5419  &249.538  &$-$46.185  &&16344$-$4605~0  &1  &&16344$-$4605~0  &60  \\
16363$-$4645  &\phn3  &250.018  &$-$46.854  &&G338.0008$-$00.1498  &250.018  &$-$46.854  &&16363$-$4645~0  &2  &&                &    \\
16396$-$4429  &\phn1  &250.817  &$-$44.589  &&G340.0708$+$00.9267  &250.817  &$-$44.588  &&16396$-$4429~0  &1  &&                &    \\
16435$-$4515  &\phn2  &251.790  &$-$45.350  &&G339.9426$-$00.0940  &251.793  &$-$45.351  &&16435$-$4515~0  &1  &&                &    \\
16435$-$4515  &\phn1  &251.762  &$-$45.359  &&G339.9267$-$00.0837  &251.767  &$-$45.356  &&                &   &&16435$-$4515~0  &62  \\
16482$-$4443  &\phn1  &252.942  &$-$44.816  &&G340.8781$-$00.3752  &252.949  &$-$44.814  &&16482$-$4443~0  &2  &&16482$-$4443~0  &64  \\
16501$-$4314  &\phn2  &253.415  &$-$43.321  &&G342.2495$+$00.3082  &253.423  &$-$43.320  &&16501$-$4314~0  &1  &&                &    \\
16501$-$4314  &\phn1  &253.427  &$-$43.325  &&G342.2477$+$00.3056  &253.424  &$-$43.323  &&                &   &&                &    \\
16574$-$4225  &\phn2  &255.244  &$-$42.503  &&G343.7217$-$00.2231  &255.244  &$-$42.500  &&16574$-$4225~0  &1  &&                &    \\
16579$-$4245  &\phn1  &255.392  &$-$42.837  &&G343.5213$-$00.5171  &255.392  &$-$42.839  &&                &   &&                &    \\
17036$-$4033  &\phn5  &256.790  &$-$40.619  &&G345.9184$+$00.0017  &256.789  &$-$40.619  &&17036$-$4033~0  &2  &&                &    \\
17118$-$3909  &\phn1  &258.824  &$-$39.225  &&G347.9676$-$00.4321  &258.826  &$-$39.223  &&17118$-$3909~0  &1  &&                &    \\
17149$-$3916  &\phn1  &259.599  &$-$39.322  &&G348.2362$-$00.9809  &259.605  &$-$39.322  &&17149$-$3916~0  &1  &&                &    \\
17184$-$3638  &\phn1  &260.458  &$-$36.686  &&G350.7833$-$00.0273  &260.461  &$-$36.687  &&17184$-$3638~0  &1  &&17184$-$3638~0  &71  \\
17195$-$3811  &\phn1  &260.754  &$-$38.232  &&G349.6433$-$01.0957  &260.756  &$-$38.232  &&                &   &&17195$-$3811~2  &74  \\
17200$-$3658  &\phn1  &260.869  &$-$37.027  &&G350.6879$-$00.4909  &260.871  &$-$37.029  &&17200$-$3658~0  &1  &&17200$-$3658~0  &77  \\
17210$-$3646  &\phn1  &261.122  &$-$36.813  &&G350.9780$-$00.5384  &261.125  &$-$36.816  &&17210$-$3646~0  &1  &&                &    \\
17221$-$3619  &\phn2  &261.381  &$-$36.357  &&G351.4768$-$00.4525  &261.385  &$-$36.355  &&                &   &&                &    \\
17221$-$3619  &\phn1  &261.381  &$-$36.368  &&G351.4661$-$00.4582  &261.383  &$-$36.367  &&17221$-$3619~1  &1  &&17221$-$3619~1  &78  \\
17225$-$3426  &\phn7  &261.445  &$-$34.497  &&G353.0473$+$00.5558  &261.439  &$-$34.489  &&                &   &&                &    \\
17242$-$3513  &\phn3  &261.872  &$-$35.268  &&G352.5975$-$00.1762  &261.874  &$-$35.271  &&17242$-$3513~0  &2  &&                &    \\
17242$-$3513  &\phn2  &261.869  &$-$35.288  &&G352.5844$-$00.1810  &261.870  &$-$35.284  &&17242$-$3513~0  &1  &&                &    \\
17249$-$3501  &\phn1  &262.074  &$-$35.069  &&G352.8577$-$00.2040  &262.079  &$-$35.070  &&17249$-$3501~0  &1  &&17249$-$3501~0  &81  \\
17256$-$3631  &\phn1  &262.257  &$-$36.556  &&G351.6976$-$01.1477  &262.255  &$-$36.558  &&17256$-$3631~0  &1  &&                &    \\
17279$-$3350  &\phn2  &262.822  &$-$33.881  &&G354.1882$-$00.0598  &262.823  &$-$33.881  &&17279$-$3350~0  &1  &&                &    \\
17296$-$3236  &\phn1  &263.217  &$-$32.643  &&G355.4049$+$00.3419  &263.217  &$-$32.642  &&17296$-$3236~0  &1  &&                &    \\
17352$-$3153  &\phn1  &264.620  &$-$31.911  &&G356.6620$-$00.2638  &264.623  &$-$31.911  &&17352$-$3153~0  &1  &&                &    \\
17355$-$3241  &\phn1  &264.708  &$-$32.724  &&G356.0124$-$00.7616  &264.710  &$-$32.726  &&                &   &&                &    \\
17377$-$3109  &\phn1  &265.238  &$-$31.183  &&G357.5571$-$00.3208  &265.238  &$-$31.183  &&17377$-$3109~0  &1  &&17377$-$3109~0  &83  \\
18014$-$2428  &\phn2  &271.219  &$-$24.445  &&G006.0485$-$01.4468  &271.221  &$-$24.445  &&18018$-$2426~9  &2  &&18018$-$2426~9  &84  \\
18144$-$1723  &\phn1  &274.349  &$-$17.368  &&G013.6554$-$00.5949  &274.347  &$-$17.369  &&                &   &&                &    \\
18198$-$1429  &\phn1  &275.673  &$-$14.462  &&G016.8217$-$00.3484  &275.674  &$-$14.461  &&18198$-$1429~0  &1  &&                &    \\
\end{longtable}
\tablefoot{
\tablefoottext{a}{Millimeter source: columns as in Table~\ref{t:type1}.}
\tablefoottext{b}{Infrared source: MSX name as listed in the MSX Point Source Catalogue (Price \et\ 1999), $\alpha_\mathrm{J2000}$ and $\delta_\mathrm{J2000}$ right ascension (in degrees) and declination (in degrees).}
\tablefoottext{c}{Centimeter source: columns as in Table~\ref{t:type1}.}
\tablefoottext{d}{H$_2$ maser spot: columns as in Table~\ref{t:type1}.}
}
\end{scriptsize}
}

\longtab{9}{
\begin{scriptsize}
\begin{longtable}{l c c c l c c l c c c}
\caption{\label{t:type3} Sources classified as \emph{type~3} objects: infrared-only sources}\\
\hline\hline
\multicolumn{3}{c}{infrared\taba}
&
&\multicolumn{2}{c}{centimeter\tabb}
&
&\multicolumn{2}{c}{H$_2$O maser\tabc}
\\\cline{1-3}\cline{5-6}\cline{8-9}
MSX name
&$\alpha_\mathrm{J2000}$
&$\delta_\mathrm{J2000}$
&
&Region
&\#
&
&Region
&\#
\\
\hline
\endfirsthead
\caption{continued.}\\
\hline\hline
\multicolumn{3}{c}{infrared\taba}
&
&\multicolumn{2}{c}{centimeter\tabb}
&
&\multicolumn{2}{c}{H$_2$O maser\tabc}
\\\cline{1-3}\cline{5-6}\cline{8-9}
MSX name
&$\alpha_\mathrm{J2000}$
&$\delta_\mathrm{J2000}$
&
&Region
&\#
&
&Region
&\#
\\
\hline
\endhead
\hline
\endfoot
G254.0567$-$00.5580  &123.997  &$-$36.140  &&                &   &&                &    \\
G263.5994$-$00.5236  &131.389  &$-$43.831  &&                &   &&                &    \\
G263.6200$-$00.5308  &131.399  &$-$43.851  &&                &   &&                &    \\
G263.6385$-$00.5217  &131.425  &$-$43.860  &&                &   &&                &    \\
G263.2493$+$00.5153  &132.200  &$-$42.906  &&08470$-$4243~0  &1  &&08470$-$4243~0  &\phn2  \\
G264.3225$-$00.1857  &132.386  &$-$44.181  &&08477$-$4359~0  &1  &&                &    \\
G268.0519$-$00.8095  &135.158  &$-$47.432  &&                &   &&                &    \\
G268.0440$-$00.7994  &135.161  &$-$47.419  &&                &   &&                &    \\
G268.0640$-$00.8162  &135.162  &$-$47.445  &&                &   &&                &    \\
G268.0799$-$00.8067  &135.188  &$-$47.451  &&                &   &&                &    \\
G268.6162$-$00.7389  &135.790  &$-$47.808  &&09014$-$4736~0  &1  &&                &    \\
G268.6184$-$00.7302  &135.802  &$-$47.804  &&                &   &&                &    \\
G270.8164$+$00.6893  &139.605  &$-$48.440  &&                &   &&                &    \\
G281.1656$-$01.6451  &149.878  &$-$57.065  &&                &   &&                &    \\
G281.1782$-$01.6519  &149.888  &$-$57.078  &&                &   &&                &    \\
G281.1834$-$01.6414  &149.908  &$-$57.073  &&                &   &&                &    \\
G282.7848$-$01.2869  &152.695  &$-$57.732  &&                &   &&10088$-$5730~2  &\phn9  \\
G283.1263$-$00.9887  &153.538  &$-$57.681  &&                &   &&                &    \\
G284.0079$-$00.8485  &155.063  &$-$58.054  &&10184$-$5748~0  &1  &&                &    \\
G284.7306$+$00.3213  &157.375  &$-$57.446  &&                &   &&10276$-$5711~0  &11  \\
G284.9025$+$00.0540  &157.392  &$-$57.764  &&10277$-$5730~0  &1  &&                &    \\
G284.9007$+$00.0653  &157.400  &$-$57.753  &&10277$-$5730~0  &2  &&                &    \\
G285.2504$-$00.0701  &157.834  &$-$58.049  &&                &   &&                &    \\
G285.2416$-$00.0505  &157.839  &$-$58.028  &&                &   &&                &    \\
G285.2412$-$00.0464  &157.842  &$-$58.024  &&                &   &&                &    \\
G285.2568$-$00.0430  &157.870  &$-$58.029  &&                &   &&                &    \\
G286.4203$-$01.4801  &158.402  &$-$59.858  &&10317$-$5936~0  &1  &&                &    \\
G286.4345$-$01.4907  &158.416  &$-$59.874  &&10317$-$5936~0  &2  &&                &    \\
G286.3931$-$01.3484  &158.487  &$-$59.730  &&                &   &&                &    \\
G287.8149$-$00.8332  &161.460  &$-$59.966  &&                &   &&                &    \\
G289.1062$-$01.1087  &163.543  &$-$60.793  &&10521$-$6031~0  &1  &&                &    \\
G288.8524$-$00.1088  &163.955  &$-$59.782  &&10537$-$5930~0  &1  &&                &    \\
G288.9583$-$00.0283  &164.213  &$-$59.755  &&10548$-$5929~0  &1  &&                &    \\
G288.9865$+$00.2533  &164.501  &$-$59.512  &&10559$-$5914~0  &1  &&10559$-$5914~0  &18  \\
G289.5787$-$00.6388  &164.825  &$-$60.570  &&                &   &&                &    \\
G288.9744$+$00.7910  &164.921  &$-$59.019  &&10575$-$5844~0  &1  &&                &    \\
G289.8953$-$00.8074  &165.271  &$-$60.855  &&                &   &&                &    \\
G289.8981$-$00.8131  &165.271  &$-$60.861  &&                &   &&                &    \\
G289.8985$-$00.7986  &165.284  &$-$60.848  &&                &   &&                &    \\
G289.4993$+$00.1231  &165.316  &$-$59.844  &&10591$-$5934~0  &1  &&                &    \\
G289.5051$+$00.1161  &165.321  &$-$59.853  &&                &   &&                &    \\
G291.0511$-$00.7784  &167.487  &$-$61.287  &&                &   &&                &    \\
G291.0515$-$00.7748  &167.491  &$-$61.284  &&                &   &&                &    \\
G294.6466$-$01.1777  &174.449  &$-$62.859  &&11354$-$6234~0  &2  &&                &    \\
G295.1013$-$01.6833  &175.107  &$-$63.470  &&                &   &&                &    \\
G295.1026$-$01.6778  &175.114  &$-$63.465  &&                &   &&                &    \\
G295.1085$-$00.7120  &175.694  &$-$62.536  &&11404$-$6215~0  &1  &&                &    \\
G295.1172$-$00.7206  &175.708  &$-$62.547  &&                &   &&                &    \\
G298.1829$-$00.7789  &182.258  &$-$63.259  &&                &   &&                &    \\
G298.8559$-$00.4623  &183.837  &$-$63.048  &&                &   &&                &    \\
G298.8700$-$00.4335  &183.877  &$-$63.022  &&                &   &&                &    \\
G298.8720$-$00.4287  &183.882  &$-$63.017  &&                &   &&                &    \\
G298.8332$+$00.1279  &183.967  &$-$62.461  &&12132$-$6211~0  &1  &&                &    \\
G299.1531$+$00.0086  &184.620  &$-$62.621  &&12157$-$6220~0  &1  &&                &    \\
G300.4007$+$00.5401  &187.422  &$-$62.225  &&                &   &&                &    \\
G300.4005$+$00.5500  &187.423  &$-$62.215  &&                &   &&                &    \\
G300.5030$-$00.1905  &187.510  &$-$62.962  &&                &   &&                &    \\
G300.5027$-$00.1833  &187.511  &$-$62.954  &&12272$-$6240~0  &1  &&                &    \\
G300.5035$-$00.1698  &187.515  &$-$62.941  &&                &   &&                &    \\
G300.5047$-$00.1745  &187.517  &$-$62.946  &&12272$-$6240~0  &2  &&                &    \\
G300.7486$+$00.1016  &188.097  &$-$62.689  &&                &   &&                &    \\
G304.9184$+$00.5645  &197.128  &$-$62.241  &&                &   &&                &    \\
G304.9275$+$00.5585  &197.148  &$-$62.247  &&                &   &&                &    \\
G304.9289$+$00.5524  &197.152  &$-$62.253  &&13054$-$6159~0  &1  &&                &    \\
G304.9406$+$00.5557  &197.176  &$-$62.249  &&                &   &&                &    \\
G305.5190$+$00.3485  &198.453  &$-$62.411  &&                &   &&                &    \\
G305.5302$+$00.3662  &198.473  &$-$62.392  &&                &   &&                &    \\
G305.5313$+$00.3505  &198.479  &$-$62.408  &&                &   &&                &    \\
G305.5345$+$00.3499  &198.486  &$-$62.408  &&13107$-$6208~0  &1  &&                &    \\
G305.5389$+$00.3672  &198.492  &$-$62.391  &&                &   &&                &    \\
G305.5388$+$00.3604  &198.493  &$-$62.397  &&                &   &&                &    \\
G308.9323$+$00.1289  &205.785  &$-$62.139  &&                &   &&                &    \\
G309.3767$-$00.1377  &206.837  &$-$62.308  &&                &   &&                &    \\
G310.8902$+$00.0089  &209.899  &$-$61.805  &&13560$-$6133~2  &1  &&                &    \\
G311.1779$-$00.0960  &210.546  &$-$61.830  &&                &   &&                &    \\
G311.8742$+$00.0890  &211.844  &$-$61.457  &&                &   &&                &    \\
G311.8810$+$00.1002  &211.851  &$-$61.445  &&                &   &&                &    \\
G311.8772$+$00.0776  &211.857  &$-$61.467  &&                &   &&                &    \\
G311.8928$+$00.0931  &211.879  &$-$61.448  &&                &   &&                &    \\
G312.3063$+$00.6647  &212.350  &$-$60.780  &&                &   &&                &    \\
G313.6668$-$00.1023  &215.508  &$-$61.072  &&                &   &&                &    \\
G313.6833$-$00.1088  &215.544  &$-$61.072  &&                &   &&                &    \\
G313.6870$-$00.1996  &215.616  &$-$61.156  &&                &   &&                &    \\
G313.7132$-$00.1766  &215.651  &$-$61.126  &&                &   &&                &    \\
G313.9140$-$00.0745  &215.967  &$-$60.960  &&14201$-$6044~0  &1  &&                &    \\
G313.9710$-$00.0741  &216.076  &$-$60.940  &&                &   &&                &    \\
G314.2269$+$00.4099  &216.218  &$-$60.397  &&                &   &&                &    \\
G314.2437$+$00.4231  &216.240  &$-$60.378  &&                &   &&                &    \\
G314.2495$+$00.4338  &216.243  &$-$60.366  &&                &   &&                &    \\
G314.2462$+$00.4174  &216.249  &$-$60.383  &&                &   &&                &    \\
G314.2004$+$00.2900  &216.253  &$-$60.518  &&                &   &&                &    \\
G314.2014$+$00.2864  &216.258  &$-$60.521  &&14214$-$6017~0  &1  &&                &    \\
G314.2015$+$00.2834  &216.260  &$-$60.524  &&                &   &&                &    \\
G314.2124$+$00.2939  &216.273  &$-$60.510  &&                &   &&                &    \\
G314.2605$+$00.4027  &216.286  &$-$60.392  &&                &   &&                &    \\
G314.2125$+$00.2632  &216.295  &$-$60.539  &&                &   &&                &    \\
G314.2161$+$00.2546  &216.308  &$-$60.546  &&                &   &&                &    \\
G314.2354$+$00.2913  &216.319  &$-$60.505  &&                &   &&                &    \\
G316.3627$-$00.3631  &220.804  &$-$60.287  &&14394$-$6004~0  &1  &&                &    \\
G320.1488$+$00.8145  &226.300  &$-$57.516  &&                &   &&                &    \\
G320.1515$+$00.8109  &226.308  &$-$57.518  &&                &   &&                &    \\
G320.1542$+$00.7976  &226.324  &$-$57.528  &&15015$-$5720~0  &1  &&                &    \\
G320.1524$+$00.7924  &226.326  &$-$57.533  &&                &   &&                &    \\
G320.1655$+$00.8126  &226.329  &$-$57.509  &&                &   &&                &    \\
G320.1697$+$00.8097  &226.338  &$-$57.510  &&15015$-$5720~0  &2  &&                &    \\
G320.1750$+$00.8001  &226.356  &$-$57.516  &&                &   &&                &    \\
G320.3144$-$00.1845  &227.507  &$-$58.300  &&                &   &&                &    \\
G320.3333$-$00.1790  &227.533  &$-$58.285  &&15061$-$5806~0  &2  &&                &    \\
G320.2698$-$01.2678  &228.504  &$-$59.253  &&15100$-$5903~0  &1  &&                &    \\
G322.3904$+$00.1957  &230.433  &$-$56.887  &&                &   &&                &    \\
G323.6814$+$00.6298  &231.944  &$-$55.812  &&15239$-$5538~0  &1  &&                &    \\
G323.6934$+$00.6328  &231.958  &$-$55.803  &&                &   &&                &    \\
G323.9169$+$00.3984  &232.523  &$-$55.870  &&15262$-$5541~1  &1  &&                &    \\
G323.7410$-$00.2552  &232.933  &$-$56.507  &&                &   &&                &    \\
G325.1158$+$00.0078  &234.656  &$-$55.492  &&                &   &&                &    \\
G325.5964$+$00.0912  &235.246  &$-$55.138  &&                &   &&                &    \\
G327.4012$+$00.4488  &237.326  &$-$53.751  &&                &   &&                &    \\
G326.4477$-$00.7485  &237.327  &$-$55.282  &&15454$-$5507~0  &1  &&15454$-$5507~0  &40  \\
G326.7831$-$00.5548  &237.573  &$-$54.921  &&                &   &&                &    \\
G326.7899$-$00.5485  &237.576  &$-$54.912  &&15464$-$5445~0  &1  &&                &    \\
G326.7878$-$00.5526  &237.577  &$-$54.916  &&                &   &&                &    \\
G326.7896$-$00.5594  &237.587  &$-$54.920  &&                &   &&                &    \\
G326.7859$-$00.5644  &237.588  &$-$54.927  &&                &   &&                &    \\
G328.1202$+$00.1115  &238.626  &$-$53.559  &&                &   &&                &    \\
G327.7555$-$00.3324  &238.630  &$-$54.134  &&                &   &&                &    \\
G327.7579$-$00.3515  &238.654  &$-$54.147  &&15507$-$5359~0  &1  &&                &    \\
G327.5761$-$00.8497  &238.961  &$-$54.647  &&                &   &&                &    \\
G328.9480$+$00.5709  &239.194  &$-$52.676  &&15530$-$5231~0  &1  &&                &    \\
G328.9483$+$00.5520  &239.215  &$-$52.691  &&                &   &&                &    \\
G329.4506$+$00.5124  &239.886  &$-$52.395  &&15557$-$5215~0  &1  &&                &    \\
G329.4533$+$00.5034  &239.899  &$-$52.401  &&                &   &&                &    \\
G328.5759$-$00.5285  &239.910  &$-$53.755  &&15557$-$5337~0  &2  &&                &    \\
G329.4626$+$00.5037  &239.911  &$-$52.394  &&                &   &&                &    \\
G328.5739$-$00.5483  &239.929  &$-$53.771  &&                &   &&                &    \\
G331.6218$+$00.5208  &242.493  &$-$50.946  &&16061$-$5048~0  &1  &&                &    \\
G331.7095$+$00.6034  &242.506  &$-$50.825  &&16061$-$5048~5  &1  &&                &    \\
G331.3292$-$00.3350  &243.083  &$-$51.772  &&                &   &&                &    \\
G331.3334$-$00.3376  &243.091  &$-$51.771  &&                &   &&                &    \\
G332.1420$+$00.0431  &243.613  &$-$50.939  &&16106$-$5048~0  &1  &&                &    \\
G332.7506$+$00.6682  &243.629  &$-$50.066  &&16107$-$4956~0  &1  &&                &    \\
G332.9793$+$00.7771  &243.769  &$-$49.830  &&                &   &&                &    \\
G332.9623$+$00.7541  &243.775  &$-$49.858  &&                &   &&                &    \\
G332.9879$+$00.7693  &243.788  &$-$49.829  &&                &   &&                &    \\
G332.9872$+$00.7659  &243.790  &$-$49.832  &&                &   &&                &    \\
G333.0221$-$00.0377  &244.701  &$-$50.385  &&                &   &&                &    \\
G332.8014$-$00.7017  &245.189  &$-$51.013  &&                &   &&                &    \\
G332.8143$-$00.6977  &245.199  &$-$51.001  &&                &   &&                &    \\
G335.1572$-$00.2172  &247.217  &$-$48.992  &&                &   &&                &    \\
G337.9758$-$00.1370  &249.979  &$-$46.864  &&                &   &&                &    \\
G337.9662$-$00.1507  &249.985  &$-$46.880  &&                &   &&                &    \\
G337.9732$-$00.1458  &249.986  &$-$46.872  &&16363$-$4645~0  &1  &&                &    \\
G337.9858$-$00.1442  &249.997  &$-$46.861  &&                &   &&                &    \\
G337.9684$-$00.1599  &249.997  &$-$46.885  &&                &   &&                &    \\
G338.0012$-$00.1337  &250.000  &$-$46.843  &&                &   &&                &    \\
G339.1707$-$00.3880  &251.396  &$-$46.129  &&16419$-$4602~0  &1  &&                &    \\
G339.1704$-$00.3927  &251.400  &$-$46.132  &&                &   &&                &    \\
G343.0290$+$02.6199  &251.695  &$-$41.242  &&16428$-$4109~2  &1  &&                &    \\
G343.0500$+$02.6094  &251.724  &$-$41.233  &&                &   &&                &    \\
G343.0574$+$02.5615  &251.778  &$-$41.259  &&                &   &&                &    \\
G339.9429$-$00.1187  &251.820  &$-$45.367  &&                &   &&                &    \\
G339.9689$-$00.1507  &251.879  &$-$45.367  &&                &   &&                &    \\
G341.2366$+$00.3361  &252.502  &$-$44.084  &&16464$-$4359~0  &1  &&                &    \\
G340.8832$-$00.3663  &252.943  &$-$44.804  &&16482$-$4443~0  &1  &&                &    \\
G342.2464$+$00.2842  &253.446  &$-$43.337  &&                &   &&                &    \\
G342.8211$-$00.0307  &254.277  &$-$43.088  &&16535$-$4300~0  &1  &&                &    \\
G343.8354$-$00.1058  &255.213  &$-$42.338  &&                &   &&                &    \\
G343.8569$-$00.0986  &255.223  &$-$42.317  &&16573$-$4214~0  &1  &&                &    \\
G345.8197$+$00.0341  &256.677  &$-$40.678  &&                &   &&                &    \\
G345.8486$+$00.0193  &256.715  &$-$40.664  &&                &   &&                &    \\
G345.8677$-$01.1011  &257.923  &$-$41.316  &&17082$-$4114~0  &1  &&                &    \\
G345.8869$-$01.1042  &257.941  &$-$41.303  &&17082$-$4114~0  &2  &&                &    \\
G348.1485$+$00.2549  &258.246  &$-$38.675  &&17095$-$3837~0  &1  &&                &    \\
G347.9779$-$00.4321  &258.834  &$-$39.215  &&                &   &&                &    \\
G347.9752$-$00.4364  &258.836  &$-$39.220  &&                &   &&                &    \\
G347.9706$-$00.4443  &258.841  &$-$39.228  &&                &   &&                &    \\
G347.9747$-$00.4435  &258.844  &$-$39.224  &&                &   &&                &    \\
G347.9742$-$00.4508  &258.851  &$-$39.229  &&                &   &&                &    \\
G347.9721$-$00.4591  &258.858  &$-$39.236  &&                &   &&                &    \\
G350.7082$+$00.9911  &259.370  &$-$36.165  &&                &   &&                &    \\
G350.8074$-$00.8267  &261.303  &$-$37.119  &&                &   &&                &    \\
G350.8040$-$00.8331  &261.307  &$-$37.125  &&17218$-$3704~0  &1  &&                &    \\
G351.4677$-$00.4512  &261.377  &$-$36.362  &&                &   &&                &    \\
G351.4880$-$00.4642  &261.405  &$-$36.352  &&                &   &&                &    \\
G351.4658$-$00.4812  &261.407  &$-$36.380  &&                &   &&                &    \\
G353.0587$+$00.5337  &261.469  &$-$34.492  &&17225$-$3426~0  &1  &&                &    \\
G352.2436$-$00.1617  &261.617  &$-$35.557  &&                &   &&                &    \\
G352.6006$-$00.1674  &261.867  &$-$35.263  &&                &   &&                &    \\
G352.6023$-$00.1983  &261.899  &$-$35.279  &&                &   &&                &    \\
G352.6159$-$00.1948  &261.905  &$-$35.266  &&                &   &&                &    \\
G351.6947$-$01.1490  &262.255  &$-$36.561  &&                &   &&                &    \\
G351.7052$-$01.1666  &262.281  &$-$36.563  &&                &   &&                &    \\
G354.2042$-$00.0419  &262.816  &$-$33.857  &&                &   &&                &    \\
G354.2041$-$00.0456  &262.819  &$-$33.859  &&                &   &&                &    \\
G354.2130$-$00.0524  &262.832  &$-$33.856  &&                &   &&                &    \\
G358.8387$-$00.7420  &266.438  &$-$30.314  &&17425$-$3017~0  &1  &&                &    \\
G013.6562$-$00.5997  &274.351  &$-$17.371  &&                &   &&18144$-$1723~0  &85  \\
G016.8182$-$00.3308  &275.657  &$-$14.456  &&                &   &&                &    \\
\end{longtable}
\tablefoot{
\tablefoottext{a}{Infrared source: columns as in Table~\ref{t:type2}.}
\tablefoottext{b}{Centimeter source: columns as in Table~\ref{t:type1}.}
\tablefoottext{c}{H$_2$O maser spot: columns as in Table~\ref{t:type1}.}
}
\end{scriptsize}
}

\Online
\begin{appendix}
\section{Figures}

\begin{figure*}[]
\begin{center}
\includegraphics[scale=0.9]{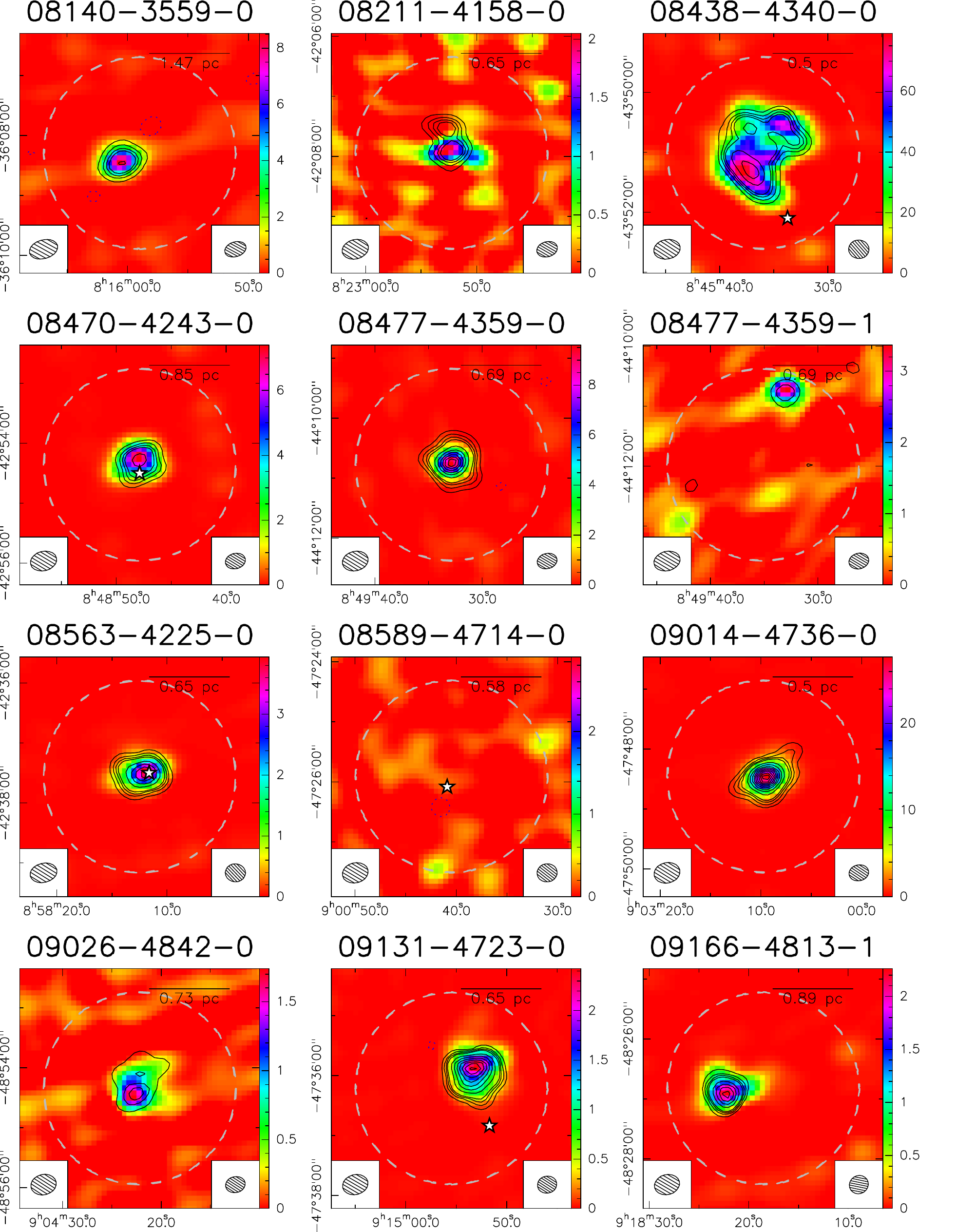}
\end{center}
\caption{\label{f:ATCAmaps}ATCA continuum images at 18.0~GHz (color scale, units m\jpb) and 22.8~GHz (contours). Synthesized beams are indicated at the bottom left (18.0~GHz) and right (22.8~GHz) corners (see Table~\ref{t:observations} for values). The primary beam at 22.8~GHz is indicated with the dashed grey circle. When a distance is available the spatial scale corresponding to 80\arcsec\ is indicated at the top of the panel. White stars show the position of water masers (see Table~\ref{t:masers}).}
\end{figure*}
\begin{figure*}[]
\ContinuedFloat
\begin{center}
\includegraphics[scale=0.9]{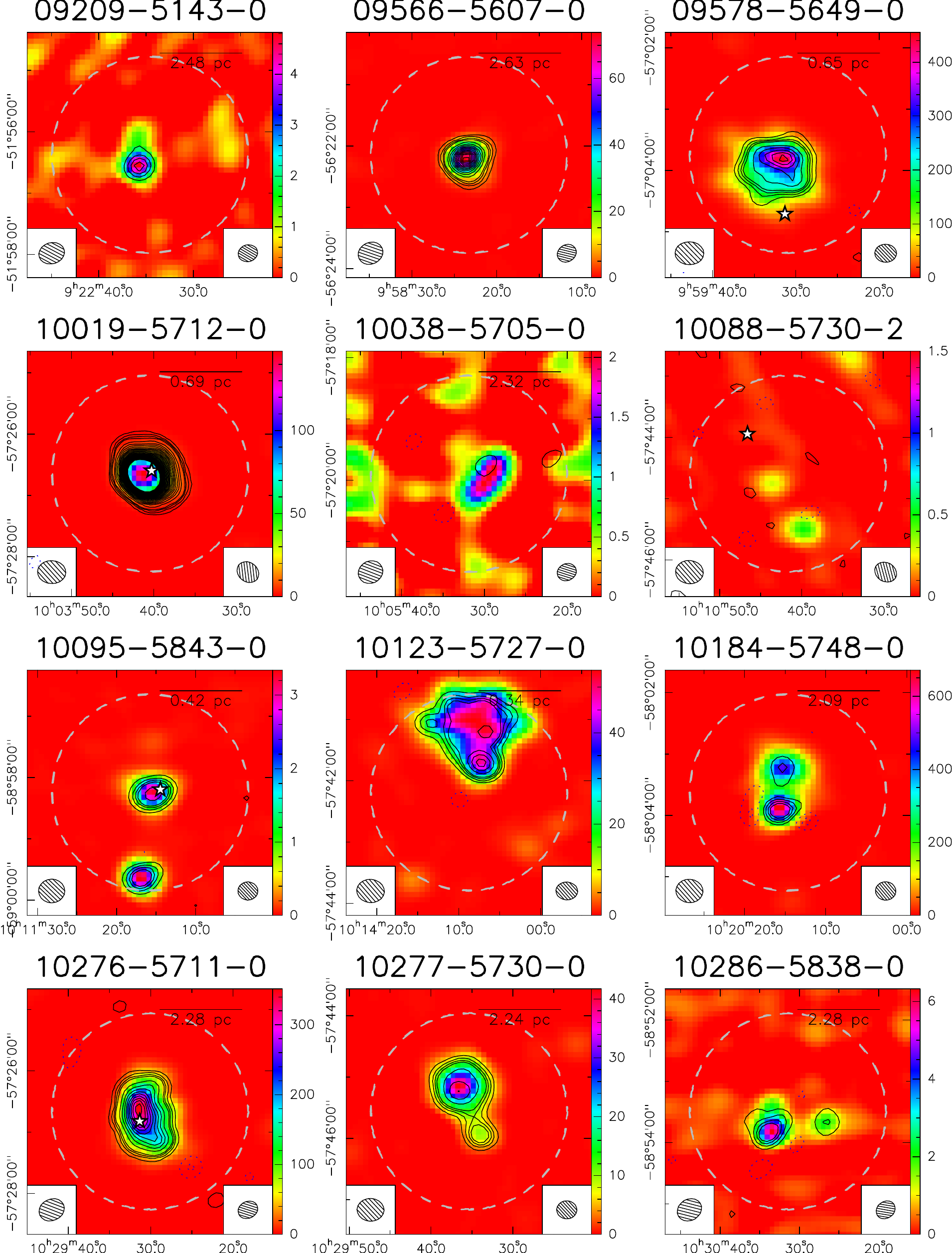}
\end{center}
\caption{continued.}
\end{figure*}
\begin{figure*}[]
\ContinuedFloat
\begin{center}
\includegraphics[scale=0.9]{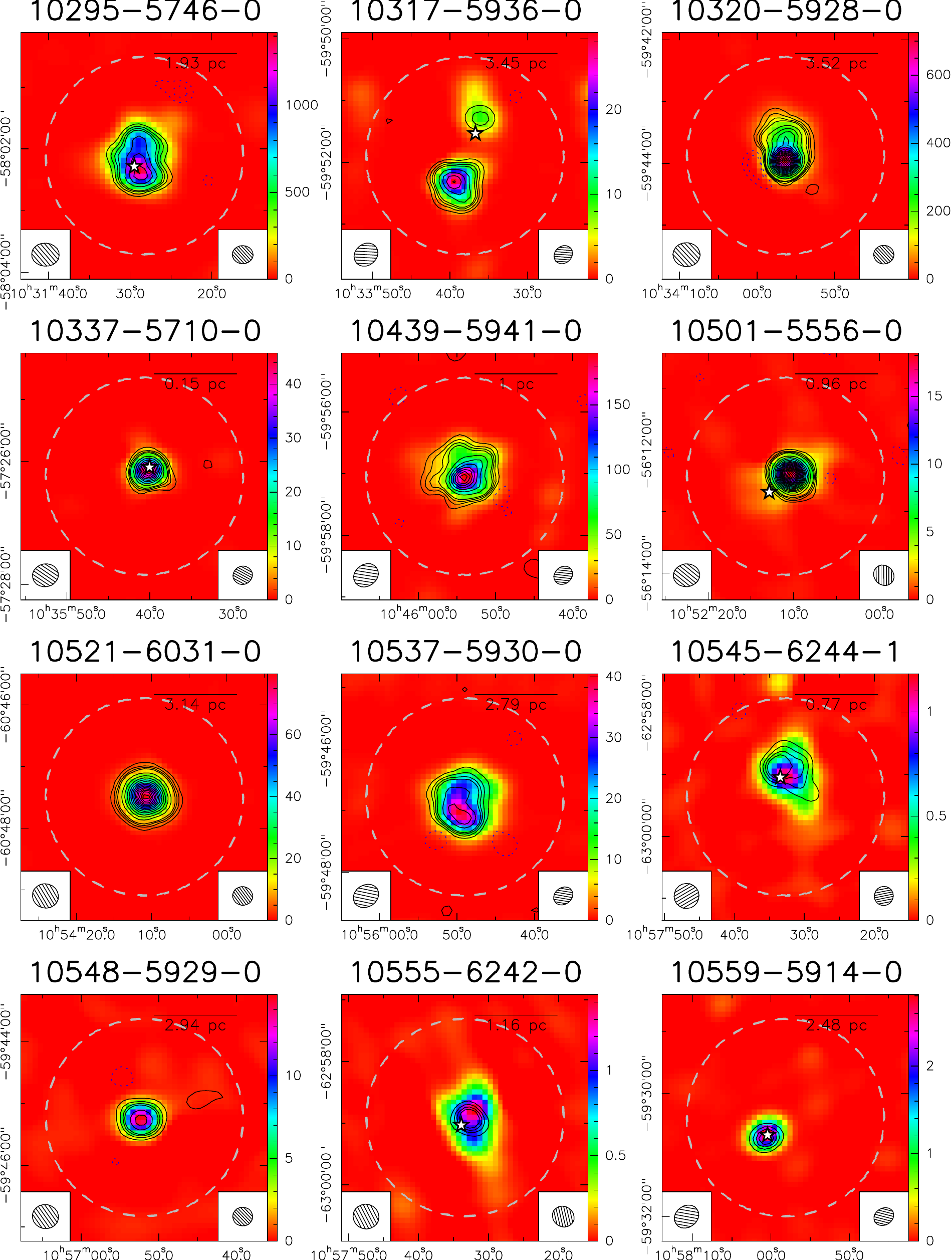}
\end{center}
\caption{continued.}
\end{figure*}
\begin{figure*}[]
\ContinuedFloat
\begin{center}
\includegraphics[scale=0.9]{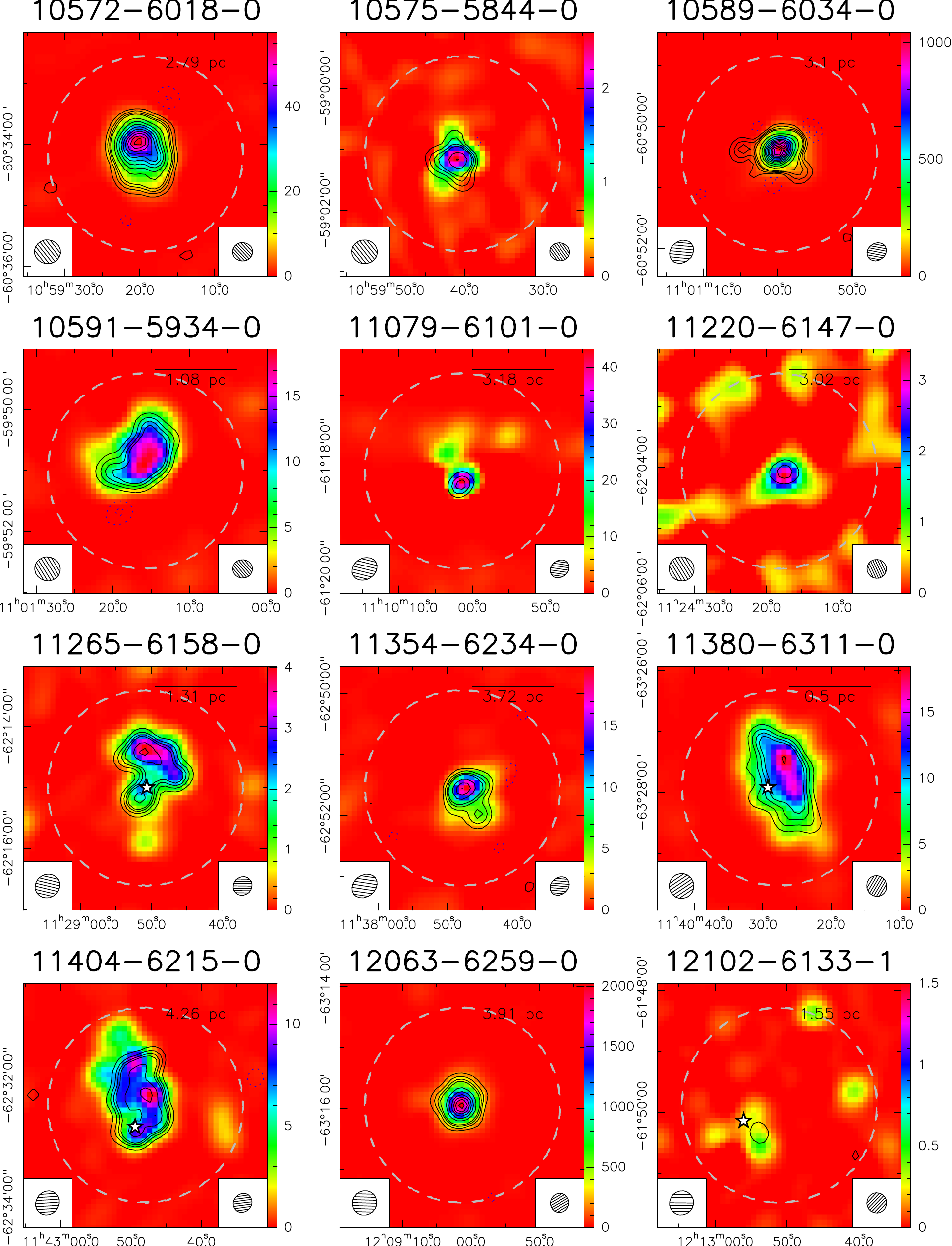}
\end{center}
\caption{continued.}
\end{figure*}
\begin{figure*}[]
\ContinuedFloat
\begin{center}
\includegraphics[scale=0.9]{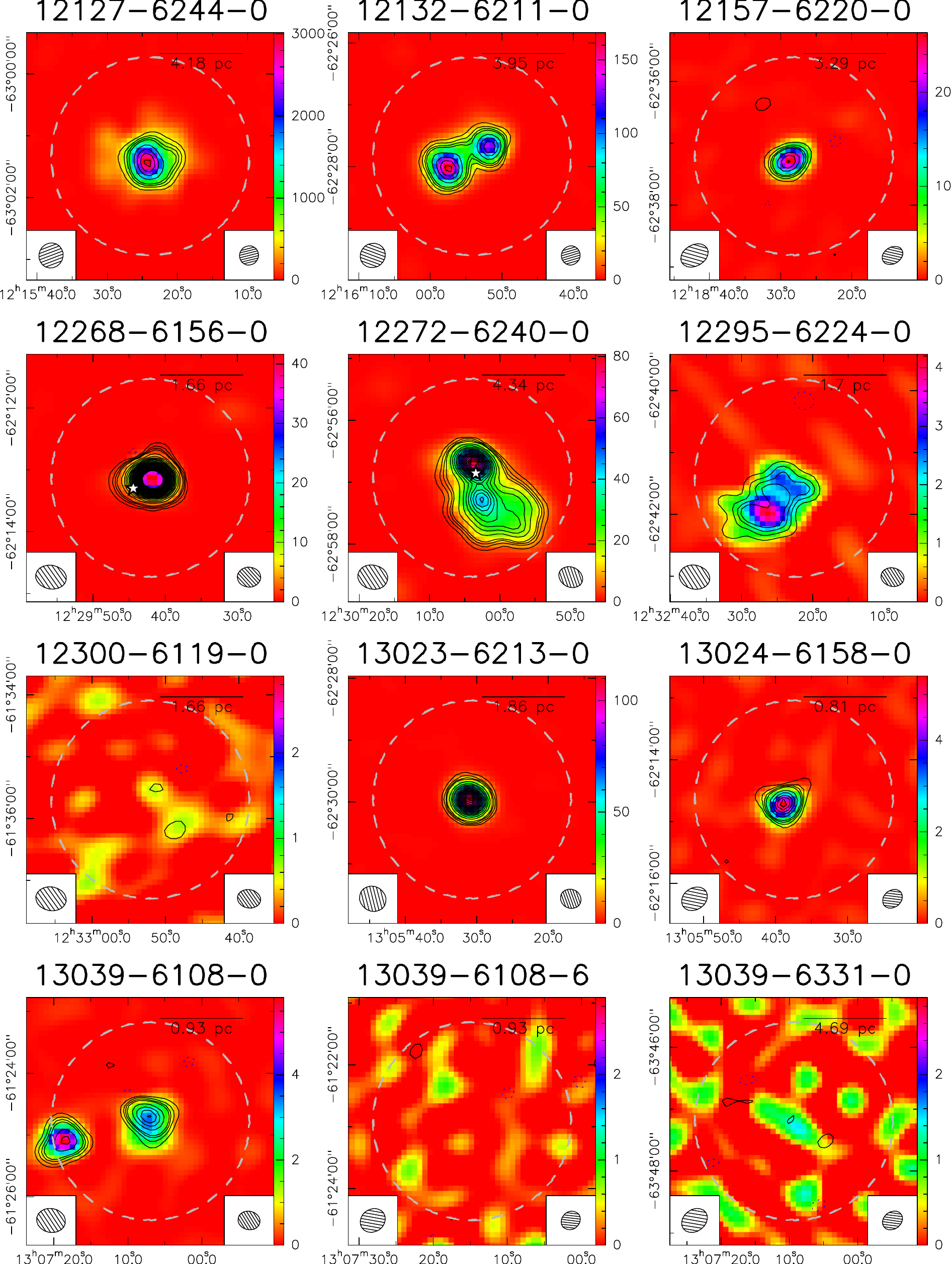}
\end{center}
\caption{continued.}
\end{figure*}
\begin{figure*}[]
\ContinuedFloat
\begin{center}
\includegraphics[scale=0.9]{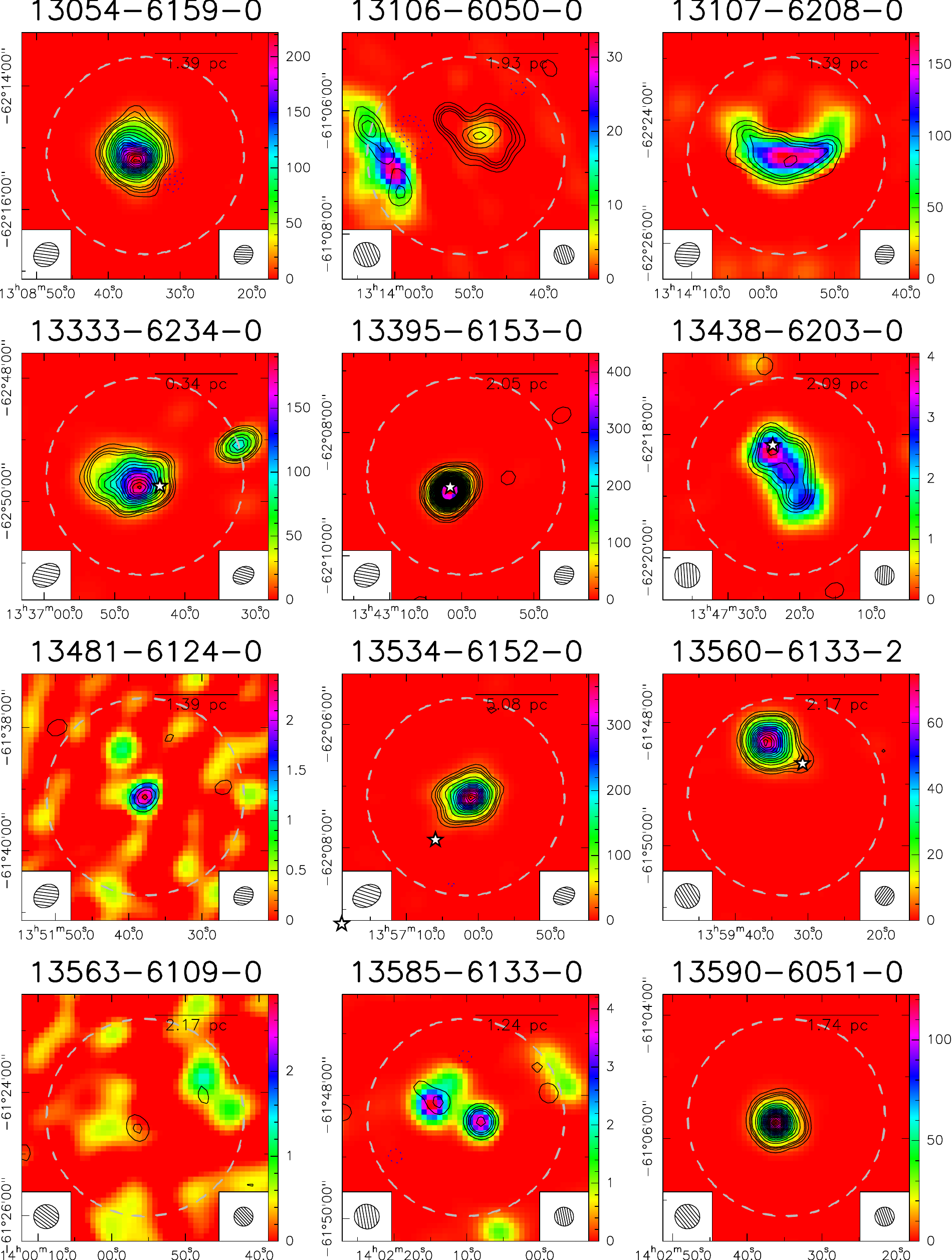}
\end{center}
\caption{continued.}
\end{figure*}
\begin{figure*}[]
\ContinuedFloat
\begin{center}
\includegraphics[scale=0.9]{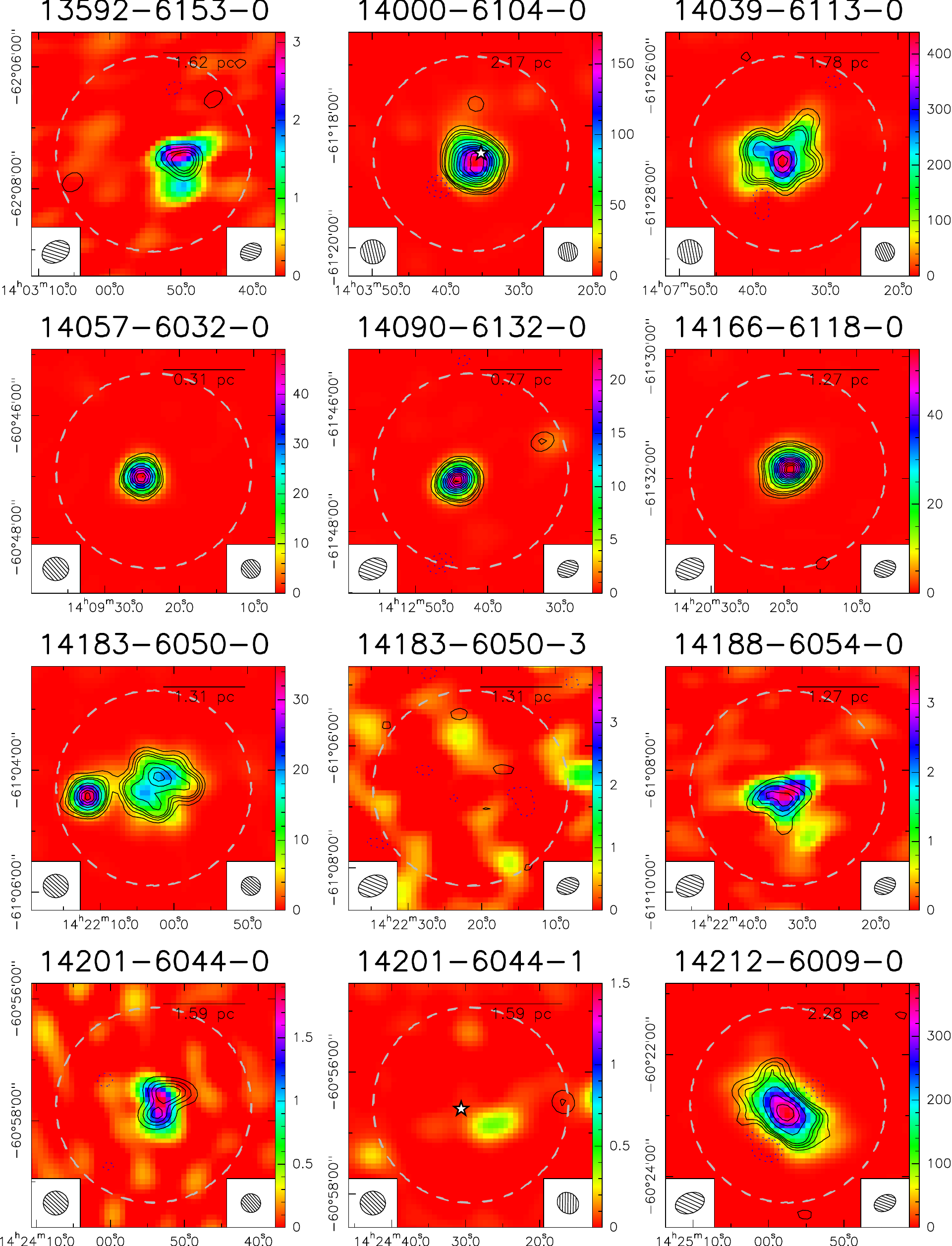}
\end{center}
\caption{continued.}
\end{figure*}
\begin{figure*}[]
\ContinuedFloat
\begin{center}
\includegraphics[scale=0.9]{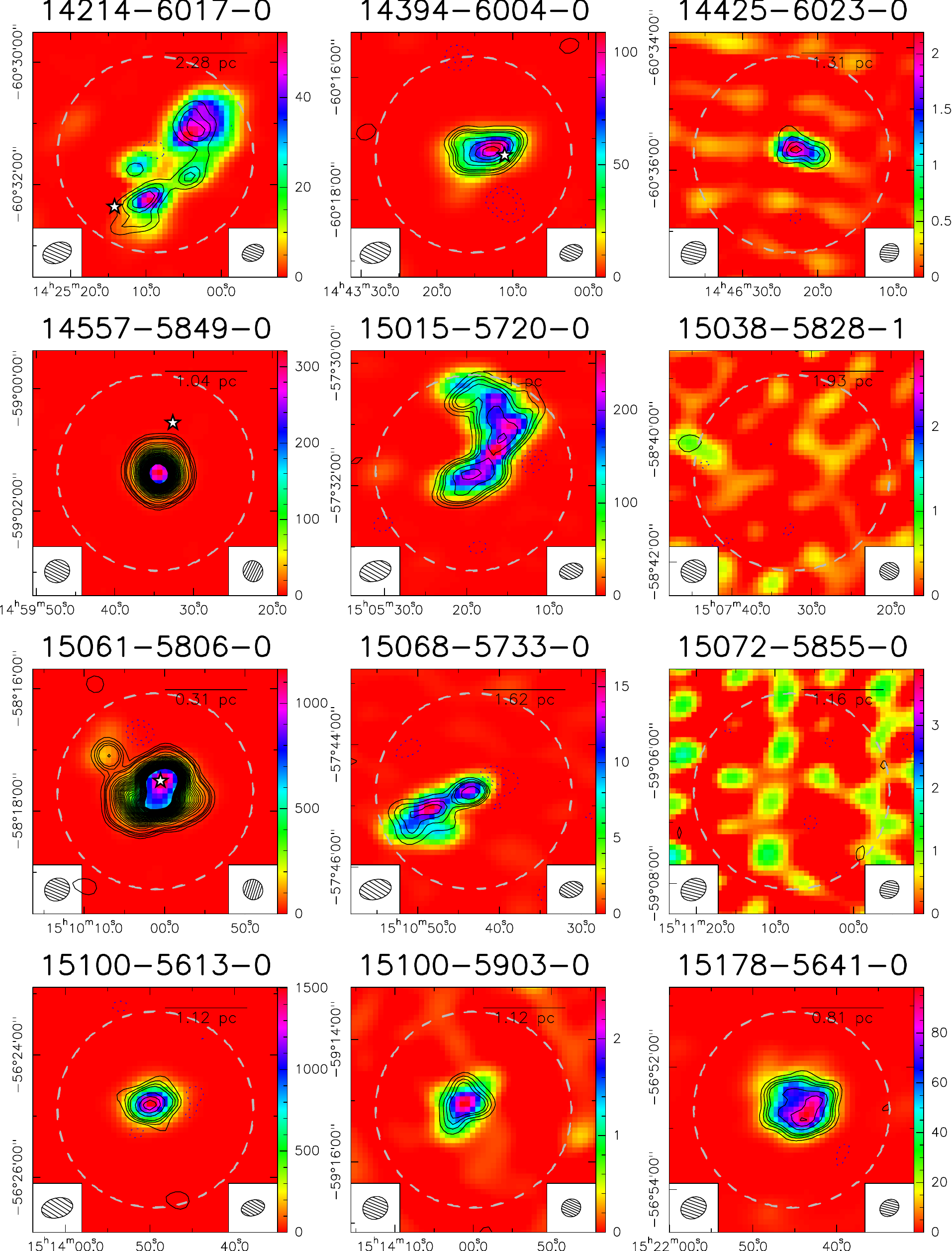}
\end{center}
\caption{continued.}
\end{figure*}
\begin{figure*}[]
\ContinuedFloat
\begin{center}
\includegraphics[scale=0.9]{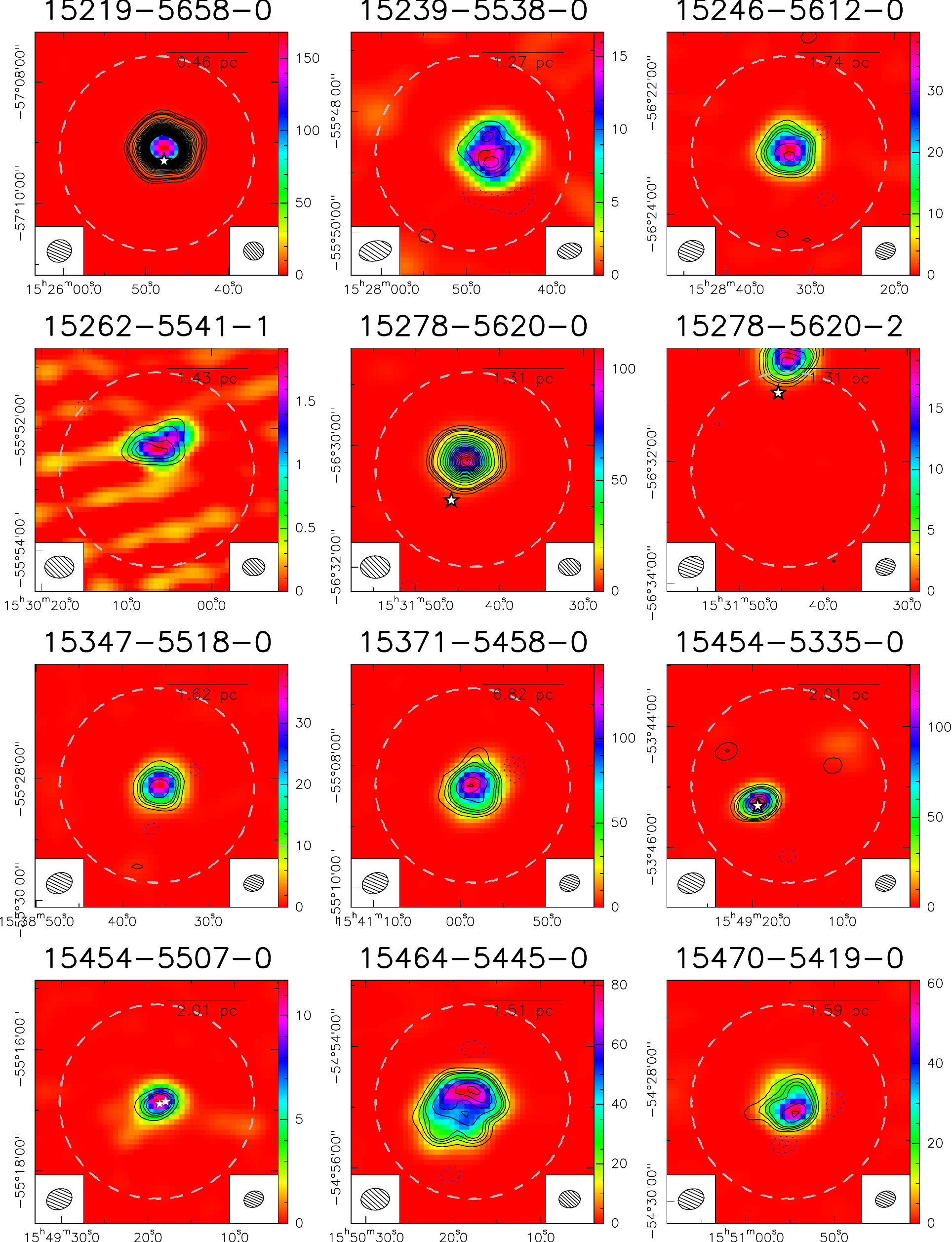}
\end{center}
\caption{continued.}
\end{figure*}
\clearpage
\begin{figure*}[]
\ContinuedFloat
\begin{center}
\includegraphics[scale=0.9]{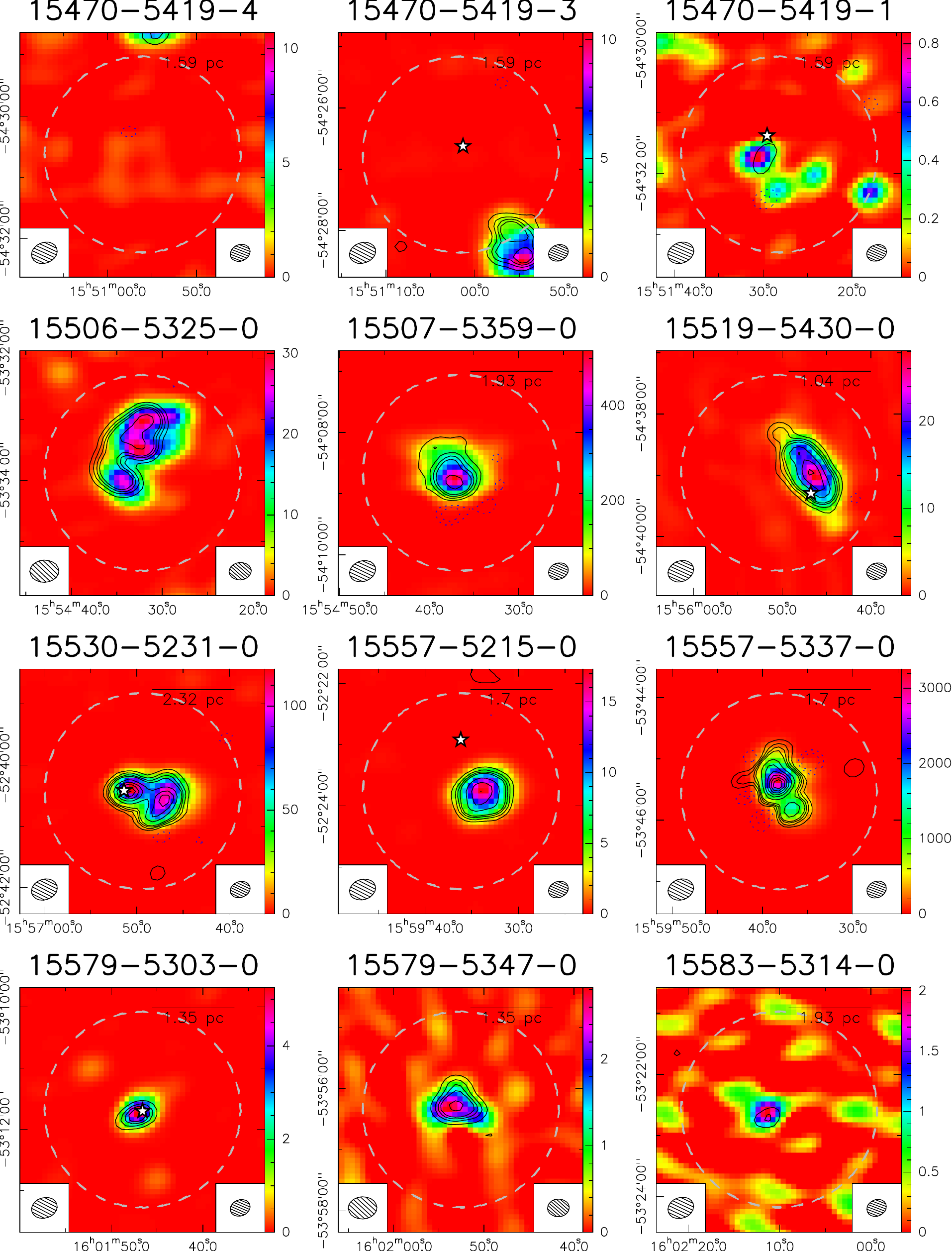}
\end{center}
\caption{continued.}
\end{figure*}
\begin{figure*}[]
\ContinuedFloat
\begin{center}
\includegraphics[scale=0.9]{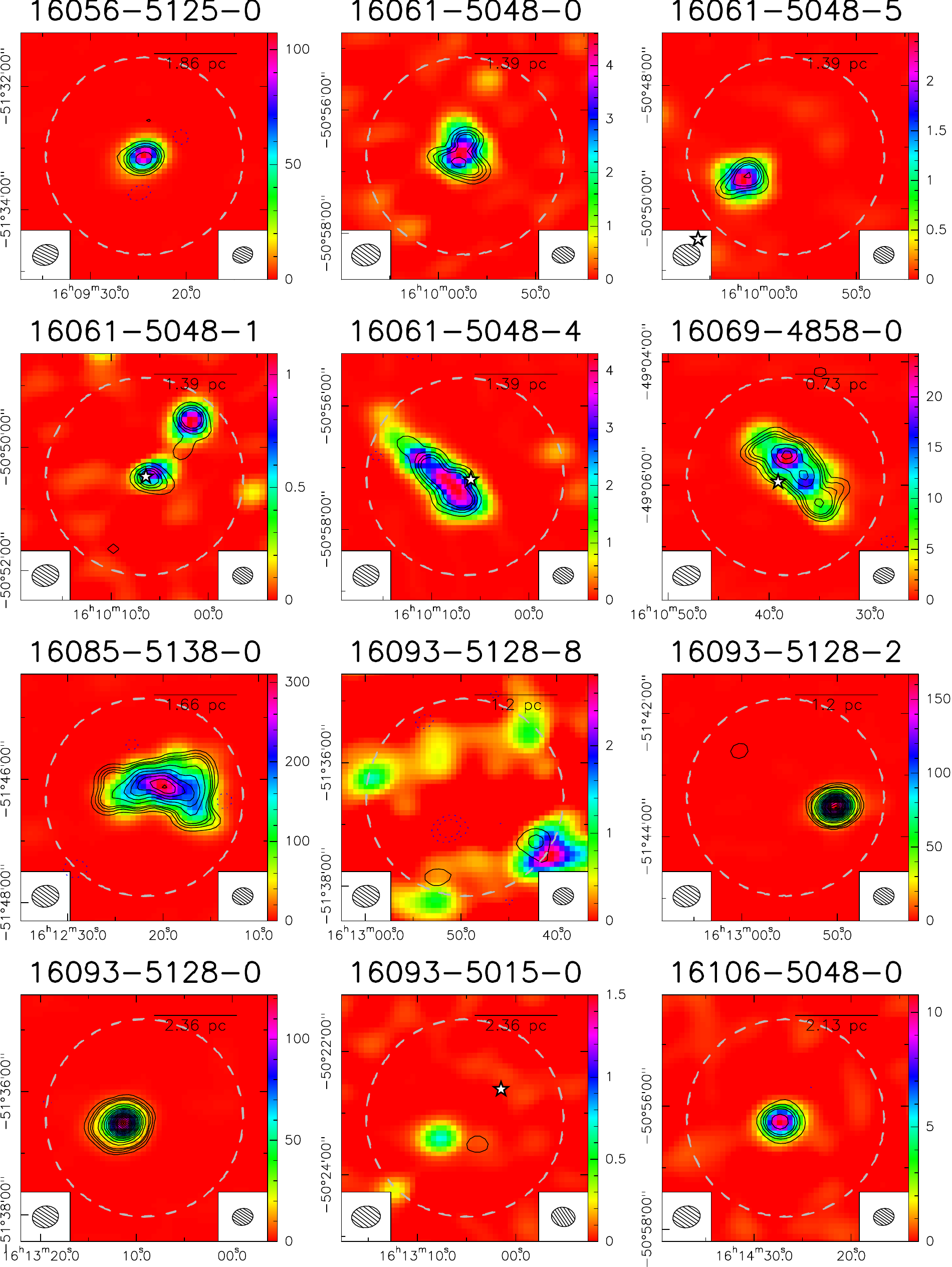}
\end{center}
\caption{continued.}
\end{figure*}
\begin{figure*}[]
\ContinuedFloat
\begin{center}
\includegraphics[scale=0.9]{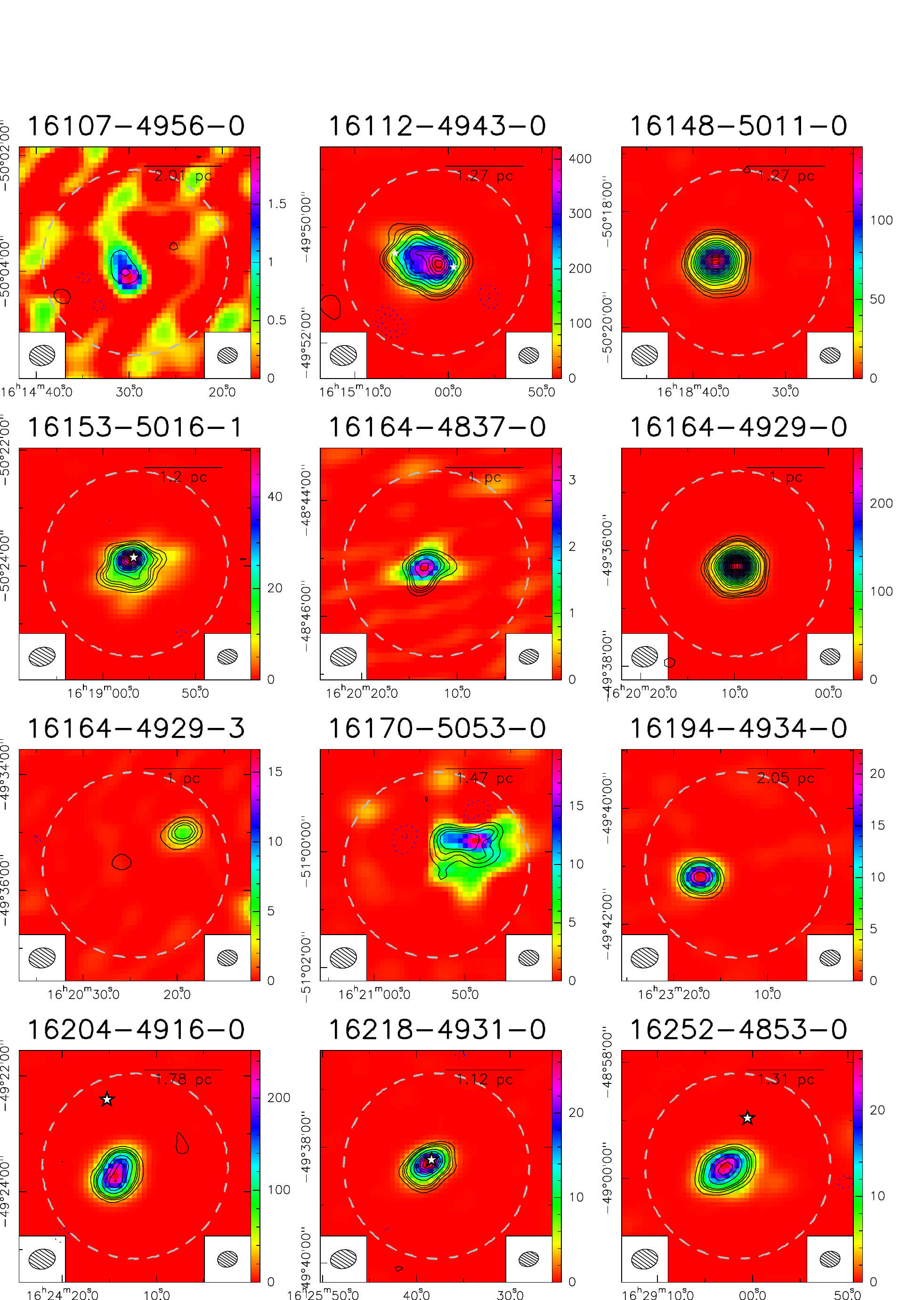}
\end{center}
\caption{continued.}
\end{figure*}
\begin{figure*}[]
\ContinuedFloat
\begin{center}
\includegraphics[scale=0.9]{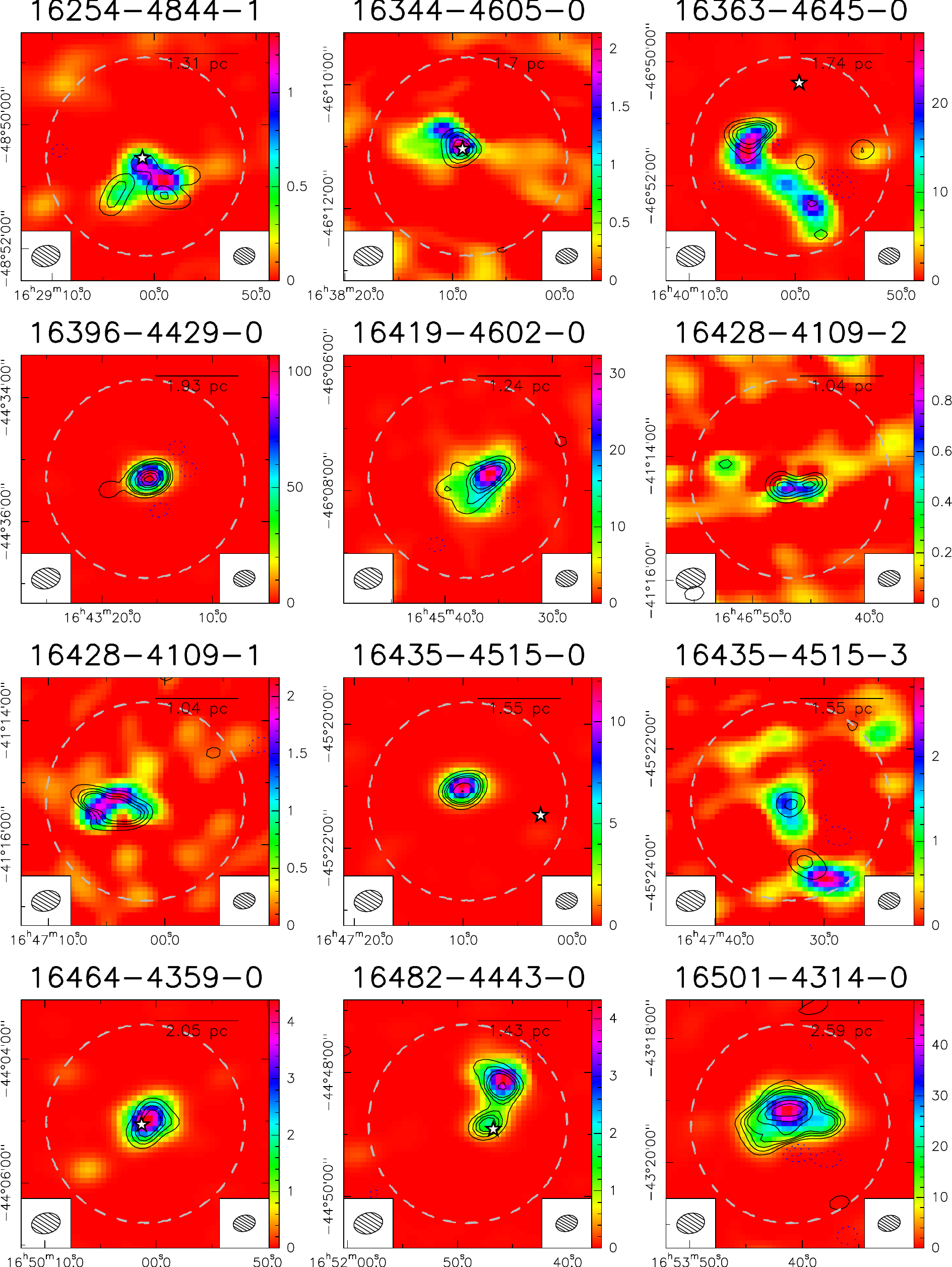}
\end{center}
\caption{continued.}
\end{figure*}
\begin{figure*}[]
\ContinuedFloat
\begin{center}
\includegraphics[scale=0.9]{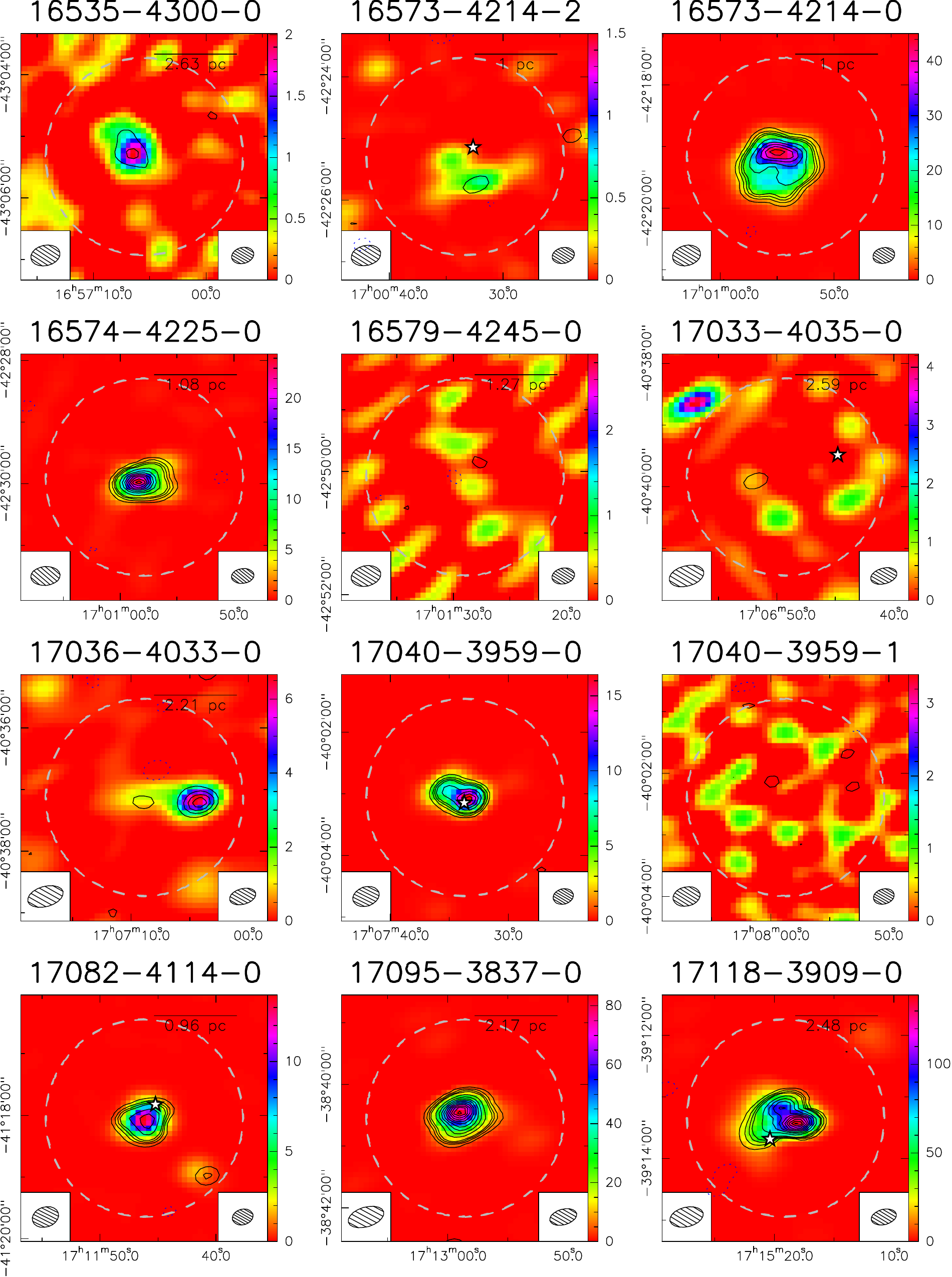}
\end{center}
\caption{continued.}
\end{figure*}
\begin{figure*}[]
\ContinuedFloat
\begin{center}
\includegraphics[scale=0.9]{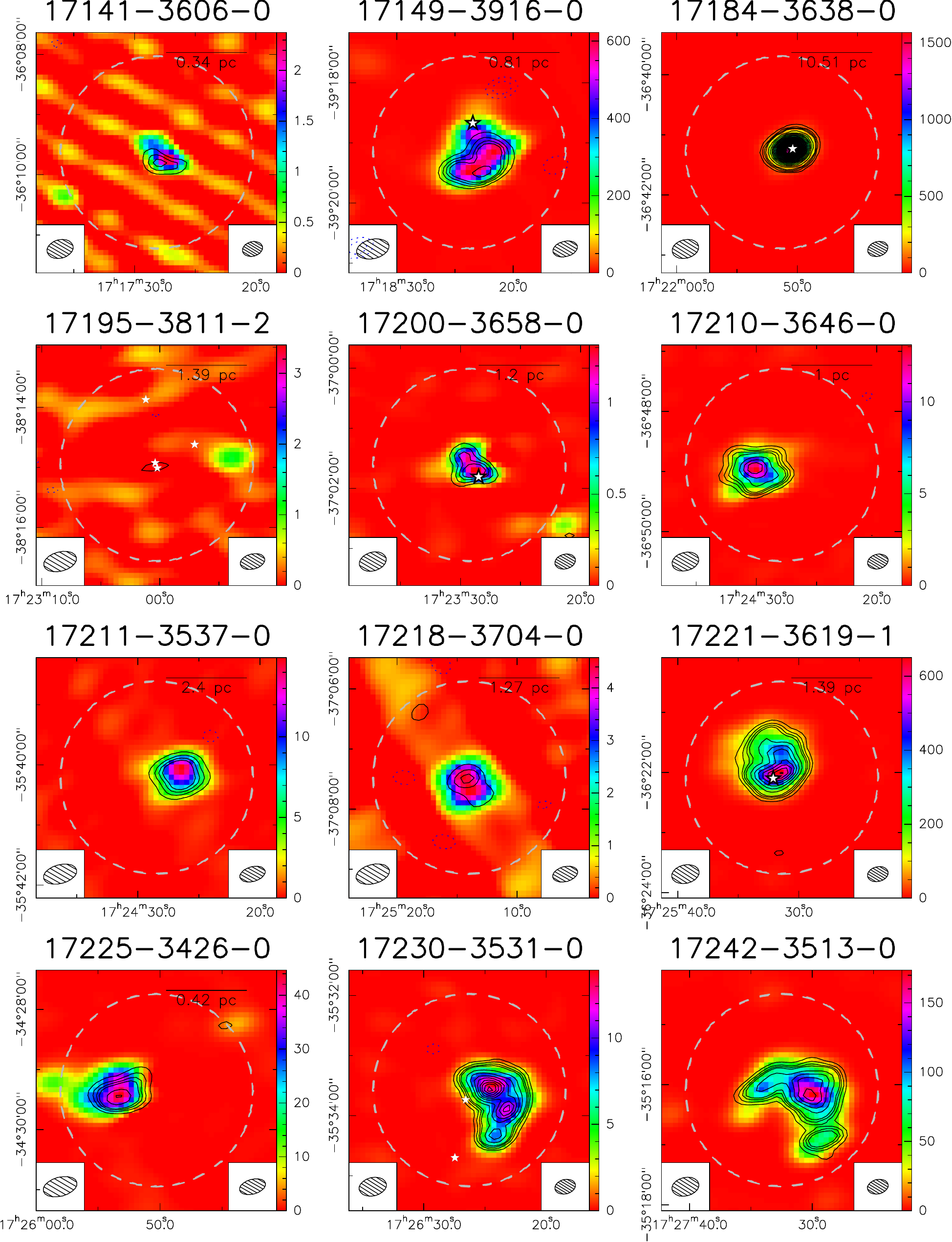}
\end{center}
\caption{continued.}
\end{figure*}
\begin{figure*}[]
\ContinuedFloat
\begin{center}
\includegraphics[scale=0.9]{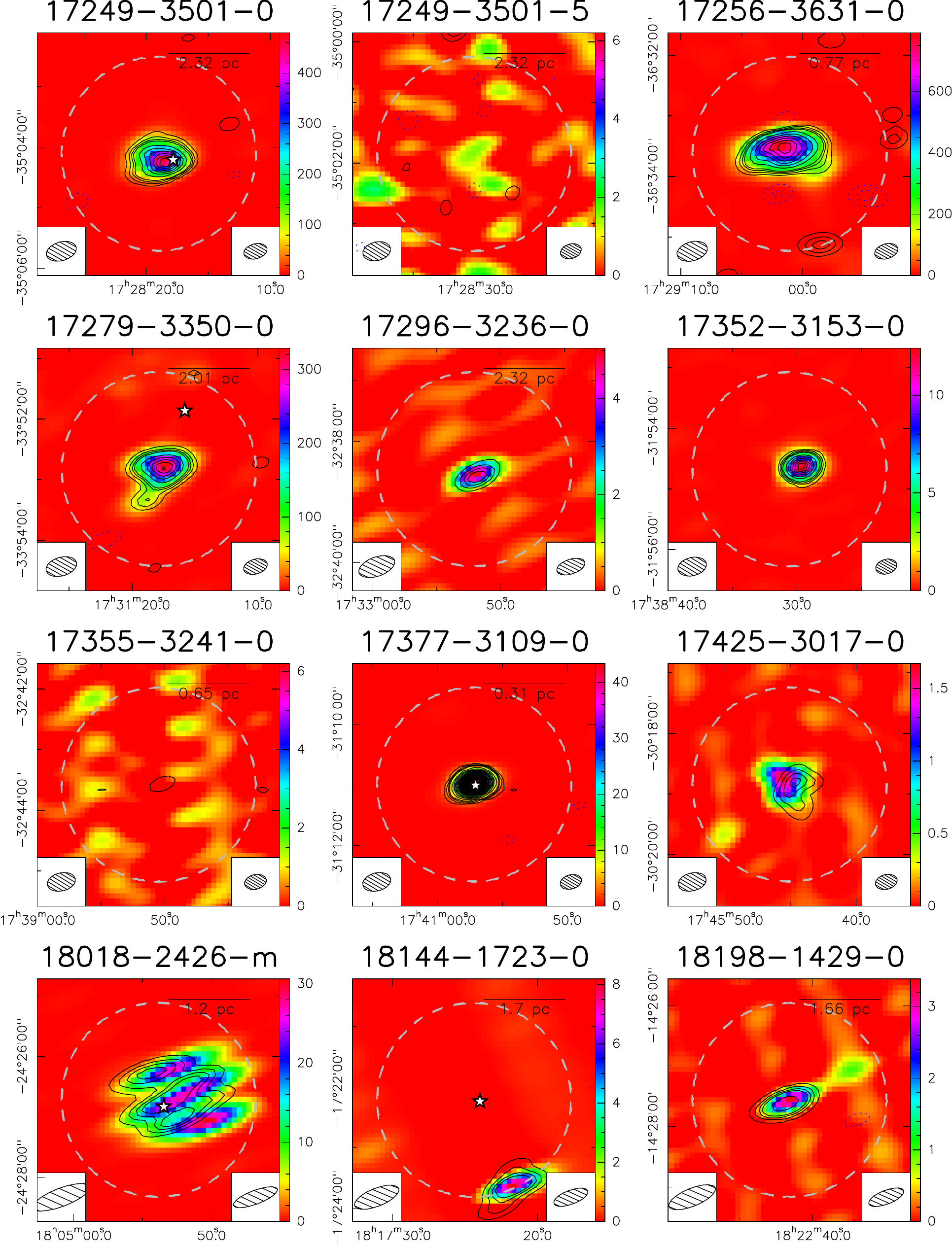}
\end{center}
\caption{continued.}
\end{figure*}

\clearpage
\begin{figure*}[]
\begin{center}
\includegraphics[scale=0.8]{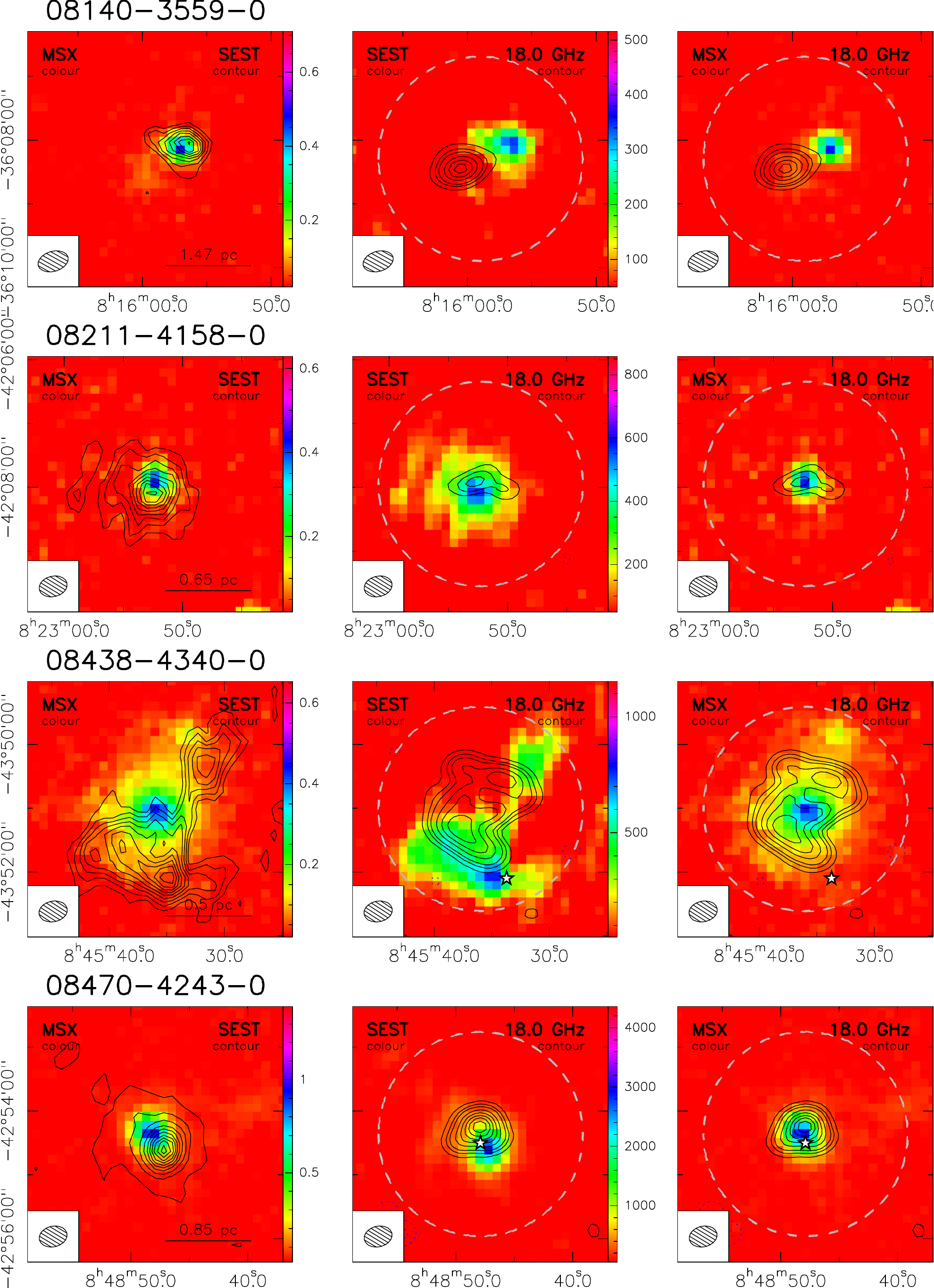}
\end{center}
\caption{\label{f:globalATCAmaps}ATCA, SEST and MSX maps. For each source we show three panels: {\bf left} 21~$\mu$m MSX image (color scale) and 1.2~mm SEST image (contours), {\bf middle} 1.2~mm SEST image (color scale) and 18.0~GHz ATCA image (contours), and {\bf right} 21~$\mu$m MSX image (color scale, as in left panel) and 18.0~GHz ATCA image (contours). The spatial scale is shown in sources with distance determination. The primary beam of the ATCA image is shown as a white dashed circle. Water masers are shown as star symbols.}
\end{figure*}
\begin{figure*}[]
\ContinuedFloat
\begin{center}
\includegraphics[scale=0.8]{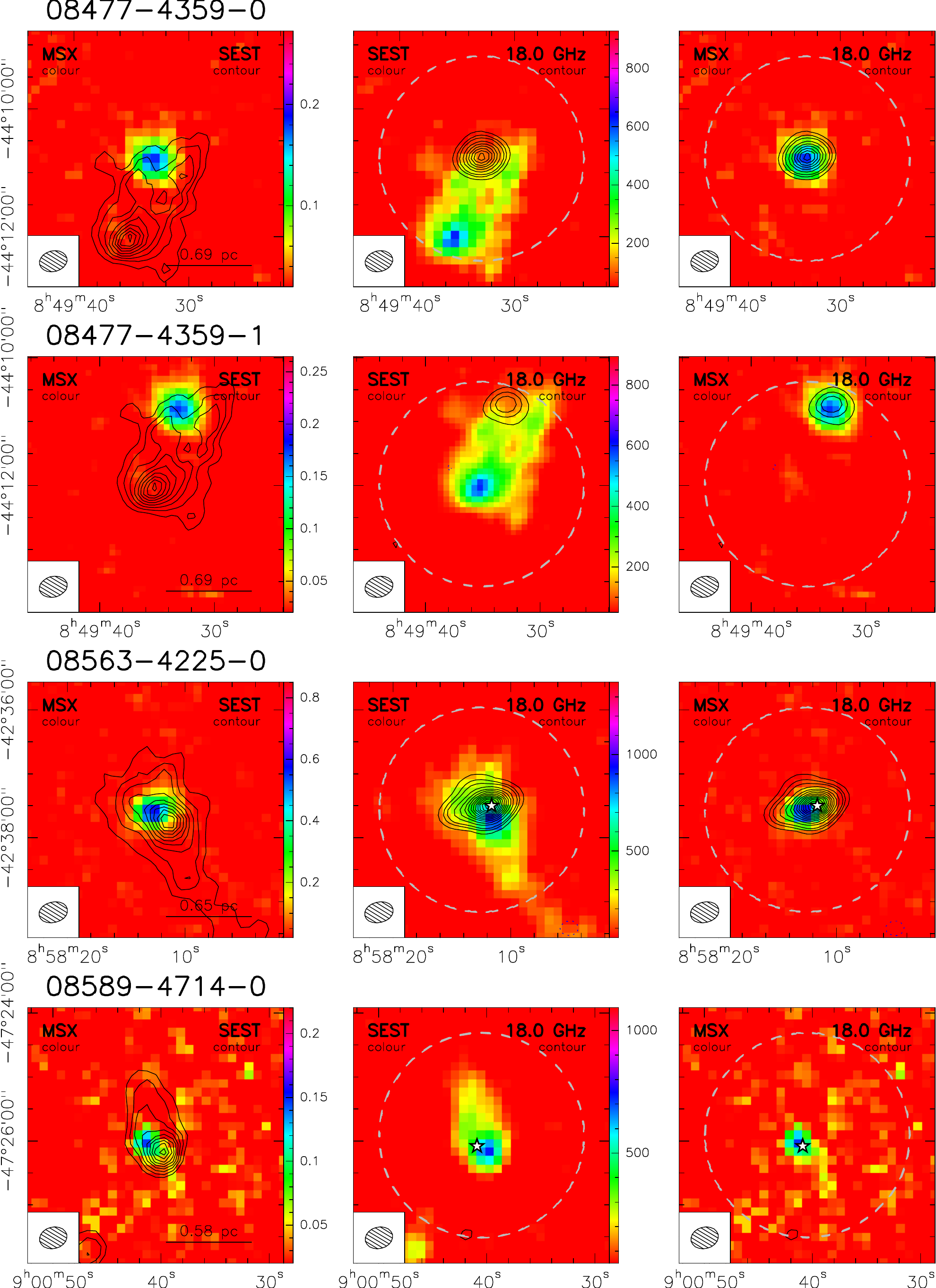}
\end{center}
\caption{continued.}
\end{figure*}
\begin{figure*}[]
\ContinuedFloat
\begin{center}
\includegraphics[scale=0.8]{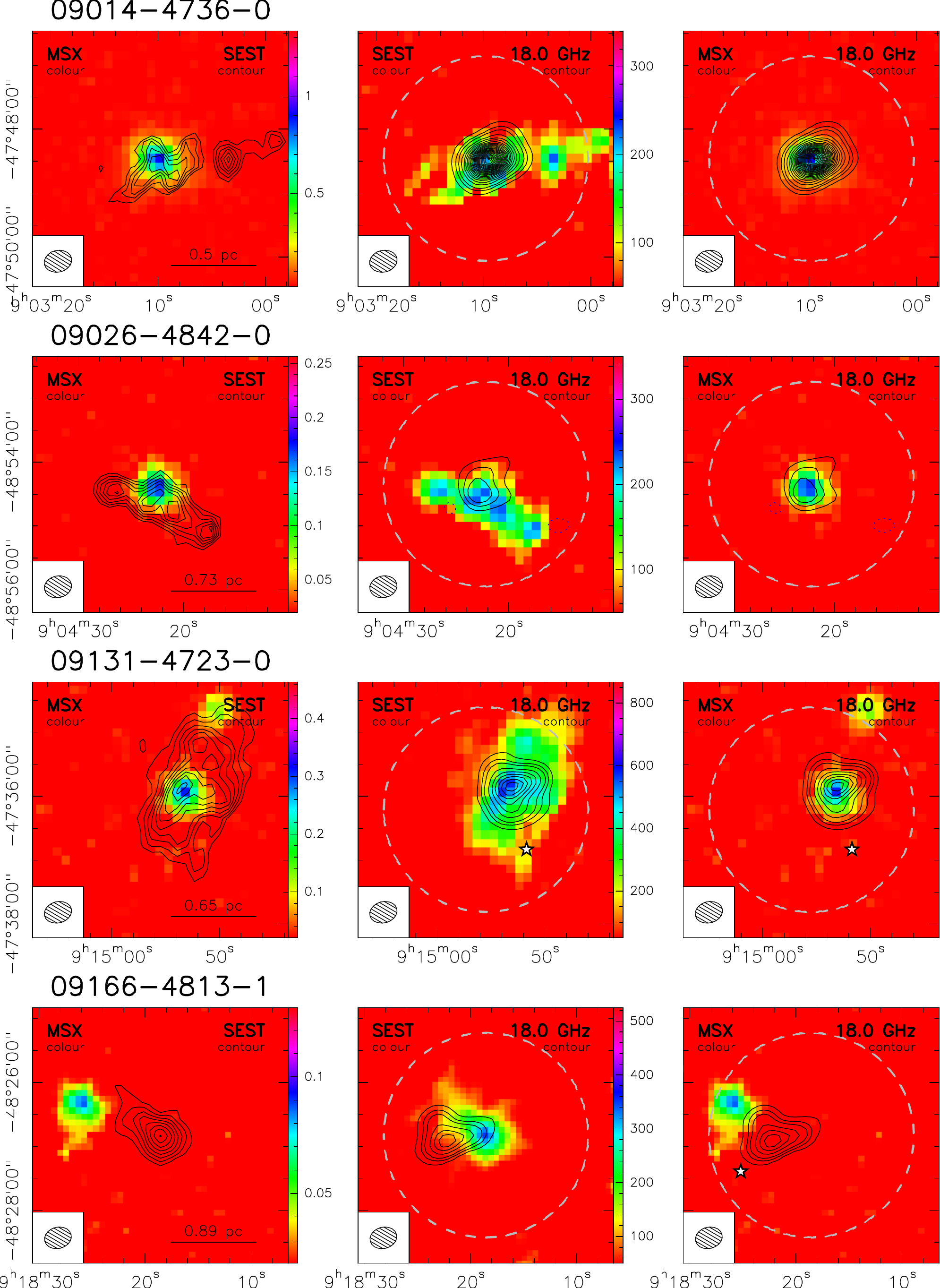}
\end{center}
\caption{continued.}
\end{figure*}
\begin{figure*}[]
\ContinuedFloat
\begin{center}
\includegraphics[scale=0.8]{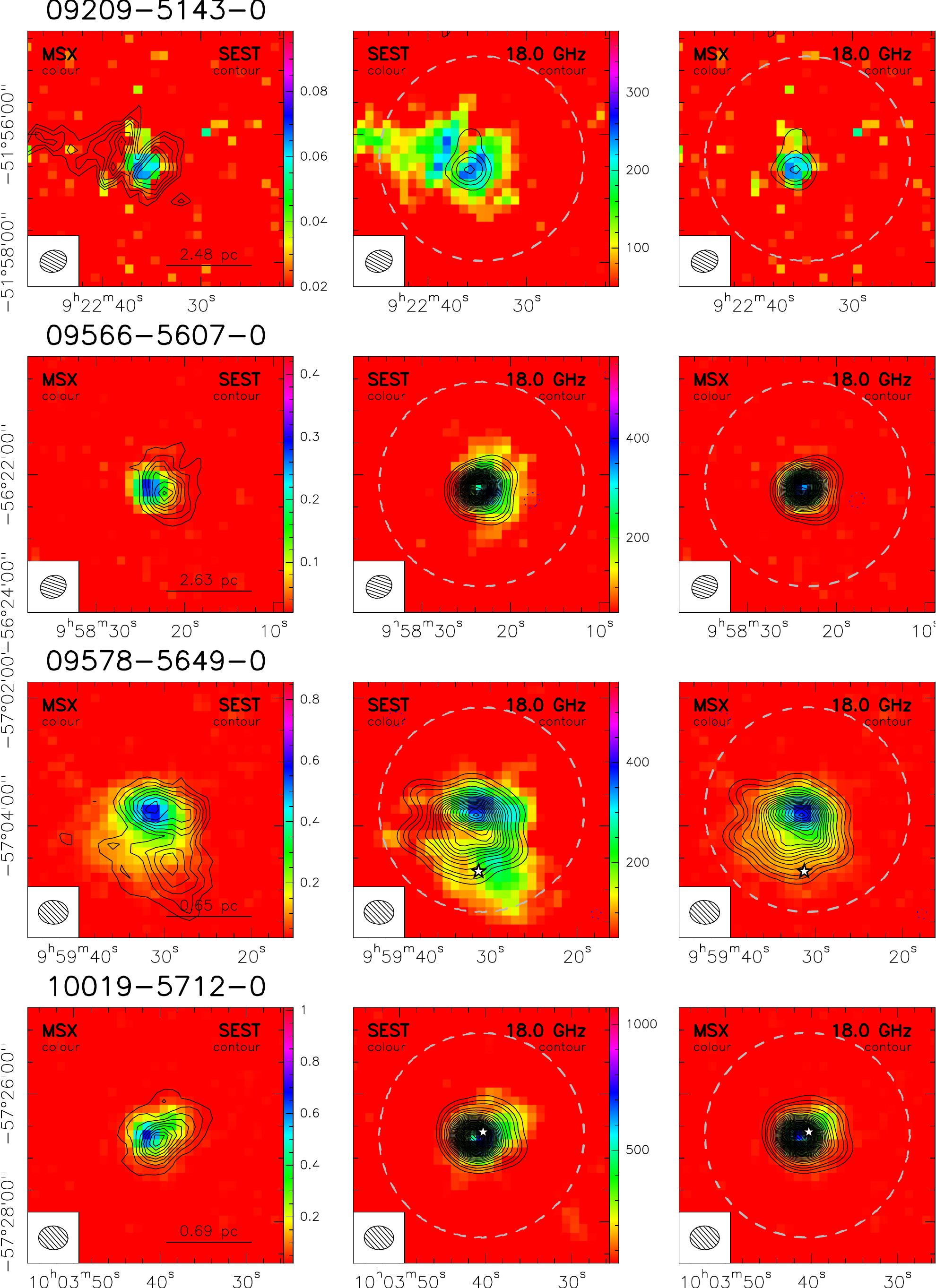}
\end{center}
\caption{continued.}
\end{figure*}
\begin{figure*}[]
\ContinuedFloat
\begin{center}
\includegraphics[scale=0.8]{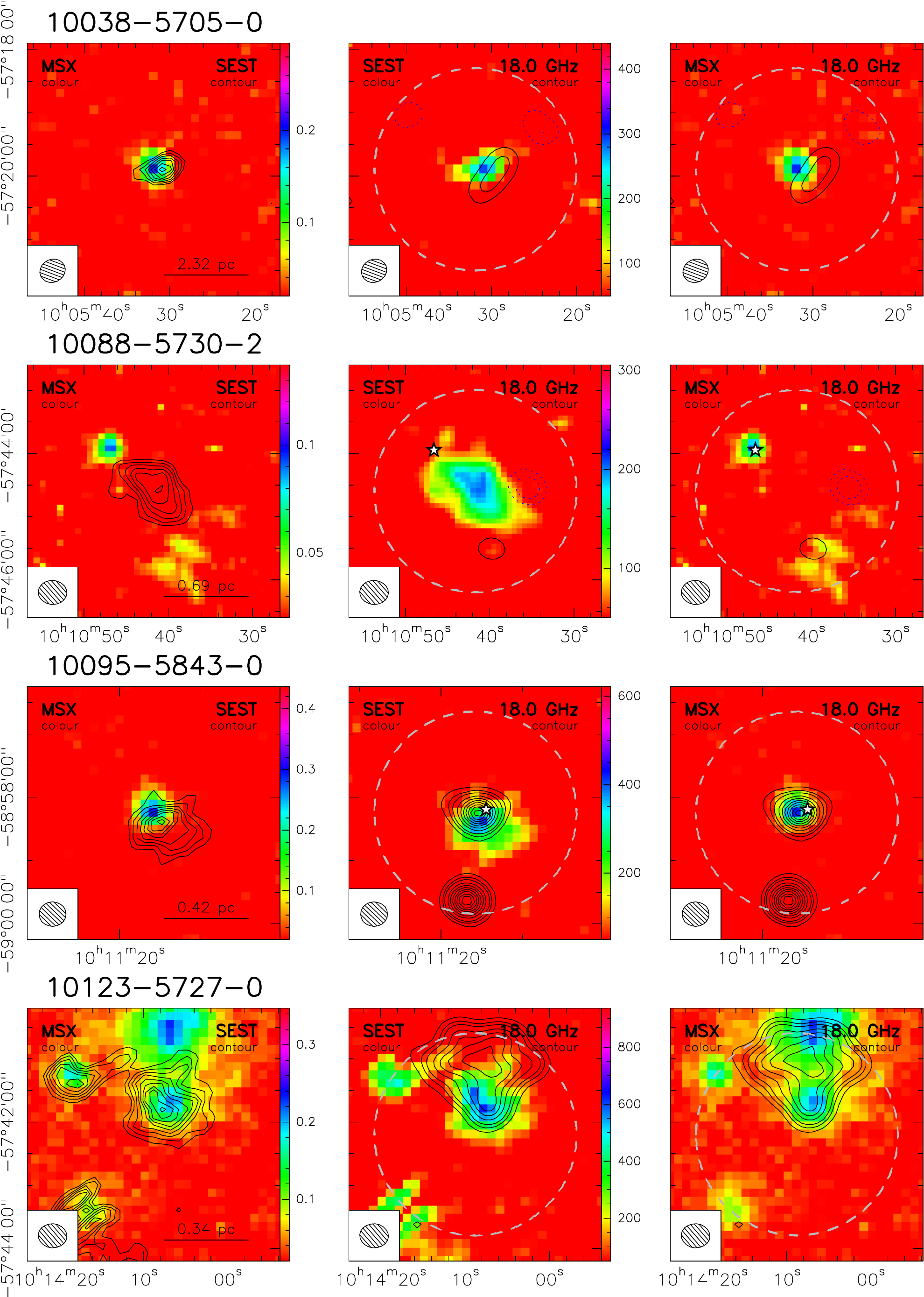}
\end{center}
\caption{continued.}
\end{figure*}
\begin{figure*}[]
\ContinuedFloat
\begin{center}
\includegraphics[scale=0.8]{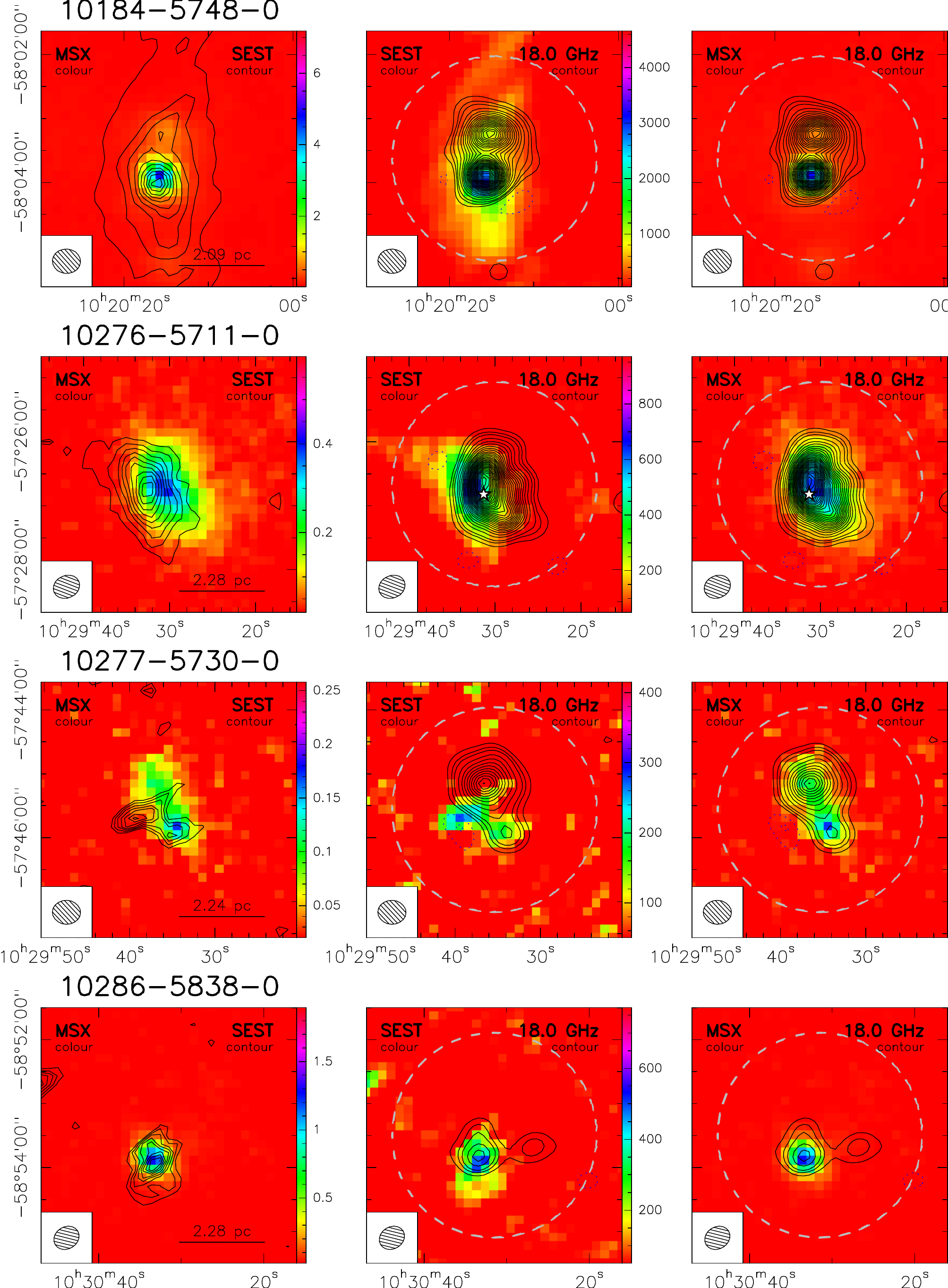}
\end{center}
\caption{continued.}
\end{figure*}
\begin{figure*}[]
\ContinuedFloat
\begin{center}
\includegraphics[scale=0.8]{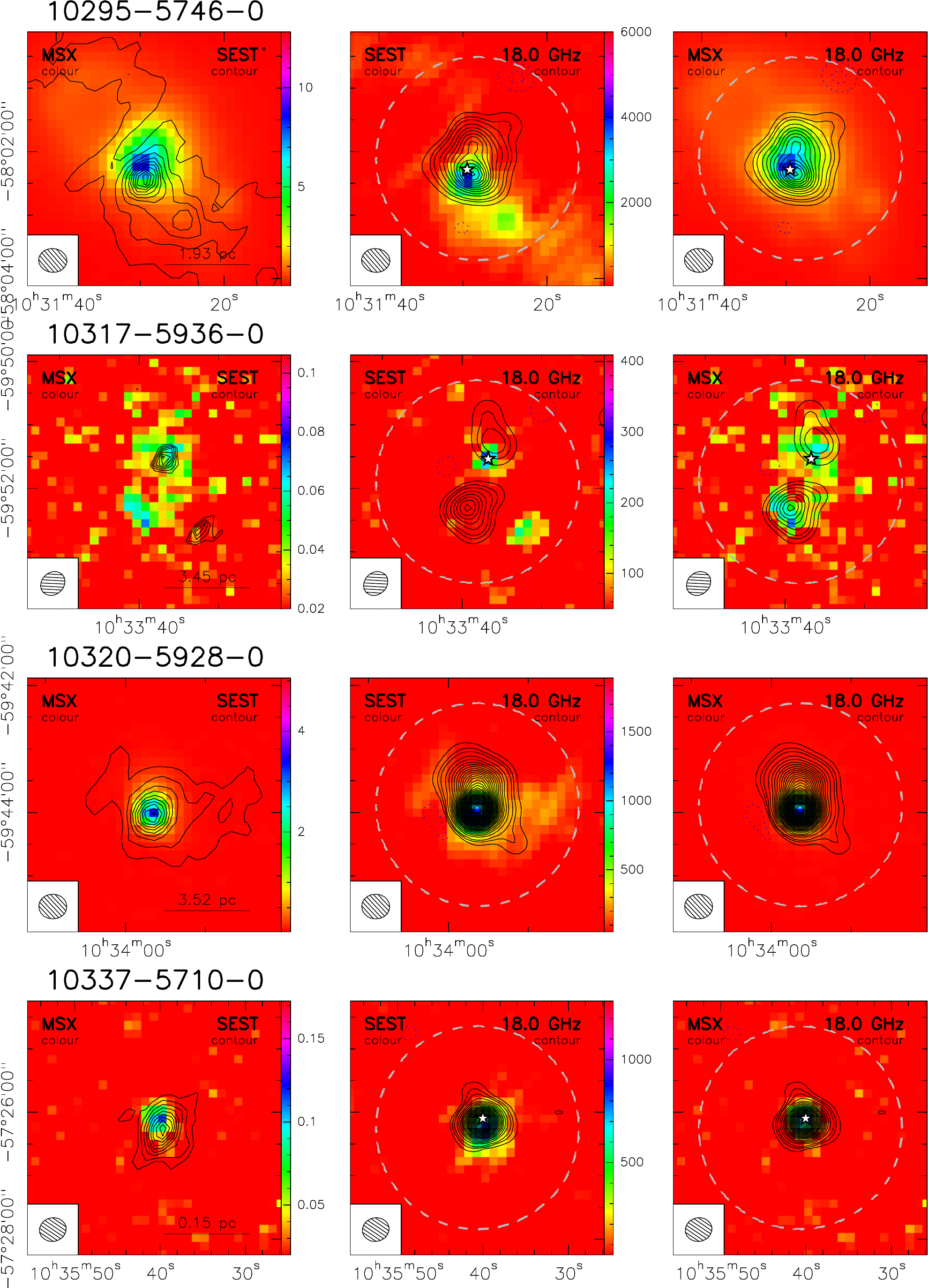}
\end{center}
\caption{continued.}
\end{figure*}
\begin{figure*}[]
\ContinuedFloat
\begin{center}
\includegraphics[scale=0.8]{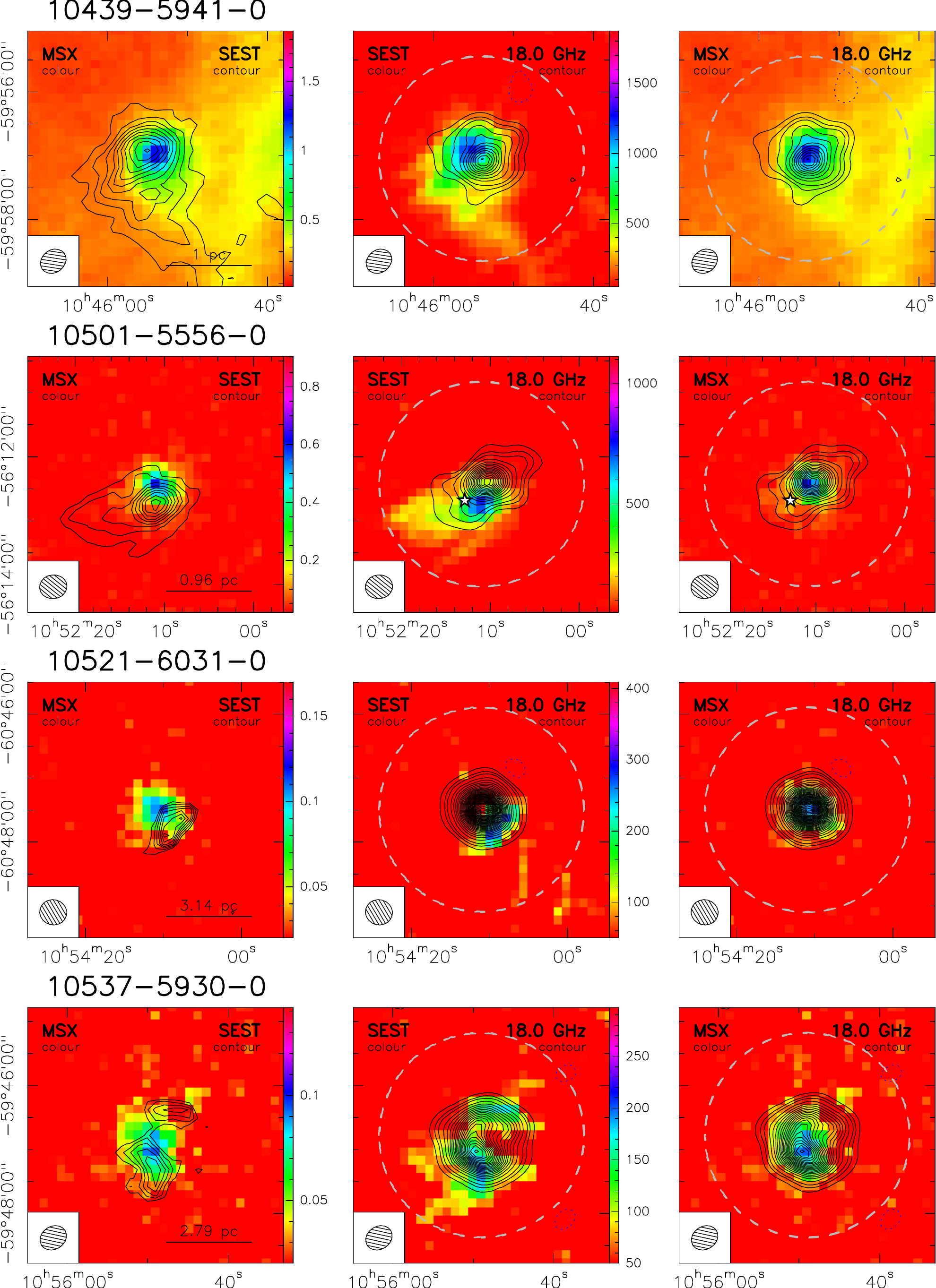}
\end{center}
\caption{continued.}
\end{figure*}
\begin{figure*}[]
\ContinuedFloat
\begin{center}
\includegraphics[scale=0.8]{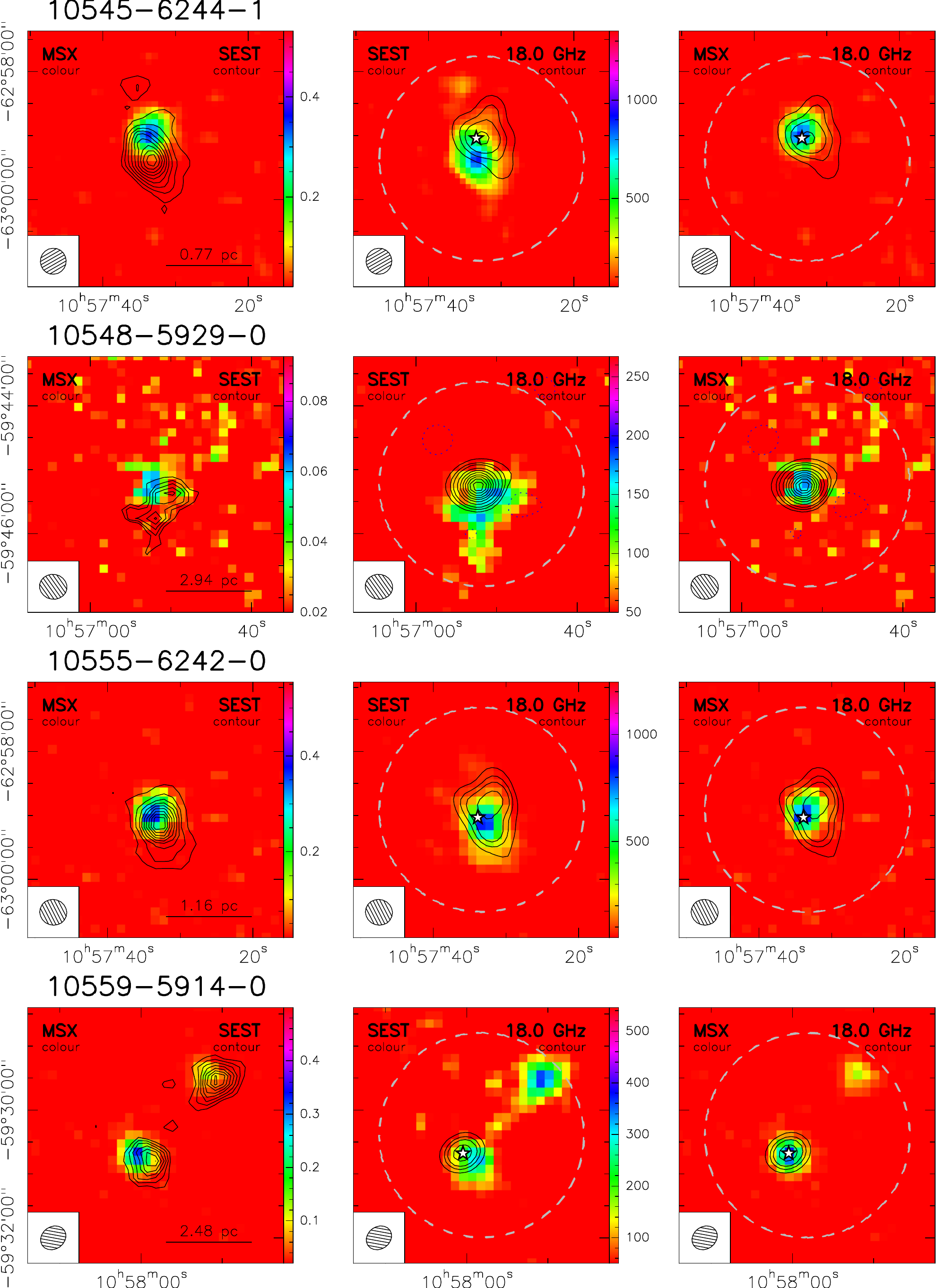}
\end{center}
\caption{continued.}
\end{figure*}
\clearpage
\begin{figure*}[]
\ContinuedFloat
\begin{center}
\includegraphics[scale=0.8]{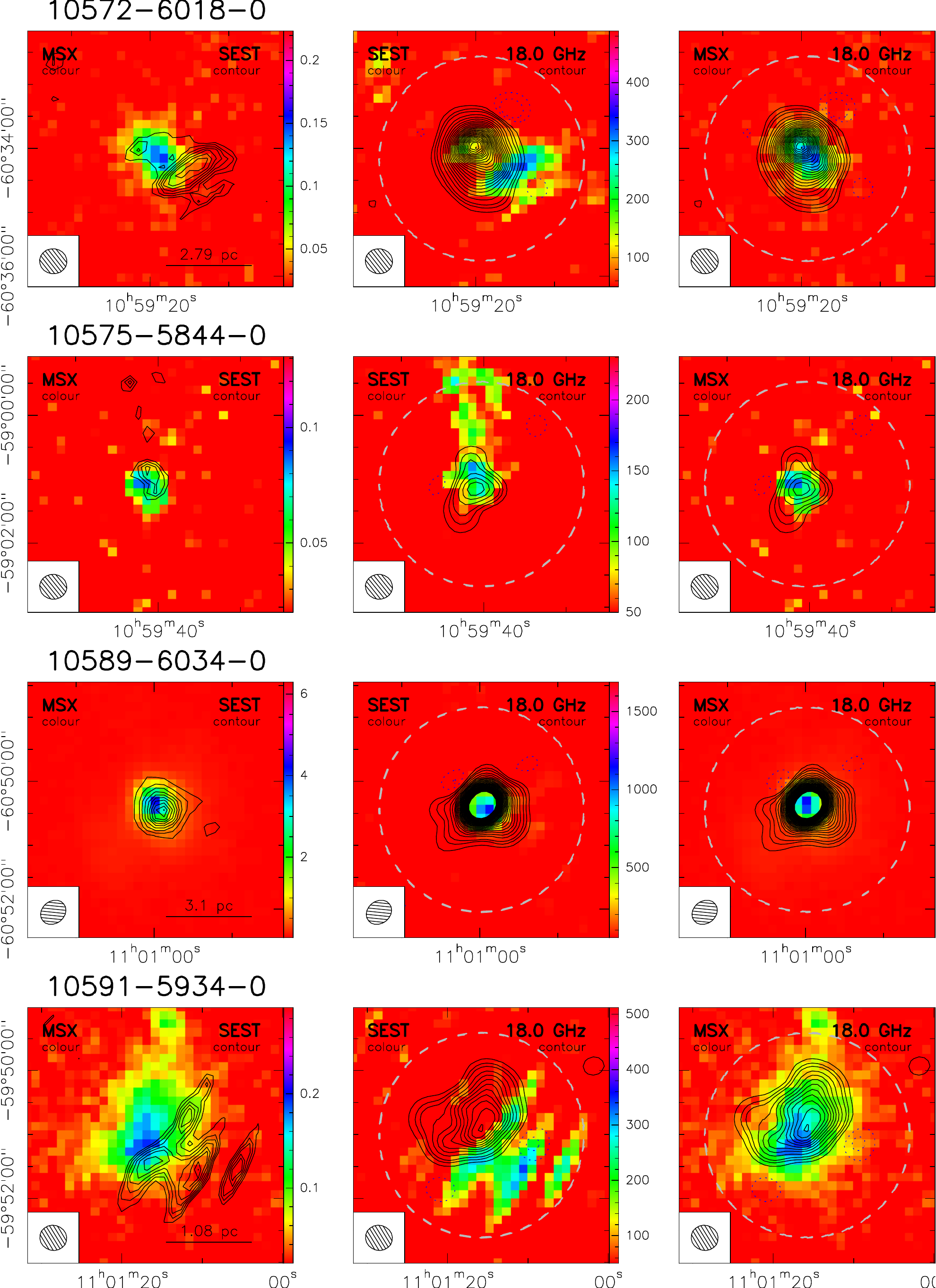}
\end{center}
\caption{continued.}
\end{figure*}
\begin{figure*}[]
\ContinuedFloat
\begin{center}
\includegraphics[scale=0.8]{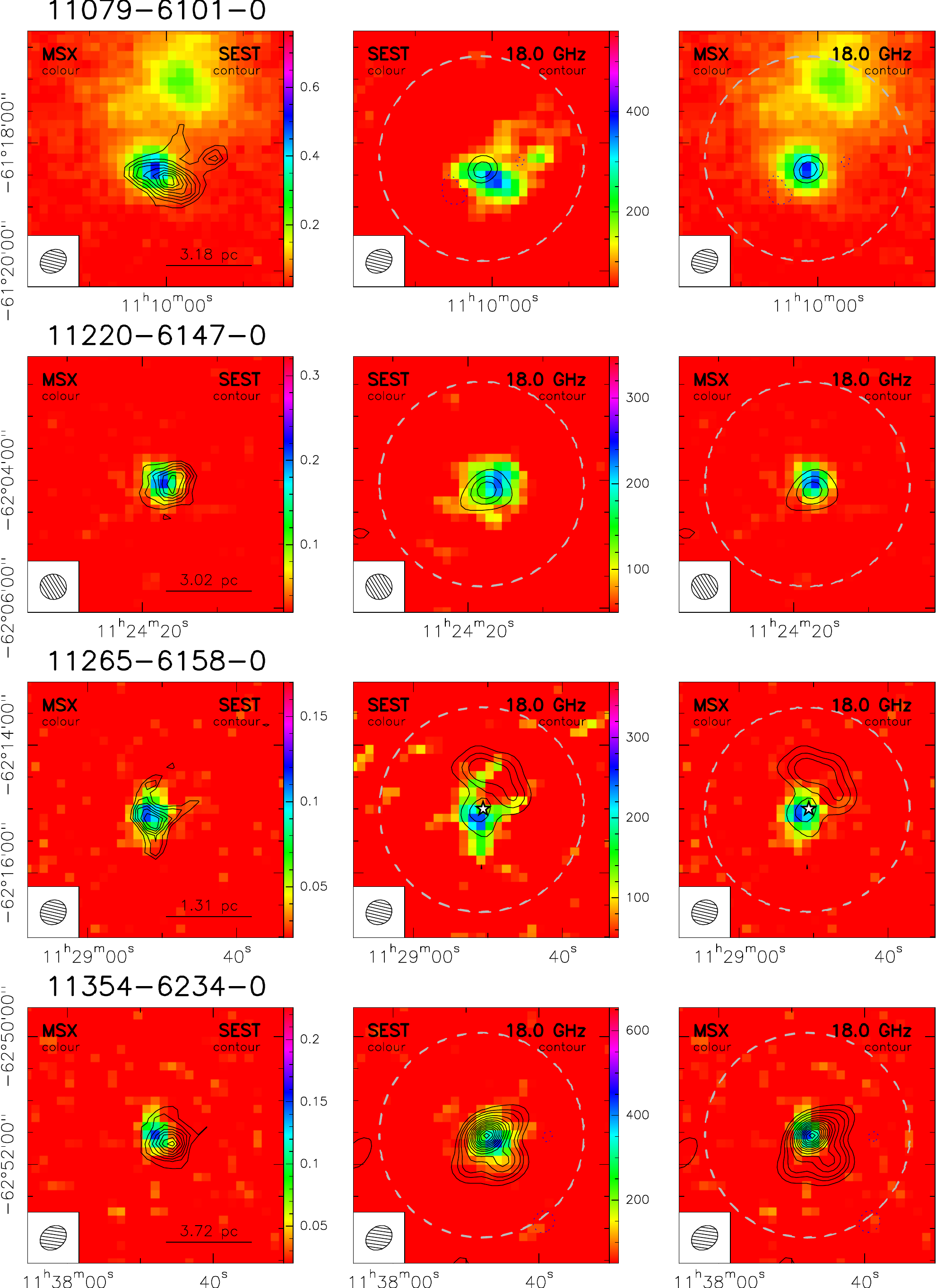}
\end{center}
\caption{continued.}
\end{figure*}
\begin{figure*}[]
\ContinuedFloat
\begin{center}
\includegraphics[scale=0.8]{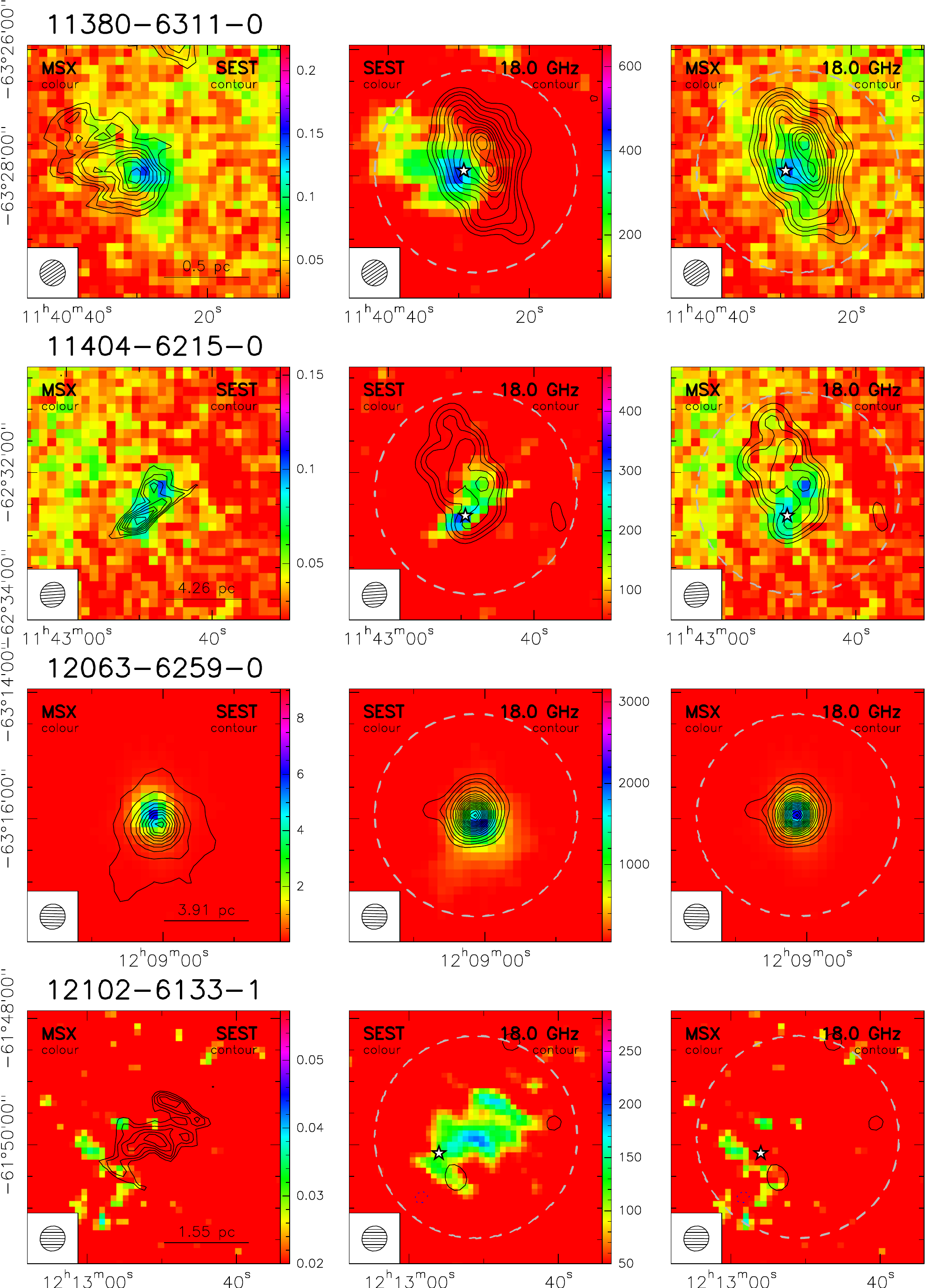}
\end{center}
\caption{continued.}
\end{figure*}
\begin{figure*}[]
\ContinuedFloat
\begin{center}
\includegraphics[scale=0.8]{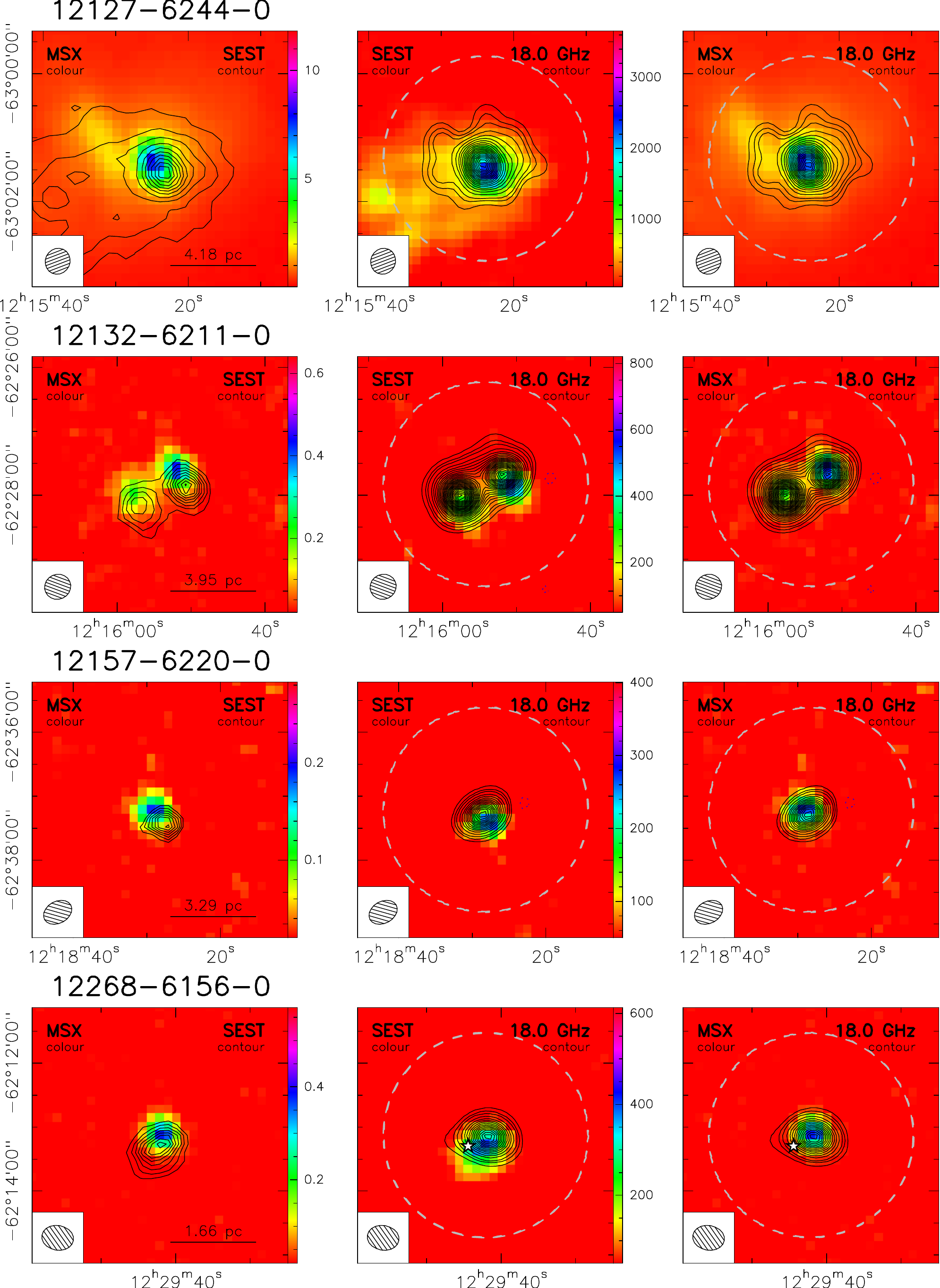}
\end{center}
\caption{continued.}
\end{figure*}
\begin{figure*}[]
\ContinuedFloat
\begin{center}
\includegraphics[scale=0.8]{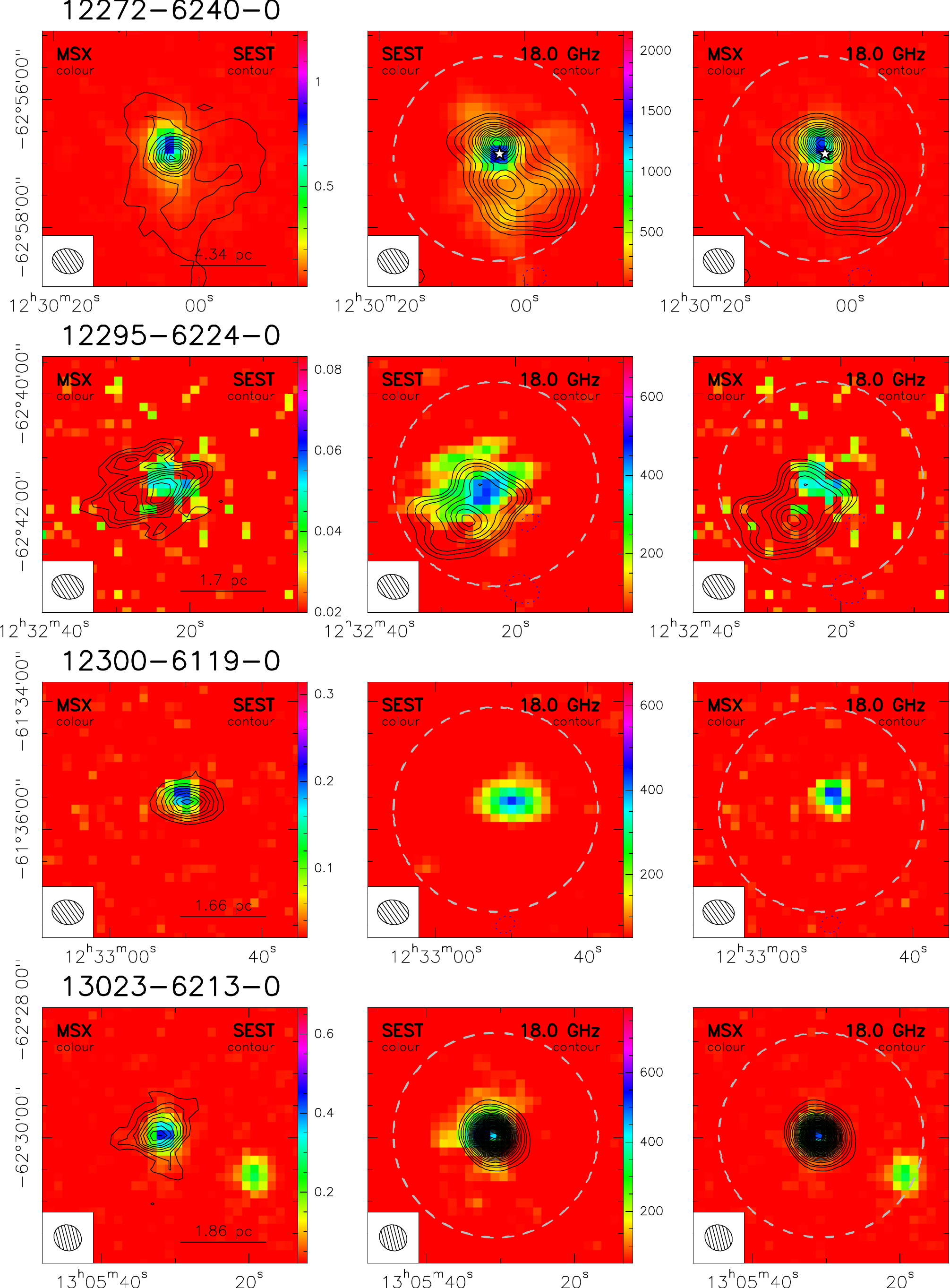}
\end{center}
\caption{continued.}
\end{figure*}
\begin{figure*}[]
\ContinuedFloat
\begin{center}
\includegraphics[scale=0.8]{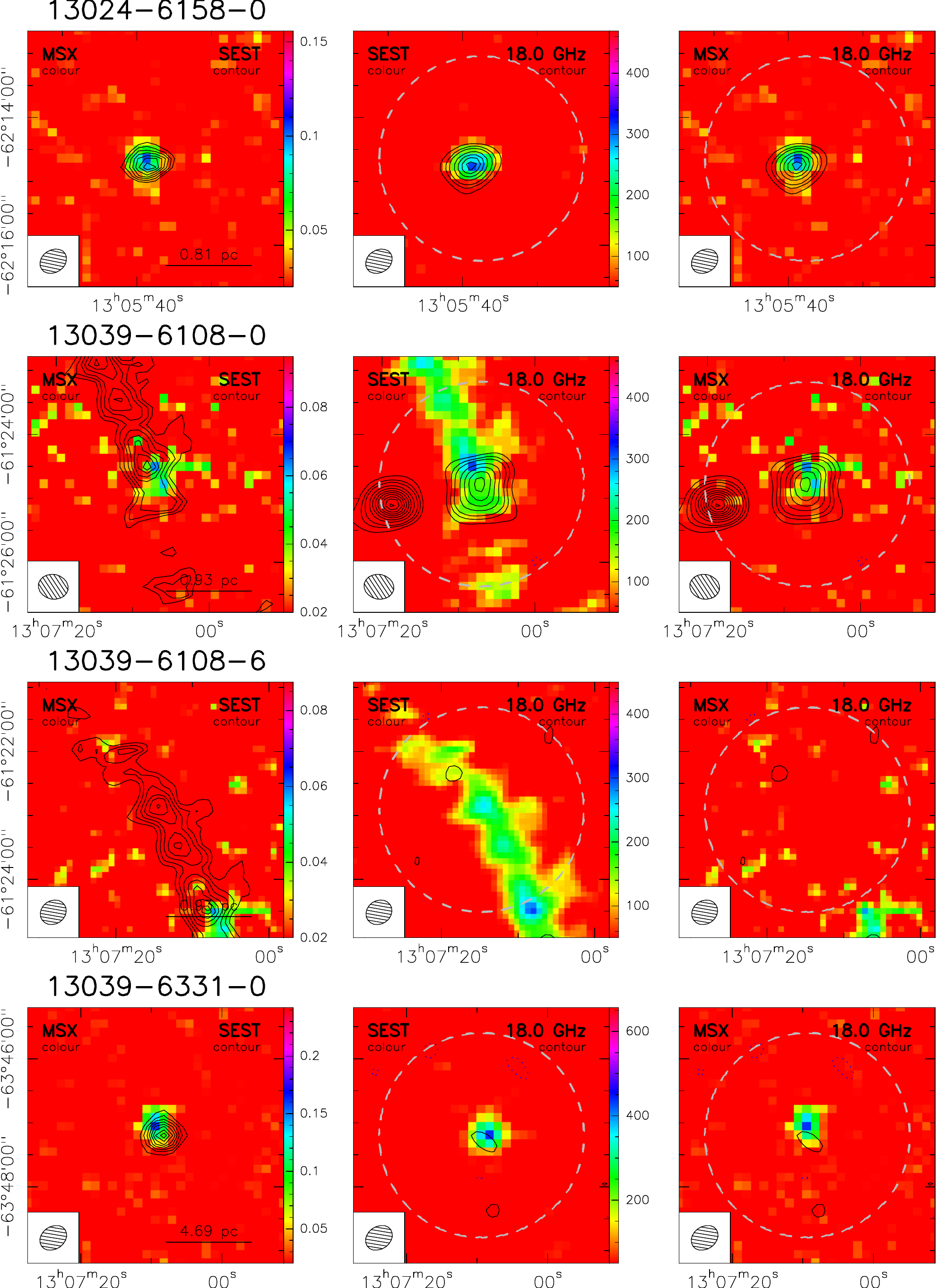}
\end{center}
\caption{continued.}
\end{figure*}
\begin{figure*}[]
\ContinuedFloat
\begin{center}
\includegraphics[scale=0.8]{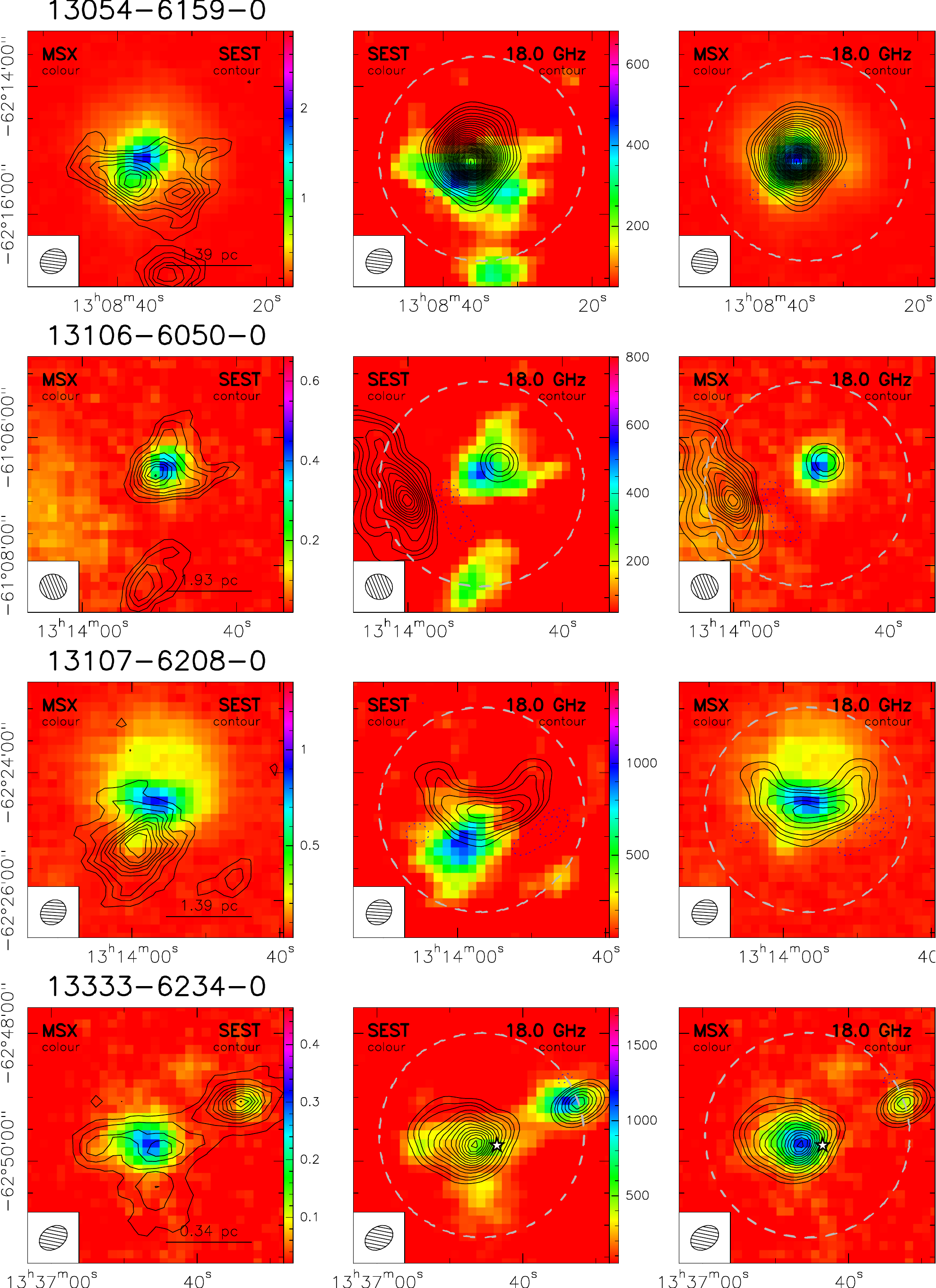}
\end{center}
\caption{continued.}
\end{figure*}
\clearpage
\begin{figure*}[]
\ContinuedFloat
\begin{center}
\includegraphics[scale=0.8]{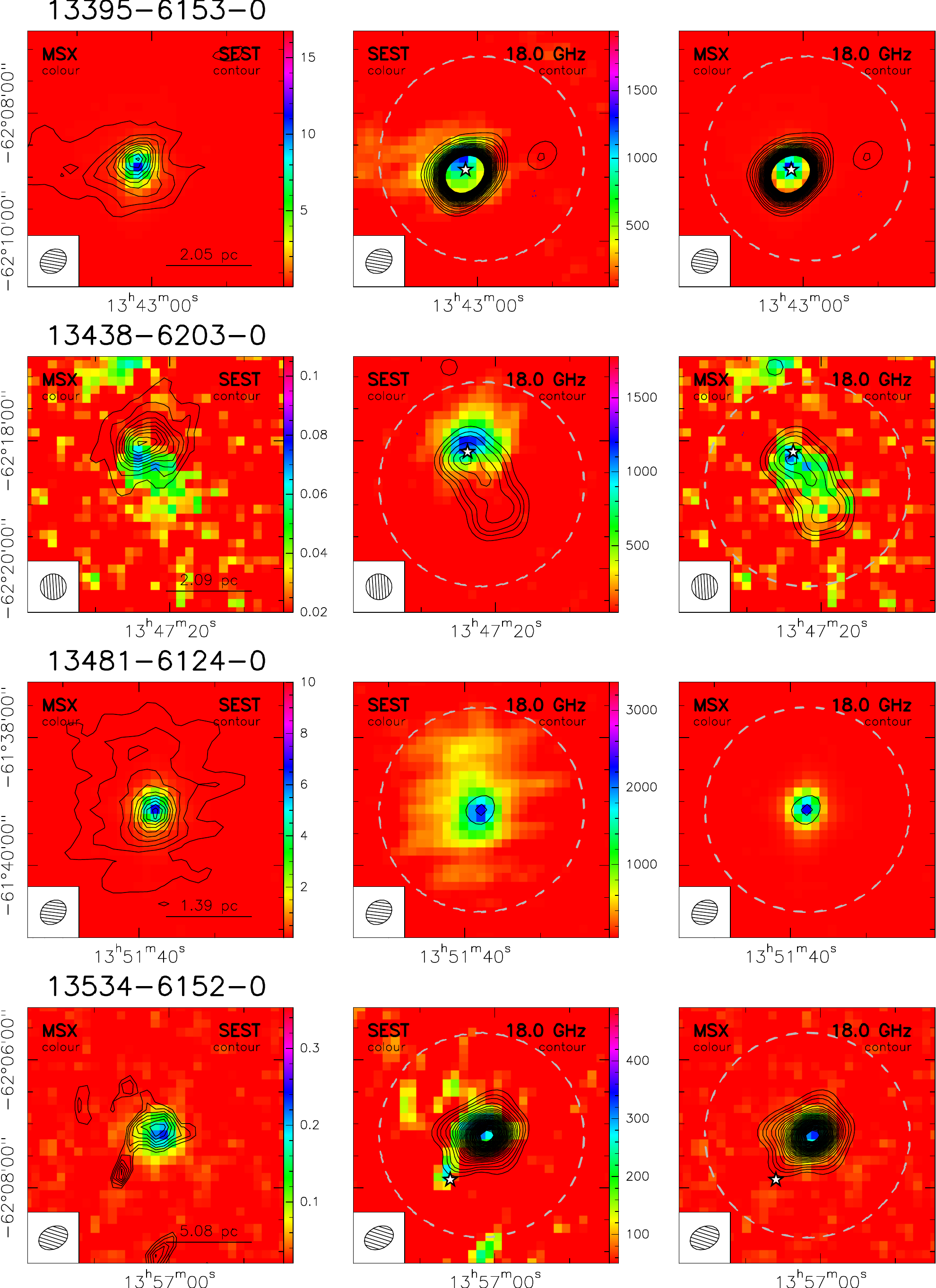}
\end{center}
\caption{continued.}
\end{figure*}
\begin{figure*}[]
\ContinuedFloat
\begin{center}
\includegraphics[scale=0.8]{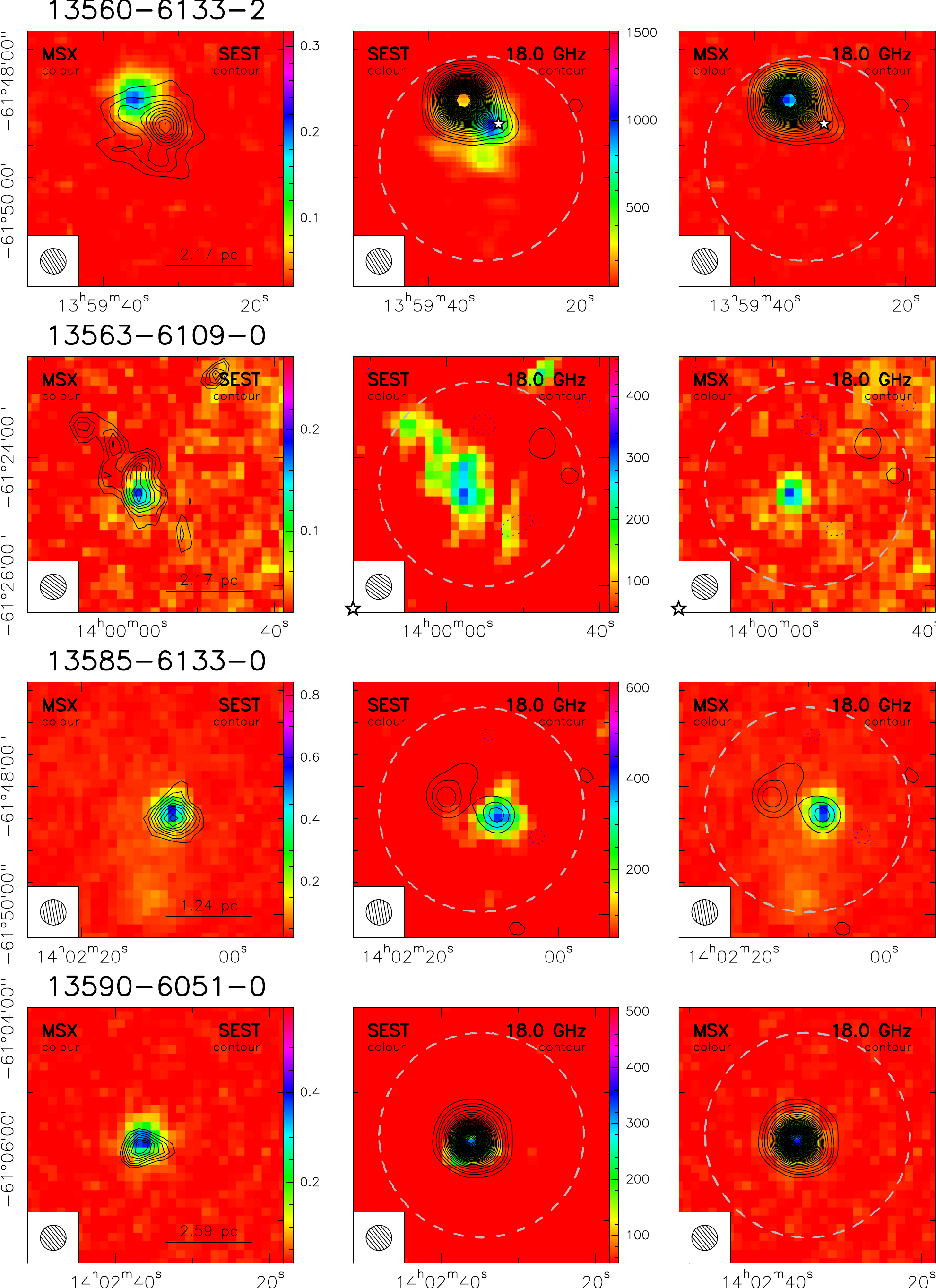}
\end{center}
\caption{continued.}
\end{figure*}
\begin{figure*}[]
\ContinuedFloat
\begin{center}
\includegraphics[scale=0.8]{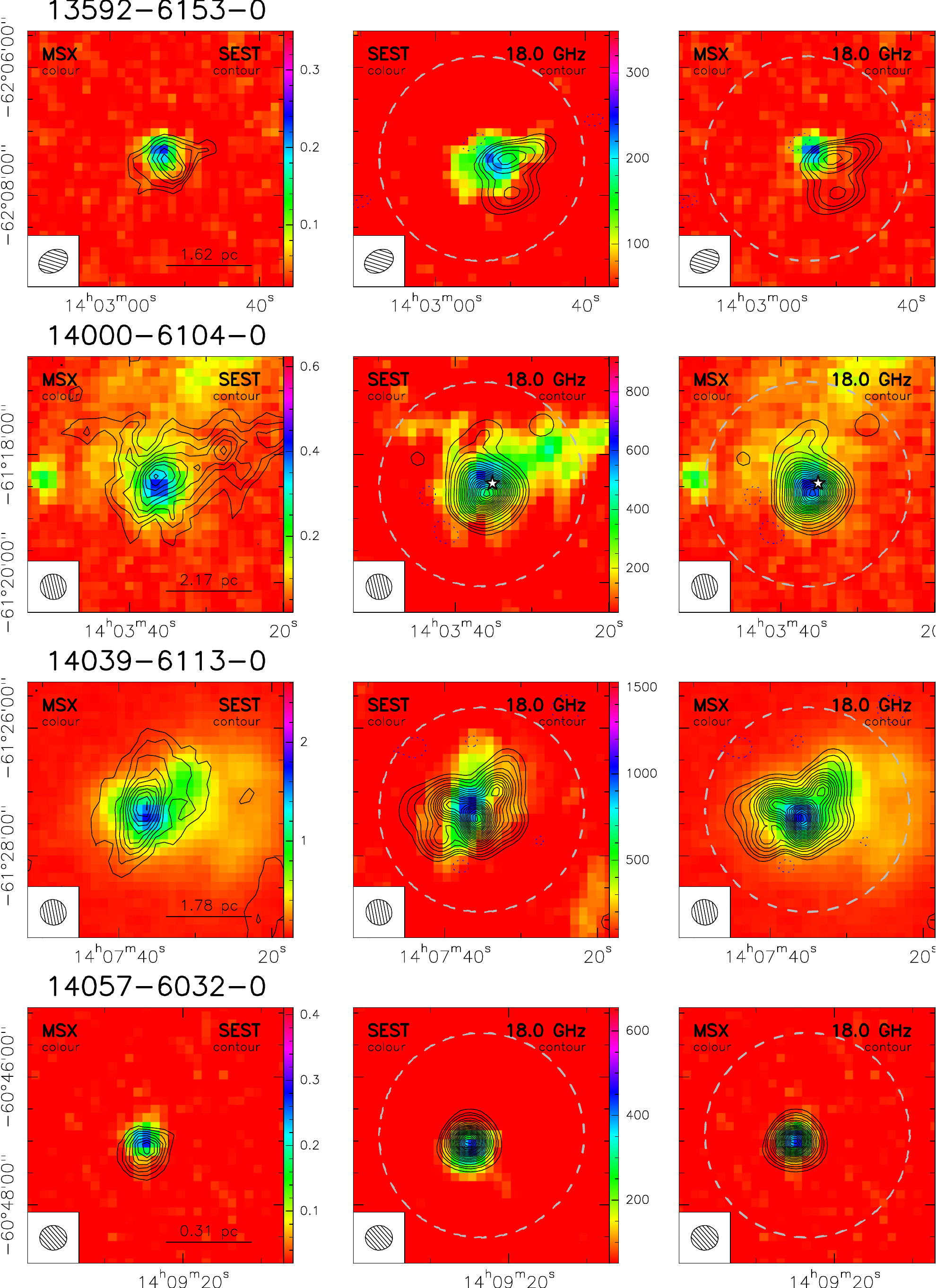}
\end{center}
\caption{continued.}
\end{figure*}
\clearpage
\begin{figure*}[t!]
\ContinuedFloat
\begin{center}
\includegraphics[scale=0.8]{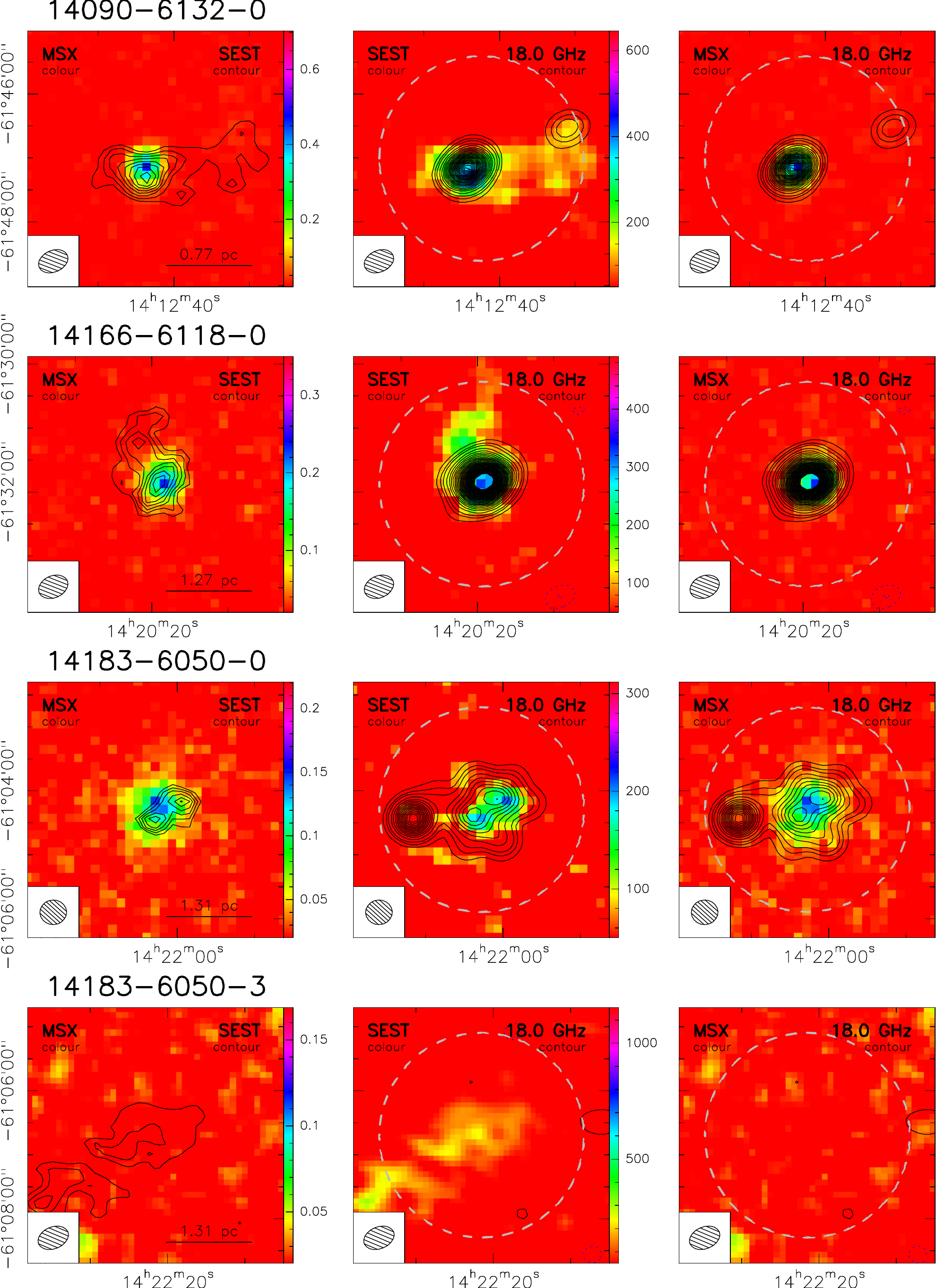}
\end{center}
\caption{continued.}
\end{figure*}
\begin{figure*}[t!]
\ContinuedFloat
\begin{center}
\includegraphics[scale=0.8]{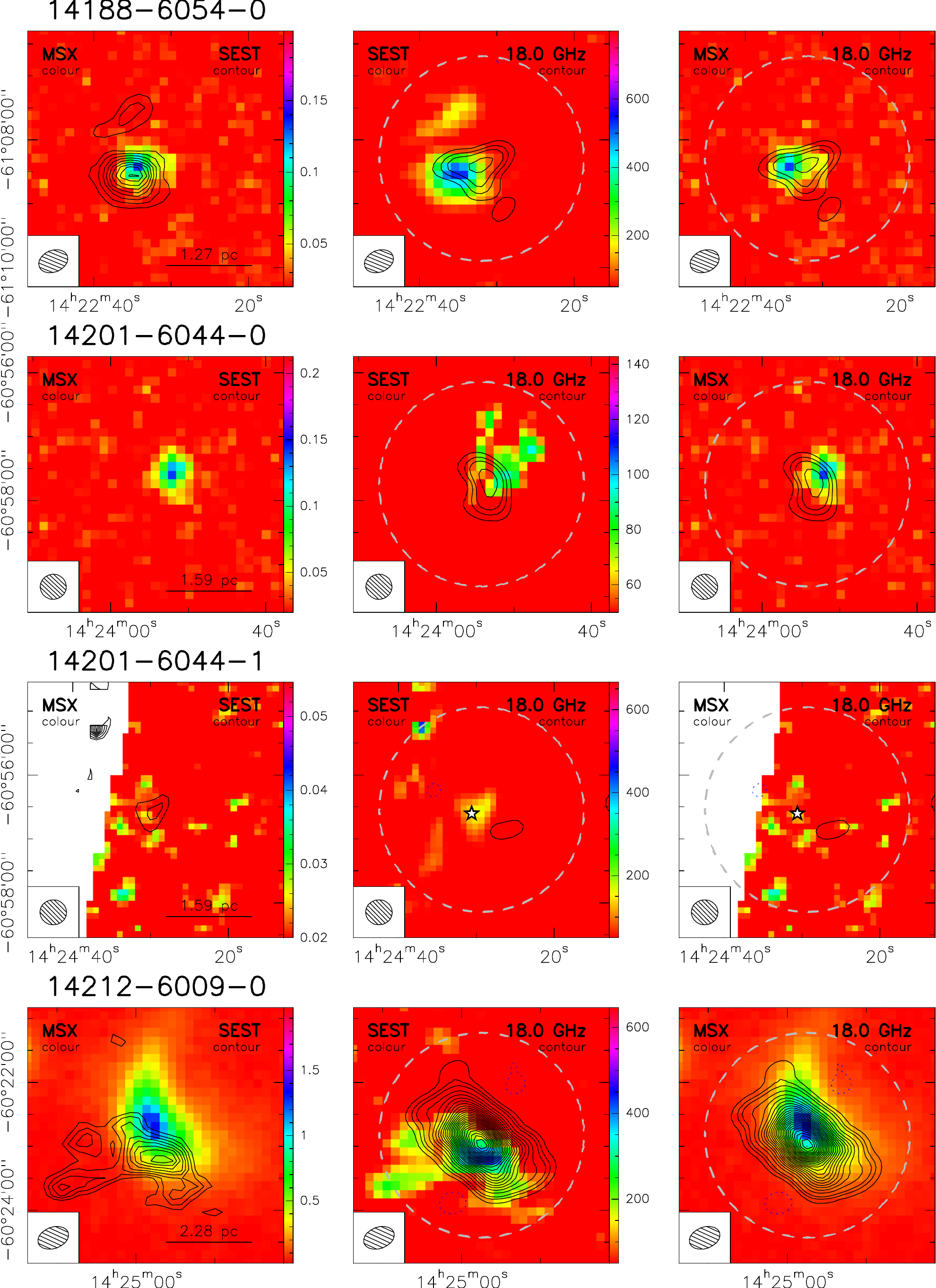}
\end{center}
\caption{continued.}
\end{figure*}
\begin{figure*}[t!]
\ContinuedFloat
\begin{center}
\includegraphics[scale=0.8]{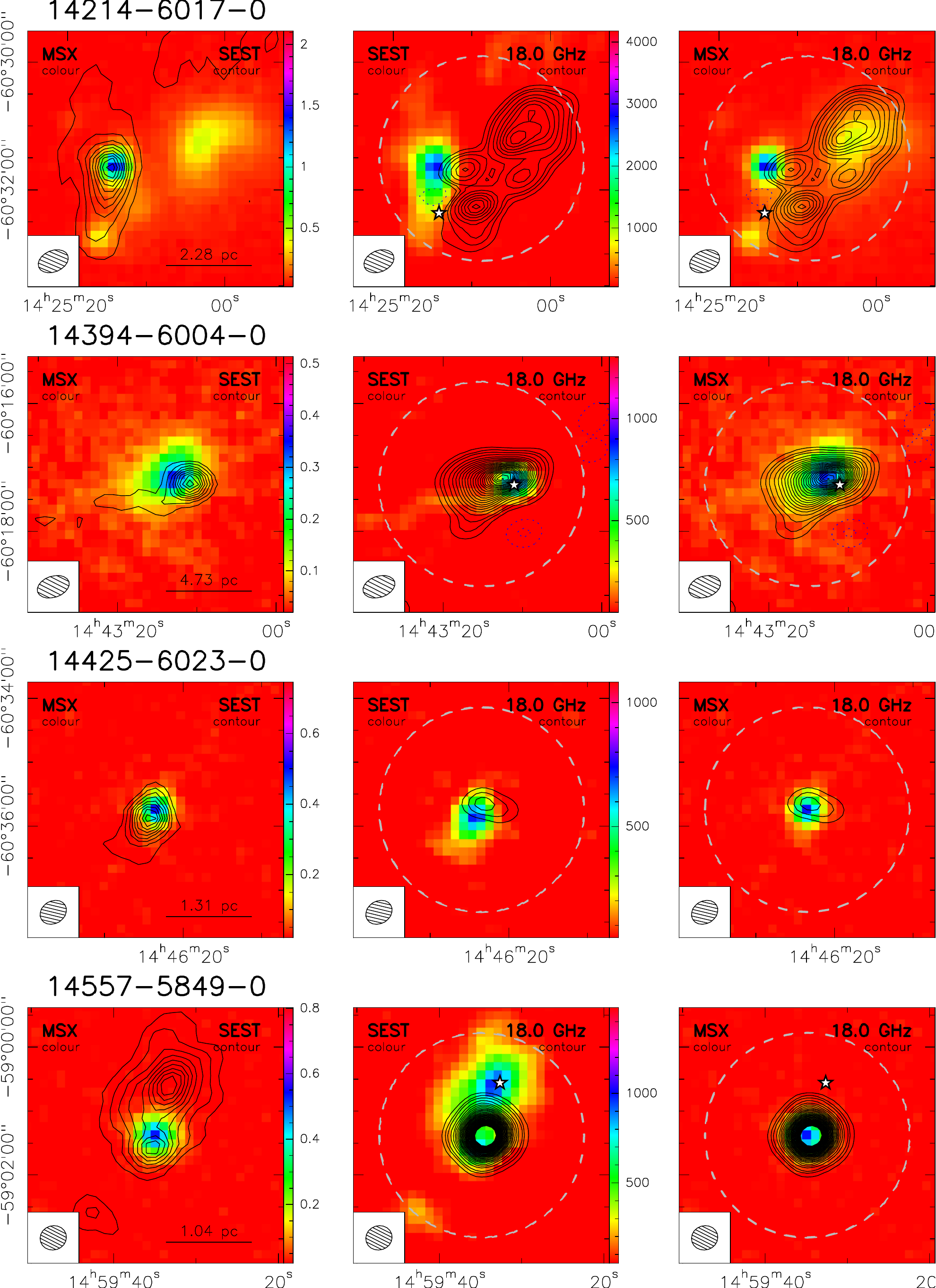}
\end{center}
\caption{continued.}
\end{figure*}
\begin{figure*}[t!]
\ContinuedFloat
\begin{center}
\includegraphics[scale=0.8]{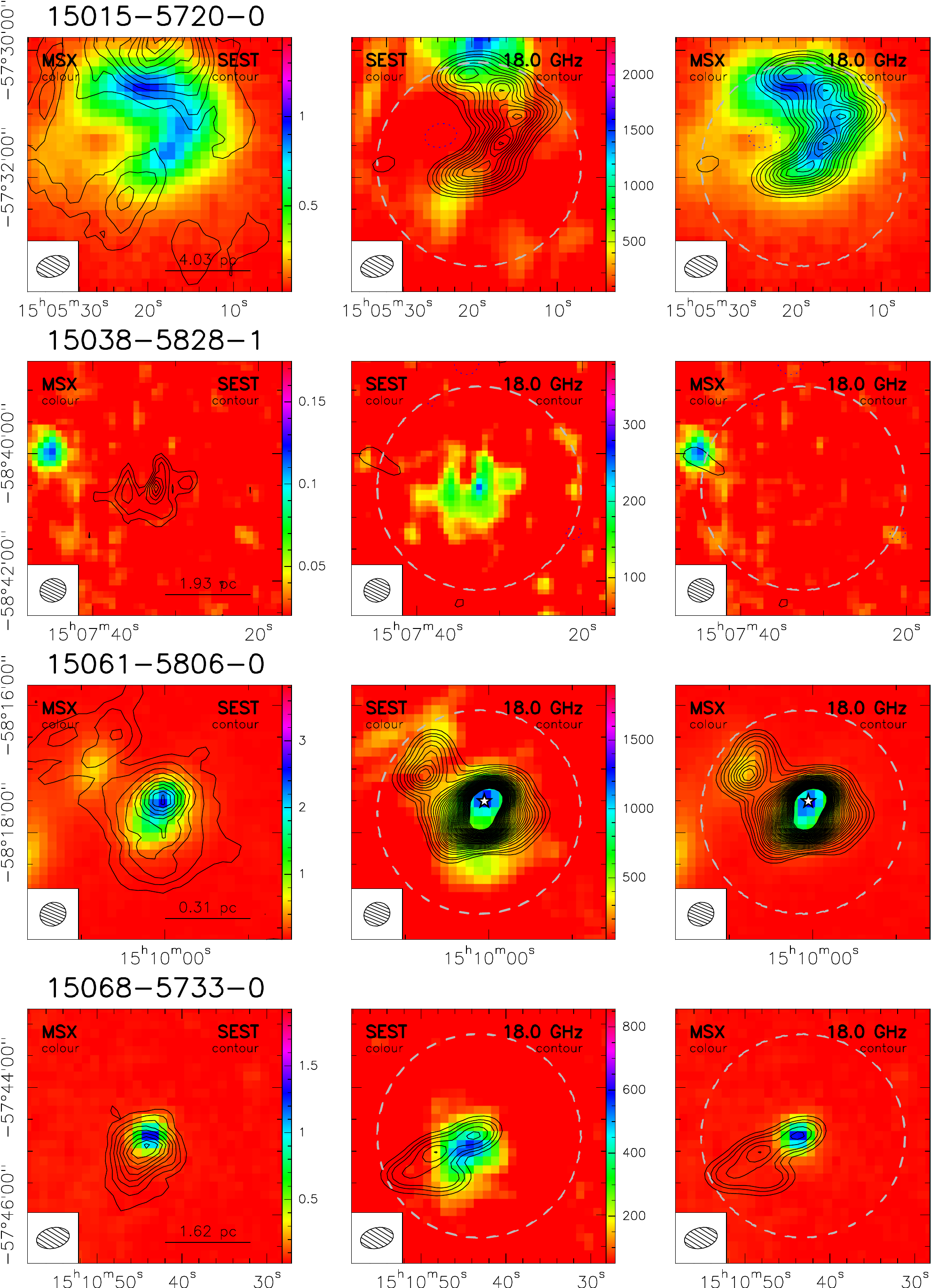}
\end{center}
\caption{continued.}
\end{figure*}
\begin{figure*}[t!]
\ContinuedFloat
\begin{center}
\includegraphics[scale=0.8]{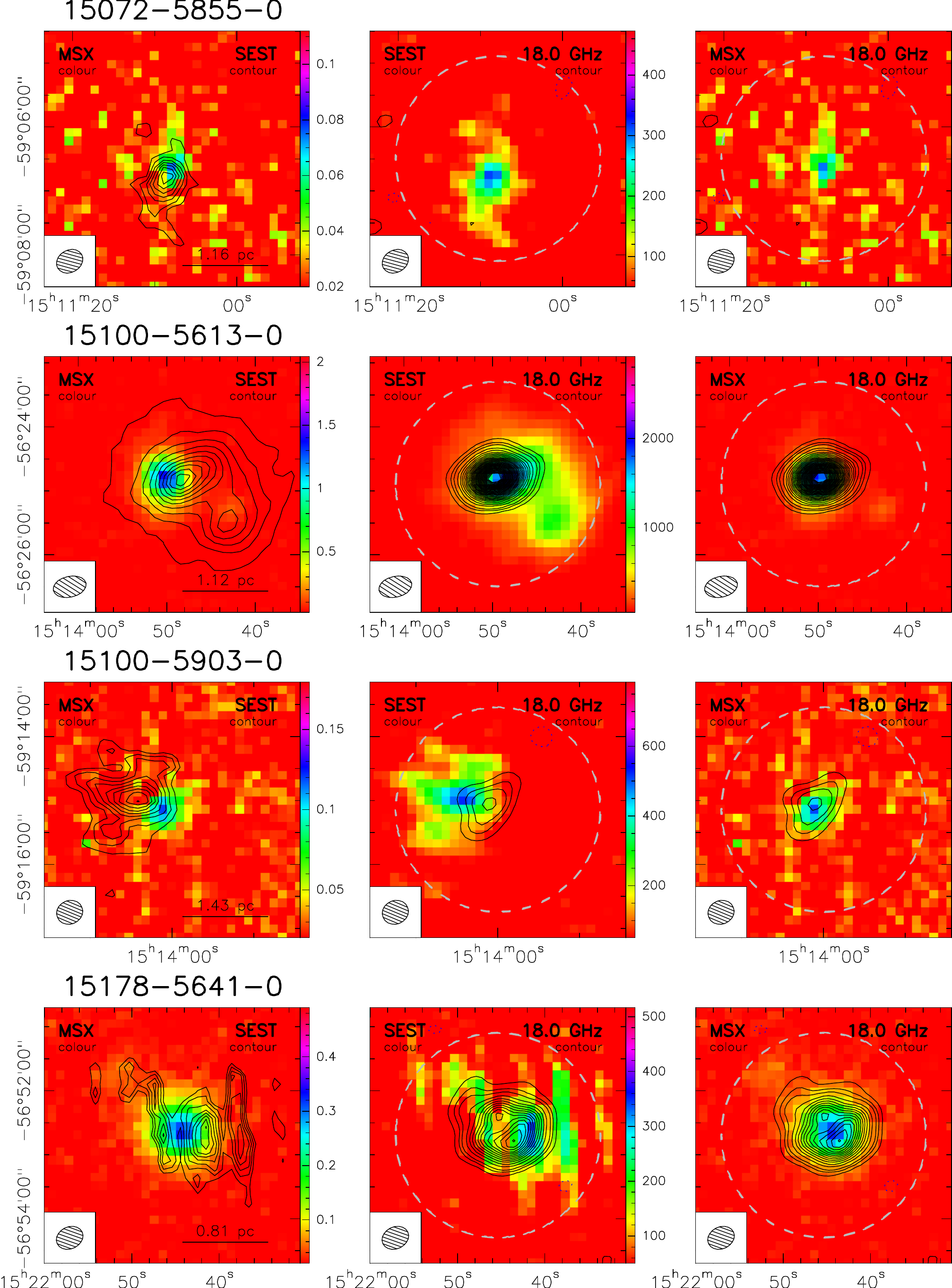}
\end{center}
\caption{continued.}
\end{figure*}
\begin{figure*}[t!]
\ContinuedFloat
\begin{center}
\includegraphics[scale=0.8]{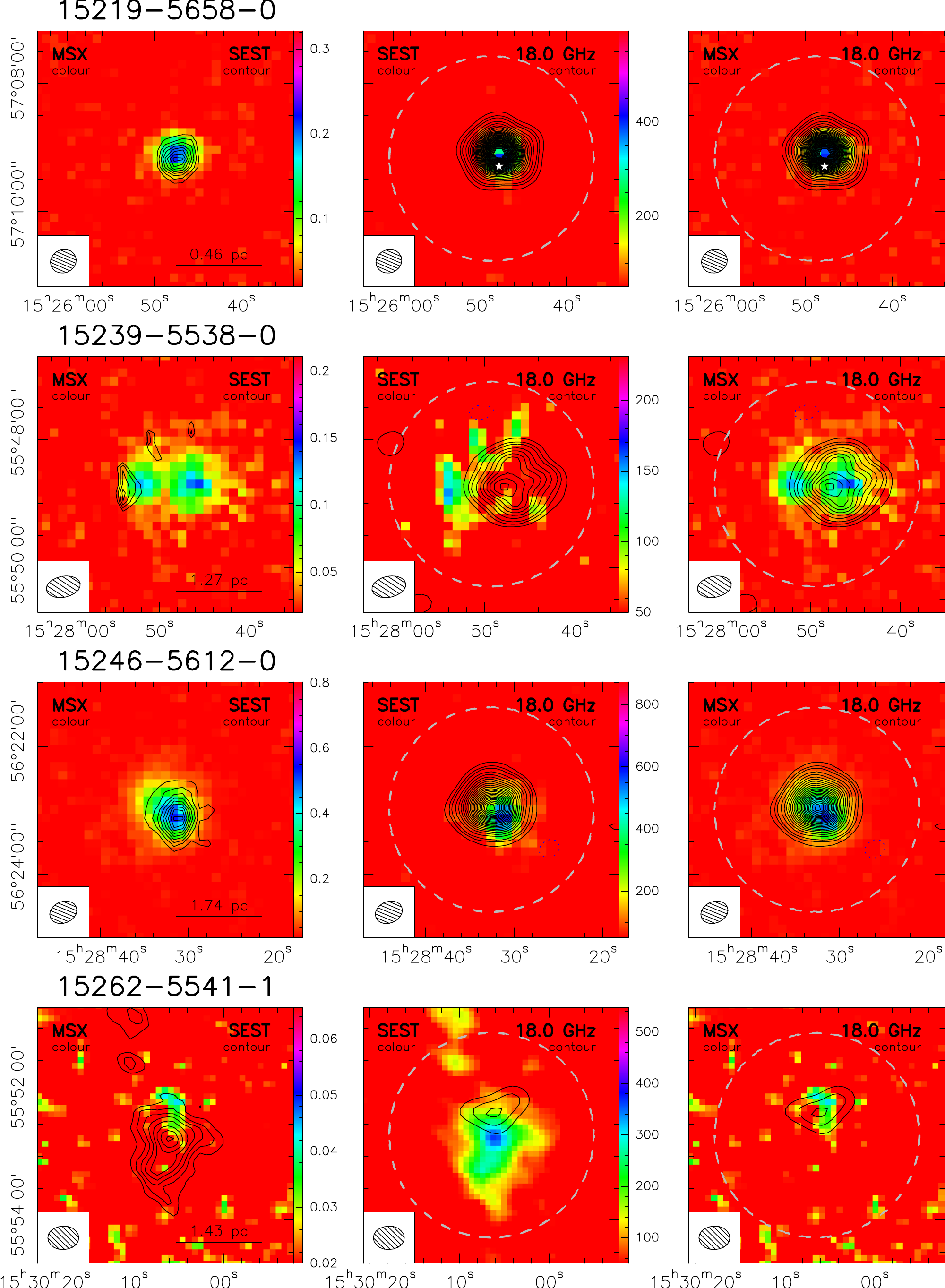}
\end{center}
\caption{continued.}
\end{figure*}
\begin{figure*}[t!]
\ContinuedFloat
\begin{center}
\includegraphics[scale=0.8]{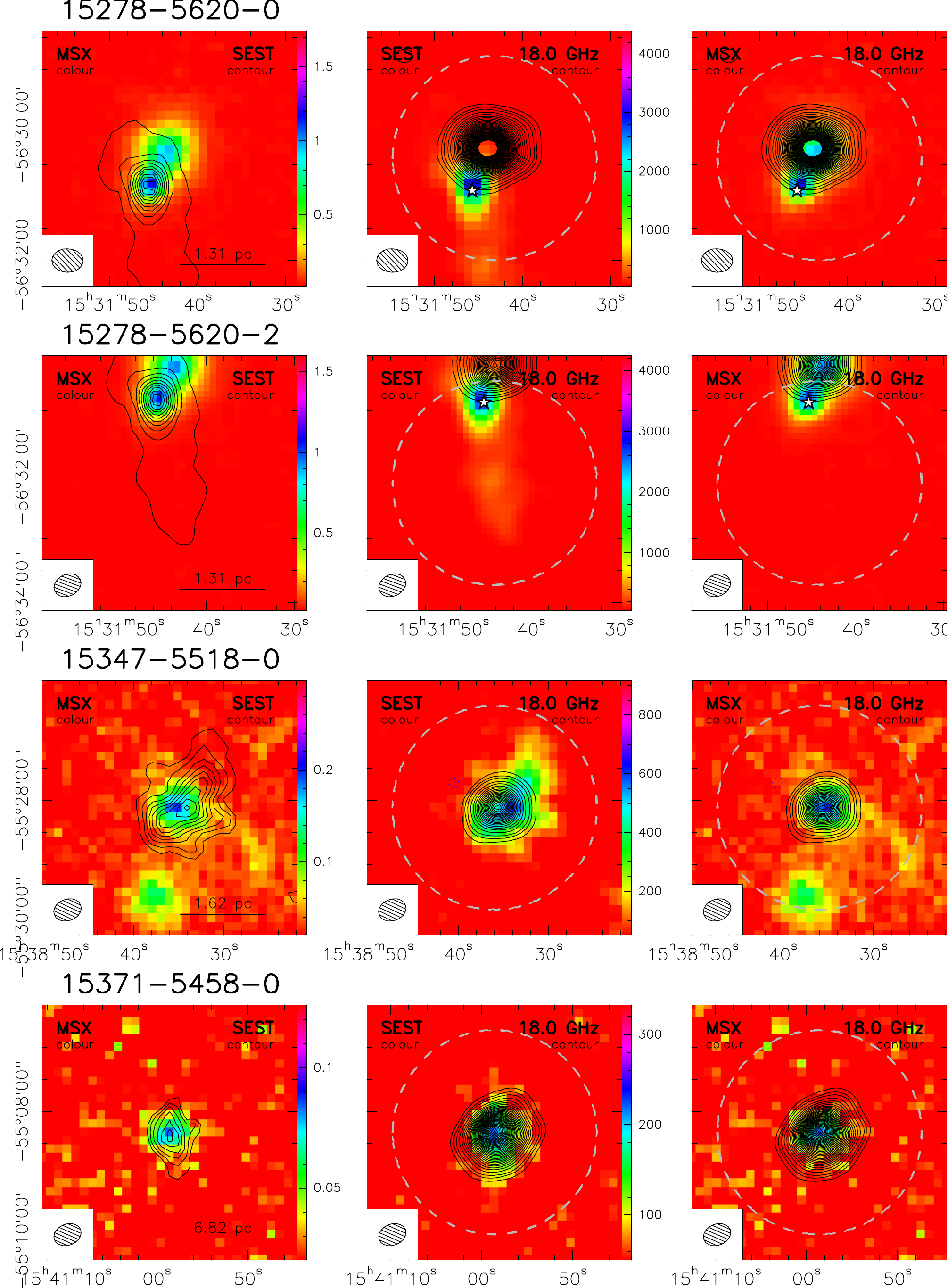}
\end{center}
\caption{continued.}
\end{figure*}
\begin{figure*}[t!]
\ContinuedFloat
\begin{center}
\includegraphics[scale=0.8]{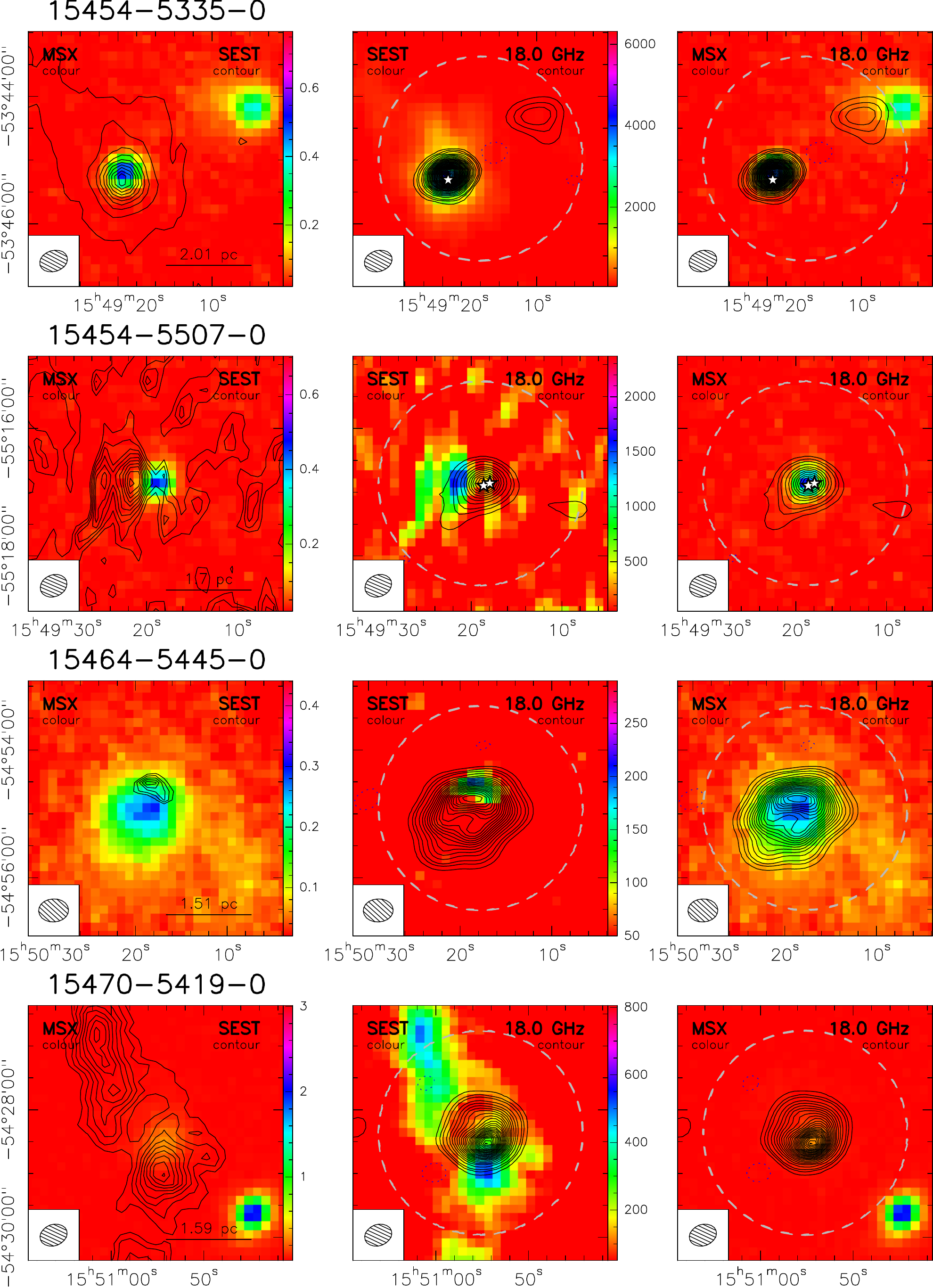}
\end{center}
\caption{continued.}
\end{figure*}
\begin{figure*}[t!]
\ContinuedFloat
\begin{center}
\includegraphics[scale=0.8]{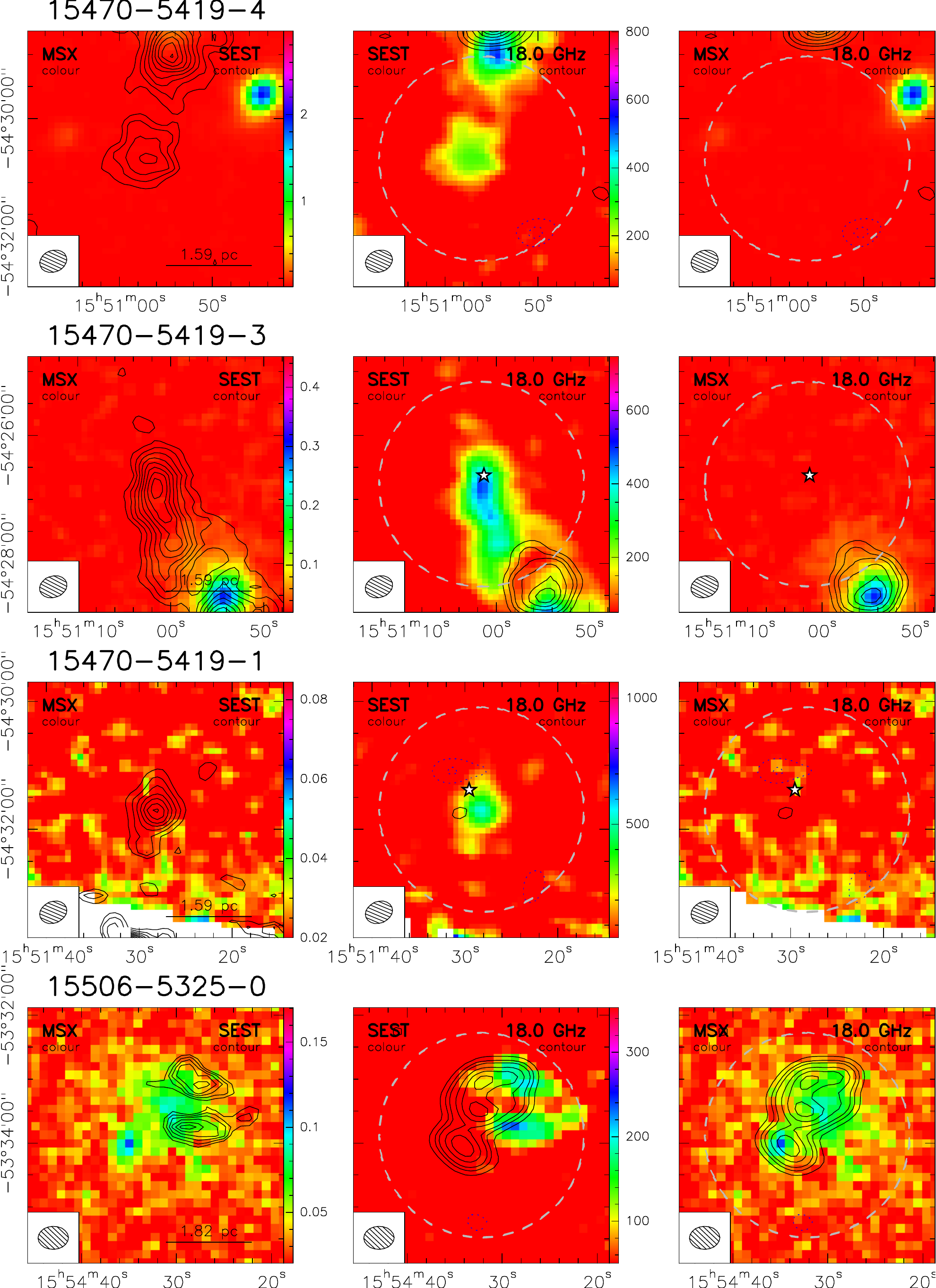}
\end{center}
\caption{continued.}
\end{figure*}
\begin{figure*}[t!]
\ContinuedFloat
\begin{center}
\includegraphics[scale=0.8]{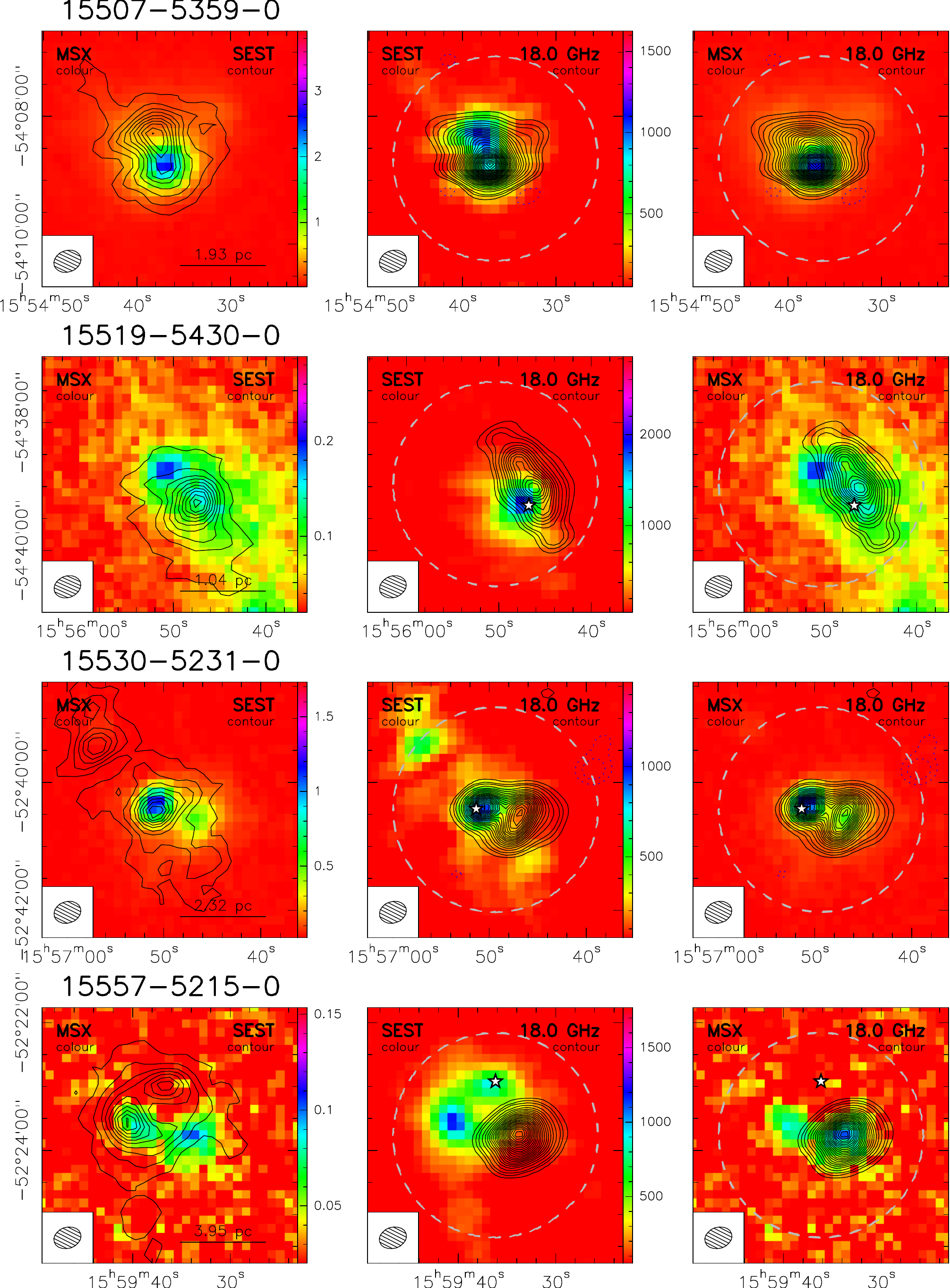}
\end{center}
\caption{continued.}
\end{figure*}
\clearpage
\begin{figure*}[t!]
\ContinuedFloat
\begin{center}
\includegraphics[scale=0.8]{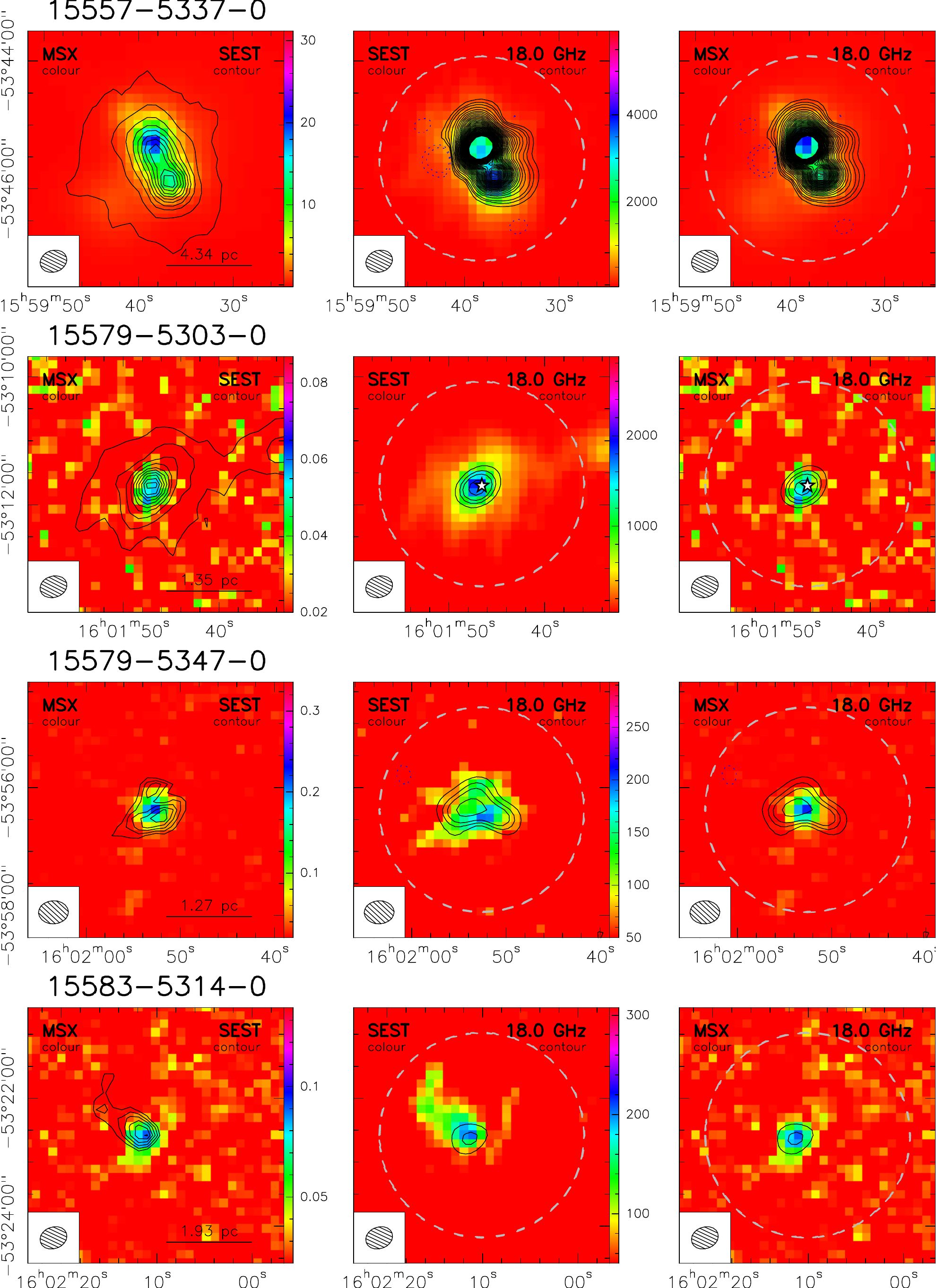}
\end{center}
\caption{continued.}
\end{figure*}
\begin{figure*}[t!]
\ContinuedFloat
\begin{center}
\includegraphics[scale=0.8]{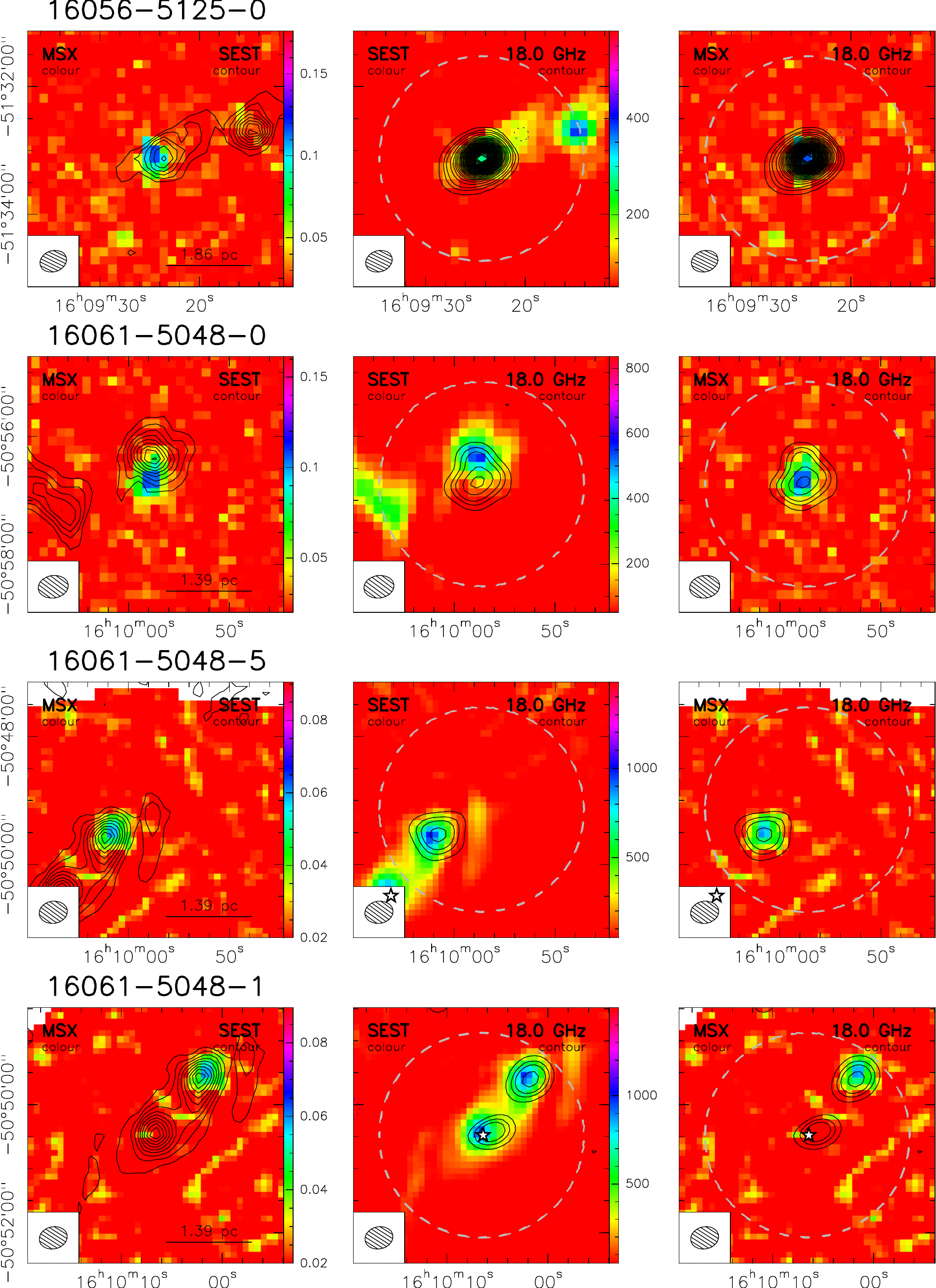}
\end{center}
\caption{continued.}
\end{figure*}
\begin{figure*}[t!]
\ContinuedFloat
\begin{center}
\includegraphics[scale=0.8]{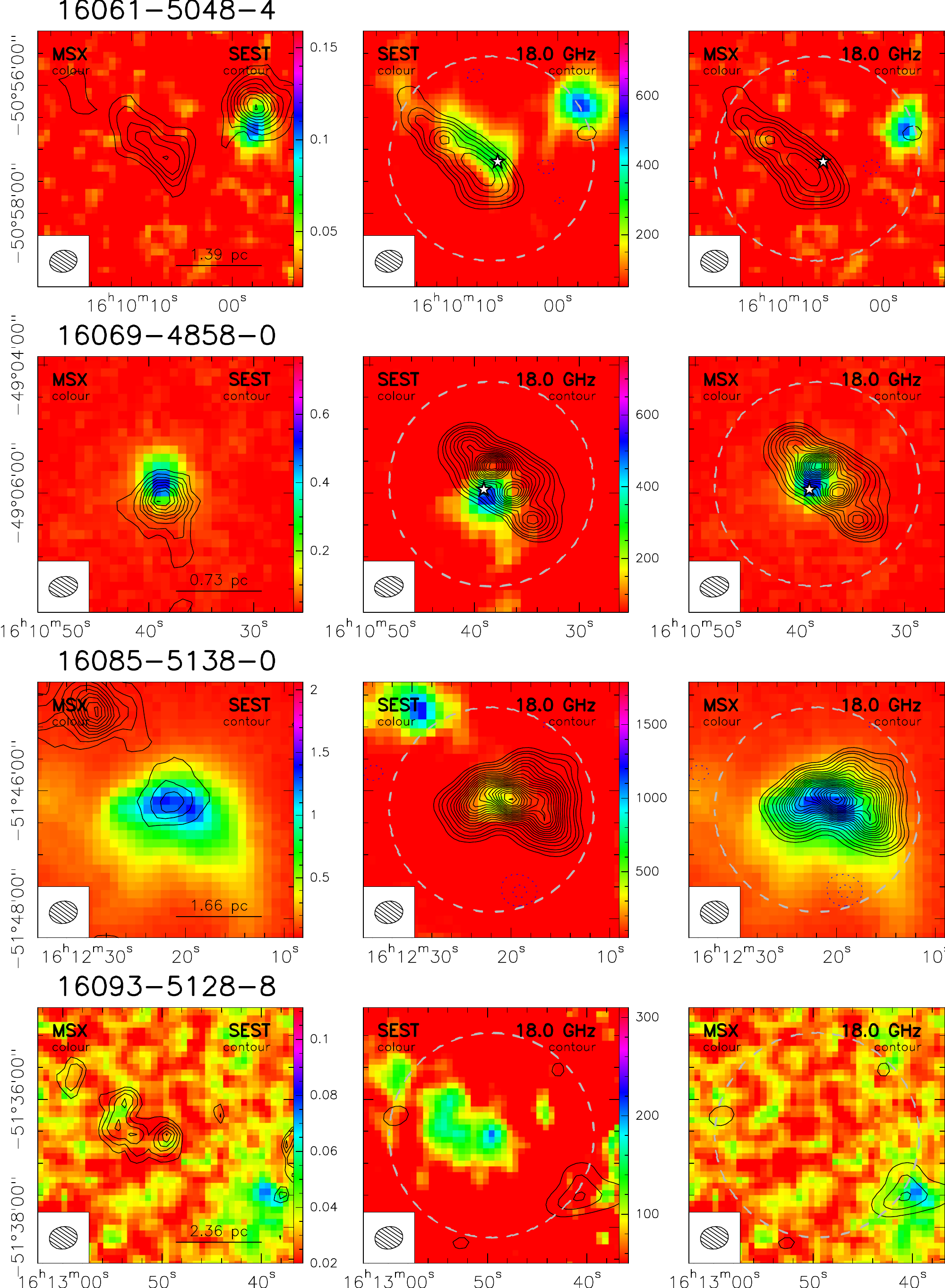}
\end{center}
\caption{continued.}
\end{figure*}
\begin{figure*}[t!]
\ContinuedFloat
\begin{center}
\includegraphics[scale=0.8]{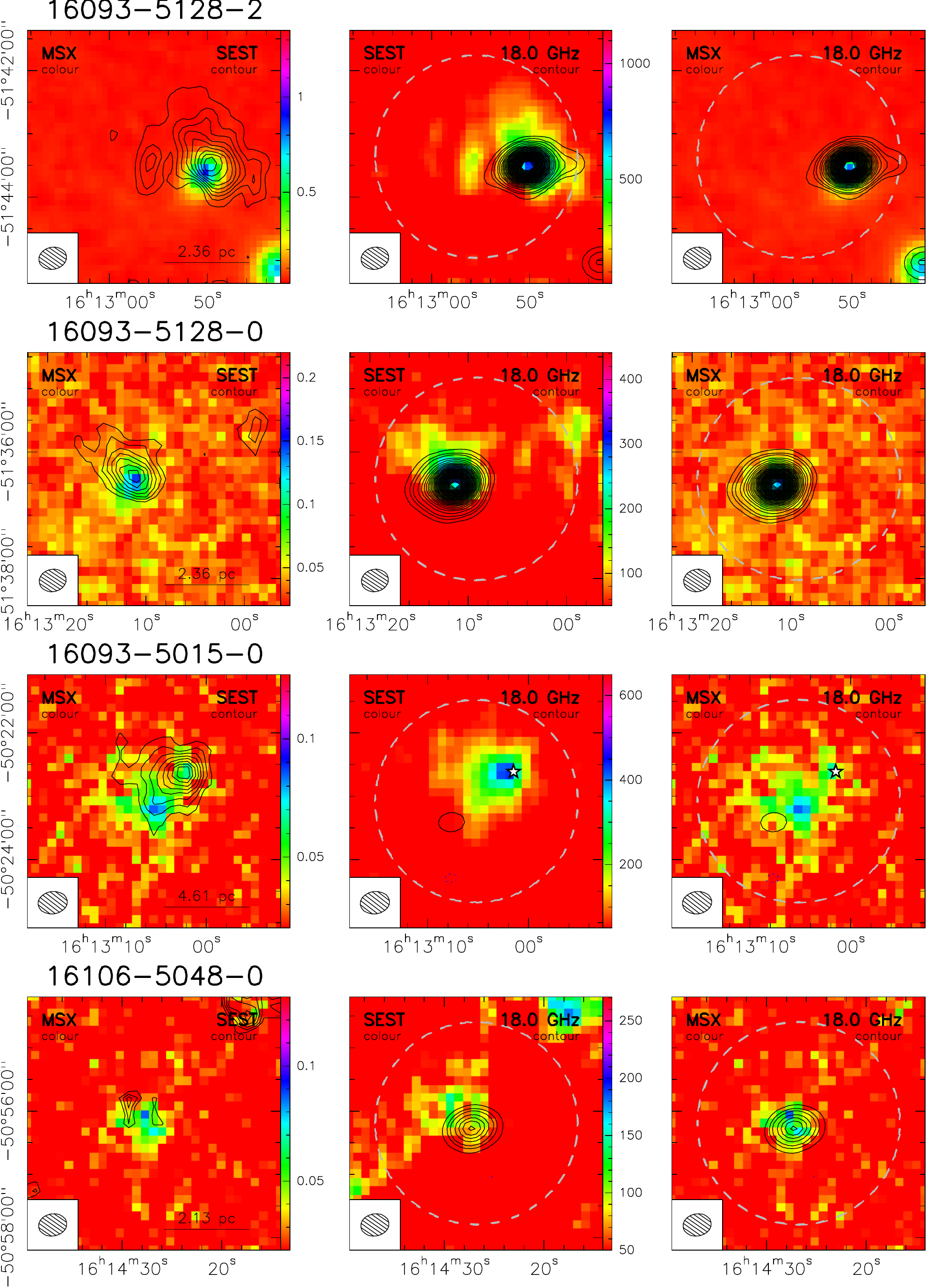}
\end{center}
\caption{continued.}
\end{figure*}
\begin{figure*}[t!]
\ContinuedFloat
\begin{center}
\includegraphics[scale=0.8]{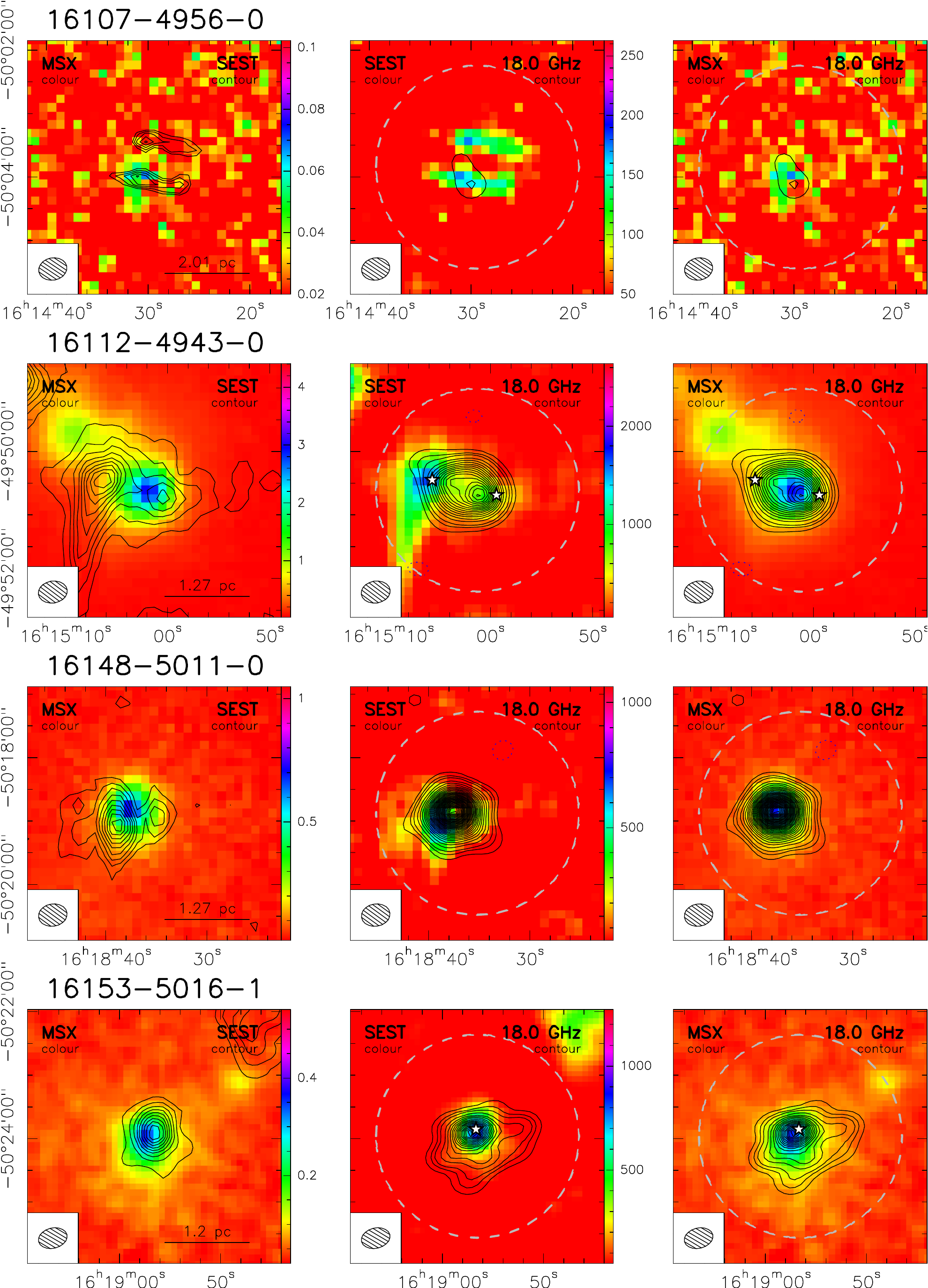}
\end{center}
\caption{continued.}
\end{figure*}
\begin{figure*}[t!]
\ContinuedFloat
\begin{center}
\includegraphics[scale=0.8]{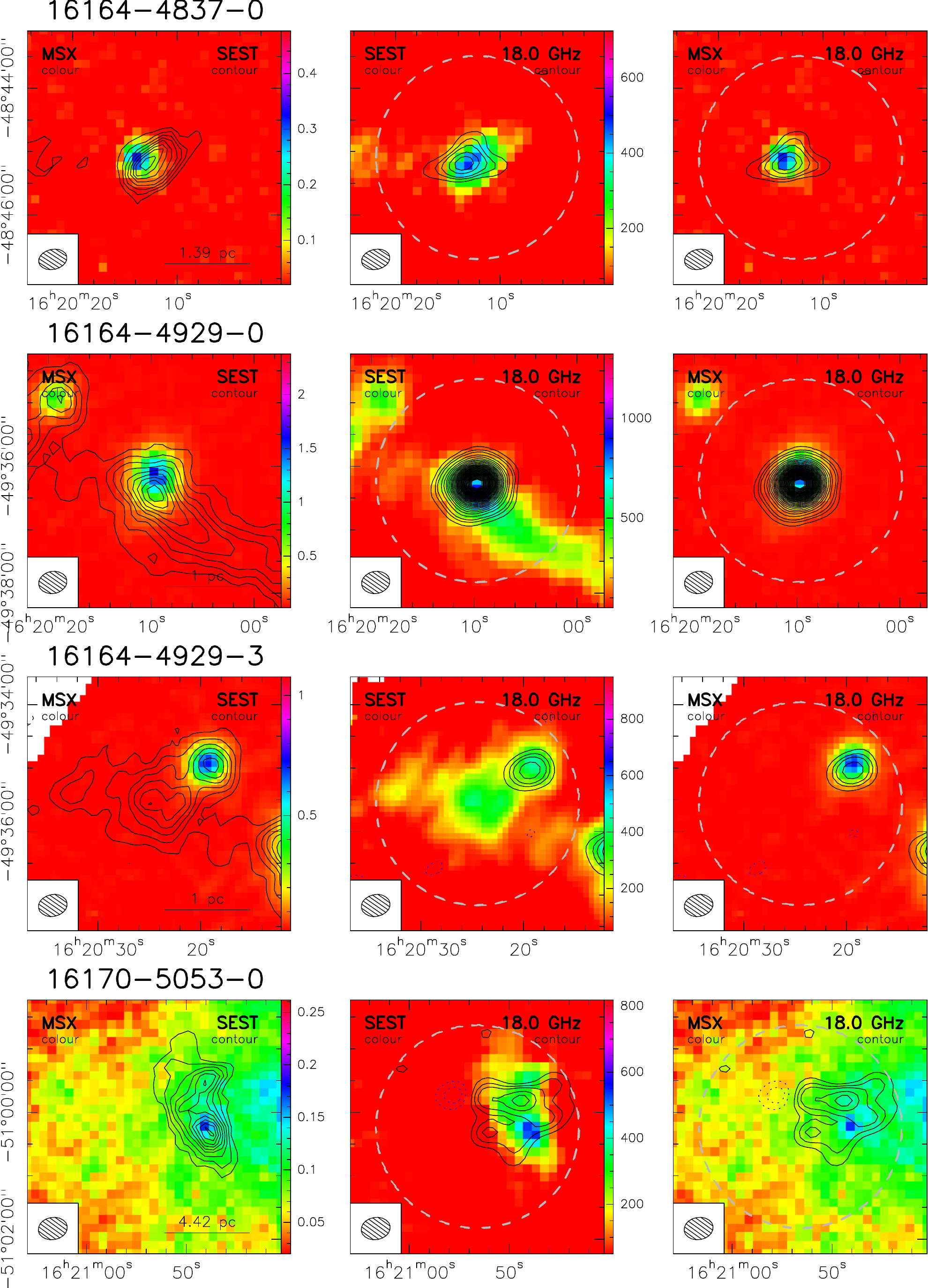}
\end{center}
\caption{continued.}
\end{figure*}
\begin{figure*}[t!]
\ContinuedFloat
\begin{center}
\includegraphics[scale=0.8]{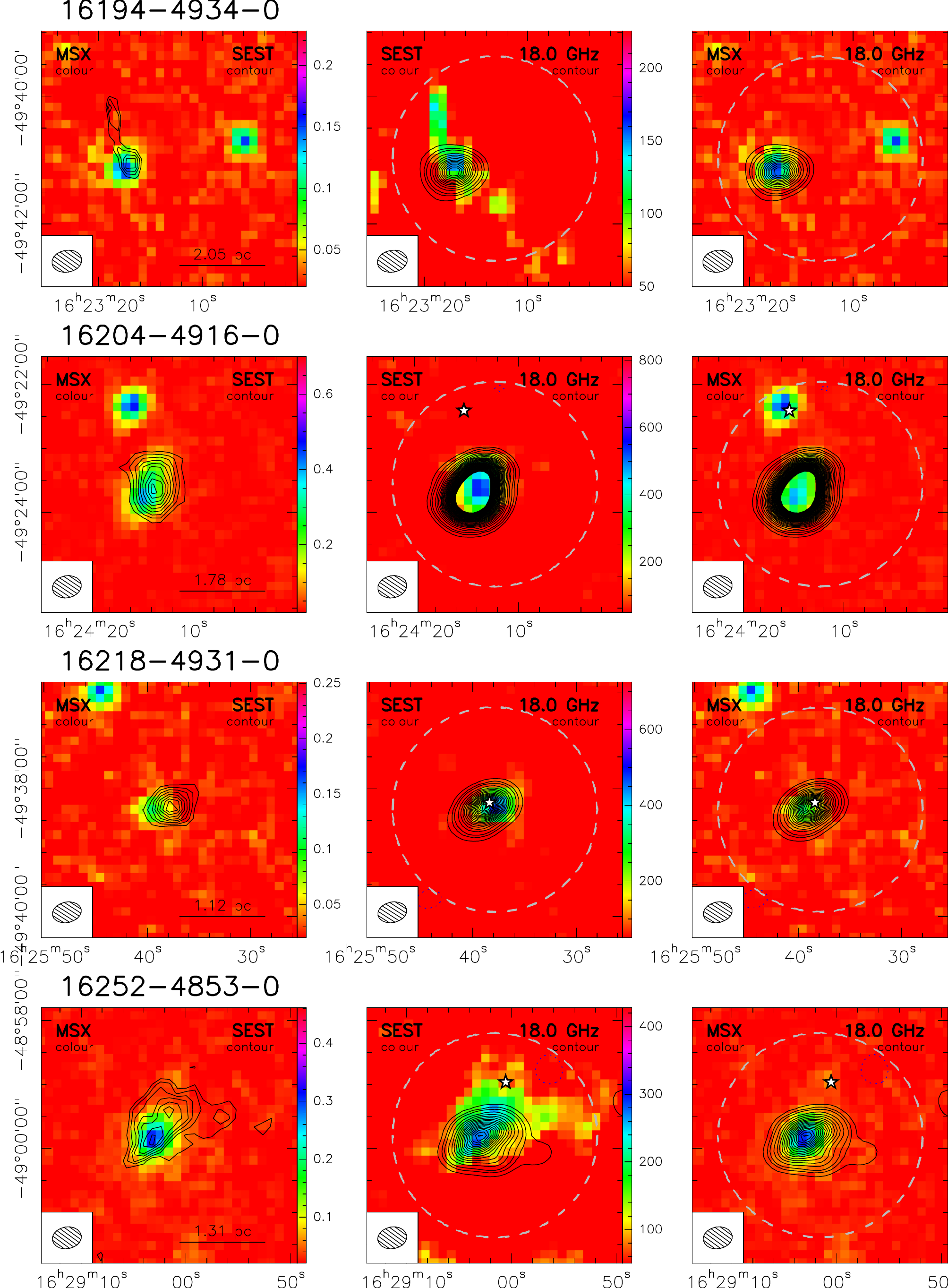}
\end{center}
\caption{continued.}
\end{figure*}
\begin{figure*}[t!]
\ContinuedFloat
\begin{center}
\includegraphics[scale=0.8]{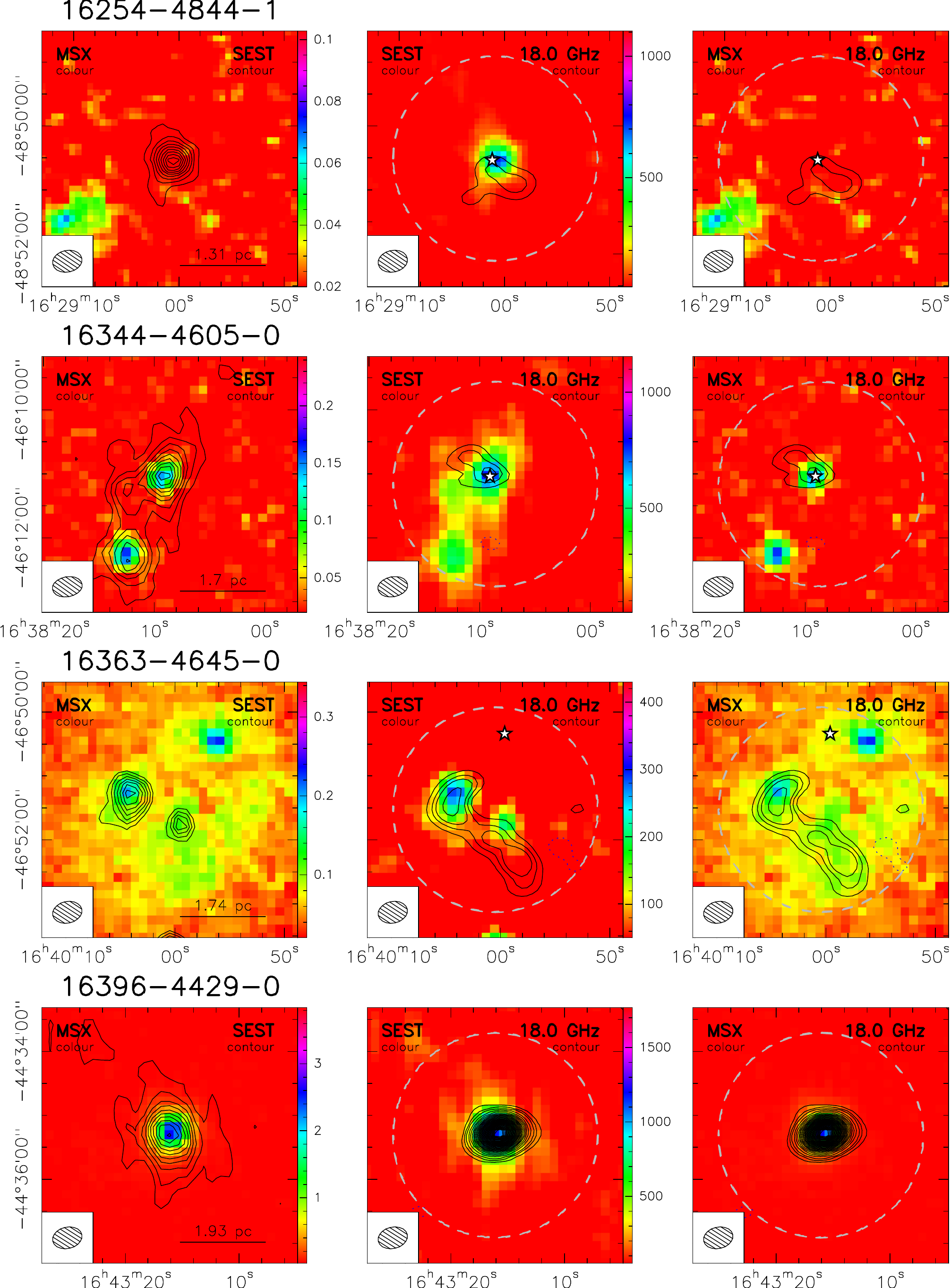}
\end{center}
\caption{continued.}
\end{figure*}
\begin{figure*}[t!]
\ContinuedFloat
\begin{center}
\includegraphics[scale=0.8]{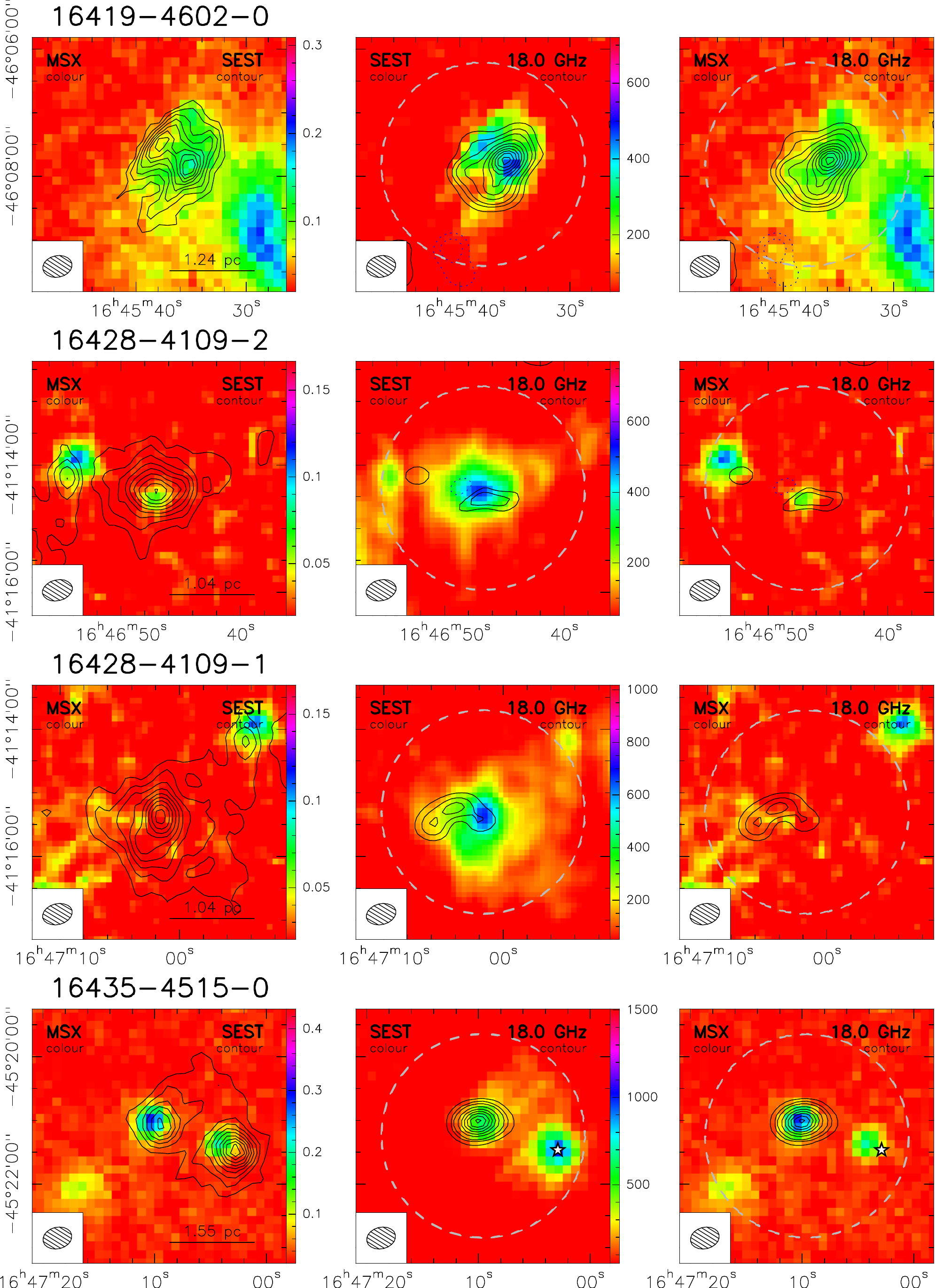}
\end{center}
\caption{continued.}
\end{figure*}
\begin{figure*}[t!]
\ContinuedFloat
\begin{center}
\includegraphics[scale=0.8]{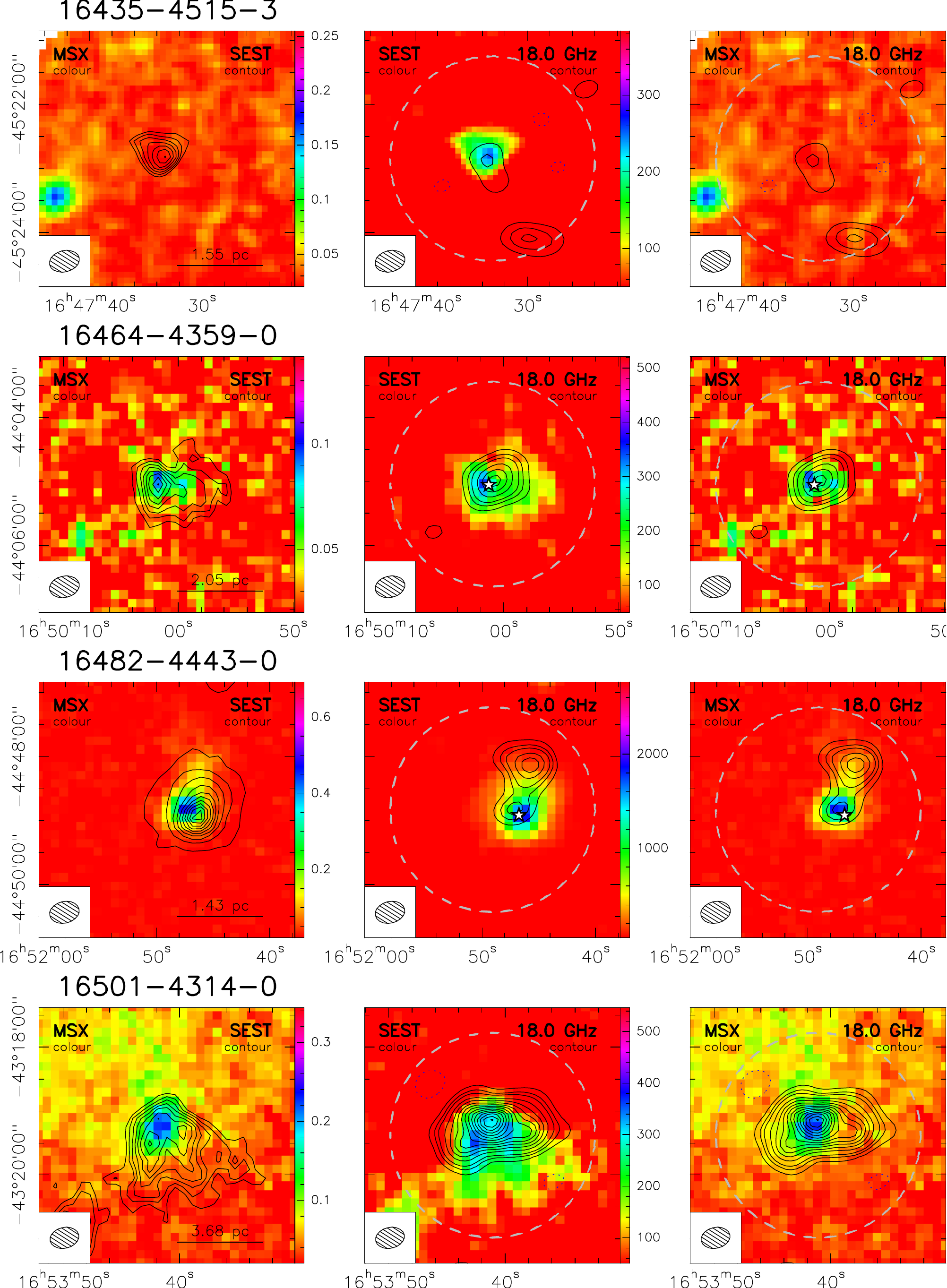}
\end{center}
\caption{continued.}
\end{figure*}
\clearpage
\begin{figure*}[t!]
\ContinuedFloat
\begin{center}
\includegraphics[scale=0.8]{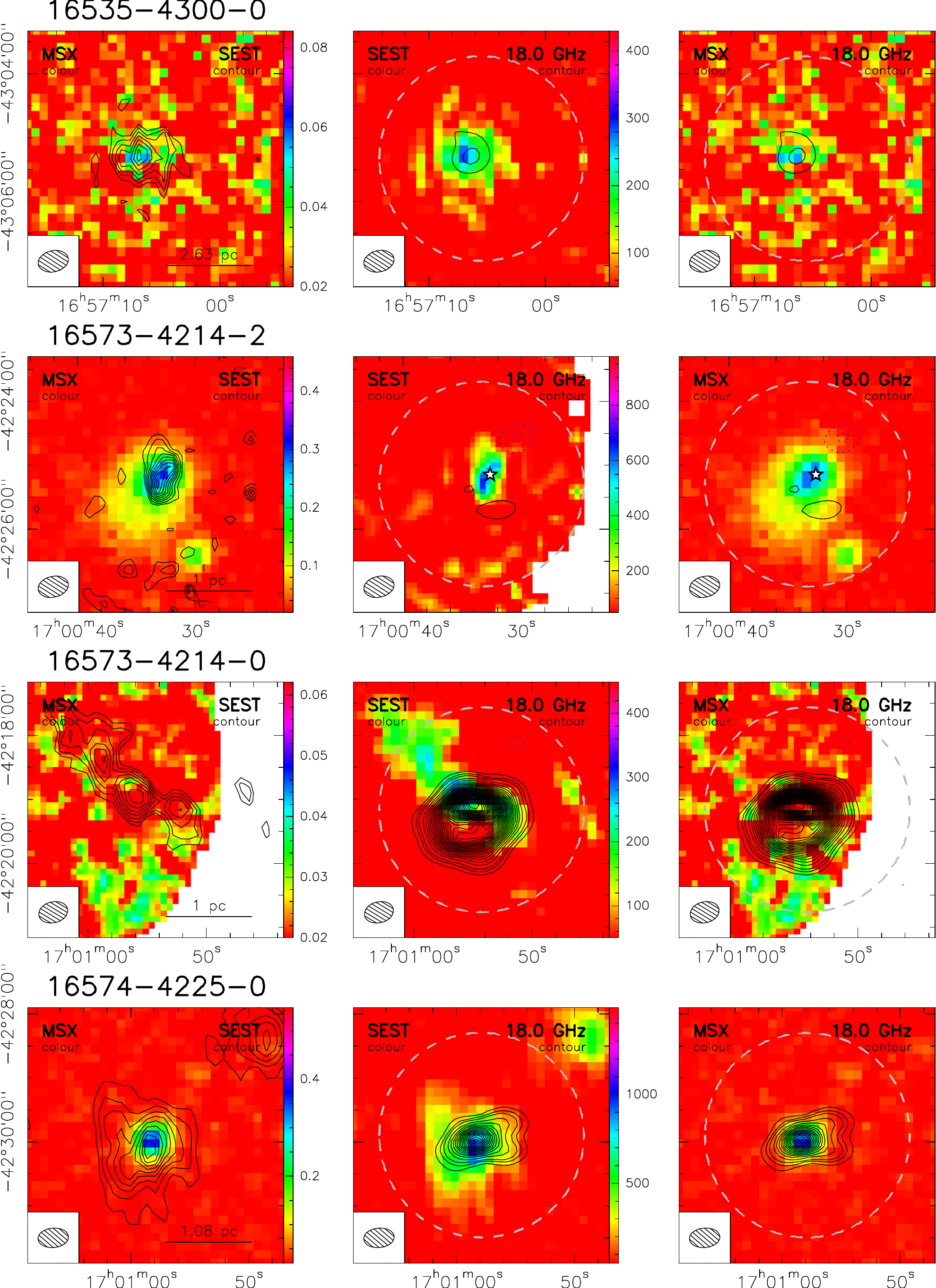}
\end{center}
\caption{continued.}
\end{figure*}
\begin{figure*}[t!]
\ContinuedFloat
\begin{center}
\includegraphics[scale=0.8]{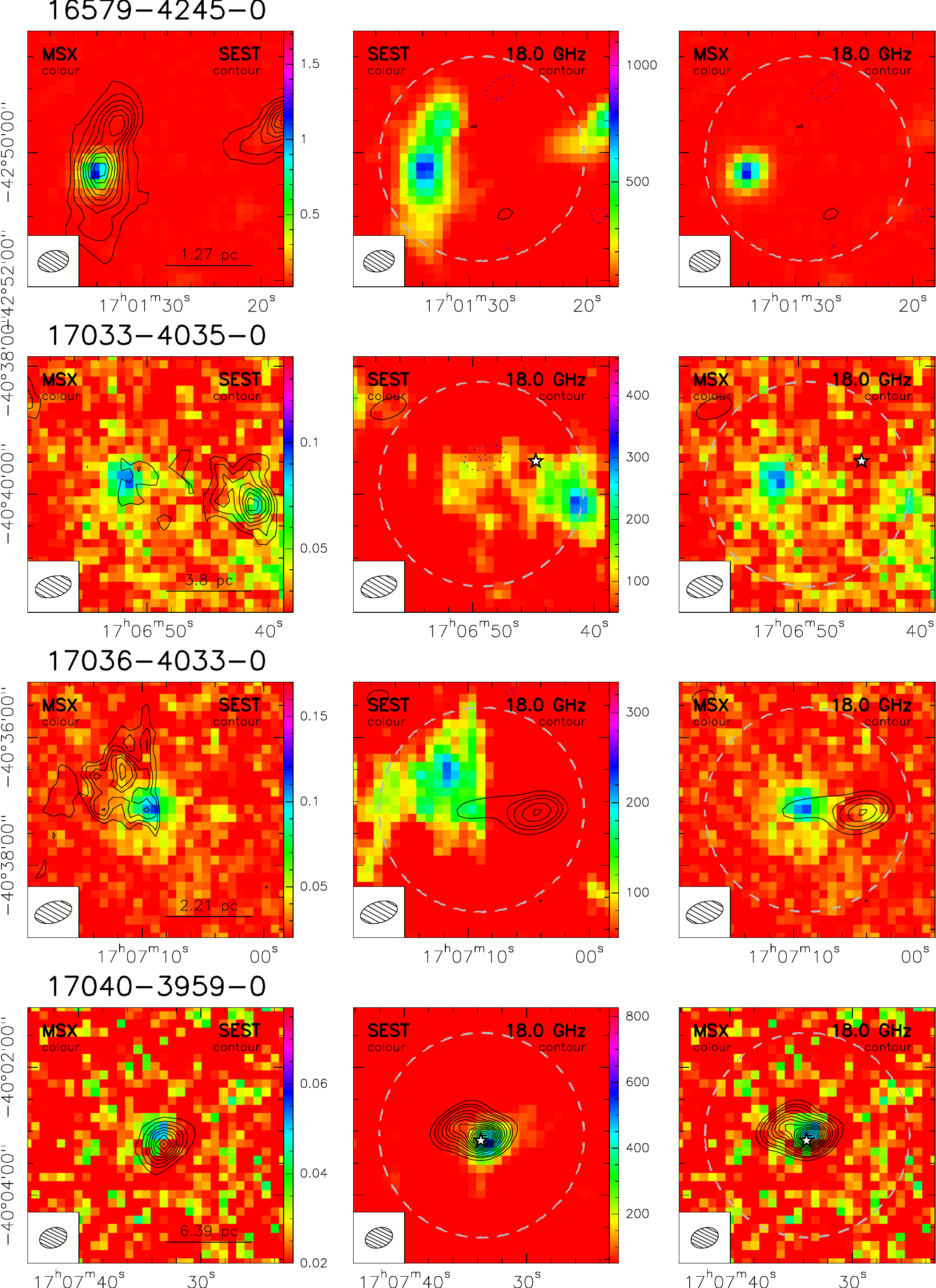}
\end{center}
\caption{continued.}
\end{figure*}
\begin{figure*}[t!]
\ContinuedFloat
\begin{center}
\includegraphics[scale=0.8]{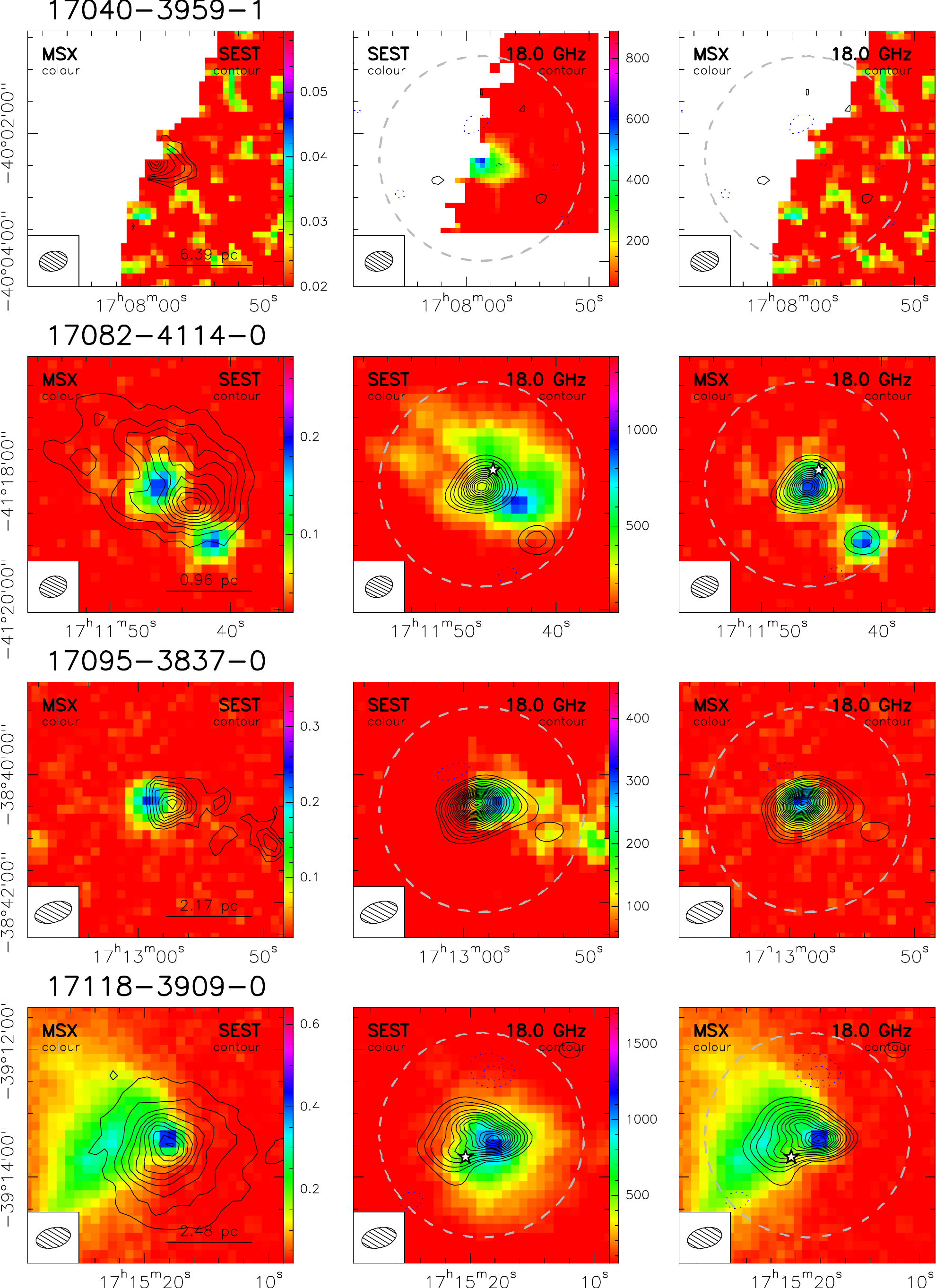}
\end{center}
\caption{continued.}
\end{figure*}
\begin{figure*}[t!]
\ContinuedFloat
\begin{center}
\includegraphics[scale=0.8]{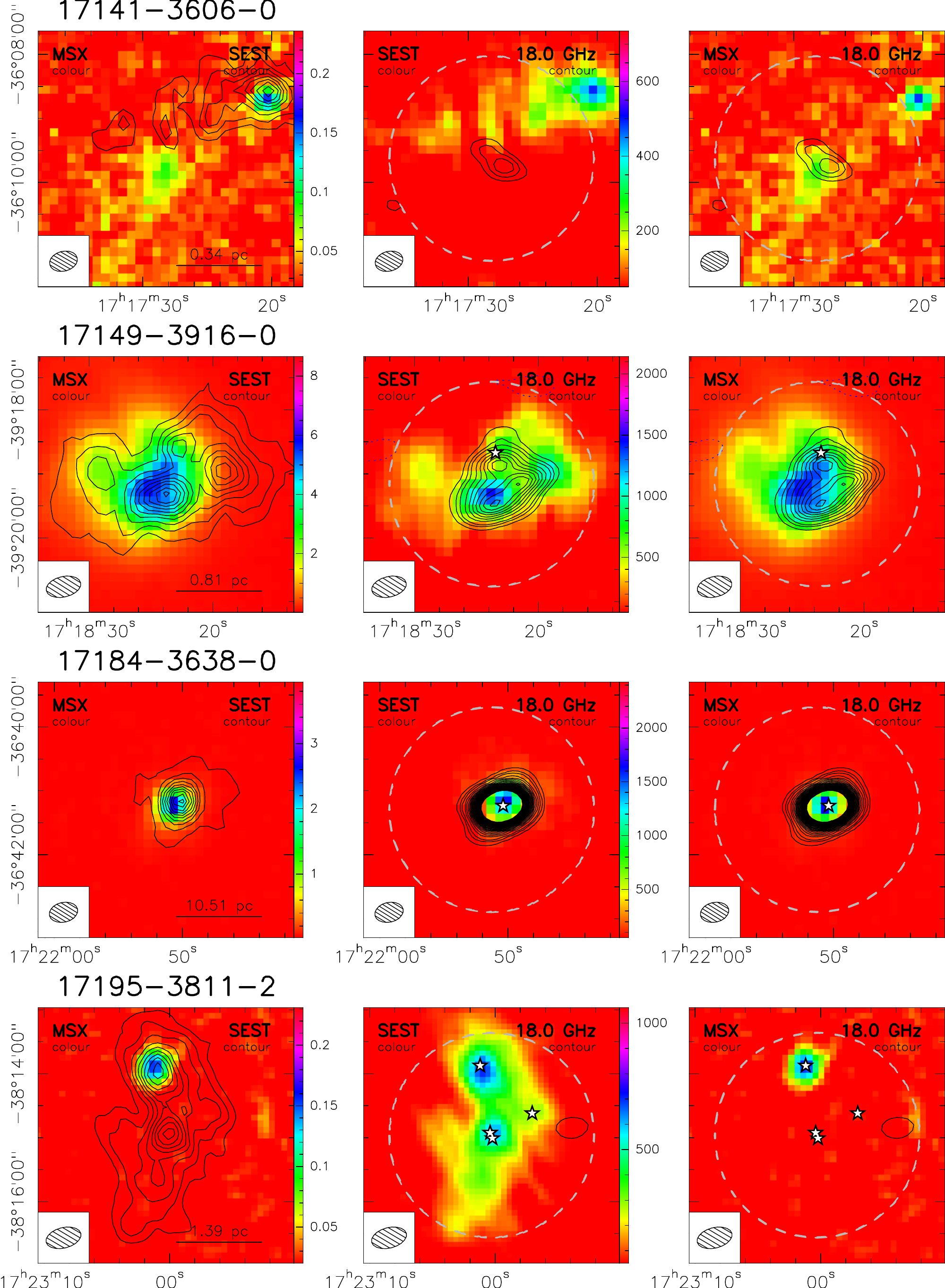}
\end{center}
\caption{continued.}
\end{figure*}
\begin{figure*}[t!]
\ContinuedFloat
\begin{center}
\includegraphics[scale=0.8]{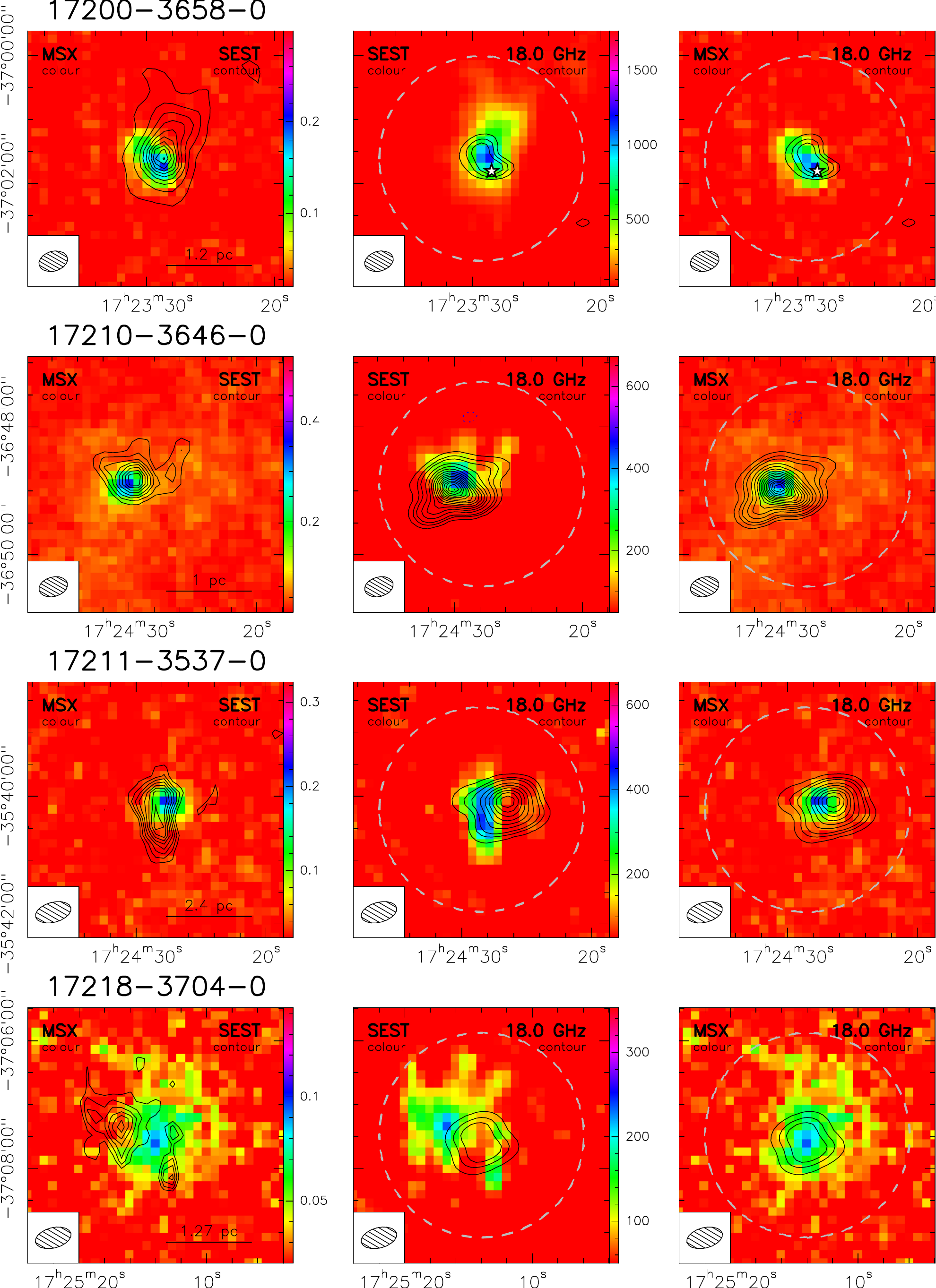}
\end{center}
\caption{continued.}
\end{figure*}
\begin{figure*}[t!]
\ContinuedFloat
\begin{center}
\includegraphics[scale=0.8]{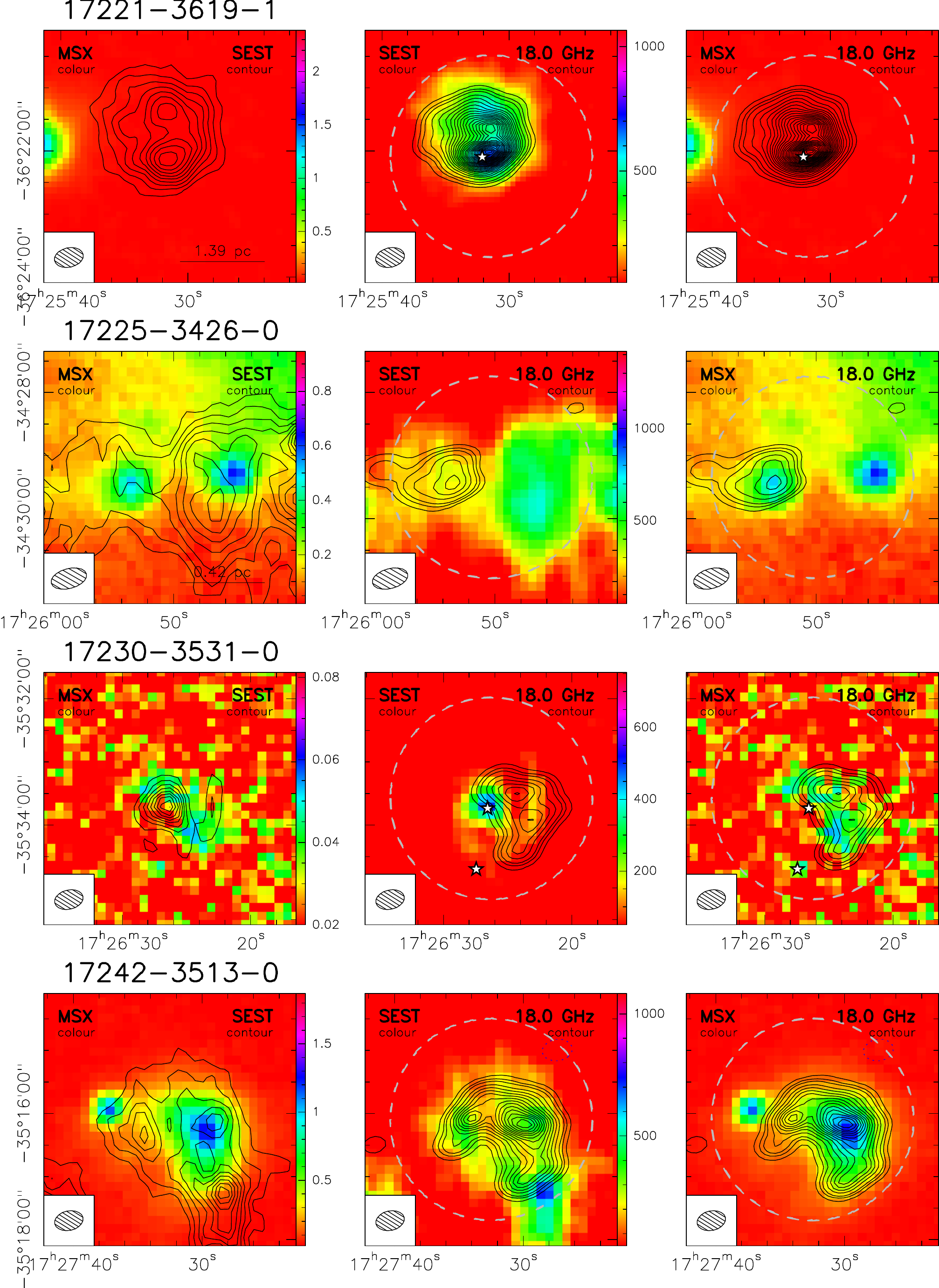}
\end{center}
\caption{continued.}
\end{figure*}
\begin{figure*}[t!]
\ContinuedFloat
\begin{center}
\includegraphics[scale=0.8]{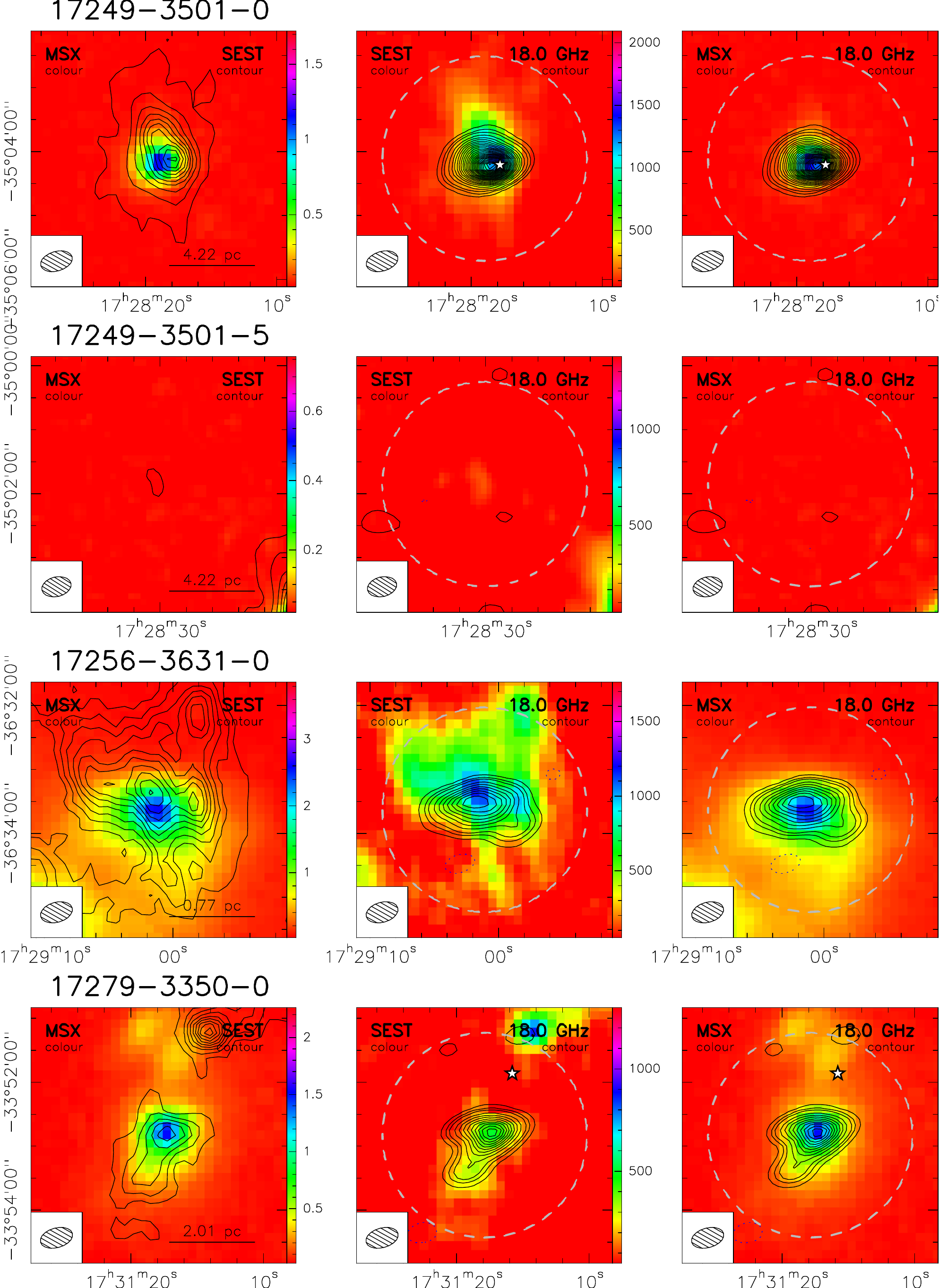}
\end{center}
\caption{continued.}
\end{figure*}
\begin{figure*}[t!]
\ContinuedFloat
\begin{center}
\includegraphics[scale=0.8]{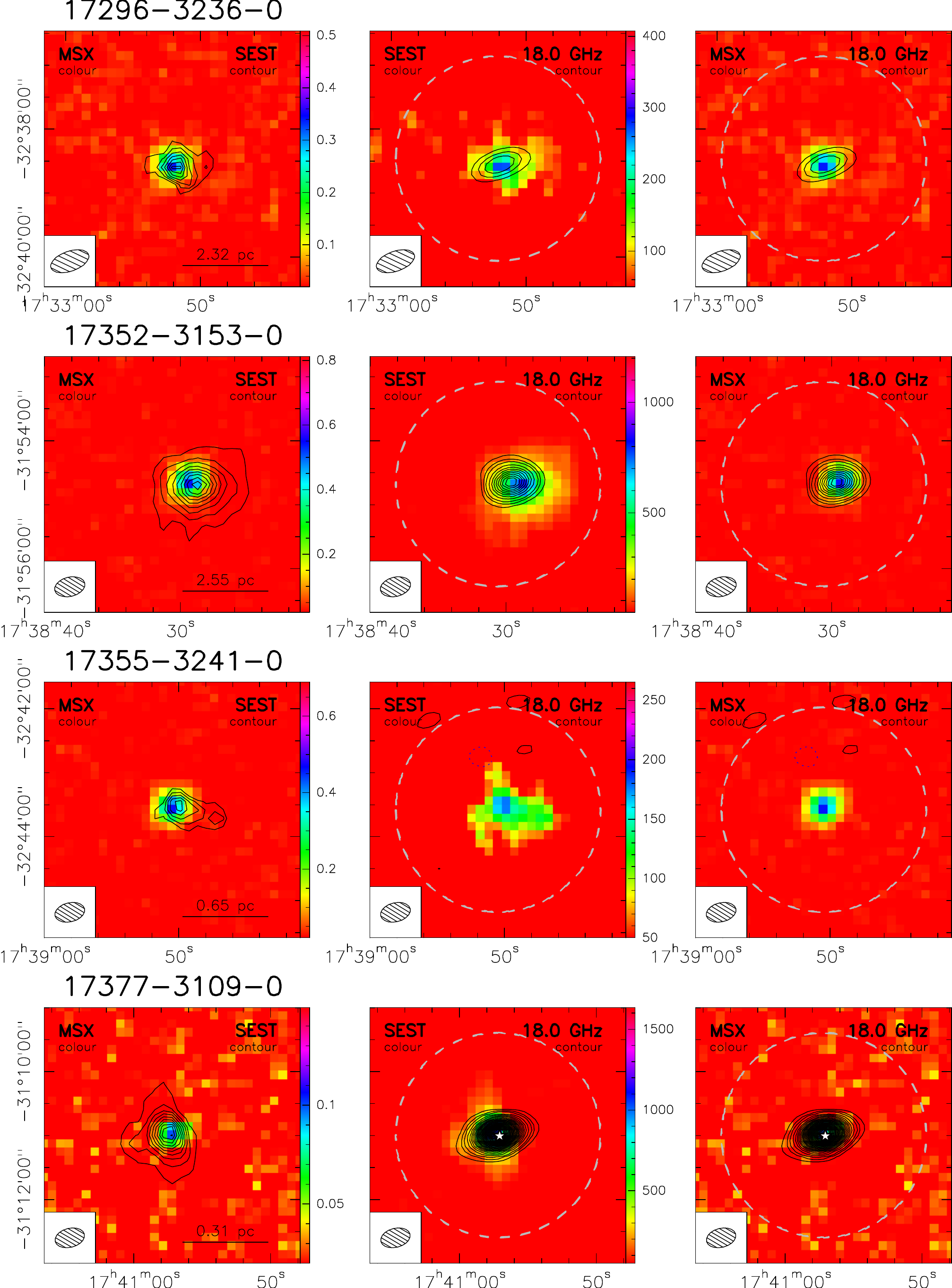}
\end{center}
\caption{continued.}
\end{figure*}
\begin{figure*}[t!]
\ContinuedFloat
\begin{center}
\includegraphics[scale=0.8]{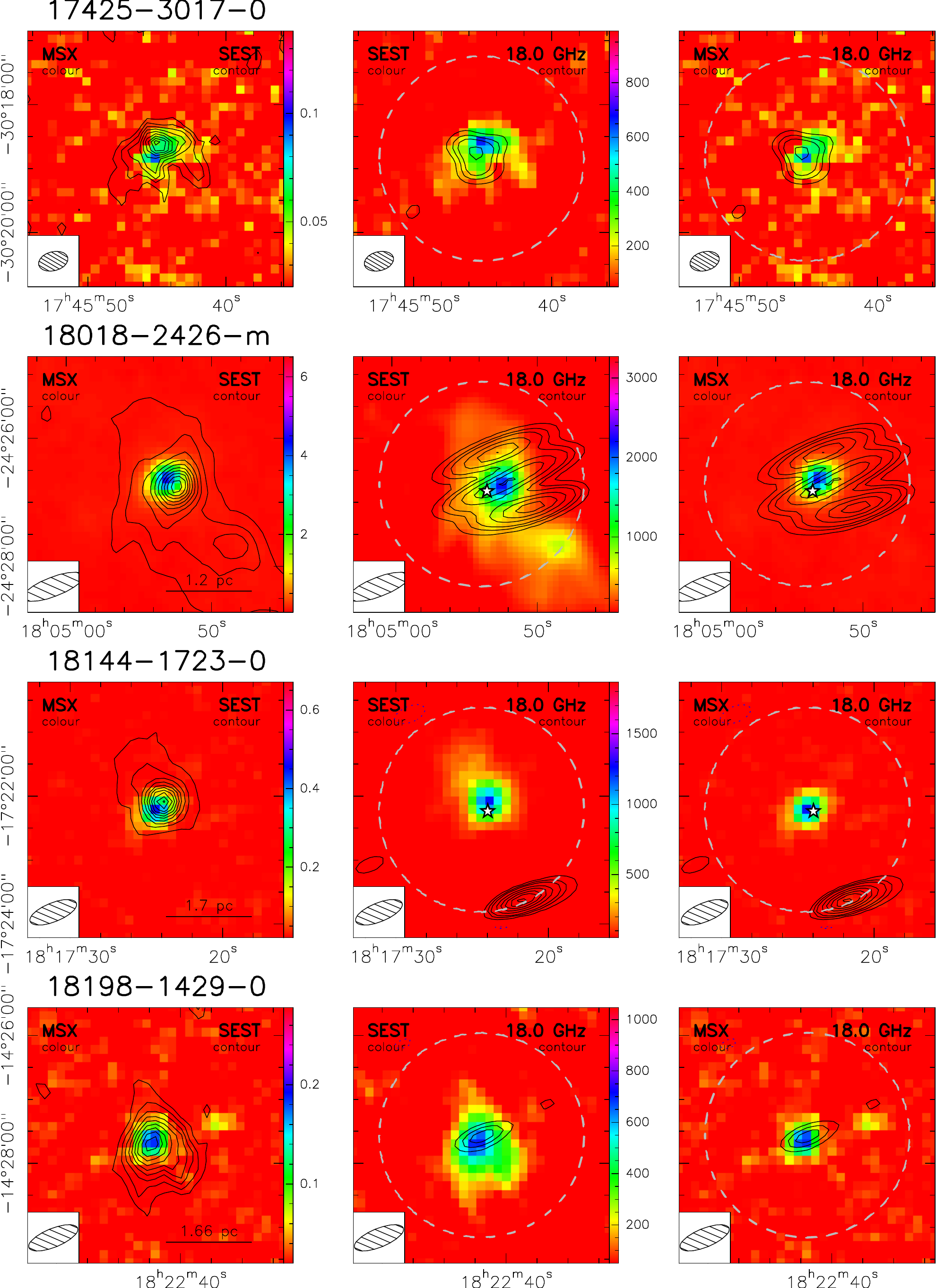}
\end{center}
\caption{continued.}
\end{figure*}
\end{appendix}

\end{document}